\documentclass[acmtog]{acmart}

\renewcommand\footnotetextcopyrightpermission[1]{} %
\usepackage{amsmath}
\usepackage[LGR,T1]{fontenc} %
\usepackage[cal = cm]{mathalpha}
\usepackage{caption,graphicx,adjustbox,subcaption}
\usepackage{algorithm,algorithmicx,algpseudocode}
\usepackage[bottom]{footmisc}
\usepackage{enumitem}
\setlist[itemize]{noitemsep,topsep=0pt,parsep=4pt,partopsep=0pt,leftmargin=*}
\setlist[enumerate]{noitemsep,topsep=0pt,parsep=4pt,partopsep=0pt,leftmargin=*}
\usepackage{comment}
\usepackage{overpic,pbox,tikz}
\usepackage{multirow,colortbl}
\usepackage{afterpage,url}
\usepackage{wrapfig}
\usepackage{booktabs}
\usepackage{makecell}
\usepackage{svg}

\usepackage{mathtools}
\usepackage[many]{tcolorbox}
\usepackage{xcolor}
\usepackage{tikz}
\usetikzlibrary{calc}
\usepackage{bm}
\usepackage{tablefootnote}
\usepackage{booktabs}
\usepackage{dsfont}

\usepackage{letltxmacro}
\LetLtxMacro{\originaleqref}{\eqref}
\renewcommand{\eqref}{Eq.~\originaleqref}

\title{Efficient Scene Appearance Aggregation for Level-of-Detail Rendering}

\author{Yang Zhou}
\email{yzhou426@cs.ucsb.edu}
\affiliation{%
  \institution{University of California, Santa Barbara}
  \city{Santa Barbara}
  \state{California}
  \country{USA}
}

\author{Tao Huang}
\email{tao_huang@ucsb.edu}
\affiliation{%
  \institution{University of California, Santa Barbara}
  \city{Santa Barbara}
  \state{California}
  \country{USA}
}

\author{Ravi Ramamoorthi}
\email{ravir@cs.ucsd.edu}
\affiliation{%
  \institution{University of California, San Diego}
  \city{La Jolla}
  \state{California}
  \country{USA} 
}

\author{Pradeep Sen}
\email{psen@ece.ucsb.edu}
\affiliation{%
  \institution{University of California, Santa Barbara}
  \city{Santa Barbara}
  \state{California}
  \country{USA}
}

\author{Ling-Qi Yan}
\email{lingqi@cs.ucsb.edu}
\affiliation{%
  \institution{University of California, Santa Barbara}
  \city{Santa Barbara}
  \state{California}
  \country{USA}
}



\definecolor{ogreen}{rgb}{0.2,0.7,0.2}
\definecolor{revision_blue}{rgb}{0.2,0.2,0.7}

\newcommand{\rev}[1]{{#1}}

\algnewcommand{\IfThen}[2]{%
	\State \algorithmicif\ #1\ \algorithmicthen\ #2}
\algnewcommand{\IfThenElse}[3]{%
	\State \algorithmicif\ #1\ \algorithmicthen\ #2\ \algorithmicelse\ #3}

\newcommand{\sLNM}{\begin{linenomath}}
\newcommand{\tLNM}{\end{linenomath}}

\newcommand{\D}{\mathrm{d}}

\definecolor{gray}{rgb}{0.8,0.8,0.8}

\newcommand{\tabincell}[2]{\begin{tabular}{@{}#1@{}}#2\end{tabular}}

\definecolor{grayL}{rgb}{.95, .95, .95}
\definecolor{purpleL}{rgb}{.9735, .95, .9761}
\definecolor{purpleD}{rgb}{.8941, .8, .9043}
\definecolor{greenL}{rgb}{.9643, .9906, .9750}
\definecolor{greenD}{rgb}{.7145, .9249, .7999}
\definecolor{orangeLL}{rgb}{0.9991, 0.9846, 0.9759}
\definecolor{orangeL}{rgb}{.9982, .9692, .9518}
\definecolor{orangeD}{rgb}{.9929, .8766, .8071}

\definecolor{redL}{rgb}{1.0, 0.95, 0.95}
\definecolor{redD}{rgb}{1.0, 0.8, 0.8}
\definecolor{yellowL}{rgb}{1.0, 1.0, 0.95}
\definecolor{yellowD}{rgb}{0.95, 0.95, 0.6}
\definecolor{blueLL}{rgb}{0.98, 0.98, 1.0}
\definecolor{blueL}{rgb}{0.95, 0.95, 1.0}
\definecolor{blueD}{rgb}{0.8, 0.8, 1.0}

\newcommand{\pluseq}{\mathrel{+}=}

\definecolor{lightgreen}{HTML}{12B503} 
\definecolor{lightblue}{HTML}{00A3FF} 
\definecolor{grassgreen}{HTML}{638F3D}
\definecolor{lightorange}{HTML}{EAAC35} 
\definecolor{lightred}{HTML}{DD672B}
\definecolor{lightteal}{HTML}{44809E}

\tcbset{
highlight math/.append style={colback=white, colframe=lightorange, boxrule=1pt, boxsep=0pt, top=2pt, bottom=2pt, left=2pt, right=2pt}
}

\makeatletter
\def\sectionautorefname{\S\@gobble}
\makeatother
\makeatletter
\def\subsectionautorefname{\S\@gobble}
\makeatother

\setcitestyle{nosort,square} %

\begin{document}

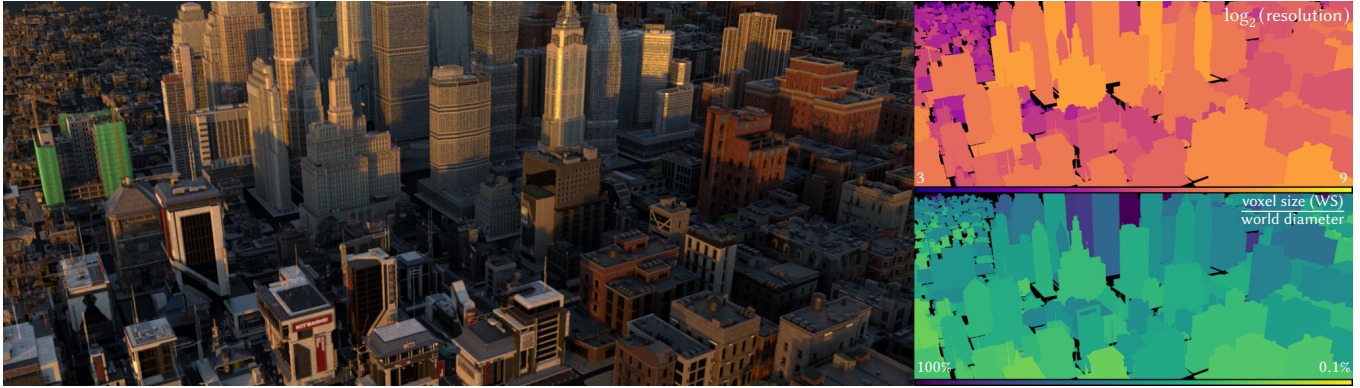
\begin{teaserfigure}
	\newlength{\lenTeaser}
	\setlength{\lenTeaser}{\linewidth}	
	\addtolength{\tabcolsep}{-5pt}
    \center
	\begin{tabular}{c}
		\begin{overpic}[width=\lenTeaser]{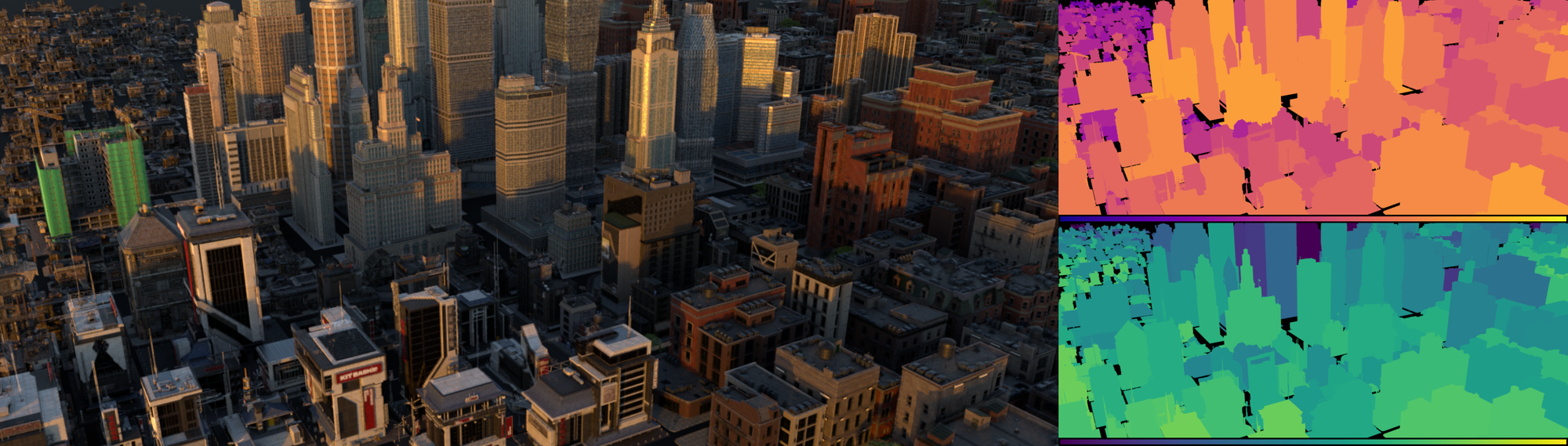}
			\put(90.5, 27){\footnotesize \color{white}  $\log_2(\textsf{resolution})$}
			\put(67.7, 15){\scriptsize \color{white}  $3$}
			\put(99, 15){\scriptsize \color{white}  $9$}
			\put(91.5, 12.5){\small \color{white}  $\frac{\textsf{voxel size (WS)}}{\textsf{world diameter}}$}
			\put(67.7, 1){\scriptsize \color{white}  $100\%$}
			\put(97.5, 1){\scriptsize \color{white}  $0.1\%$}
		\end{overpic}
	\end{tabular}		
	\caption{\label{fig:teaser}
	\rev{
		\textbf{Left}: \emph{Metropolis}, a cityscape rendered with our scene aggregation approach. The scene includes 82 unique buildings and 270 
		instances and originally requires 46.9 GB to store, making it challenging to render while stay in in-core memory. Our representation drastically reduces 
		the size to 5.33 GB while preserving the detailed appearance. \textbf{Right Top}: Each instance selects the appropriate LoD resolution where the 
		projected voxel size matches the pixel footprint (rounded to the nearest power of two). \textbf{Right Bottom}: As a result, close-view instances are 
		rendered with finer voxels while distant instances are rendered with coarser voxels.
		}}
\end{teaserfigure}

\begin{abstract}
Creating an appearance-preserving level-of-detail (LoD) representation for arbitrary 3D scenes is a challenging problem.
The appearance of a scene is an intricate combination of both geometry and material models, and is further complicated by correlation due to the spatial
configuration of scene elements.
We present a novel \rev{volumetric} representation for the aggregated appearance of complex scenes and an efficient pipeline for LoD generation and rendering.
The core of our representation is the \emph{Aggregated Bidirectional Scattering Distribution Function} (ABSDF) that summarizes the far-field appearance of all
surfaces inside a voxel. We propose a closed-form factorization of the ABSDF that accounts for spatially varying and orientation-varying material parameters.
We tackle the challenge of capturing the correlation existing locally within a voxel and globally across different parts of the scene.
Our method faithfully reproduces appearance and achieves higher quality than existing scene filtering methods while being inherently
efficient to render. The memory footprint and rendering cost of our representation are independent of the original scene complexity.
\end{abstract}

\maketitle

\section{Introduction}

Modern physically based rendering is widely adopted to synthesize photorealistic images, animations, and immersive 3D experiences. Generating content at such
level of realism requires large-scale assets with extremely detailed geometry, textures, and sophisticated material models. This presents significant
challenges to the rendering process both in terms of storage and speed. Among them, one prominent issue comes from the mismatch between scene complexity and
image resolution. In an open-world environment, it is typical to only have a small portion of the scene contribute to the foreground, while the majority of the
scene is minified in the background. It is wasteful to load and render the entirety of the scene when the image resolution is not even enough to resolve the
details. Moreover, the rendering cost of different pixels can be highly uneven. Some pixels may cover a drastically more complex part of the scene than others
and thus require an excessive sampling budget for convergence. \rev{On the other hand, ignoring such complexity often leads to aliasing, artifacts or incorrect appearance.}

Level-of-detail (LoD) techniques reduce the heavy, unbalanced rendering cost by converting, or prefiltering, the original scene to a multi-scale
representation in a precomputation step. Depending on how much detail is required for each pixel, only an appropriate scale of the representation is
accessed and used for rendering. In this way, LoD techniques are able to decouple rendering cost from the original scene complexity and distribute the cost
evenly among pixels.

In order to improve efficiency, LoD techniques usually perform simplification to the original geometry. A key challenge for
any LoD technique is that it should preserve the original appearance after the simplification. When viewed at distance, the appearance is a compound phenomenon
of both geometry and material models. Simply discarding or averaging geometry would result in appearance mismatch and artifacts~\citep{luebke2003level}.
Instead, the technique should condense
the effect of the original geometry into the simplified representation. This process can be called appearance aggregation. It is important to realize that the
aggregated appearance can be more complex than, say, the original material models because it describes more information. However, it can still be advantageous
performance-wise compared to tracing the explicit geometry.

\rev{
Many existing LoD solutions converts geometry to volumes for filtering or downsampling. The recurring difficulty for these solutions is the loss of 
geometric correlation. 
}
We lose track of how the geometry is distributed locally within a
volume when it is abstracted away. Furthermore, if multiple regions are simplified separately, we lose track of the long-range visibility caused by the specific
spatial configuration between geometry of different regions. Correlation exists ubiquitously in different types of scenes, such as those containing large,
connected surface or regularly organized structures. Ignoring correlation leads to incorrect appearance for the LoD representation.

\begin{figure*}[tb]
	\newlength{\lenCoralreef}
	\setlength{\lenCoralreef}{2.4in}
	\addtolength{\tabcolsep}{-5pt}
    \center
	\begin{tabular}{cccc}
		\begin{overpic}[height=\lenCoralreef]{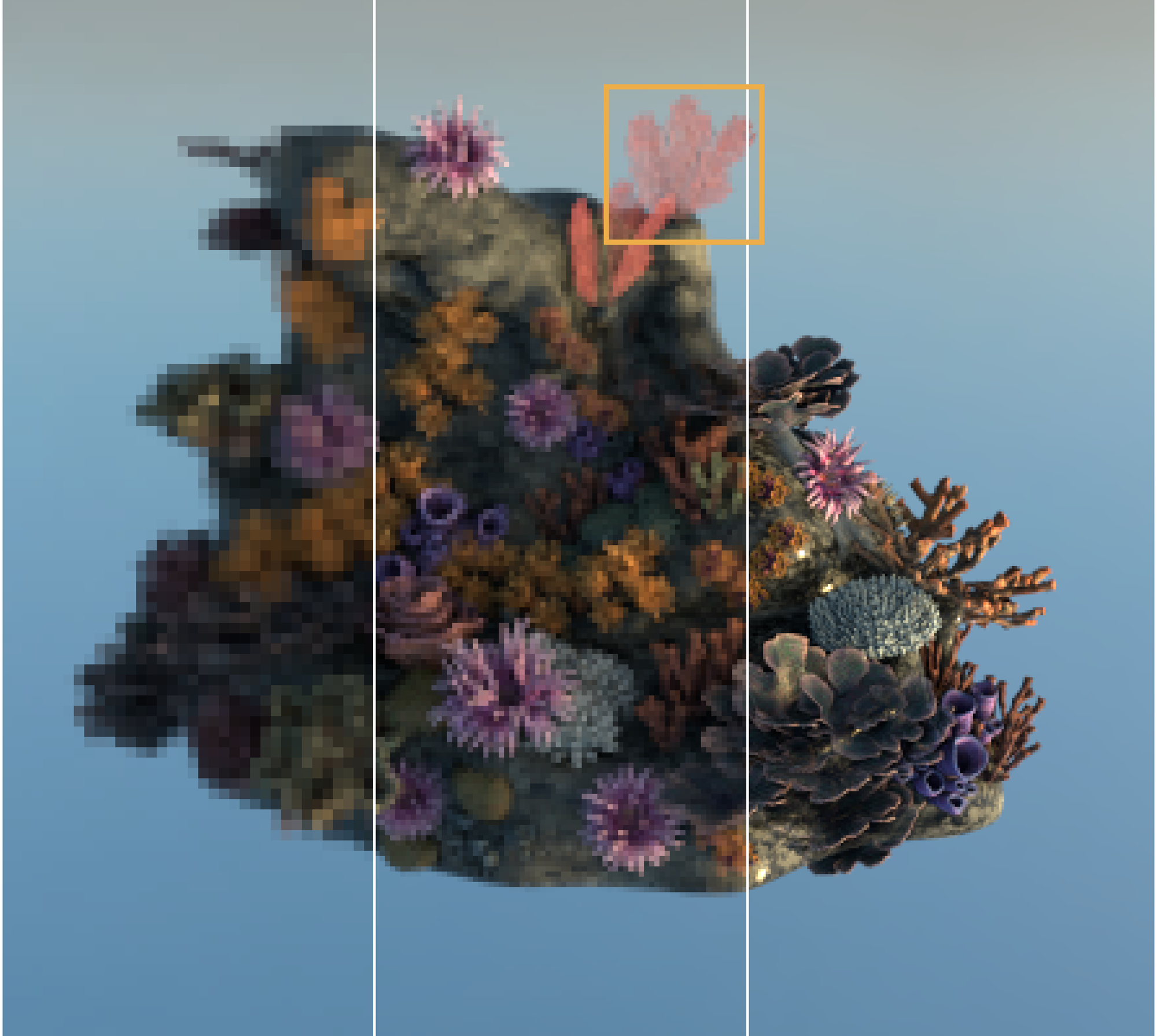}
			\put(1, 2){\normalsize \color{white}  $64^3$}
			\put(33, 2){\normalsize \color{white} $128^3$}
			\put(65.5, 2){\normalsize \color{white} $256^3$}
		\end{overpic}
		&
		\begin{overpic}[height=\lenCoralreef]{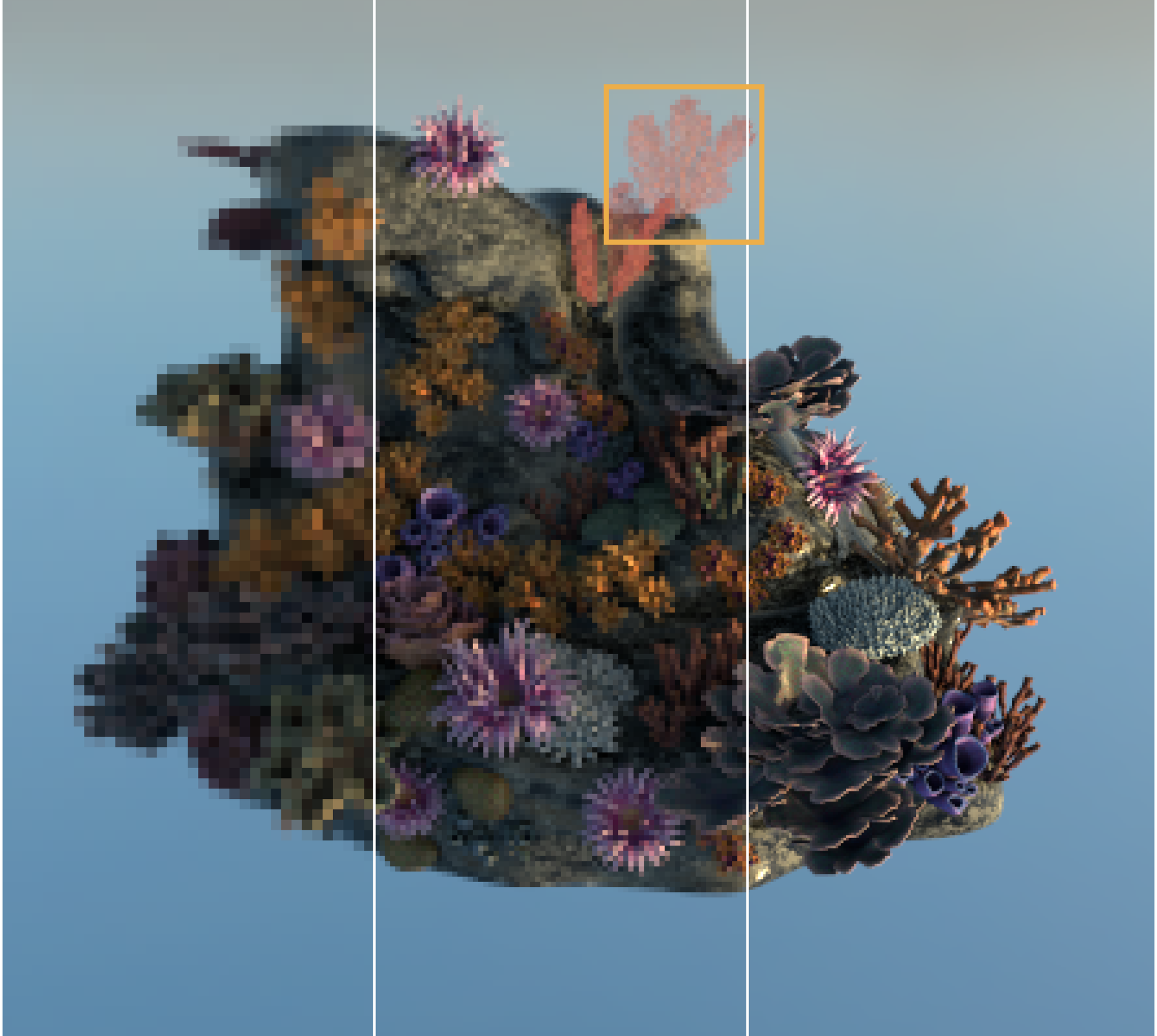}
			\put(1, 2){\normalsize \color{white}  $128^2$}
			\put(33, 2){\normalsize \color{white} $256^2$}
			\put(65.5, 2){\normalsize \color{white} $512^2$}
		\end{overpic}
		&
		\multicolumn{2}{c}{\begin{overpic}[height=\lenCoralreef]{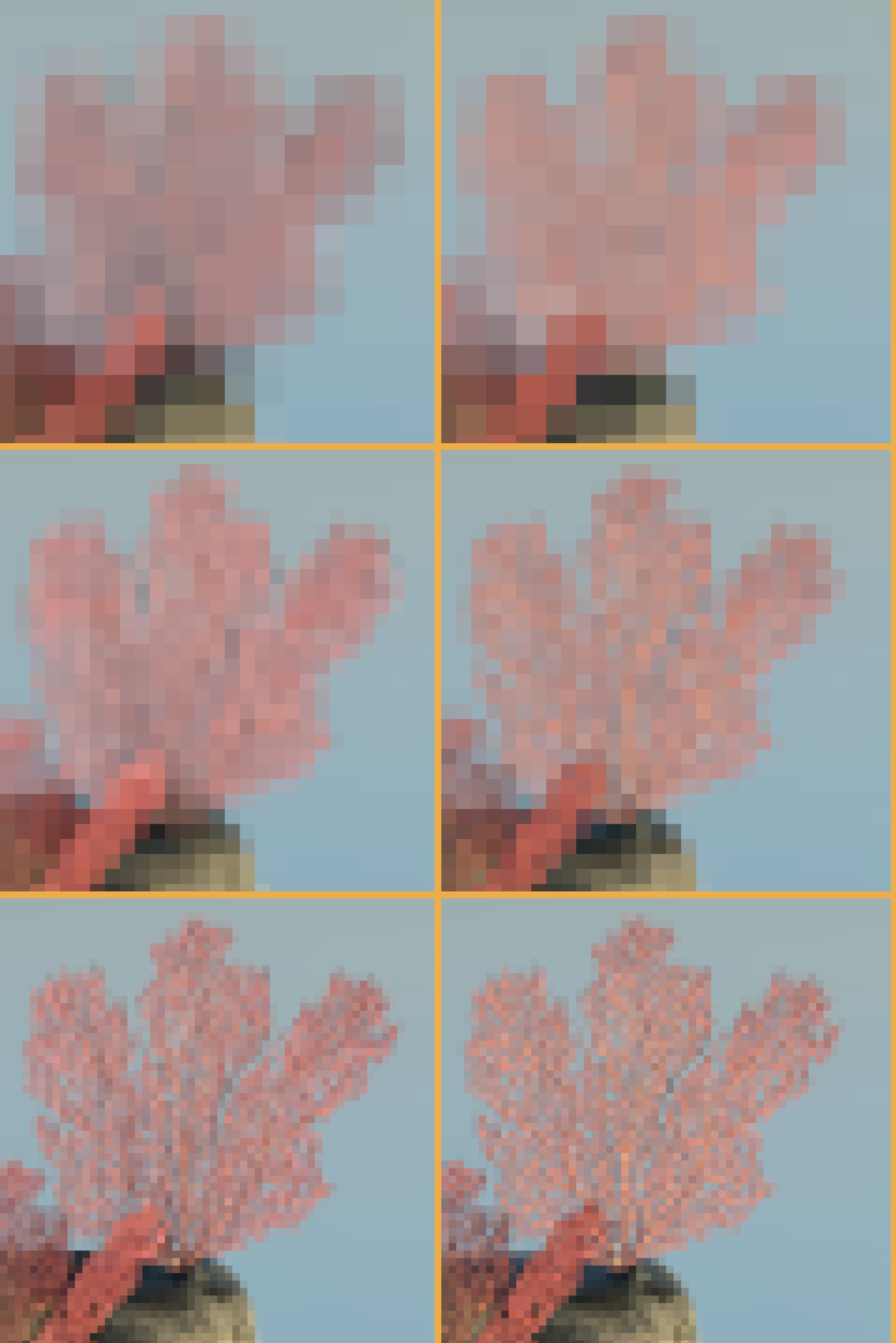}
			\put(21.5, 2){\normalsize \color{white}  $256^3$}
			\put(21.5, 36){\normalsize \color{white}  $128^3$}
			\put(21.5, 69){\normalsize \color{white}  $64^3$}
			\put(54.5, 2){\normalsize \color{white}  $512^2$}
			\put(54.5, 36){\normalsize \color{white}  $256^2$}
			\put(54.5, 69){\normalsize \color{white}  $128^2$}
		\end{overpic}}
		\\
		\textsf{(a) Ours} & \textsf{(b) PT Reference} & \;\quad \textsf{(c) Ours} & \;\quad \textsf{(d) PT Ref.}
	\end{tabular}
	\caption{\label{fig:coralreef}
		Our method captures the complex visual appearance of the \emph{Coral Reef} scene that consists of a variety of geometry and materials. We show (a) the
		renders with our representation at 3 scales ($64^3$, $128^3$, and $256^3$) and compare them to (b) path-traced (PT) references. Image resolutions are
		chosen such that one voxel approximately projects to the footprint of a single pixel ($128^2$, $256^2$, and $512^2$). We highlight a challenging part
		that features partial transparency and a glossy material on the right.}
	\end{figure*}

In this work, we propose an efficient \rev{volumetric} scene appearance aggregation method for LoD rendering. Our representation supports arbitrary types of scenes geometry
from completely opaque surfaces to stochastically distributed structures, and a wide range of appearance from glossy to diffuse. At the heart of our
representation is the Aggregated Bidirectional Scattering Distribution Function (ABSDF) that summarizes the appearance of all surfaces inside a voxel.
\rev{Contrary to existing volume-based methods,}
our method inherently keeps track of long-range correlation by recording the global visibility originated from a voxel and from the scene boundary.
Simultaneously, we propose a novel truncated ellipsoid primitive to better handle the local correlation within a voxel.
We focus on the appearance of a scene at far field, as is the case when an LoD representation gains the most benefit.
Similar to \citet{bako2023deep}, we focus on the appearance with direct illumination, which is arguably the more challenging part compared to
the indirectly illuminated counterpart as it is subject to more visible artifacts such as leaking and bloating.
Our method achieves high rendering fidelity by preserving the complex visual appearance caused by both geometry and materials
(\autoref{fig:teaser}, \autoref{fig:coralreef}, and \autoref{fig:botanic}).

To summarize, our contributions include:
\begin{itemize}
    \item A novel formulation for representing and rendering far-field scene aggregates for arbitrary scenes with the \emph{Aggregated Bidirectional Scattering
            Distribution Function} (ABSDF).
    \item A closed-form factorization of the aggregated appearance that captures all-frequency and view-dependent effects.
          The resulting model supports efficient evaluation and importance sampling.
    \item A practical solution that handles local correlation by truncated ellipsoid primitives and long-range correlation by recording global visibility.
    \item An efficient scene aggregation pipeline that is scalable to large, complex assets and offers asymptotic memory saving and rendering speed boost.
\end{itemize}

\section{Related work}
Representing scenes and appearance at multiple scales to improve rendering efficiency and quality is a long-standing problem in computer graphics. We draw
inspiration from various previous work ranging from surface-based approaches to volume-based approaches, together with hybrid approaches in between. In addition,
the recent advances of neural representations provide a set of new tools proven to be effective in certain graphics applications.

\paragraph{Mesh simplification} Polygon meshes are by far the most common representation of 3D models in computer graphics. A large amount of study has been focused on
algorithms that simplify a complex mesh by collapsing edges and merging vertices~\citep{hoppe1996progressive, garland1997surface}. Some
attempts have been made to extend mesh simplification to consider appearance to a limited extent~\citep{cohen1998appearance, cook2007stochastic}. Mesh
simplification techniques are widely employed in movie and video game production~\citep{karis2021deep}. However, they are fundamentally unable to preserve the
complex appearance that is a combination of both detailed geometry and material models.
More recently, \citet{hasselgren2021appearance} jointly optimize triangle
meshes and material parameters to minimize the image-space difference to the target scene by a differentiable rasterizer. However, the shading model is limited
to be the same before and after optimization. The optimization also ignores global effects such as shadows.

\paragraph{Surface appearance filtering} Surface-based filtering techniques focus on filtering the spatially-varying material attributes and microscale geometric
details while keeping the original macro-scale surface geometry. Normal map filtering, for example, converts the normal directions inside a footprint to a normal
distribution function (NDF) to preserve highlights when viewed from afar~\citep{toksvig2005mipmapping, han2007frequency, olano2010lean, kaplanyan2016filtering}.
Xu et al. propose to jointly mipmap BRDF and normal maps~\citep{xu2017real}.
Glints rendering~\citep{yan2014rendering, yan2016position} focuses on resolving the highlight from specular micro-geometry, which is essentially the same problem.
However, both the spatial resolution of the normal maps and the angular resolution of the NDFs are much higher. The source normal maps can also be procedurally
generated to alleviate the high memory cost~\citep{jakob2014discrete, zirr2016real, wang2020example}. Displacement map filtering
incorporates the microscale geometric details provided by displacement maps inside a footprint into a shading model~\citep{dupuy2013linear, wu2019accurate}.
Bi-scale material design models the macro-scale appearance of an object by designing its microscale details and aggregates their appearance
~\citep{wu2011physically, iwasaki2012interactive}.
Bidirectional texture functions (BTFs) represent non-parametric 6D spatially-varying surface appearance. Filtering BTFs offers significant memory savings and
a performance boost~\citep{jarabo2014effects}.
Surface-based techniques successfully simplify microscale details by prefiltering them into an appearance model.
However, they do not alter macro-scale geometry, thus they are not helpful when macro-scale geometry is the dominant factor in scene complexity.

\paragraph{Volumetric appearance models and filtering}
Using volumes to represent complex geometry has been explored extensively since first introduced by \citet{kajiya1989rendering}.
Volumes are traditionally used to accelerate the rendering of dense, unstructured geometry such as fur, hair, and foliage~\citep{neyret1998modeling, moon2008efficient}.
\citet{jakob2010radiative} proposes the microflake theory that extends the radiative transfer equation
(RTE)~\citep{Chandrasekhar:1960:Radiative} to anisotropic participating media, enabling volumes to represent a wider range of appearance such as fabric and
cloth~\citep{zhao2011building, zhao2012structure}. \citet{heitz2015sggx} further proposes the SGGX distribution to construct efficient
microflake phase functions that support linear interpolation and closed-form importance sampling. As a high-resolution volume can be very memory-intensive,
several works consider the problem of downsampling microflake volumes while preserving the important self-shadowing effect~\citep{zhao2016downsampling, loubet2018new}.
The classic volumetric light transport theory that builds on the RTE assumes independently distributed scatterers and thus does not support spatial correlation,
limiting its expressiveness for general scene representation.
More recently, it has been further extended to support spatially-correlated participating media through different formulations
~\citep{jarabo2018radiative,bitterli18framework}. \citet{vicini2021non} proposes an empirical non-exponential transmittance model that, while not
physically-based, improves the ability to model correlation and opaque surfaces when combined with data-driven optimization. While
volume-based techniques are able to simplify macro-scale geometry, volumetric light transport itself is significantly harder to solve than surface light transport
and typically takes a longer time to converge for Monte Carlo path tracing.

Another line of works focuses on building efficient voxel-based data structures. \citet{crassin2009gigavoxels} and
\citet{laine2010efficient} propose different variants of a sparse voxel octree (SVO) to render massive volumes at interactive rates. The SVO data
structure can be further specialized to support even higher resolution~\citep{kampe2013high}. Building on top of SVO, \citet{heitz2012representing}
proposes a representation to filter the appearance of detailed surfaces with the ability to reproduce view-dependent effects and account for correlation of
occlusion and attributes with visibility. However, they only support opaque surfaces modeled by a boundary representation. Thus, their work is not applicable
to a wide variety of subjects consisting of dense, unstructured geometry.

\paragraph{Hybrid approaches}
A number of works attempt to combine the advantages of surface-based techniques and volume-based techniques. \citet{dupuy2016additional} draws
a theoretical connection between microfacet and microflake theories. Granular material rendering techniques achieve acceleration by switching representation at
different scales of light transport~\citep{moon2007rendering, meng15granular,muller16efficient, zhang2020multi}. Grains are only explicitly traced during initial
bounces. For longer-scale light transport and multiple scattering, grains are replaced with a volumetric representation that is rendered by volumetric path
tracing and eventually diffusion methods.
\citet{loubet2017hybrid} proposes a hybrid LoD technique that performs a binary classification on the input scene to divide it into a mesh
part and a volume part at each scale. Subsequently, the mesh part undergoes mesh simplification and the volume part is represented by a microflake participating
medium. While the idea sounds straightforward, the classification unfortunately suffers from ambiguity and the technique produces artifacts when misclassification
happens. Additionally, mesh simplification may drastically alter surface curvature that results in incorrect glossy appearance, as shown in
\autoref{fig:main_comparison}.

\paragraph{Neural representations}
Neural implicit representations are shown to be particularly effective at compactly reconstructing signals in low-dimensional spaces such as radiance
fields and shapes~\citep{mildenhall2020nerf, lombardi2021mixture, martel2021acorn, muller2022instant, park2019deepsdf}. While most works focus on point-wise
query and inference, some techniques build multi-scale representations that support range queries for anti-aliasing~\citep{barron2021mip, takikawa2021neural}.
However, most neural implicit representations are unable to model full appearance, with limited capability for relighting~\citep{bi2020neural, lyu2022neural,
baatz2022nerf}. For a more comprehensive review, we refer readers to two recent surveys on the subject~\citep{tewari2022advances, xie2022neural}.

\rev{
Traditional surface-based techniques can be enhanced by neural components. \citet{kuznetsov2021neumip,kuznetsov2022rendering} achieve BTF compression and 
filtering by simultaneously training a latent texture pyramid and a small multilayer perceptron (MLP) decoder that supports isotropic range queries.
\citet{gauthier2022mipnet} improve normal map filtering by using a MLP cascade to learn downsampling kernels.
}

Recently, \citet{bako2023deep} propose a deep learning based \\appearance-prefiltering framework. An input scene is converted to a volumetric
representation where each voxel records a monochromatic phase function, an average albedo, and a 4D view-dependent coverage mask. To reduce the otherwise
infeasible memory requirement, each type of data is compressed by a separate encoder-decoder network that produces per-voxel latent vectors. The volume is
rendered by a beam tracer that traverses the voxels and decodes them for shading and transmittance computation.
The method preserves accurate appearance but at a heavy cost in both precomputation and rendering. The typical precomputation time is reported
to be 0.5 to 2 days on a GPU cluster with 256 NVIDIA Volta GPUs. The compressed per-voxel size is still large with 256 floats. In addition, the beam tracer must
traverse voxels ordered by distance to correctly compute transmittance by accumulating the coverage masks from each voxel. In contrast, our method only requires
a much lighter precomputation pass, a smaller memory cost, offers much faster rendering speed, and results in similar rendering quality.
\citet{weier2023neural} propose a neural prefiltering pipeline by learning a compressed
representation of the intra-voxel light transport. Two independent networks for appearance and visibility are trained with a multi-level feature grid. The
method handles diffuse-like appearance well and supports indirect lighting. However, it struggles at preserving glossy appearance and capturing all-frequency
directional signals.

\section{Far-field Appearance Aggregation and Factorization} \label{sec:factorize}

\subsection{Overview}
\begin{figure*}[tb]
	\newlength{\lenOverview}
	\setlength{\lenOverview}{2in}
    \addtolength{\tabcolsep}{-4pt}
    \renewcommand{\arraystretch}{0.5}
    \centering
    \includegraphics[height=\lenOverview]{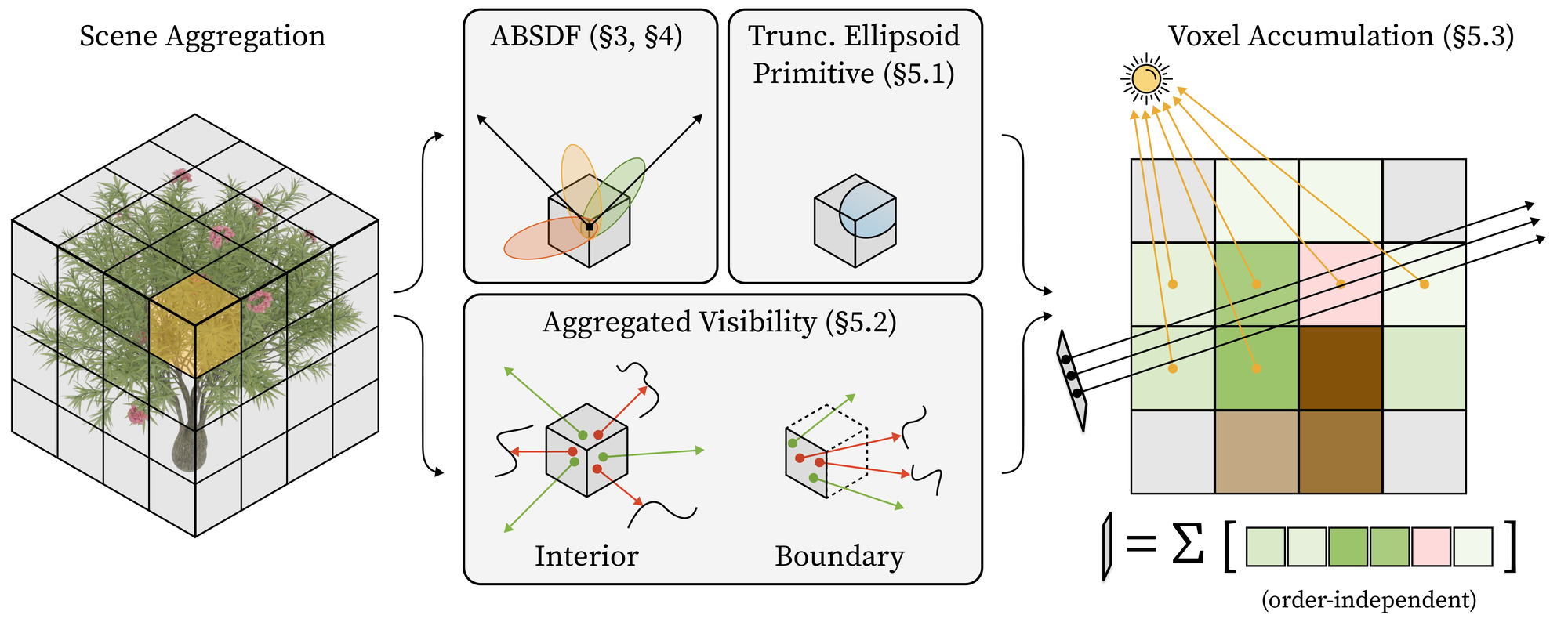}
    \caption{\label{fig:overview}
        An overview of our method. We start by voxelizing a scene such that the voxel size matches the given pixel footprint. For each voxel, we model its
        aggregated appearance by its ABSDF. To preserve the local spatial correlation, we use a truncated ellipsoid primitive that describes the intra-voxel
        geometric distribution. To preserve the long-range correlation, we record global aggregated visibility. Both lead to accurate voxel accumulation that
        is order-independent.
        }
\end{figure*}

Our goal is to develop an appearance-preserving representation of a scene that is independent of the original geometry complexity, which we term \emph{scene
aggregate}.
\autoref{fig:overview} provides an overview of our method.
When measured externally, the general light transport of a scene aggregate can be characterized as an 8D function of incident/outgoing positions (on
a suitable bound of the scene) and directions, similar to a BSSRDF. However, directly computing and storing such a function is impractical due to the
prohibitive memory requirement. It is also not necessary as one might as well simply switch back to the original representation for near-field appearance.
Therefore, our formulation is based on the far-field assumption. When a scene is sufficiently far from the measuring sensor and
emitters that the sensor can no longer distinguish the internal spatial structure, we may drop the positional dependency by integrating the 8D light transport
function over positions. The resulting 4D function of incident/outgoing directions describes the far-field appearance of the scene aggregate and we name it
\emph{Aggregated Bidirectional Scattering Distribution Function} (ABSDF), again due to its similarity to a BSDF.
In \autoref{sec:factorize}, \autoref{sec:factorize_diffuse}, and
\autoref{sec:factorize_specular}, we define the ABSDF and present an efficient, closed-from factorization of it.

In practice, an entire scene is usually too large to be considered far-field all together. We apply spatial subdivision to the scene at a suitable resolution
given a certain pixel footprint such that the subset of the scene included in each voxel satisfies the far-field assumption. This introduces the subsequent
problem of accumulating the outgoing radiance from voxels and eventually measuring the pixel intensities. Crucially, the accumulation problem is non-trivial
because the spatial configuration of voxels is not independent. Traditional volumetric representations model a scene as independently distributed particles,
which is incorrect because a scene made of surfaces typically exhibits spatial correlation and ignoring such correlation leads to artifacts or
inaccurate appearance.
In \autoref{sec:accumulation}, we analyze the problem of spatial correlation in detail, discuss our strategies to preserve correlation, and derive the formulation
for voxel accumulation. We provide a summary for commonly used symbols throughout the paper in \autoref{tab:symbols}.

\begin{table}[tb]
	\centering
	\caption{Table of notation.}
	\begin{tabular}{lll}
        \Xhline{1pt}
        \textbf{Symbol} & \textbf{Explanation} & \textbf{Def.} \\
        \hline
        $A$                         & A set of surfaces (in a voxel)                                                        & \\
        \hline
        $|A|$                       & Surface area of $A$                                                                   &  \\
        \hline
        $|A|_{\omega}$              & Projected surface area of $A$ along $\omega$                                          & \autoref{subsec:define_absdf}, \autoref{eq:average_Lo} \\
        \hline
        $\hat{f}$                   & \makecell[cl]{Aggregated Bidirectional Scattering \\ Distribution Function (ABSDF)}   & \autoref{subsec:define_absdf}, \autoref{eq:absdf_def} \\
        \hline
        $\hat{f}_{\mathrm{novis}}$  & ABSDF without visibility                                                              & \autoref{subsec:factorize_absdf}, \autoref{eq:absdf_distrib} \\
        \hline
        $n$                         & Surface normal                                                                        & \\
        \hline
        $p_N(n)$                    & Surface normal distribution function                                                  & \autoref{subsec:factorize_absdf}, \autoref{eq:multi_lobe_ndf} \\
        \hline
        $D_{\mathrm{sggx}}(n)$      & \rev{\makecell[cl]{SGGX distribution parameterized by \\
                                                    eigenbasis $R$ and roughness \\ 
                                                    $\bm{\alpha} \coloneq (\alpha_x, \alpha_y)$}}                           & \autoref{subsec:factorize_absdf} \\
        \hline
        $\beta$                     & \rev{\makecell[cl]{Material parameters at point $x$, \\
                                        including roughness $\alpha$, basecolor $\beta^c$ \\ 
                                        (spectral), metallic $\beta^m$ (scalar), and \\
                                        specular intensity $\beta^s$ (scalar)}}                                             & \autoref{subsec:factorize_absdf} \\
        \hline
        $\gamma$                    & Concatenation of $n$ and $\beta$                                                      & \autoref{subsec:factorize_absdf} \\
        \hline
        $p_Y(\gamma)$               & Joint distribution of $\gamma$                                                        & \autoref{subsec:factorize_absdf}\\
        \hline
        $B$                         & Truncated ellipsoid primitive                                                         & \autoref{subsec:trunc_ellipsoid} \\
        \hline
        $c(\omega)$                 & Primitive coverage                                                                    & \autoref{subsec:trunc_ellipsoid}, \autoref{eq:primitive_coverage} \\
        \hline
        $\hat{V}$                   & Aggregated interior visibility                                                        & \autoref{subsec:agn_vis}, \autoref{eq:define_aiv} \\
        \hline
        $\hat{V}_b$                 & Aggregated boundary visibility                                                        & \autoref{subsec:agn_vis}, \autoref{eq:define_abv} \\
        \hline
        $\langle - \cdot - \rangle$     & Clamped dot product                                                               & \\
        \Xhline{1pt}
	\end{tabular}
	\label{tab:symbols}
\end{table}

\subsection{Defining ABSDF} \label{subsec:define_absdf}
We consider a subset of a scene $A$ that consists of a set of surfaces. From a point $x \in A$, the outgoing radiance given some direct incident radiance
$L_i(x, \omega_i)$ is calculated by the following equation~\citep{cohen1993radiosity}:
\begin{align} \label{eq:rendering_eq}
\begin{split}
    &L_o(x, \omega_o) = \int_{\mathbb{S}^2} f(x, \omega_i, \omega_o) L_i(x, \omega_i) \langle n_x \cdot \omega_i \rangle  V(x, \omega_i) \,\D{\omega_i}, \\
\end{split}
\end{align}
where $f(x, \omega_i, \omega_o)$ is the surface \rev{BRDF} at $x$, and we explicitly write the visibility term $V(x, \omega_i)$.
We are interested in the average outgoing radiance from $A$ when viewed from  $\omega_o$, which can be written as a weighted average of per-point outgoing
radiance masked by another visibility term along $\omega_o$:
\begin{align} \label{eq:average_Lo}
\begin{split}
    L_o(\omega_o) &= \frac{1}{|A|_{\omega_o}} \int_{A}  L_o(x ,\omega_o) \langle n_x \cdot \omega_o \rangle V(x, \omega_o) \,\D{x}, \\
    |A|_{\omega_o} &= \int_A \langle n_x\cdot\omega_o \rangle \,\D{x},
\end{split}
\end{align}
where $|A|_{\omega_o}$ is the projected area of $A$ along $\omega_o$. We can apply the far-field assumption such that the
incident radiance is independent of positions $L_i(x, \omega_i) \approx L_i(\omega_i)$ and rearrange \autoref{eq:average_Lo} by reordering the integrations:
\begin{align} \label{eq:absdf_def}
\begin{split}
    L_o(\omega_o) &= \int_{\mathbb{S}^2} \hat{f}(\omega_i, \omega_o) L_i(\omega_i) \,\D{\omega_i}, \\
    \hat{f}(\omega_i, \omega_o) &= \frac{1}{|A|_{\omega_o}} \int_A \Big( f(x,\omega_i, \omega_o) \langle n_x\cdot\omega_i \rangle \langle n_x\cdot\omega_o \rangle \\
    &\qquad \qquad \qquad V(x,\omega_i)V(x,\omega_o) \Big) \,\D{x}. \\
\end{split}
\end{align}
We define $\hat{f}(\omega_i, \omega_o)$ as the ABSDF of $A$ as it captures the intrinsic geometrical and material characteristics of $A$ and is independent of
external sensors or emitters. It is not hard to see that the ABSDF satisfies energy conservation as long as $f(x, \omega_i, \omega_o)$ is energy-conserving.
\rev{The ABSDF can be interpreted as an extension of the effective BRDF~\citep{wu2011physically} where surfaces are not longer confined to a macrosurface or 
a heightfield, but allowed to be arranged arbitrarily in free space.
}
It also satisfies a generalized form of reciprocity that is similar to the situation in the microflake theory~\citep{jakob2010radiative}. To summarize:
\begin{align}
    &\int_{\mathbb{S}^2} \hat{f}(\omega_i, \omega_o) \,\D{\omega_i} \leq 1 \\
    &|A|_{\omega_o} \hat{f}(\omega_i, \omega_o) = |A|_{\omega_i} \hat{f}(\omega_o, \omega_i)
\end{align}

We note that it is possible to use the visible projected area\\
 $|A_v(\omega_o)| = \int_A \langle n_x \cdot \omega_o \rangle V(x, \omega_o) \,\D{x}$ as the normalization term in \autoref{eq:absdf_def}. In fact, the choice
 is not critical as the term will eventually be cancelled out (\autoref{eq:primitive_coverage}, \autoref{eq:accumulate_single_voxel}). We choose to use $|A|_{\omega_o}$
 for simplicity. In practice, neither $|A_v(\omega_o)|$ nor $|A|_{\omega_o}$ needs to be computed or stored.

\subsection{A Closed-form Factorization of the ABSDF} \label{subsec:factorize_absdf}
According to \autoref{eq:absdf_def}, an ABSDF is defined by integrating the product of base material, foreshortening factors, and bidirectional visibility over
the underlying surfaces. This is challenging, as in general, no closed-form solution exists. However, we would also like to avoid stochastic evaluation or
numerical integration which would greatly increase the rendering cost and undermine the purpose of scene aggregation. To achieve closed-form evaluation and
importance sampling, we factor the ABSDF with the following steps.

\paragraph{Separate Visibility}
We perform a splitting approximation that separates the integration of the bidirectional visibility from the rest:
\begin{align} \label{eq:split_vis}
\begin{split}
        \hat{f}(\omega_i, \omega_o) \approx
        \frac{1}{|A|_{\omega_o}} &\int_A f(x,\omega_i, \omega_o) \langle n_x\cdot\omega_i \rangle \langle n_x\cdot\omega_o \rangle \,\D{x} \\
        \frac{1}{|A|} &\int_A V(x, \omega_i) V(x, \omega_o)  \,\D{x}. \\
\end{split}
\end{align}
\rev{
This is similar to the approximation by \citet{jimenez2016practical}.
}
We focus on the first integral in the rest of this section and further describe how to incorporate visibility into our framework in \autoref{subsec:agn_vis}.

\paragraph{The distribution form of the ABSDF}
With the visibility terms separated, we convert the ABSDF into a convolution between the base material and the joint distribution of material parameters.
Similar operations have been employed in normal map and displacement map filtering techniques~\citep{han2007frequency,olano2010lean,dupuy2013linear}, but they
usually only consider the distribution of surface normals. Our formulation can be seen as a generalization that incorporates all spatially-varying parameters.
Let $\gamma_x\coloneqq (n_x, \beta_x)$ be a vector consisting of surface normal and all material parameters at $x$. It can also be interpreted
as a value of a random vector $Y$ with joint PDF $p_{Y}(\gamma)$. We can rewrite the ABSDF (no visibility) as follows:
\begin{align} \label{eq:absdf_distrib}
\begin{split}
    \hat{f}_{\mathrm{novis}}(\omega_i, \omega_o)
    &=\frac{1}{|A|_{\omega_o}} \int_A f(\omega_i,\omega_o; \gamma_x) \langle n_x\cdot\omega_i \rangle \langle n_x\cdot\omega_o \rangle\,\D{x} \\
    &=\frac{|A|}{|A|_{\omega_o}} \int_{\Gamma} f(\omega_i,\omega_o; \gamma) \langle n\cdot\omega_i \rangle \langle n\cdot\omega_o \rangle p_{Y}(\gamma) \,\D{\gamma},
\end{split}
\end{align}
where $\Gamma$ is the product space of all parameters.

\paragraph{Surface normal distribution function}
The marginal distribution of surface normals $p_N(n)$, or the surface NDF, is important as it affects the glossiness and anisotropy of the aggregated
appearance. It can also be complex because the underlying surfaces may be arbitrarily oriented. We use a mixture of the SGGX distribution~\citep{heitz2015sggx}
as a compact yet expressive representation for the surface NDF:
\begin{equation} \label{eq:multi_lobe_ndf}
    p_N(n) = \sum_{i=1}^k w_i D^i_{\mathrm{sggx}}(n),
\end{equation}
where the weights $\{ w_i \}$ are positive and sum to $1$. We describe the fitting process for the mixture model in \autoref{subsec:precompute}.

As will be seen in \autoref{sec:factorize_diffuse} and \autoref{subsec:conv_with_ndf}, ours
factorization involves convolving $p_N(n)$ with another isotropic spherical distribution $g(\omega; n)$. Because $p_N(n)$ is a mixture of SGGX lobes, the
result is the sum of the convolution between each lobe $D^i_{\mathrm{sggx}}(n)$ and $g(\omega; n)$.
We propose to represent the per-lobe convolution as a similar but roughened SGGX. 
\rev{
We first parameterize an SGGX distribution by its eigenbasis
$R \coloneqq (\omega_1, \omega_2, \omega_3)$ and anisotropic roughness $\bm{\alpha} \coloneq (\alpha_x, \alpha_y)$. The parameterization is detailed in the supplemental document.
The convolution can then be written as
\begin{equation} \label{eq:per_sggx_lobe_conv}
    \int_{\mathbb{S}^2} g(\omega; n) D_{\mathrm{sggx}}(n; R, \bm{\alpha}) \, \D{n} \approx D_{\mathrm{sggx}}(\omega; R, \bm{\alpha}_{+}),
\end{equation}
where $R$ stays fixed but $\bm{\alpha}$ gains additional values. This is inspired by
\citet{xu2013anisotropic}, where a similar approximation is made for anisotropic Spherical Gaussians. While Xu et al. seeks a symbolic
approximation, we simply perform a nonlinear least-square fit to find the best mapping $\bm{\alpha}_{+} = M(\bm{\alpha}, g)$. Note
that this mapping is \emph{scene-independent} and typically smooth, thus only requiring precomptation once and negligible storage. We observe accurate fits for all our
target distributions $g$. We provide derivation details and numerical validation with different $g$ in the supplemental document. Following
\autoref{eq:per_sggx_lobe_conv}, the post-convolution distribution for the entire $p_N(n)$ is
\begin{equation} \label{eq:multi_sggx_lobe_conv}
    D_\mathrm{conv}(\omega) = \int_{\mathbb{S}^2} g(\omega; n) p_N(n) \, \D{n} \approx \sum_{i=i}^k w_i D^i_{\mathrm{sggx}}(\omega; R^i, \bm{\alpha}_{+}^i).
\end{equation}
}

\paragraph{Base material}
The ABSDF is dependent on the underlying surface base materials. For the widest applicability, it
is desirable to support material models used by existing assets. Therefore, we target the Disney Principled BRDF \citep{burley2012physically}, which is one of the most
commonly used models in production and capable of recreating a wide range of appearance. The Disney BRDF is a sophisticated model consisting of multiple lobes.
We preserve its core feature but make three modifications to the original model:
\begin{enumerate}
    \item For diffuse reflection, we use the simpler Lambertian model instead of the original empirical model with retro-reflection.
    \item We omit the optional sheen and clearcoat lobes.
    \item We assume surfaces are double-sided.
\end{enumerate}

The modified model can be written as
\begin{align} \label{eq:base_material}
\begin{split}
    f_{\text{disney}}(\omega_i, \omega_o) &= f_d(\omega_i, \omega_o) + f_{s}(\omega_i, \omega_o), \\
    f_{d}(\omega_i,  \omega_o) &= \frac{1}{\pi}(1 - \beta^m) \beta^c, \\
    f_{s}(\omega_i, \omega_o) &= \frac{ D(\omega_h; \alpha) G(\omega_i, \omega_o; \alpha) }{ 4 | n \cdot \omega_o | | n \cdot \omega_i | }
    \Big( \beta^m F(\omega_h, \omega_o;\beta^c) + \\
    & \quad (1 - \beta^m) F(\omega_h, \omega_o;\beta^s) \Big),
\end{split}
\end{align}
where $f_{s}(\omega_i, \omega_o)$ is the specular component that consists of both metallic and dielectric Fresnel reflection,
$f_d(\omega_i, \omega_o)$ is the diffuse component,
$D(\omega_h)$ is the Trowbridge-Reitz (GGX) distribution as the microfacet distribution, $G(\omega_i, \omega_o)$ is the shadowing-masking function, and
$F(\omega_h, \omega_o; r_0)$ is the Schlick Fresnel reflectance:
\begin{align*}
F(\omega_h, \omega_o; r_0) &= r_0(1-\mathcal{F}_c) + \mathcal{F}_c, \\
\mathcal{F}_c &= (1 - |\omega_h \cdot \omega_o|)^5,
\end{align*}
where $r_0$ is the normal incidence reflectance (either $\beta^c$ for the metallic lobe or $\beta^s$ for the dielectric lobe).
The model is controlled by a set of parameters
$\beta \coloneqq (\alpha, \beta^c, \beta^m, \beta^s) = $ (roughness, basecolor, metallic, specular intensity), which can all be spatially-varying.
We now focus on factoring of each component, starting from the simpler diffuse component.

\rev{
\section{Diffuse ABSDF Factorization} \label{sec:factorize_diffuse}
}
We substitute $f_d(\omega_i, \omega_o)$ into the integral of \autoref{eq:absdf_distrib} and perform a split by first assuming $\beta^c$ and $\beta^m$ are
orientation-independent. This means the joint parameter PDF $p_Y(\gamma)$ becomes a product of two marginal PDFs
$p_Y(\gamma) \approx p(\beta^c, \beta^m) p_N(n)$ and we have
\begin{align} \label{eq:diffuse_orient_split}
&\int_{\Gamma} f_d(\omega_i,\omega_o; \gamma) \langle n\cdot\omega_i \rangle \langle n\cdot\omega_o \rangle p_{Y}(\gamma) \,\D{\gamma} \approx \\
&\frac{1}{\pi} \int_{[0,1]^2} (1 - \beta^m) \beta^c p(\beta^c, \beta^m) \,\D{\beta^c}\D{\beta^m}
\int_{\mathbb{S}^2} \langle n\cdot\omega_i \rangle \langle n\cdot\omega_o \rangle p_N(n) \,\D{n}. \nonumber
\end{align}
We extend our formulation to handle orientation-varying material parameters in \autoref{subsec:orient}. After the splitting, the left integral can be simply
represented by the means and second-order moments of the parameters
\begin{equation} \label{eq:diffuse_split_moments}
    \int_{[0,1]^2} (1 - \beta^m) \beta^c p(\beta^c, \beta^m) \,\D{\beta^c}\D{\beta^m} = \mathbb{E}[\beta^c] - \mathbb{E}[\beta^m\beta^c].
\end{equation}
For the right integral, we follow \citet{wang2009all} and fit the clamped dot product function by a Spherical Gaussian (SG).
Because SGs are closed under multiplication with a closed-form expression~\citep{wang2009all}, we can expand the right
integral of \autoref{eq:diffuse_orient_split} as
\begin{equation} \label{eq:diffuse_sg}
    \int_{\mathbb{S}^2} \langle n\cdot\omega_i \rangle \langle n\cdot\omega_o \rangle p_{N}(n) \,\D{n} \approx
    \int_{\mathbb{S}^2} c \cdot \mathrm{SG}(\omega_h; n, \kappa) p_{N}(n) \,\D{n},
\end{equation}
where $c$ is the amplitude and $\kappa$ is the concentration for the product SG. We refer readers to \citet{wang2009all} for the full expressions
for them. Notably, the product SG becomes a function of the half vector $\omega_h$.
The problem is then reduced to the convolution between an SG and an SGGX, which we solve by employing the convolution technique described in
\autoref{eq:multi_sggx_lobe_conv}. \rev{This comes with a table $\bm{\alpha}_{+} = M_1(\bm{\alpha}, \kappa)$.} \autoref{eq:diffuse_sg}
then can be evaluated in closed form given the surface NDF $p_N(n)$.

\section{Specular ABSDF Factorization} \label{sec:factorize_specular}
Next, we describe how to factorize the more challenging specular component of the ABSDF. Together with the diffuse component, our complete factorization will
be validated at the end of the section. We start by substituting $f_s(\omega_i, \omega_o)$ into the integral of \autoref{eq:absdf_distrib} and expanding it as
\begin{align} \label{eq:factor_specular}
\begin{split}
    \int_{\Gamma} &f_s(\omega_i,\omega_o; \gamma) \langle n\cdot\omega_i \rangle \langle n\cdot\omega_o \rangle p_{Y}(\gamma) \,\D{\gamma} = \\
    &\frac{1}{4} \Big[ (1-\mathcal{F}_c) \,
    \tcbhighmath[colback=white, colframe=grassgreen, boxrule=1pt, boxsep=0pt, top=2pt, bottom=2pt, left=2pt, right=2pt]
    {\int_{\Gamma} \mathcal{R} \mathcal{D} p_{Y}(\gamma) \,\D{\gamma}} + \mathcal{F}_c
    \tcbhighmath[colback=white, colframe=lightorange, boxrule=1pt, boxsep=0pt, top=2pt, bottom=2pt, left=2pt, right=2pt]
    {\int_{\Gamma} \mathcal{D} p_{Y}(\gamma) \,\D{\gamma}}
     \Big], \\
    &\mathcal{R} = \beta^m \beta^c + (1-\beta^m)\beta^s, \\
    &\mathcal{D} = D(\omega_h; n, \alpha) G(\omega_i, \omega_o; n, \alpha) \mathds{1}(n \cdot \omega_o) \mathds{1}(n \cdot \omega_i), \\
\end{split}
\end{align}
\rev{
where $\mathds{1}(\cdot)$ is the Heaviside (step) function that evaluates to $0$ or $1$.} 
$\mathds{1}(n \cdot \omega_o)$ and $\mathds{1}(n \cdot \omega_i)$
appear due to the clamped dot products in \autoref{eq:absdf_distrib}. Similar to \autoref{sec:factorize_diffuse}, we split the green highlighted integral in
\autoref{eq:factor_specular} by assuming $p_{Y}(\gamma) \approx p_{Y_1}(\beta^c, \beta^m, \beta^s) p_{Y_2}(n, \alpha)$ and extend the formulation to handle 
orientation-varying material parameters in \autoref{subsec:orient}:
\begin{equation} \label{eq:specular_orient_split}
    \tcbhighmath[colback=white, colframe=grassgreen, boxrule=1pt, boxsep=0pt, top=2pt, bottom=2pt, left=2pt, right=2pt]
    {\int_{\Gamma} \mathcal{R} \mathcal{D} p_{Y}(\gamma) \,\D{\gamma}} \approx
    \int_{\Gamma_1} \mathcal{R} p_{Y_1}(\gamma) \,\D{\gamma} \,
    \tcbhighmath[colback=white, colframe=lightorange, boxrule=1pt, boxsep=0pt, top=2pt, bottom=2pt, left=2pt, right=2pt]
    {\int_{\Gamma_2} \mathcal{D} p_{Y_2}(\gamma) \,\D{\gamma}},
\end{equation}
where $\Gamma_1$ and $\Gamma_2$ are the product space of $(\beta^c, \beta^m, \beta^s)$ and $(n, \alpha)$, respectively.
Once again, the left integral can be simply represented by the means and second-order moments of the parameters
\begin{equation} \label{eq:specular_split_moments}
    \int_{\Gamma_1} \mathcal{R} p_{Y_1}(\gamma) \,\D{\gamma} = \mathbb{E}[\beta^m \beta^c] + \mathbb{E}[\beta^s] - \mathbb{E}[\beta^m \beta^s].
\end{equation}
For the yellow highlighted integral in \autoref{eq:factor_specular} and \autoref{eq:specular_orient_split}, we propose a closed-form solution with several small,
\emph{scene-independent} precomputed tables. The total storage for
the tables is less than 5MB in practice.
\rev{
We focus on the characteristic microfacet distribution term $D$, which we now call $D_\mathrm{mic}$ in the rest of \autoref{sec:factorize_specular} for 
better clarity. The shadowing-masking term $G$ is in general very smooth~\citep{AshikminPS00,wang2009all,kaplanyan2016filtering}.
Our strategy is based on the convolution technique described in \autoref{subsec:factorize_absdf} and include three steps:
\begin{itemize}
    \setlength{\itemsep}{0pt}
    \setlength{\parskip}{1pt}
    \item \autoref{subsec:agn_microfacet}: Identify the aggregated microfacet distribution $\hat{D}_\mathrm{mic}(\omega_h)$.
    \item \autoref{subsec:conv_with_ndf}: Convolve $\hat{D}_\mathrm{mic}(\omega_h)$ with the surface NDF $p_N(n)$ to get the post-convolution distribution 
    $D_\mathrm{conv}(\omega_h)$.
    \item \autoref{subsec:cond_angular_domain}: Apply a scaling factor $S(\omega_i, \omega_o)$ to $D_\mathrm{conv}(\omega_h)$ to correct leaking beacuse 
    surfaces can be back-facing either to view or lights.
\end{itemize}
We now proceed to describe each step in detail. 
The complete model is summarized at end of the section (\autoref{subsec:absdf_summary_and_validation}, \autoref{fig:absdf_diagram}).
Additional derivation details and numerical validation are available in the supplemental document.

\subsection{Aggregated Microfacet Distribution} \label{subsec:agn_microfacet}
As roughness $\alpha$ can vary on the surfaces of $A$, the microfacet distribution can no longer be represented by one GGX lobe.
Let $\mathcal{A}$ be the underlying random variable for the roughness and
$p_{\mathcal{A}}(\alpha)$ be its marginal density function. The aggregated microfacet distribution is the expectation of the microfacet
distribution
\begin{equation} \label{eq:rough_var}
\hat{D}_\mathrm{mic}(\omega_h) = \int_{[0,1]}  D_\mathrm{mic}(\omega_h; \alpha) p_{\mathcal{A}}(\alpha) \, \D{\alpha}.
\end{equation}
We should represent $p_{\mathcal{A}}(\alpha)$ by a parametric distribution while acknowledging that $\alpha$ is bounded in [0,1]. A Gaussian distribution is thus
not a valid choice. Instead, we use a beta distribution $\mathcal{B}(\alpha; a, b)$ which has the correct support and is reasonably
expressive. The shape parameters $a$ and $b$ can be easily estimated (see supplemental document).
\autoref{eq:rough_var} can be interpreted as the weighted average of infinite GGX lobes with different possible $\alpha$. Since there is no closed-form
solution for it, we further propose to approximate $\hat{D}_\mathrm{mic}(\omega_h) $ by a weighted average of 2 lobes:
\begin{equation} \label{eq:rough_var_two_lobe_fit}
    \hat{D}_\mathrm{mic}(\omega_h) \approx m_1D_\mathrm{mic}(\omega_h; \alpha_1) + m_2D_\mathrm{mic}(\omega_h; \alpha_2),
\end{equation}
where $m_1+m_2=1$.
The approximation can be extended to use an arbitrary number $k$ of lobes, but we find $k=2$ provides a good balance between cost and accuracy. We perform a 
nonlinear least square fit to find the best mapping $(m_1, \alpha_1, \alpha_2) = M_2(a, b)$ given the shape parameters of the beta distribution and
store it as a small 2D table.

\subsection{Convolution with Surface NDF} \label{subsec:conv_with_ndf}
The aggregated microfacet distribution $\hat{D}_\mathrm{mic}(\omega_h)$ is then convolved with the surface NDF $p_N(n)$. This is similar to normal map filtering and 
specular shading antialiasing techniques~\citep{olano2010lean, kaplanyan2016filtering}. 
We use the convolution technique described in \autoref{eq:multi_sggx_lobe_conv} for it and find the best mapping
$\bm{\alpha}_{+} = M_3(\bm{\alpha}, \alpha)$, which is stored as a small 3D table. Because both $\hat{D}_\mathrm{mic}$ and $p_N(n)$ 
are mixtures, the convolution can be carried out per pair of lobes:
\begin{align} \label{eq:normal_var}
    D_\mathrm{conv}(\omega_h) &=
    \int_{\mathbb{S}^2} \hat{D}_\mathrm{mic}(\omega_h; n) p_N(n) \, \D{n} = \sum_{i=1}^k \sum_{j=1}^2 w_i m_{ij} D^{ij}_\mathrm{conv}(\omega_h), \nonumber \\
    \tcbhighmath[colback=white, colframe=lightred, boxrule=1pt, boxsep=0pt, top=2pt, bottom=2pt, left=2pt, right=2pt]
    {D^{ij}_\mathrm{conv}(\omega_h)}
    &\approx D_{\mathrm{sggx}}(\omega; R^i, \bm{\alpha}^{ij} = M_3(\bm{\alpha}^i, M_3^j(\alpha^i))), 
\end{align}
where we denote $M_3^j(\alpha^i)$ as a shorthand for fitting a beta distribution for $\alpha^i$ and querying $M_3$ for the roughness of the $j$-th $D_\mathrm{mic}$ 
lobe (\autoref{eq:rough_var_two_lobe_fit}).
}

\rev{
\subsection{Correction for Conditioned Angular Domain} \label{subsec:cond_angular_domain}
So far, we have ignored the Heaviside function terms $\mathds{1}(n \cdot \omega_i)$ and $\mathds{1}(n \cdot \omega_o)$ in $\mathcal{D}$.
Alternatively, when integrating
$\mathcal{D}p_{Y_2}(\gamma)$, the angular domain should not be the full sphere $\mathbb{S}^2$, but only a subset conditioned on $\omega_i$ and $\omega_o$:
$\mathcal{X} = \mathcal{X}_{\omega_i, \omega_o} \coloneqq \{ n \in \mathbb{S}^2 | (n \cdot \omega_i) > 0,  (n \cdot \omega_o) > 0 \}$.
Intuitively speaking, the conditioned domain avoids the
incorrect contribution when $\omega_i$ and $\omega_o$ are from different sides of the surface Otherwise, the ABSDF will suffer from leaking.
This complicates the problem because \autoref{eq:normal_var} is not exactly a spherical convolution when the integration domain is $\mathcal{X}$. To keep the
efficient convolution-based solution while addressing the potential leaking, we rewrite \autoref{eq:per_sggx_lobe_conv} and apply the following approximation:
\begin{align} \label{eq:angular_domain}
\begin{split}
        &\int_{\mathcal{X}} D_\mathrm{mic}(\omega_h; n, \alpha) D_{\mathrm{sggx}}(n; R, \bm{\alpha}) \, \D{n} \\
        =\, &D_{\mathrm{sggx}}(\omega; R, \bm{\alpha}_{+}) \frac
            {\int_{\mathcal{X}} D_\mathrm{mic}(\omega_h; n, \alpha) D_{\mathrm{sggx}}(n; R, \bm{\alpha}) \, \D{n}}
            {D_{\mathrm{sggx}}(\omega; R, \bm{\alpha}_{+})} \\
        \approx \, &D_{\mathrm{sggx}}(\omega; R, \bm{\alpha}_{+}) 
        \tcbhighmath[colback=white, colframe=lightteal, boxrule=1pt, boxsep=0pt, top=2pt, bottom=2pt, left=2pt, right=2pt]
            {\frac
            {\int_{\mathcal{X}} D_\mathrm{mic}(\omega_h; n, \alpha) \, \D{n}}
            {\int_{\mathbb{S}^2} D_\mathrm{mic}(\omega_h; n, \alpha) \, \D{n}} \, \Big( \coloneq S(\omega_i, \omega_o; \alpha) \Big)}.\\
\end{split}
\end{align}

Effectively, we replace the $D_\mathrm{sggx}$ term in both the right numerator and denominator with a constant term of 1. We name the numerator of $S$,
$\int_{\mathcal{X}} D_\mathrm{mic}(\omega_h; n, \alpha) \, \D{n}$, the \emph{shape term}, as it reflects the geometric shape of the angular domain $\mathcal{X}$. Computing
the shape term requires integrating a GGX over $\mathcal{X}$, which is a ``spherical lune'' formed by the intersection of two hemispheres. Note that
$\mathcal{X}$ can be decomposed into two spherical triangles, and the problem reduces to integrating a GGX over a spherical triangle, which can be solved
in closed form using Linearly Transformed Cosines (LTC)~\citep{heitz2016real}. We precompute the inverse LTC transform into a 1D table
$T^{-1}_{\mathrm{LTC}} = M_4(\alpha)$ (different from \citet{heitz2016real}, there is only roughness variation in our case).
The denominator of $S$ is the normalization term for a GGX in the spherical domain and can be easily precomputed as another 1D table
$\int_{\mathbb{S}^2} D_\mathrm{mic}(\omega_h; n, \alpha) \, \D{n} = M_5(\alpha)$.

Finally, we utilize the fact that the microfacet shadowing-masking term $G$ is very smooth. Therefore, we simply multiply it to each lobe post convolution.
We arrive at the following expression for the yellow highlighted integral in \autoref{eq:specular_orient_split}:

\begin{align} \label{eq:factor_spec_final}
    \tcbhighmath[colback=white, colframe=lightorange, boxrule=1pt, boxsep=0pt, top=2pt, bottom=2pt, left=2pt, right=2pt]
    {\int_{\Gamma_2} \mathcal{D} p_{Y_2}(\gamma) \,\D{\gamma}}
    &\approx \sum_{i=1}^k \sum_{j=1}^2 w_i m_{ij} 
    \tcbhighmath[colback=white, colframe=lightred, boxrule=1pt, boxsep=0pt, top=2pt, bottom=2pt, left=2pt, right=2pt]
    {D^{ij}_{\mathrm{conv}}(\omega_h)}
    \tcbhighmath[colback=white, colframe=lightteal, boxrule=1pt, boxsep=0pt, top=2pt, bottom=2pt, left=2pt, right=2pt]
    {S^{ij}(\omega_i, \omega_o)}
     G^{ij}(\omega_i, \omega_o), \nonumber \\
    S^{ij}(\omega_i, \omega_o) &= S(\omega_i, \omega_o; \alpha^{ij}), \\ 
    G^{ij}(\omega_i, \omega_o) &= G(\omega_i, \omega_o, R^i, \bm{\alpha}^{ij}). \nonumber
\end{align}
The expression consists of contributions from all convolved NDF lobes (\autoref{eq:normal_var}). Each term is multiplied by its scaling factor to 
account for the conditioned angular domain (\autoref{eq:angular_domain}). \autoref{eq:factor_spec_final} can be evaluated in closed form given the surface NDF 
$p_N(n)$ and the first two moments of roughness $\alpha$.
}

\subsection{Orientation-varying Parameters} \label{subsec:orient}
We have previously assumed that $\beta^c$, $\beta^m$, and $\beta^s$ are independent of orientation in order to perform the split in
\autoref{eq:diffuse_orient_split} and \autoref{eq:specular_orient_split}. To lift this limitation, we notice that both
\autoref{eq:diffuse_split_moments} and \autoref{eq:specular_split_moments} collapse to simple combinations of moments (means and second-order mixed moments) of
$\beta^c$, $\beta^m$, and $\beta^s$. Therefore, we extend our formulation by augmenting the moments to be orientation-varying. As an example, when calculating the
mean of a parameter $\mathbb{E}[\beta]$ for a particular direction $\omega$, each sample on the surfaces should be weighted to reflect its influence on
$\omega$. In other words, each sample is ``splatted'' to the spherical domain with a spherical function $s(\omega)$ as the kernel, followed by normalization.
The \emph{directional moments} can thus be defined as
\begin{equation} \label{eq:directional_moments}
    \mathbb{E}_s[\beta] \Bigr\rvert_{\omega} = \frac{\int_A \beta_x s(T_x\omega) \,\D{x}}{\int_A s(T_x\omega) \,\D{x}}, \,\,
    \mathbb{E}_s[\beta\beta'] \Bigr\rvert_{\omega} = \frac{\int_A \beta_x \beta'_x s(T_x\omega) \,\D{x}}{\int_A s(T_x\omega) \,\D{x}},
\end{equation}
where $T_x$ is the local transform at $x$. The surface NDF is implicitly accounted for by transforming $\omega$ into the local coordinate system. The kernel
$s(\omega)$
is different for each component of the ABSDF: For the specular component, it is the microfacet distribution; for the diffuse component, it is the SG in
\autoref{eq:diffuse_sg}. Finally, we query the directional moments at $\omega_h$ when evaluating \autoref{eq:diffuse_split_moments} and
\autoref{eq:specular_split_moments}.

Now that the moments become orientation-varying, we can no longer store them as simple scalars. In practice, we find that it is usually sufficient to coarsely
partition the spherical domain (e.g., $3\times3$) because the angular frequency usually decreases as the scale of aggregation becomes larger. Each surface sample
can then be splatted to the partition during precomputation (see \autoref{subsec:precompute}).

\rev{
\subsection{Summary and Validation} \label{subsec:absdf_summary_and_validation}
The derivation of our factorized ABSDF is complete at this point. We conclude this section with a brief summary of the complete model with a 
schematic diagram \autoref{fig:absdf_diagram}. The ABSDF (\autoref{eq:absdf_distrib}) with the base material (\autoref{eq:base_material}) is the 
linear sum of a diffuse compoennt (\autoref{eq:diffuse_orient_split}) and a specular component (\autoref{eq:factor_specular}). The diffuse component is 
decomposed to a moment term (\autoref{eq:diffuse_split_moments}) and convolution term (\autoref{eq:diffuse_sg}). The specular component is decomposed 
similarly (\autoref{eq:specular_split_moments}), but the convolution (\autoref{eq:normal_var}) needs to be performed with care given to the base distribution 
(\autoref{eq:rough_var}) and the domain (\autoref{eq:angular_domain}). Finally, the moments can be augmented to be orientation-varying 
(\autoref{eq:directional_moments}). A total of 5 small, scene-independent precomputed tables are utilized in different components as highlighted in 
\autoref{fig:absdf_diagram}.
\begin{figure}[h]
    \addtolength{\tabcolsep}{-4pt}
    \centering
    \begin{tabular}{c}
        \includegraphics[width=\linewidth]{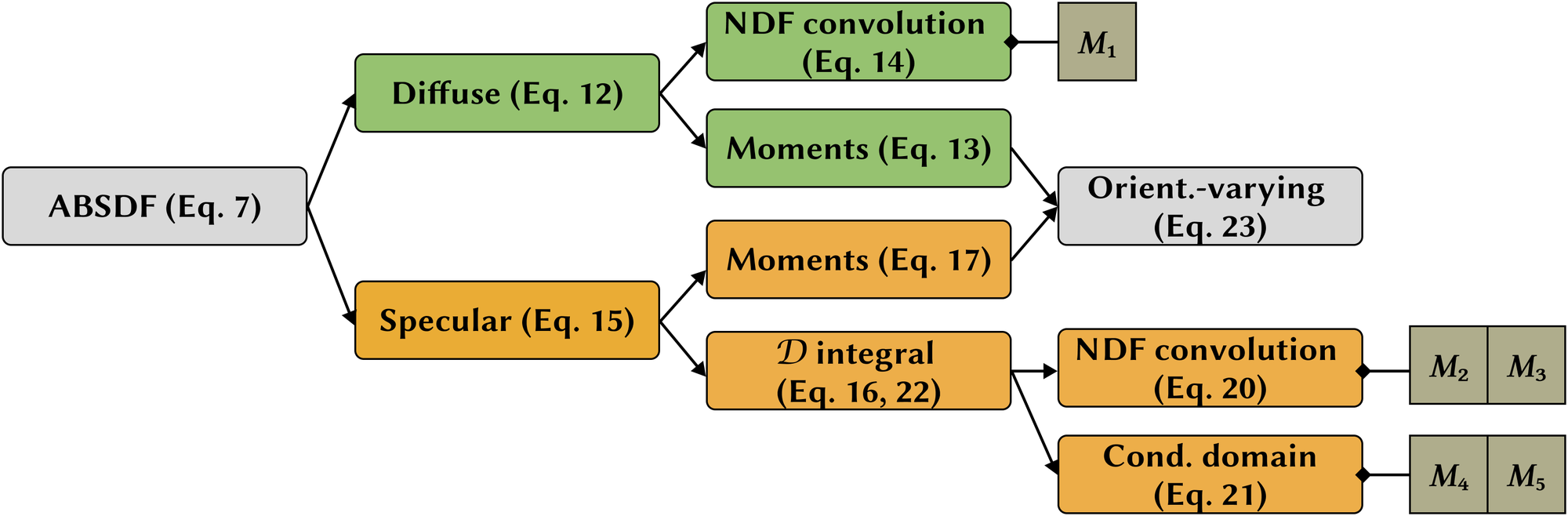}
    \end{tabular}
    \caption{\label{fig:absdf_diagram}
        Schematic diagram of our full factorized ABSDF model. The usages of precomputed tables $M_1, ..., M_5$ are highlighted.}
\end{figure}
}

\begin{figure*}[tb]
	\newlength{\lenABSDFPlot}
	\setlength{\lenABSDFPlot}{0.95in}
    \addtolength{\tabcolsep}{-4pt}
    \renewcommand{\arraystretch}{0.5}
    \centering
    \begin{tabular}{cccccccc}
        &
        \textsf{(a) Setup} & \textsf{(b) Reference} &  \textsf{(c) Ours} & \textsf{(d) Weier~\shortcite{weier2023neural}} & \textsf{(e) Reference} &
        \textsf{(f) Ours} & \textsf{(g) Weier~\shortcite{weier2023neural}} \\
        \raisebox{25pt}{\rotatebox{90}{\emph{Helmet}}}
        &
        \frame{\includegraphics[height=\lenABSDFPlot]{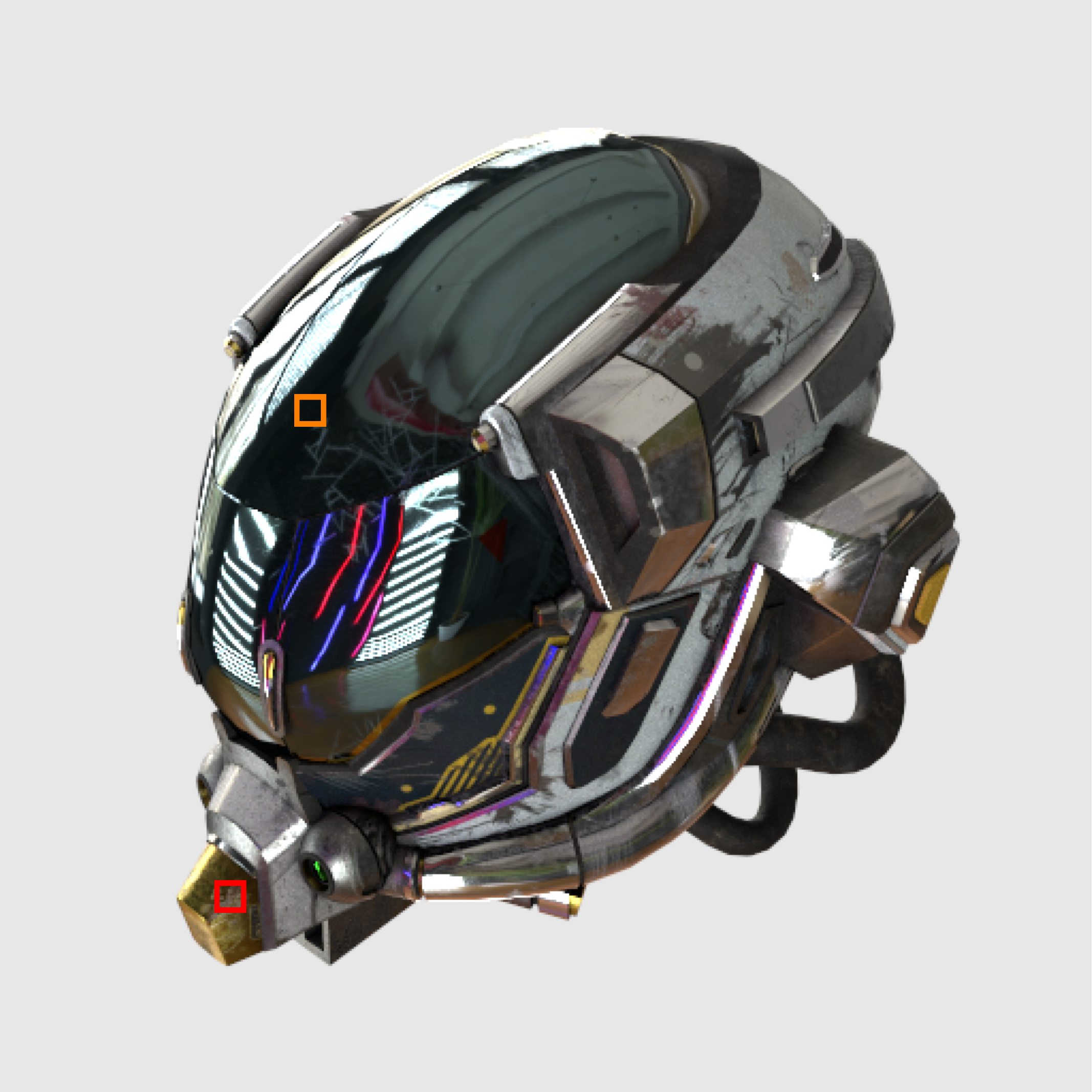}}
        &
        \begin{adjustbox}{margin=0.4pt, bgcolor=red}{\includegraphics[height=\lenABSDFPlot]{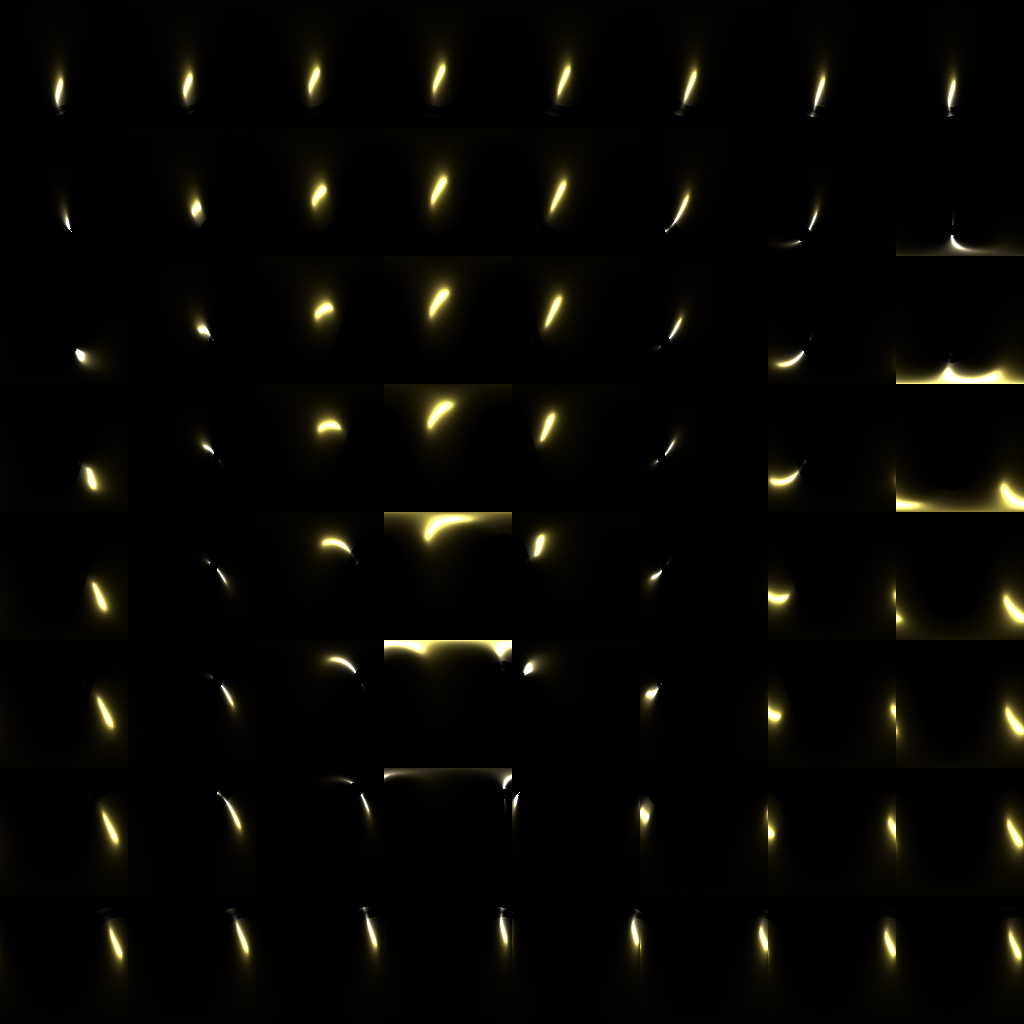}}\end{adjustbox}
        &
        \begin{adjustbox}{margin=0.4pt, bgcolor=red}{\includegraphics[height=\lenABSDFPlot]{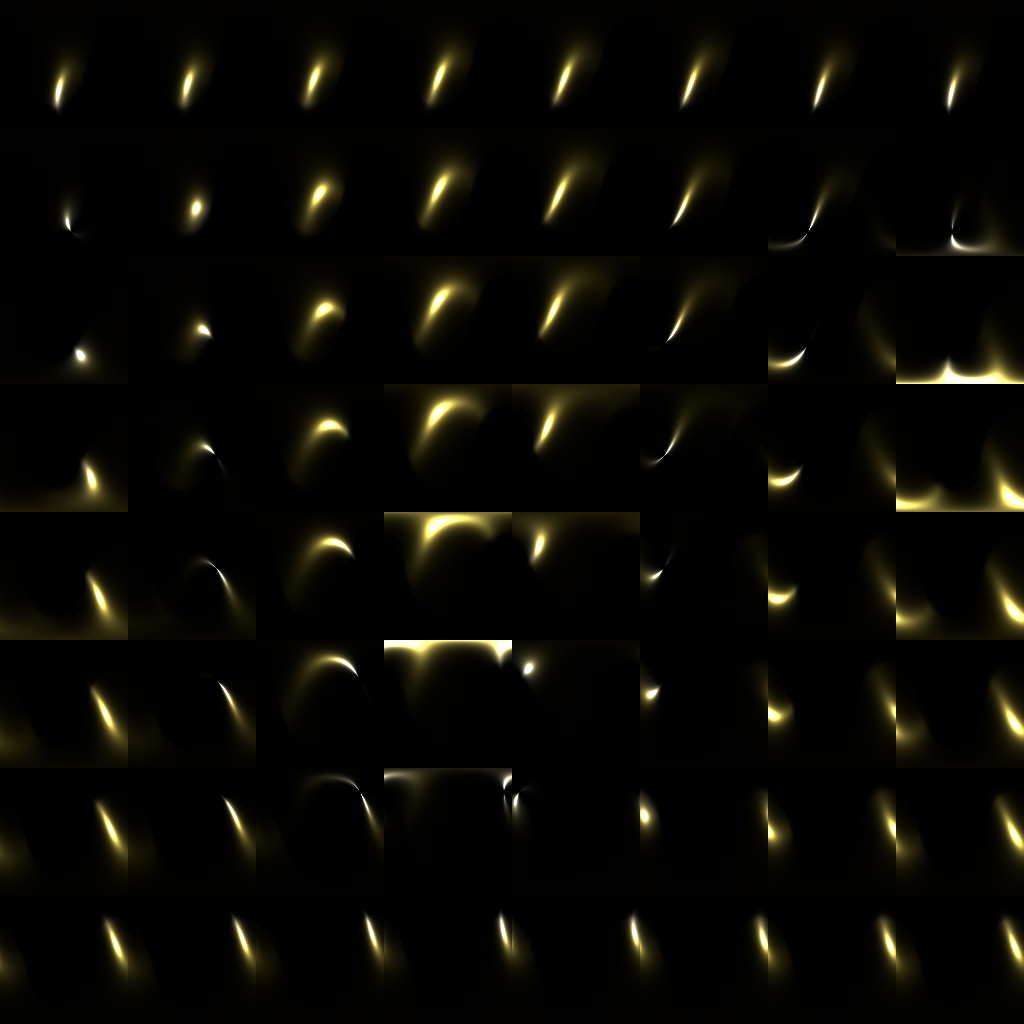}}\end{adjustbox}
        &
        \begin{adjustbox}{margin=0.4pt, bgcolor=red}{\includegraphics[height=\lenABSDFPlot]{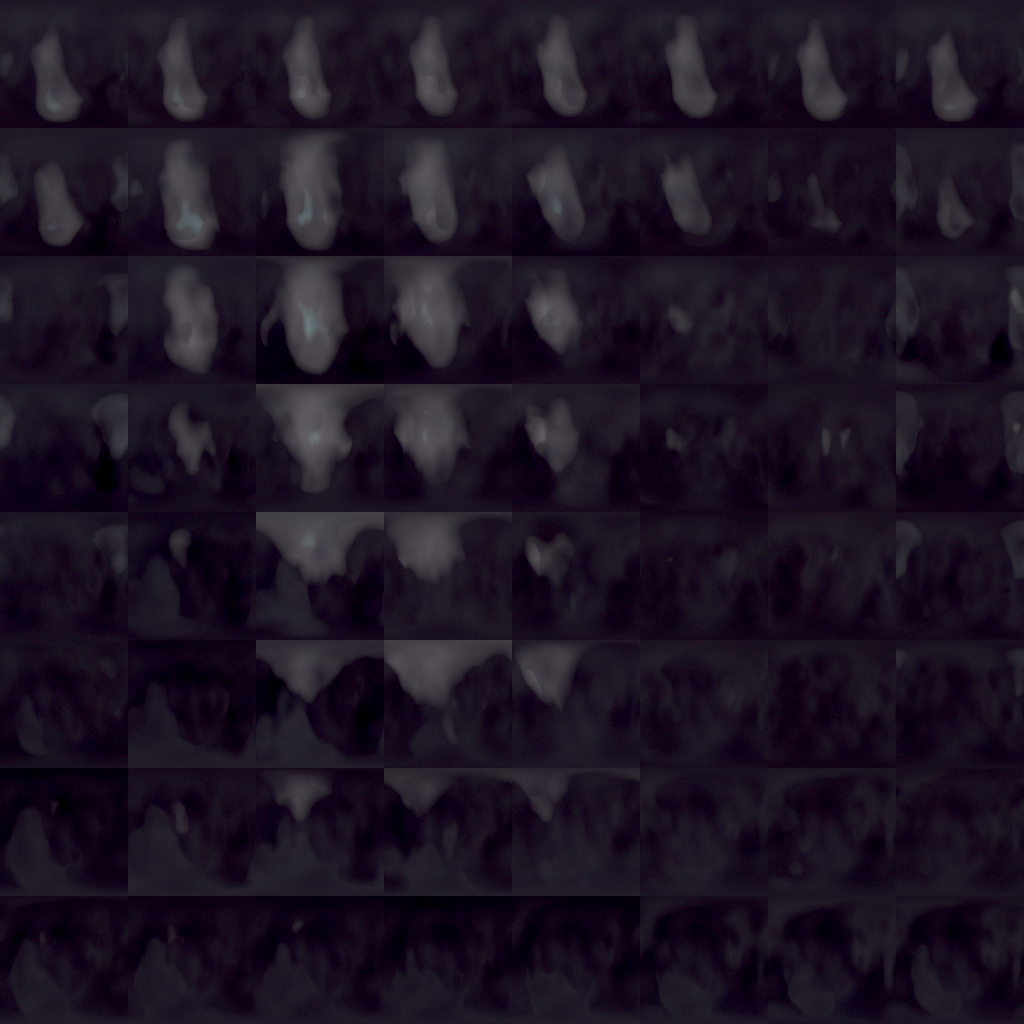}}\end{adjustbox}
        &
        \begin{adjustbox}{margin=0.4pt, bgcolor=orange}{\includegraphics[height=\lenABSDFPlot]{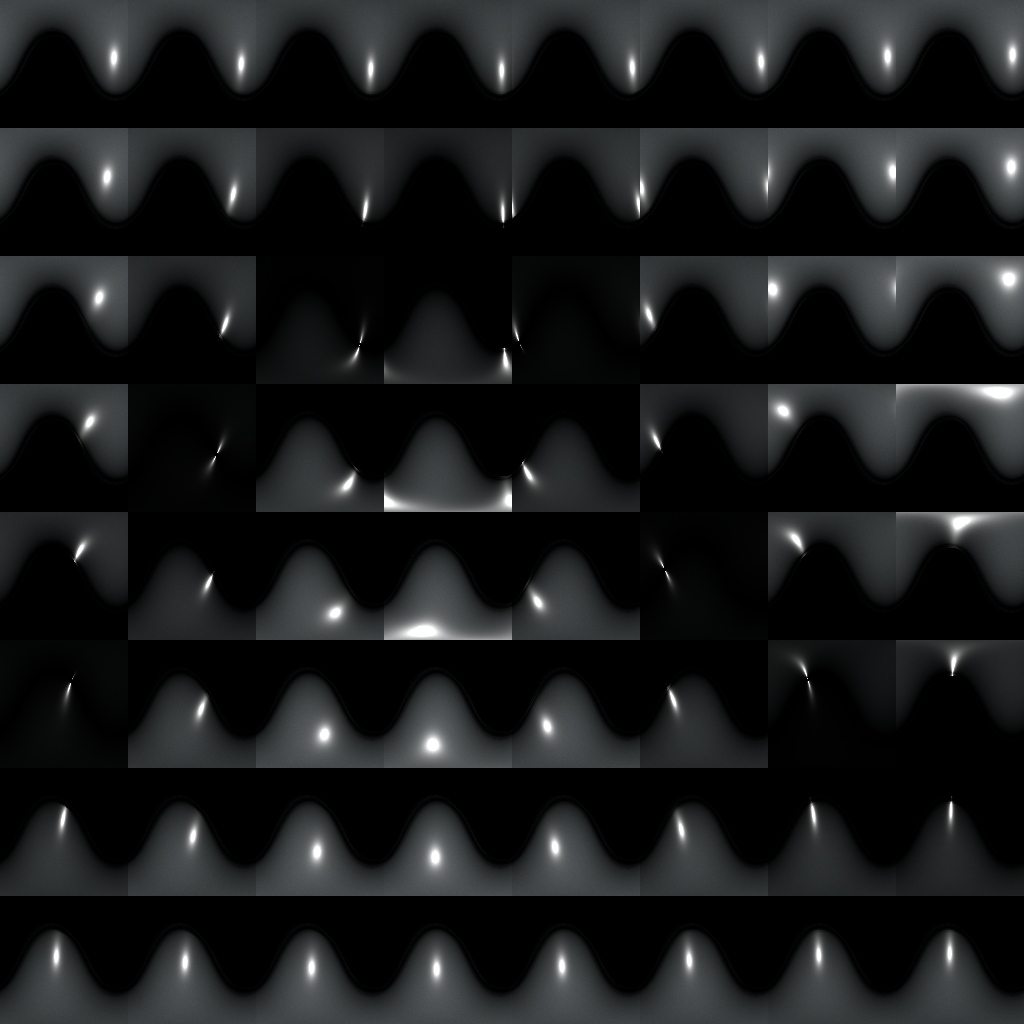}}\end{adjustbox}
        &
        \begin{adjustbox}{margin=0.4pt, bgcolor=orange}{\includegraphics[height=\lenABSDFPlot]{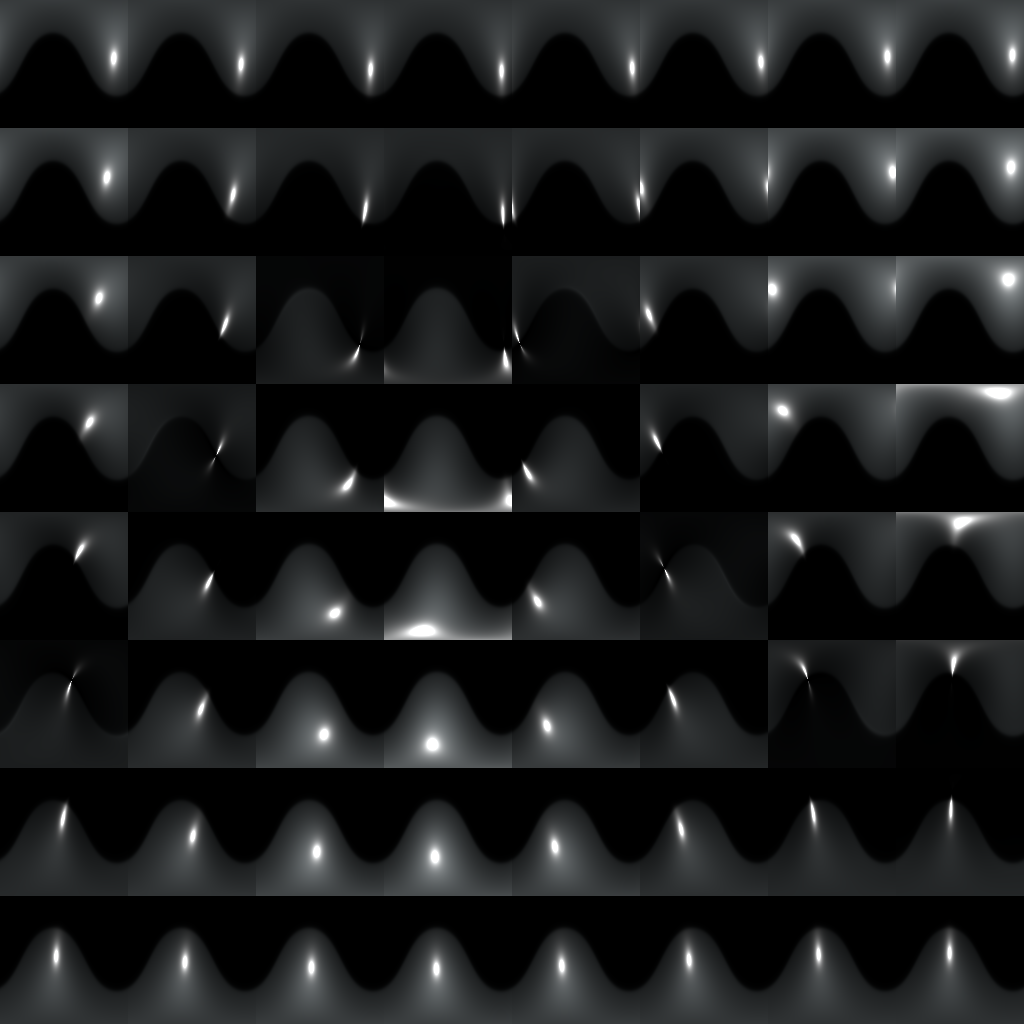}}\end{adjustbox}
        &
        \begin{adjustbox}{margin=0.4pt, bgcolor=orange}{\includegraphics[height=\lenABSDFPlot]{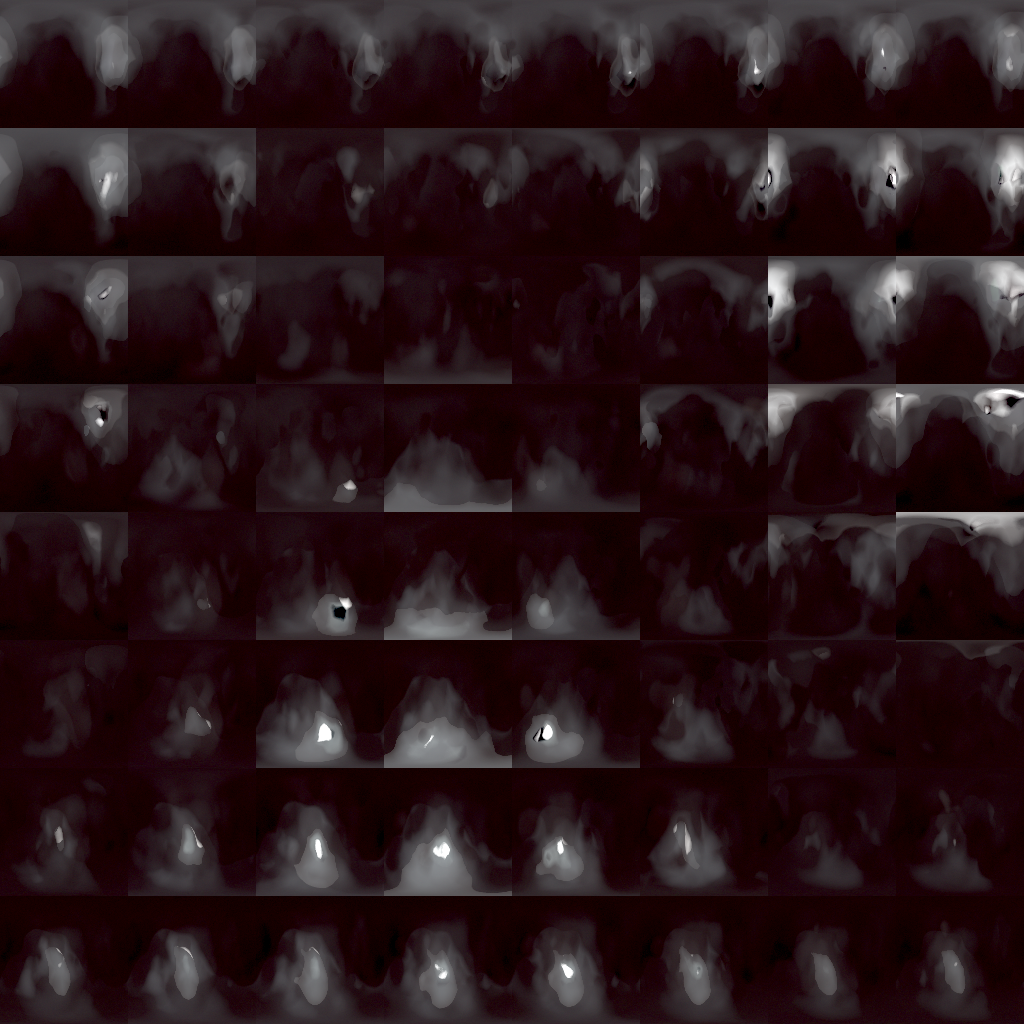}}\end{adjustbox}
        \\
        & \textsf{PSNR (dB):} & & \textbf{39.63} & 33.15 & & \textbf{55.48} & 45.27
        \\
        \raisebox{25pt}{\rotatebox{90}{\emph{Palm}}}
        &
        \frame{\includegraphics[height=\lenABSDFPlot]{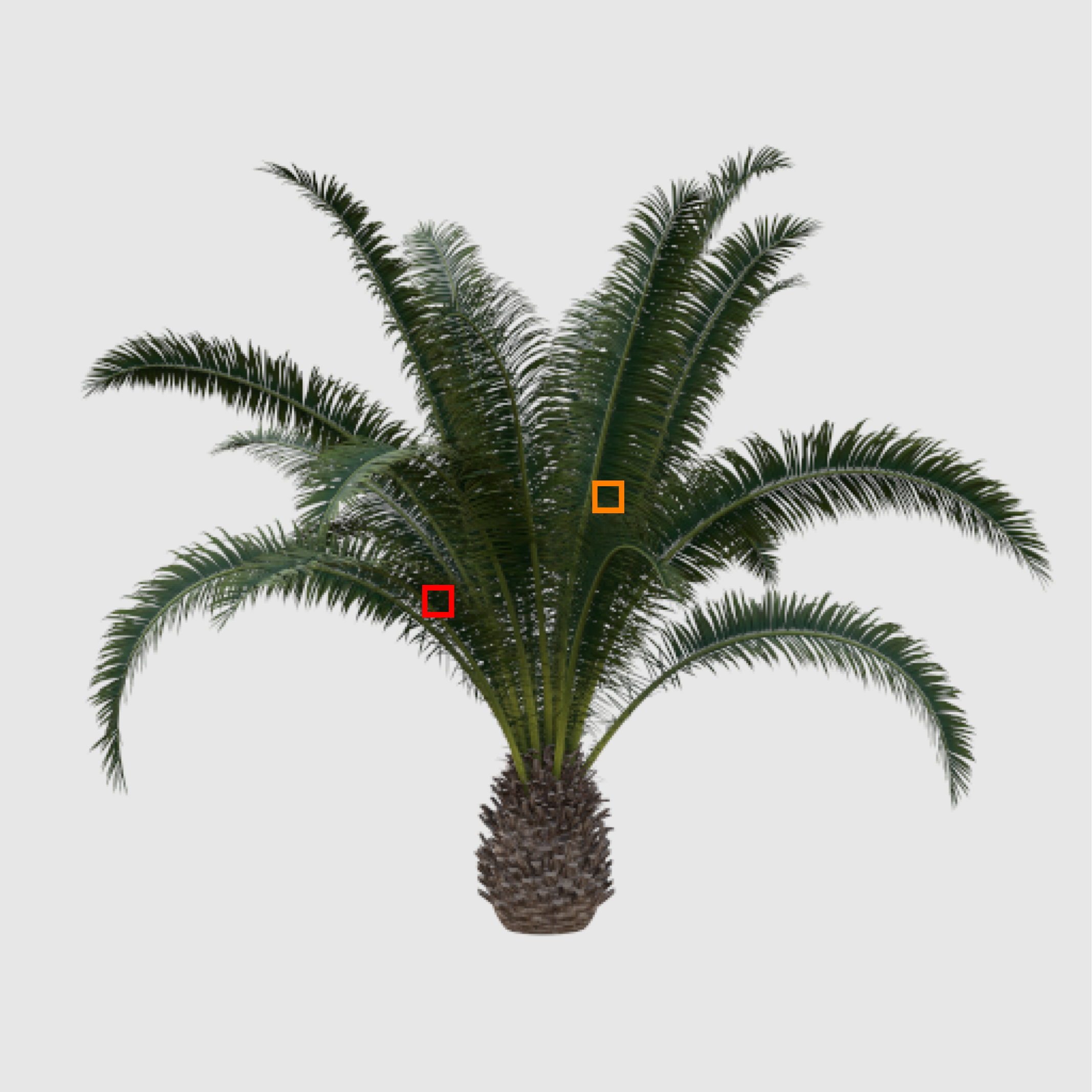}}
        &
        \begin{adjustbox}{margin=0.4pt, bgcolor=red}{\includegraphics[height=\lenABSDFPlot]{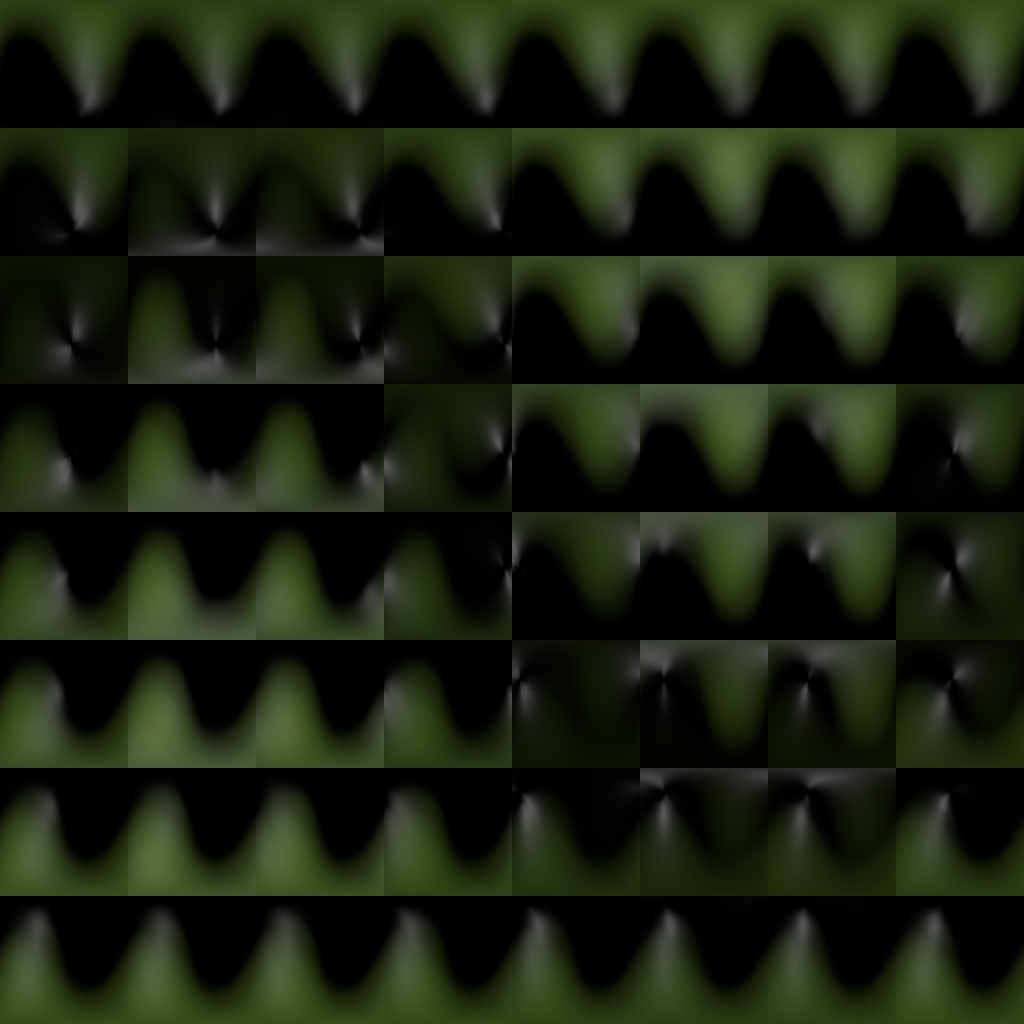}}\end{adjustbox}
        &
        \begin{adjustbox}{margin=0.4pt, bgcolor=red}{\includegraphics[height=\lenABSDFPlot]{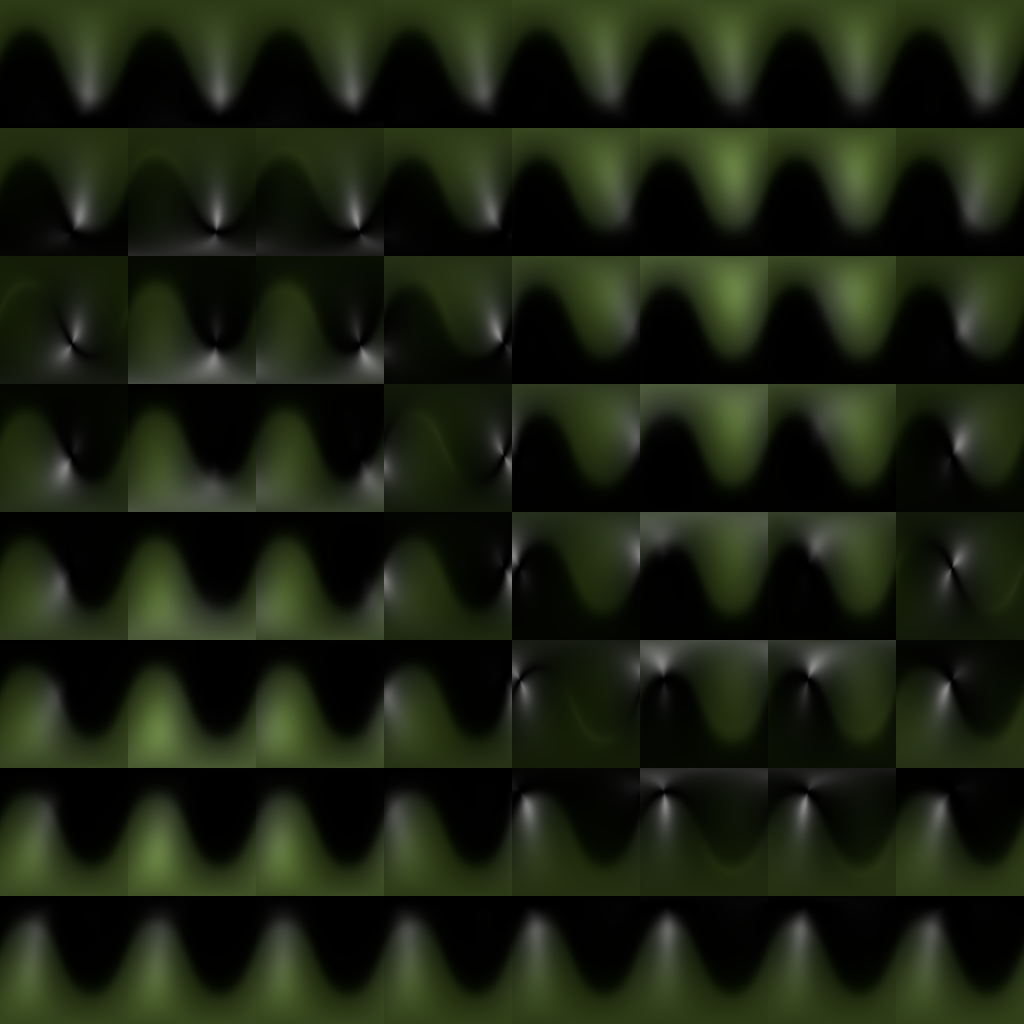}}\end{adjustbox}
        &
        \begin{adjustbox}{margin=0.4pt, bgcolor=red}{\includegraphics[height=\lenABSDFPlot]{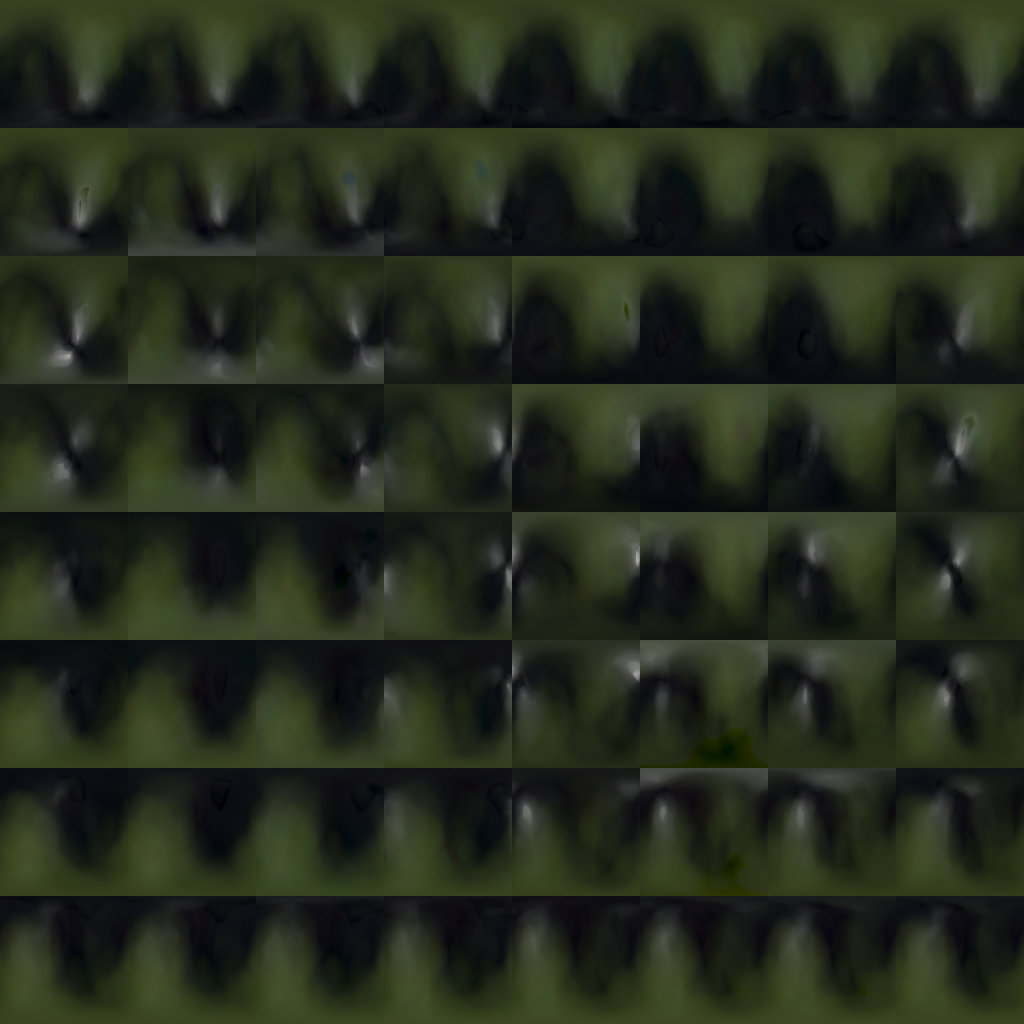}}\end{adjustbox}
        &
        \begin{adjustbox}{margin=0.4pt, bgcolor=orange}{\includegraphics[height=\lenABSDFPlot]{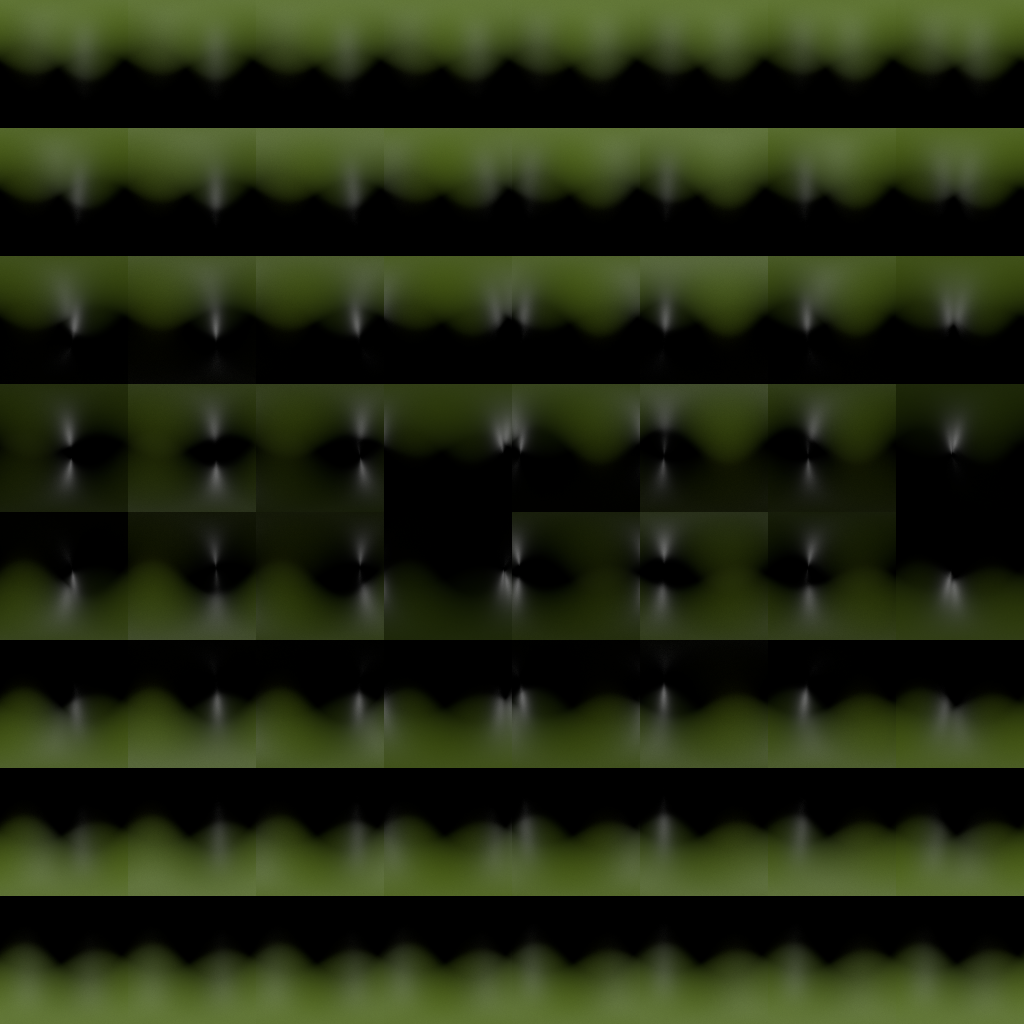}}\end{adjustbox}
        &
        \begin{adjustbox}{margin=0.4pt, bgcolor=orange}{\includegraphics[height=\lenABSDFPlot]{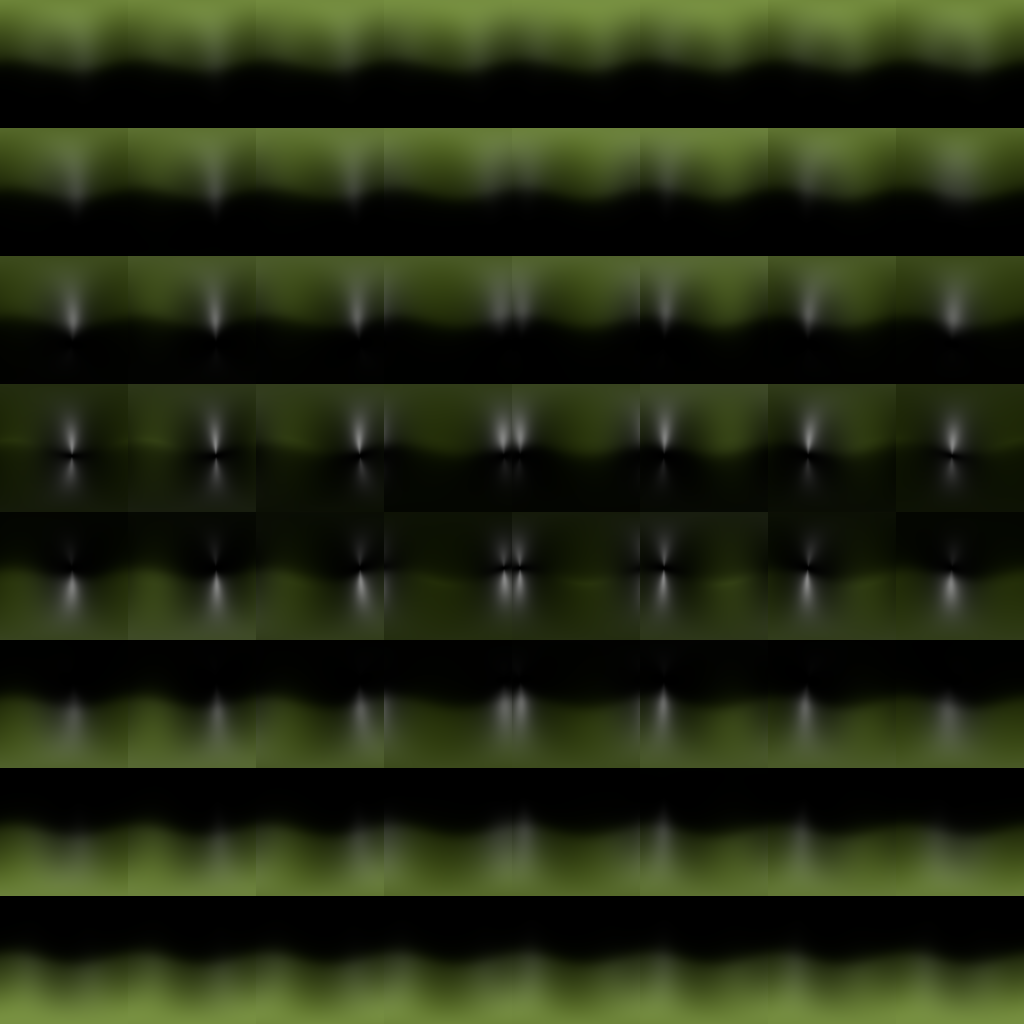}}\end{adjustbox}
        &
        \begin{adjustbox}{margin=0.4pt, bgcolor=orange}{\includegraphics[height=\lenABSDFPlot]{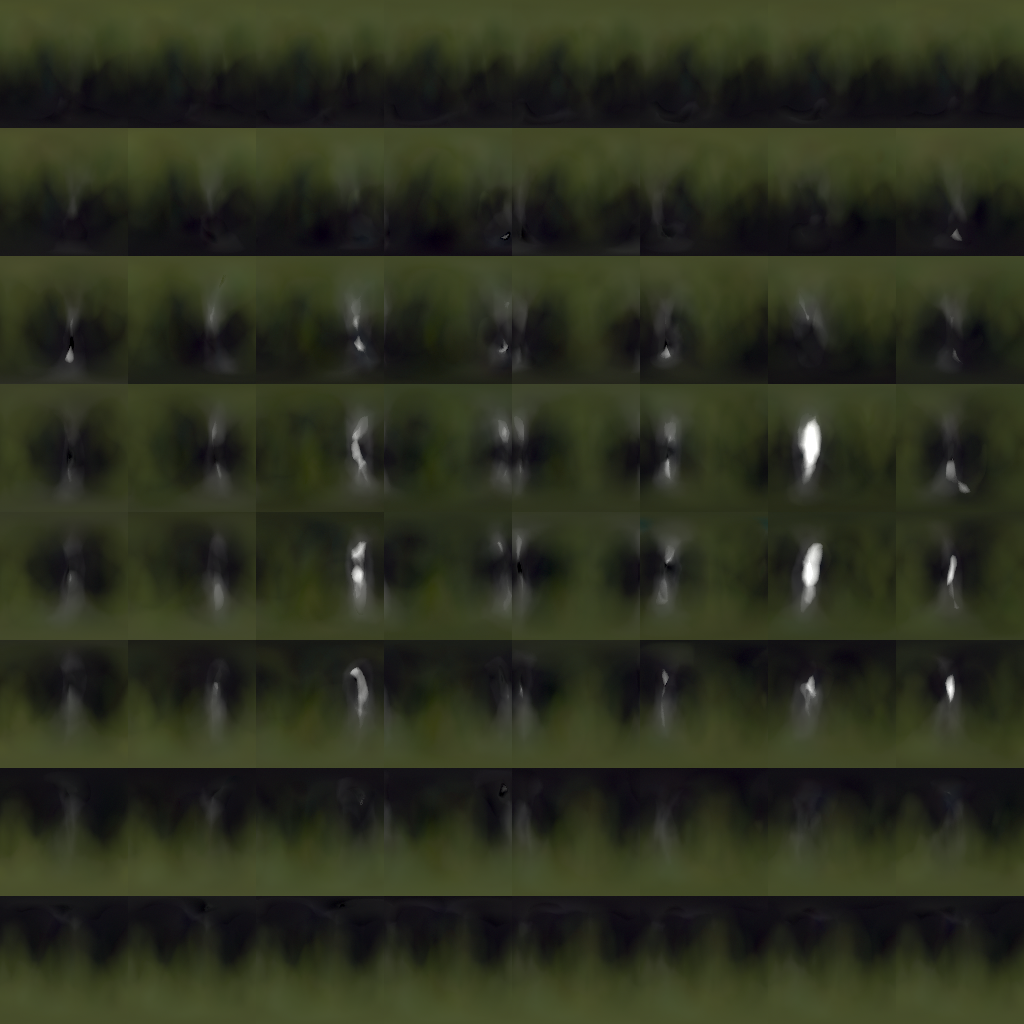}}\end{adjustbox}
        \\
        & \textsf{PSNR (dB):} & & \textbf{59.31} & 46.94 & & \textbf{56.08} & 38.82
        \\
    \end{tabular}
    \caption{\label{fig:absdf_plot}
        We select voxels from each scene, highlighted in (a), and compare the ground truth ABSDFs (b)/(e) to our factored ABSDFs (c)/(f) and the appearance network fitted
        results (d)/(g). Each plot contains $8 \times 8$ 2D outgoing slices in the lat-long coordinate system with different incident directions. Our results
        achieve better accuracy both qualitatively and quantitatively with lower RMSE. We encourage readers to zoom in for better comparison.
        Exposure is adjusted for clarity.
        }
\end{figure*}

In \autoref{fig:absdf_plot}, we compare our factored ABSDF results to the ground truth and further, the fitting results of the recent neural solution
presented by \citet{weier2023neural}. Their network (named the ``appearance network'') consists of a multi-resolution hash grid encoding
~\citep{muller2022instant} for spatial coordinates, spherical harmonics encodings for incident and outgoing directions, and a multi-layer perceptron to produce
the final output. We use the exact same architecture and hyperparameters as \citet{weier2023neural} with 8 features per level and 8 degrees of
spherical harmonics. We follow a similar training procedure by feeding a large batch of stochastic queries to the network each iteration and optimizing for
relative $L_2$ loss.

The \emph{Helmet} example presents a particularly challenging case with highly glossy anisotropic highlights, which we are able to reconstruct well. On the other
hand, the appearance network suffers from various artifacts, including color shift, ``blotchiness'', mode collapse, and perhaps most significantly, loss of highlights.
This could
be due to not enough features to capture the spatial variety and that the spherical harmonics encoding cannot handle high frequency signals. The network could
potentially benefit from more features and a better directional encoding, but will likely become much larger.
For the more diffuse \emph{Palm} example, we are able to capture the dual-mode shape reasonably well thanks to the multi-lobe surface NDF representation.
The appearance network performs relatively better on this example but still produces worse accuracy than ours.

\begin{figure}[t]
	\newlength{\heightsmall}
    \setlength{\heightsmall}{0.5in}
	\newlength{\heightbig}
    \setlength{\heightbig}{1.59in}
    \addtolength{\tabcolsep}{-4pt}
    \centering
    \begin{tabular}{ccccccc}
        \multicolumn{3}{c}{}
        &
        \includegraphics[height=\heightsmall]{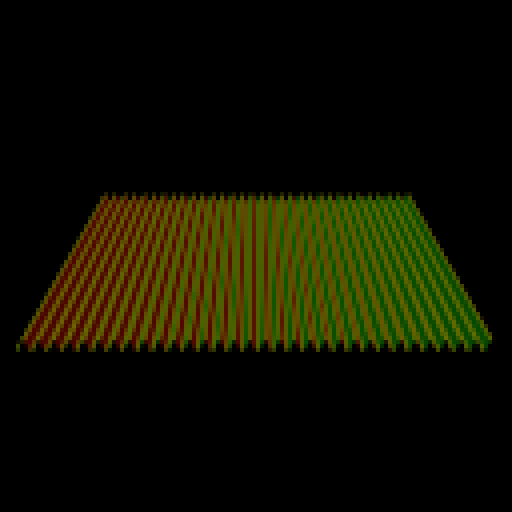}
        &
        \includegraphics[height=\heightsmall]{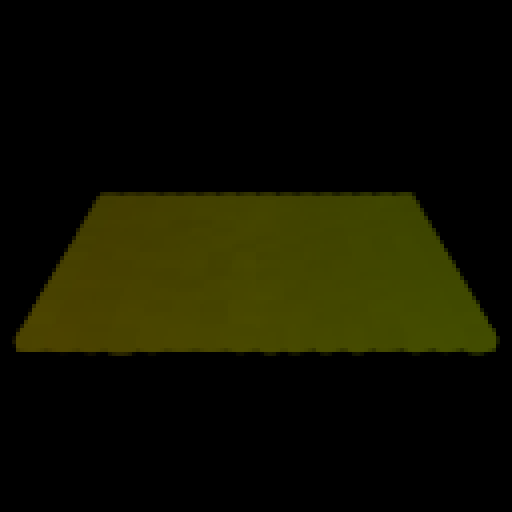}
        &
        \includegraphics[height=\heightsmall]{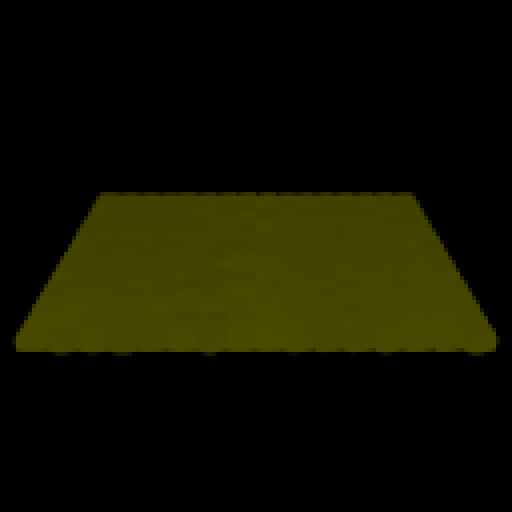}
        &
        \raisebox{35pt}{\rotatebox{-90}{\textsf{View 1}}}
        \\
        \multicolumn{3}{c}{}
        &
        \includegraphics[height=\heightsmall]{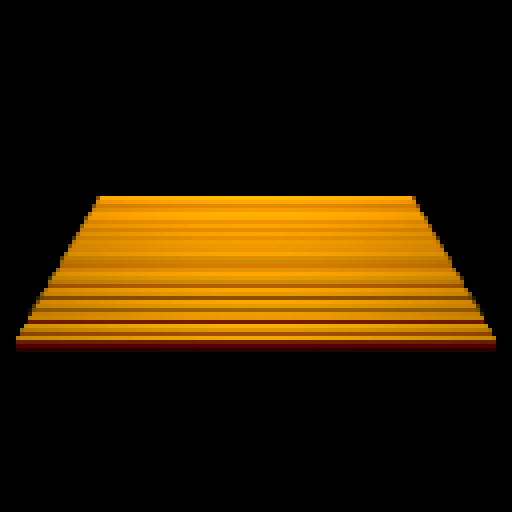}
        &
        \includegraphics[height=\heightsmall]{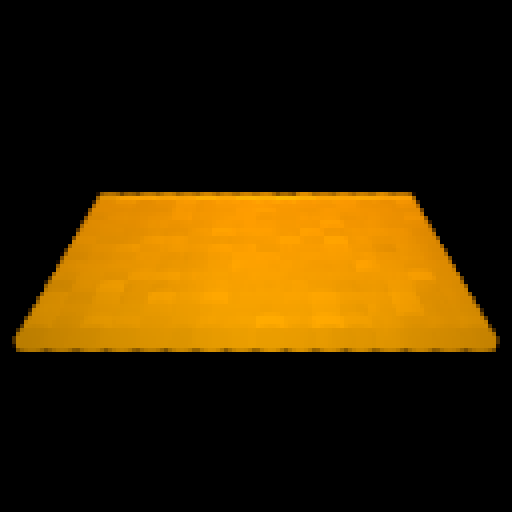}
        &
        \includegraphics[height=\heightsmall]{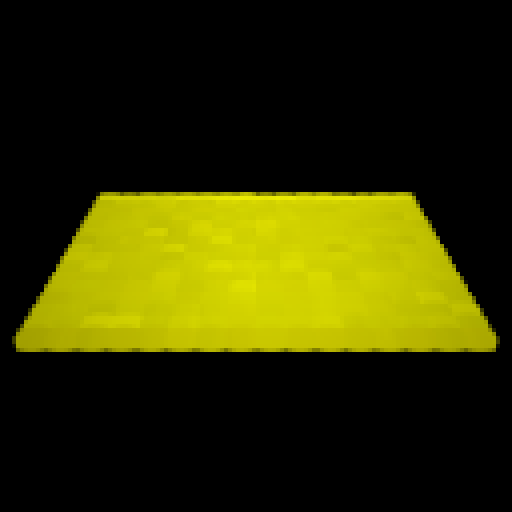}
        &
        \raisebox{35pt}{\rotatebox{-90}{\textsf{View 2}}}
        \\
        \multicolumn{3}{c}{\multirow[t]{3}{*}{\includegraphics[height=\heightbig]{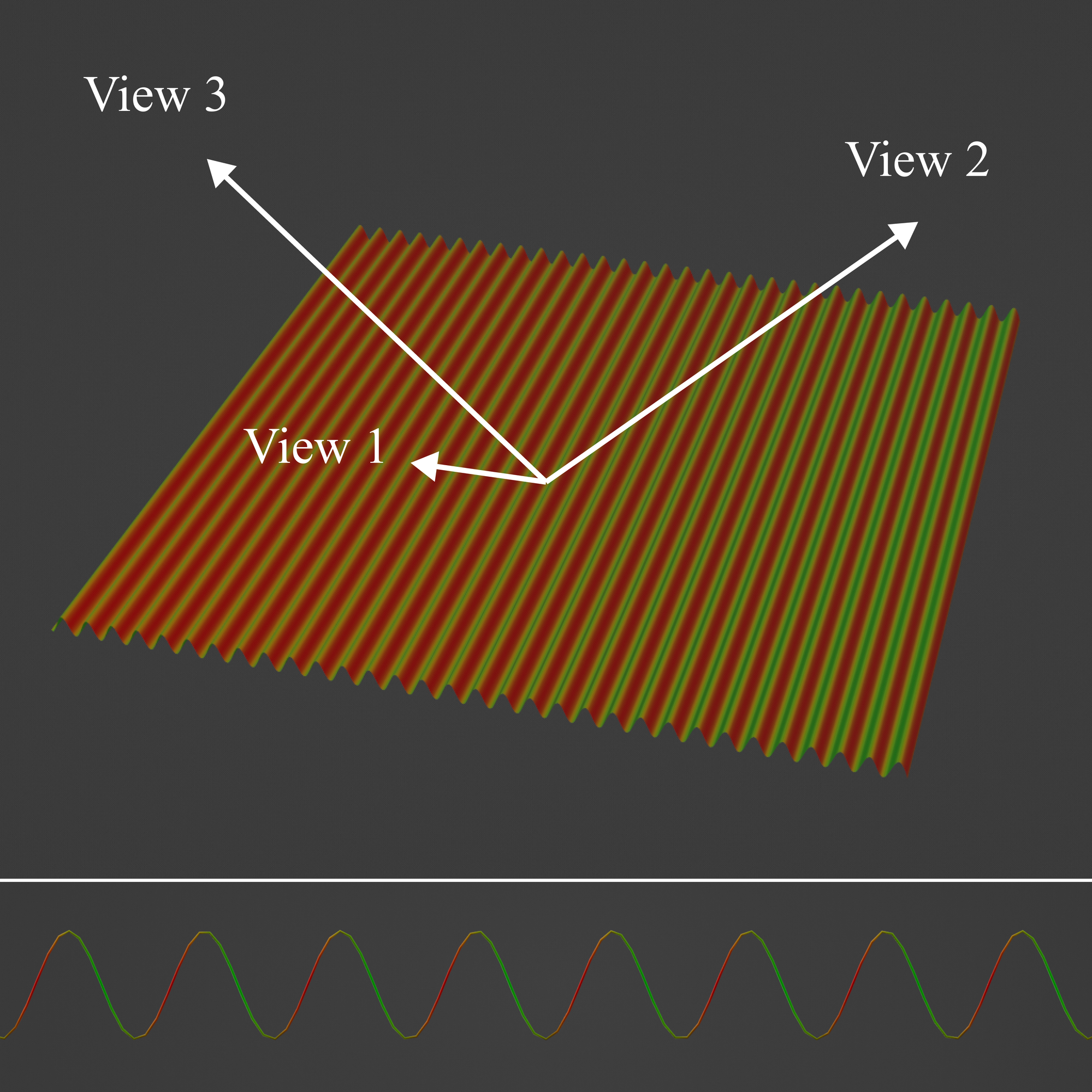}}}
        &
        \includegraphics[height=\heightsmall]{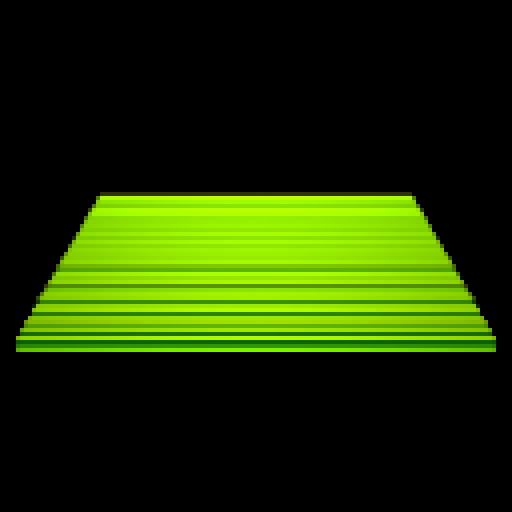}
        &
        \includegraphics[height=\heightsmall]{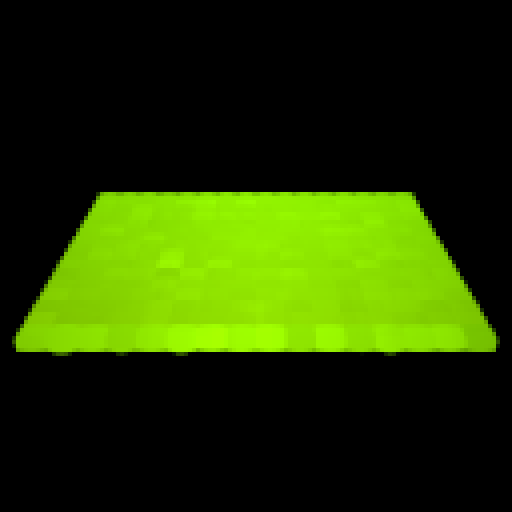}
        &
        \includegraphics[height=\heightsmall]{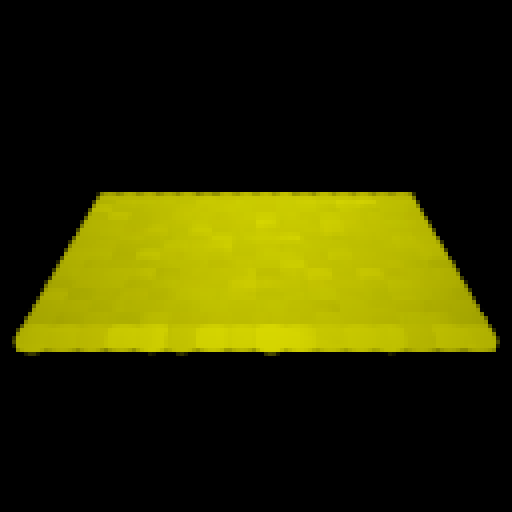}
        &
        \raisebox{35pt}{\rotatebox{-90}{\textsf{View 3}}}
        \\
        \multicolumn{3}{c}{\textsf{(a) Setup}} & \textsf{(b) Ref.} & (c) \textsf{Ours} & \tabincell{c}{\textsf{(d) Na\"ive} } &
    \end{tabular}
    \caption{\label{fig:orientation_varying_validation}
        When a scene has orientation-varying material parameters (a), our method (c) captures the view dependency and matches the reference (b), while ignoring
        it leads to incorrect results (d).}
\end{figure}

In \autoref{fig:orientation_varying_validation}, we demonstrate the necessity of supporting orientation-varying material parameters and the effectiveness of
our method. We aggregate a displaced surface with basecolor varying based on orientation and render it from 3 views. The surface exhibits drastically
different appearance from different views and our method correctly captures this view-dependent appearance.

\section{Correlation-aware Appearance Accumulation} \label{sec:accumulation}
So far, we have presented the definition and an efficient solution for the aggregated appearance of a voxel. In order to render a scene aggregate, we need to
accumulate the outgoing radiance contributions of multiple voxels for each pixel. Intuitively speaking, voxel accumulation requires two pieces of information:
(1) sub-voxel geometry distribution, and (2) the inter-occlusion across voxels. A core challenge arises from the
fact that spatial correlation generally exists in a 3D scene made of surfaces. In the following, we motivate the importance of preserving spatial correlation
and discuss our design to model the necessary information for voxel accumulation. With these components, we formulate the process of accumulating voxel
contributions into pixel intensity.

\subsection{Truncated Ellipsoid Primitive} \label{subsec:trunc_ellipsoid}
Most volumetric representations produce cube-shaped voxels. For each voxel, it is implicitly assumed that surfaces behave like uncorrelated particles and are
independently and uniformly distributed inside. This has been the \emph{de facto} choice and one may argue that with sufficient spatial subdivision, the raw
resolution could compensate for the simplicity of this assumption. However, it is important to realize that scene aggregation is more than an image-space signal
reconstruction problem. The Nyquist-Shannon sampling rate is thus not twice the image resolution, but twice the highest frequency of geometric details, which
can be much higher (if not unbounded) for a scene consisting of hard surfaces. Because it is infeasible to reach such a sampling rate, ignoring
the spatial correlation inside each voxel does negatively impact appearance. We demonstrate this by a minimal example in the following.

\begin{figure}[h]
	\newlength{\lenValidateEllipsoid}
	\setlength{\lenValidateEllipsoid}{1.0in}
    \addtolength{\tabcolsep}{-4pt}
    \renewcommand{\arraystretch}{0.5}
    \centering
    \begin{tabular}{cccc}
        \raisebox{5pt}{\rotatebox{90}{\textsf{Illustration}}}
        &
        \includegraphics[width=\lenValidateEllipsoid]{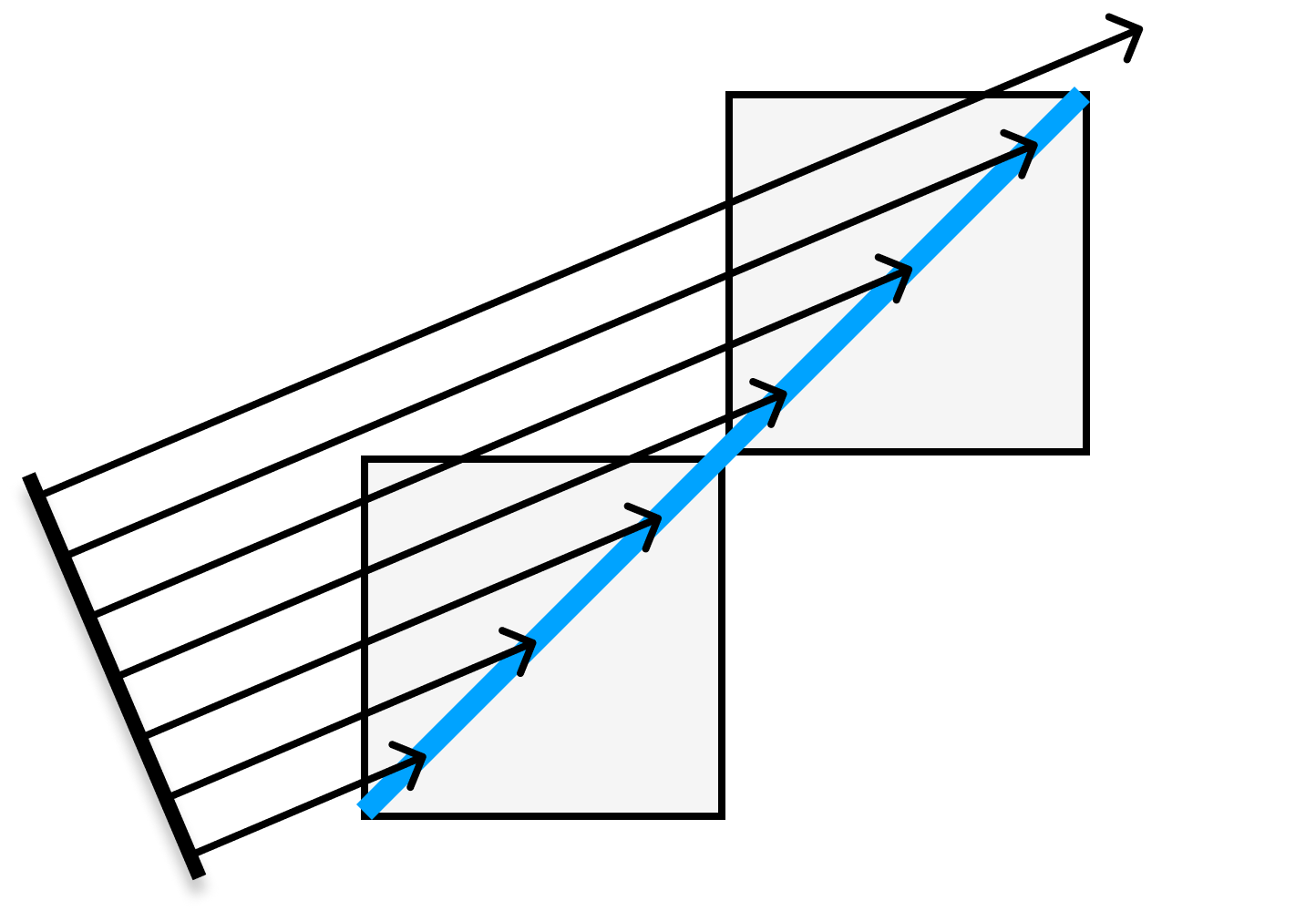}
        &
        \includegraphics[width=\lenValidateEllipsoid]{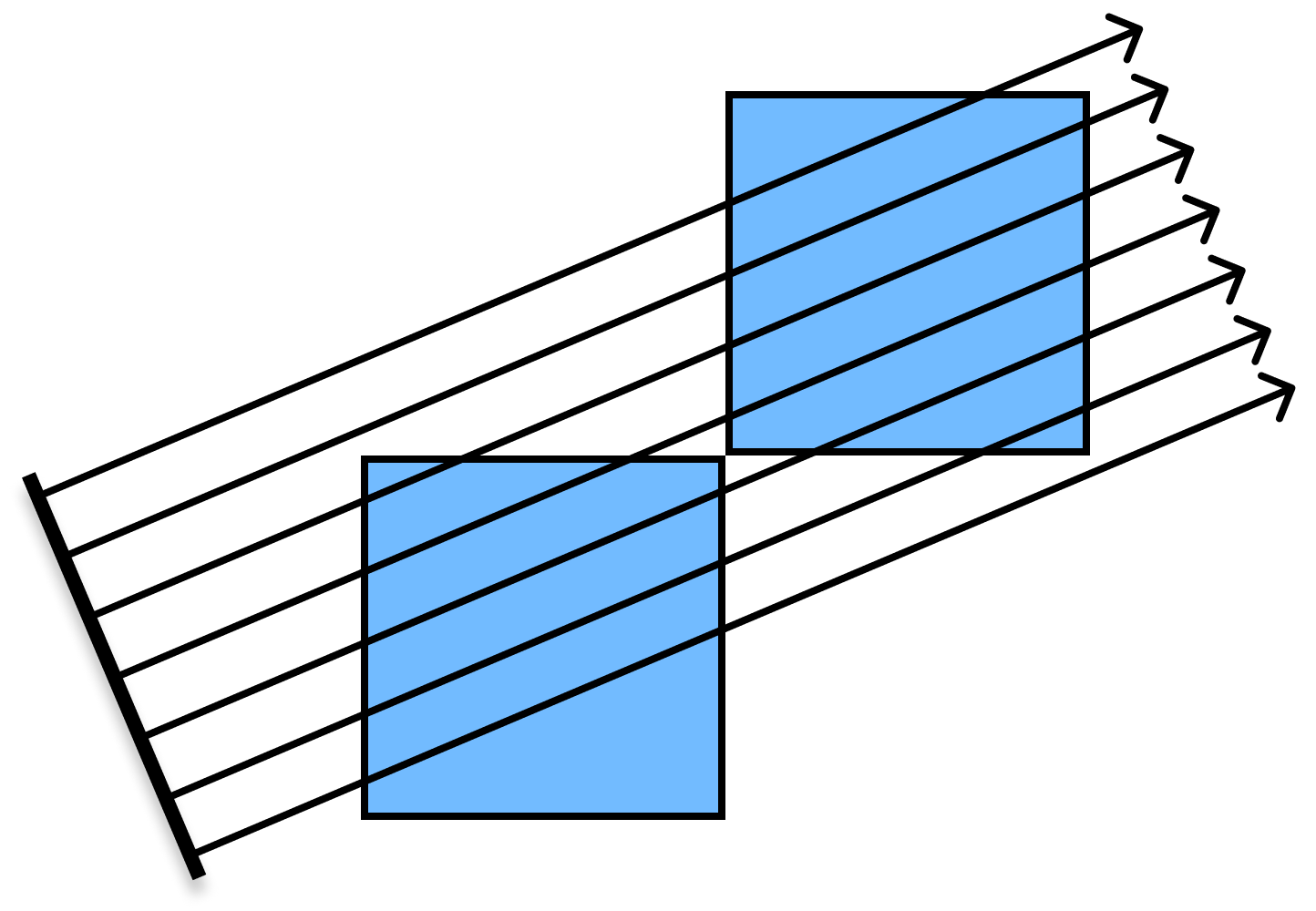}
        &
        \includegraphics[width=\lenValidateEllipsoid]{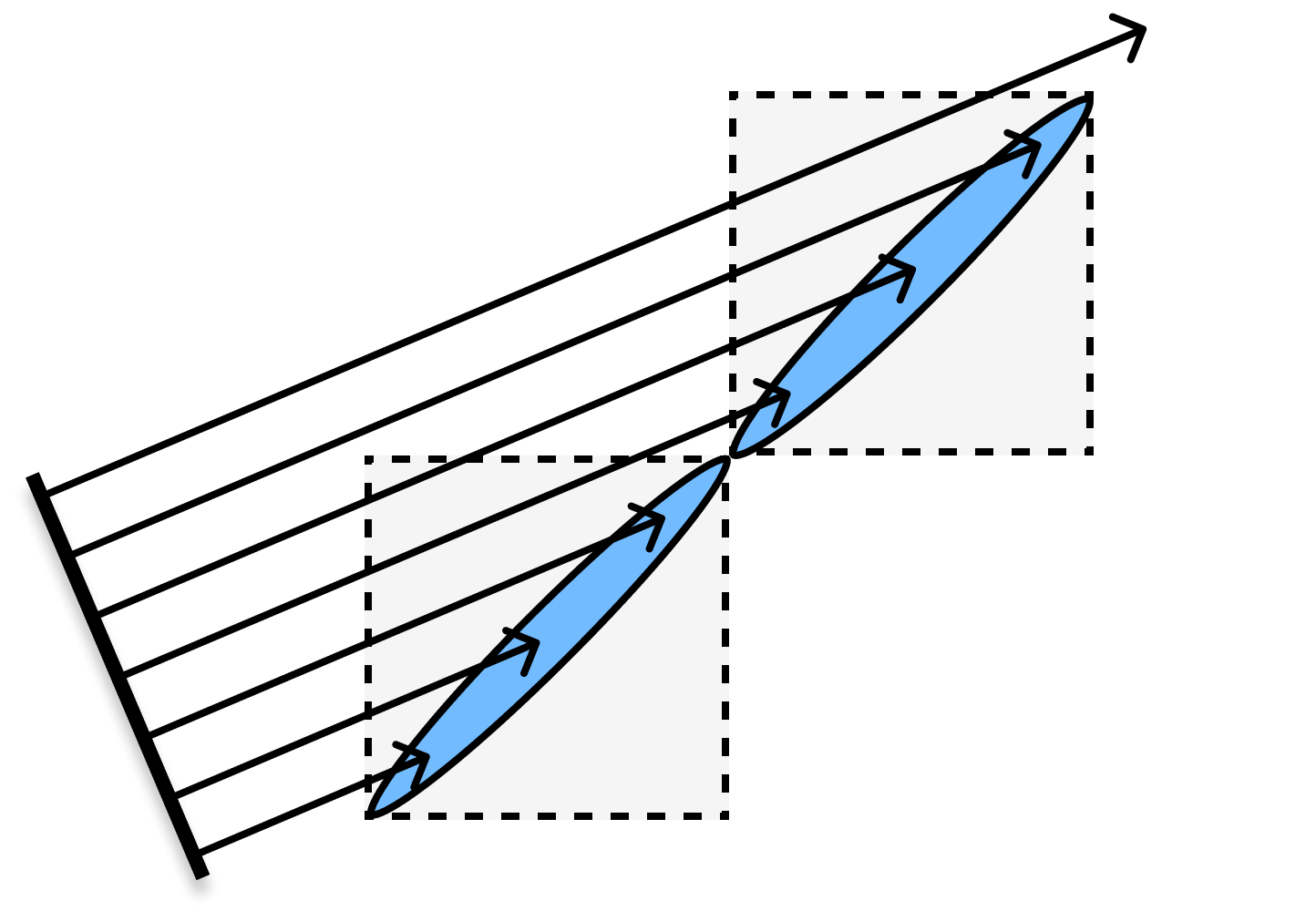}
        \\
        \raisebox{25pt}{\rotatebox{90}{\textsf{Render}}}
        &
        \frame{\includegraphics[height=\lenValidateEllipsoid]{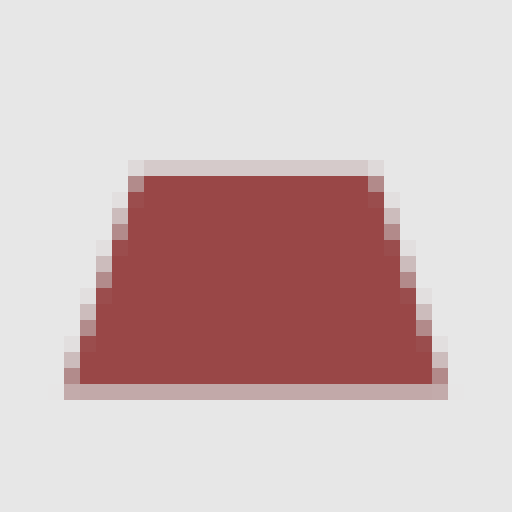}}
        &
        \frame{\includegraphics[height=\lenValidateEllipsoid]{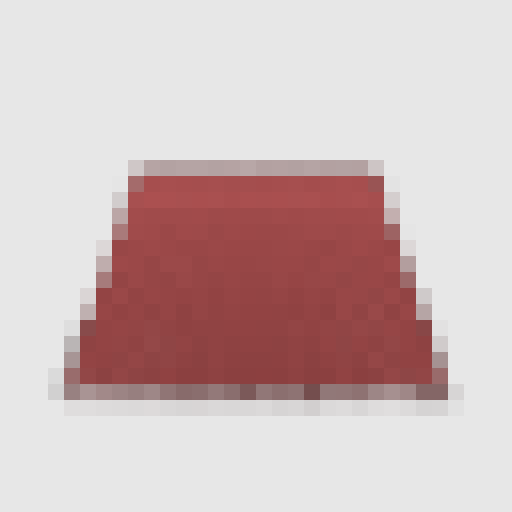}}
        &
        \frame{\includegraphics[height=\lenValidateEllipsoid]{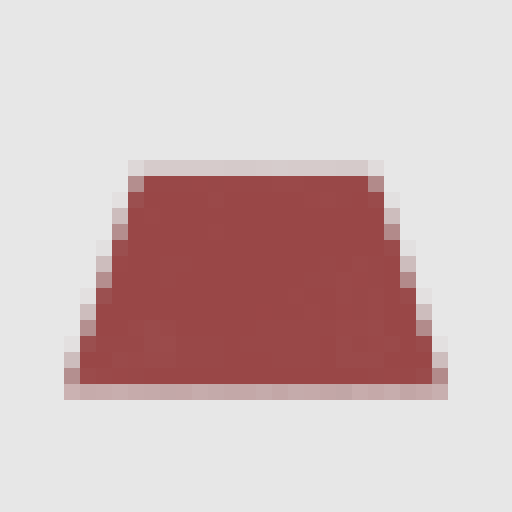}}
        \\
        \raisebox{15pt}{\rotatebox{90}{\textsf{Difference}}}
        &
		\multicolumn{1}{r}{\frame{\begin{overpic}[height=\lenValidateEllipsoid,unit=1mm]{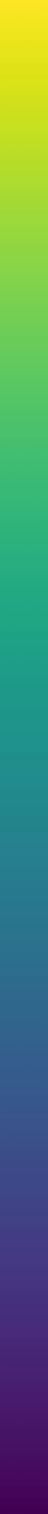}
			\put(-20, 91){\normalsize 0.1}
			\put(-20, 1){\normalsize 0.0}
		\end{overpic}}}
        &
        \frame{\includegraphics[height=\lenValidateEllipsoid]{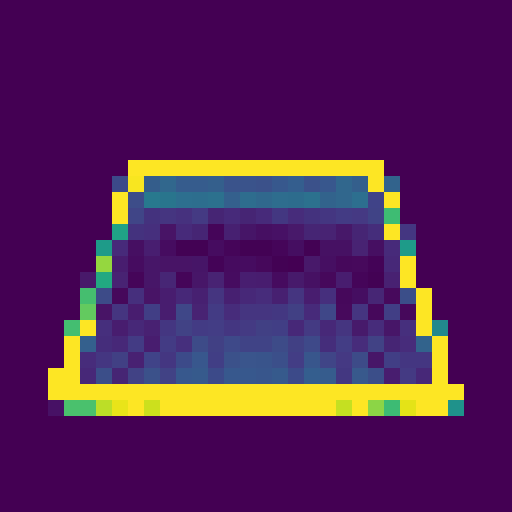}}
        &
        \frame{\includegraphics[height=\lenValidateEllipsoid]{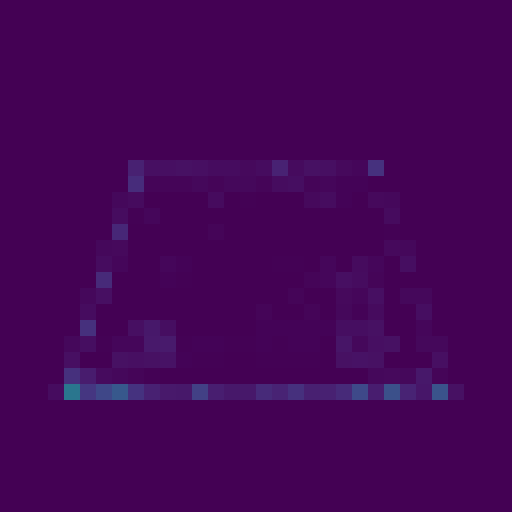}}
        \\
        \raisebox{15pt}{\rotatebox{90}{\textsf{Magnified}}}
        &
        \frame{\includegraphics[height=\lenValidateEllipsoid]{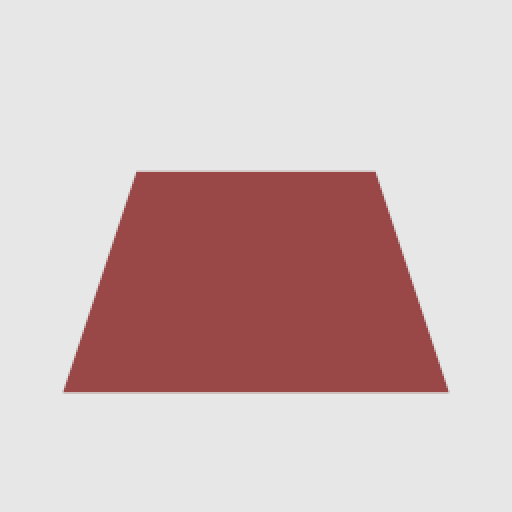}}
        &
        \frame{\includegraphics[height=\lenValidateEllipsoid]{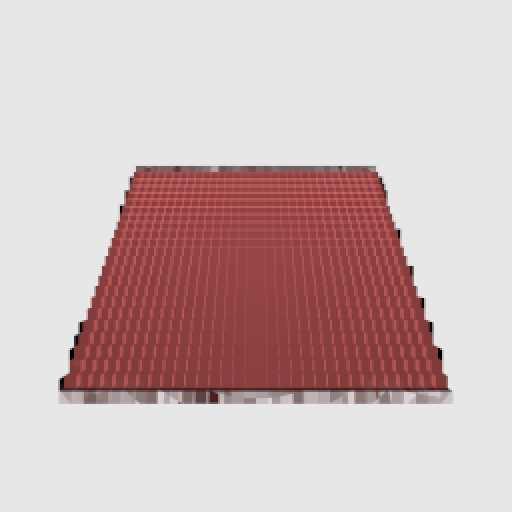}}
        &
        \frame{\includegraphics[height=\lenValidateEllipsoid]{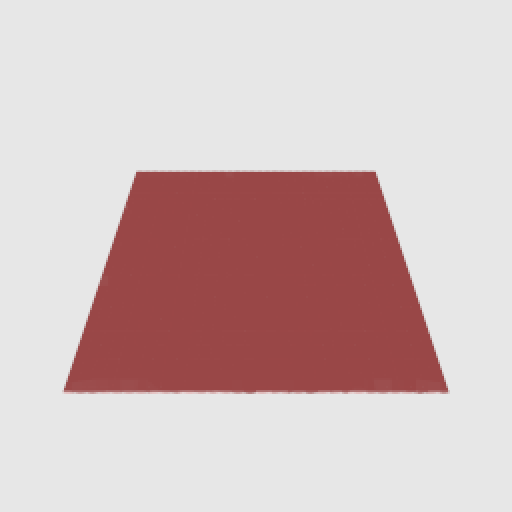}}
        \\
        &
        \textsf{(a) Reference} & \textsf{(b) Simple} & \textsf{(c) Trunc. Ellipsoid}
    \end{tabular}
    \caption{\label{fig:validate_ellipsoid}
        Compared to the reference, LoD with a simple cube primitive results in a bloated silhouette and worse, structured artifacts on the red plane.
        With the help of the truncated ellipsoid primitive, our method produces a tighter silhouette and more importantly, artifact free results.
        We encourage readers to zoom in to better identify the checkerboard-like artifact.
    }
\end{figure}

In this \emph{double-counting} example illustrated in \autoref{fig:validate_ellipsoid}, a simple plane is discretized into diagonally neighboring voxels and is
viewed from aside.
The voxel size is chosen to be half of the pixel footprint to match the image-space Nyquist-Shannon sampling rate. Because the voxels have thickness, whenever
the film plane is not axis-aligned, some points on the film plane receive contributions from more than one voxel. This is clearly wrong because if we perform
ray casting from the film plane, a ray only intersects the ground-truth geometry once. The mismatch is fundamentally because geometry is not distributed
uniformly inside a voxel. It could also be interpreted as strong correlation between different voxels: whenever a ray hits one voxel, it should never hit
another. However, simple voxels fail to capture this information and result in systematic error. In particular, the error manifests as an objectionable
checkerboard-like artifact. We also provide magnified renders with higher image resolution that better illustrate the source of this artifact.

To improve the accuracy of voxel accumulation, we consider ways to support non-uniform intra-voxel distribution. Common approaches that introduce further
subdivision within a single voxel, such as using a coverage mask, are essentially no different from brute-force supersampling. They are not cost-effective as
we discussed earlier.
Instead, we propose to fit a bounding ellipsoid for the geometry in each voxel and define the new voxel primitive as the intersection of the voxel and the
ellipsoid. The new \emph{truncated ellipsoid primitive} is much more effective at adapting to different geometry distributions:
when the voxel includes a flat surface or a fiber-like thin structure, the primitive now provides a much tighter fit;
when the voxel includes unstructured geometry, it falls back to a cube shape.
As shown in \autoref{fig:validate_ellipsoid} (c), the new primitive greatly reduces the artifacts while also producing a tighter object silhouette.
Note that we never need to explicitly store the cube-ellipsoid intersection; it is sufficient to store the separate shapes and calculate the properties of the
intersection on-the-fly. Because the primitive is a bounding volume, it should support ``semi-transparency'' to reflect the quantity of the underlying geometry as
the geometry is abstracted away (\autoref{fig:theory_illus1} (a)). We define the \emph{primitive coverage} of a truncated ellipsoid primitive as
\begin{equation} \label{eq:primitive_coverage}
    c(\omega) = \frac{|A|_\omega}{|B|_\omega},
\end{equation}
where $|A|_\omega$ and $|B|_\omega$ are the projected areas of the surfaces bounded by the primitive and the primitive itself, respectively. As will be shown
in \autoref{subsec:evaluate_pixel}, the primitive coverage is useful when accumulating the contributions from multiple voxels, though the numerator
$|A|_\omega$ will be canceled and never explicitly needed.
\rev{
For simplicity, we use an efficient Monte Carlo estimator to calculate $|B|_\omega$, the projected area of the intersection of a cube and an ellipsoid, detailed 
in the supplemental document. Alternatively, it is possible to explicitly calculate the projected area by integrating over the projected contour using 
Green's theorem.
}
The truncated ellipsoid primitive provides a good trade-off in practice as it is easy to fit and compact to store. See \autoref{subsec:precompute} for details.
Recently, 3D Gaussians have been shown to be effective at representing radiance fields~\citep{kerbl20233d}. Our primitive bears some resemblance to a
3D Gaussian but is ultimately designed for a different purpose. The truncation avoids the ambiguity in defining the inter-occlusion between the otherwise 
overlapping primitives. It also makes precomputation and rendering more straightforward in practice.  

\subsection{Aggregated Visibility} \label{subsec:agn_vis}
In order to accurately accumulate the contributions of voxels, we need to model the visibility between them when the input scene is heterogeneous with a varying
degree of spatial correlation. Existing works show that when spatial correlation exists, transmittance is no longer exponential and cannot be modeled only by
extinction coefficients~\citep{jarabo2018radiative,bitterli18framework,vicini2021non}. Hypothetically, it might be appealing to augment the traditional
volumetric models with more parameters per voxel. However, simply enhancing the local representation is unlikely to be sufficient because spatial correlation is
inherently a long-range effect and it is necessary to model the interaction between voxels. Another attempt is to record and accumulate the coverage masks
of voxels~\citep{bako2023deep}. This is again similar to brute-force supersampling and requires an impractical amount of memory.

\begin{figure}[tb]
    \centering
    \begin{tabular}{cc}
        \includegraphics[width=1.5in]{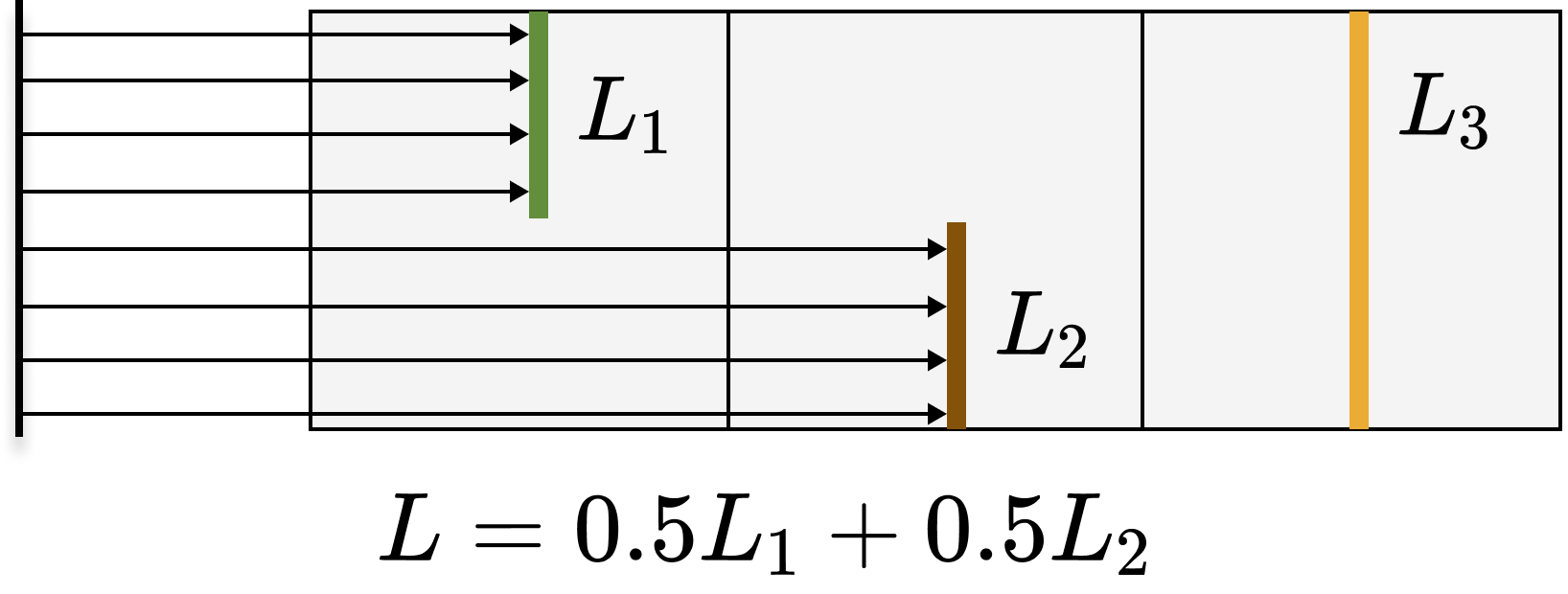}
        &
        \includegraphics[width=1.5in]{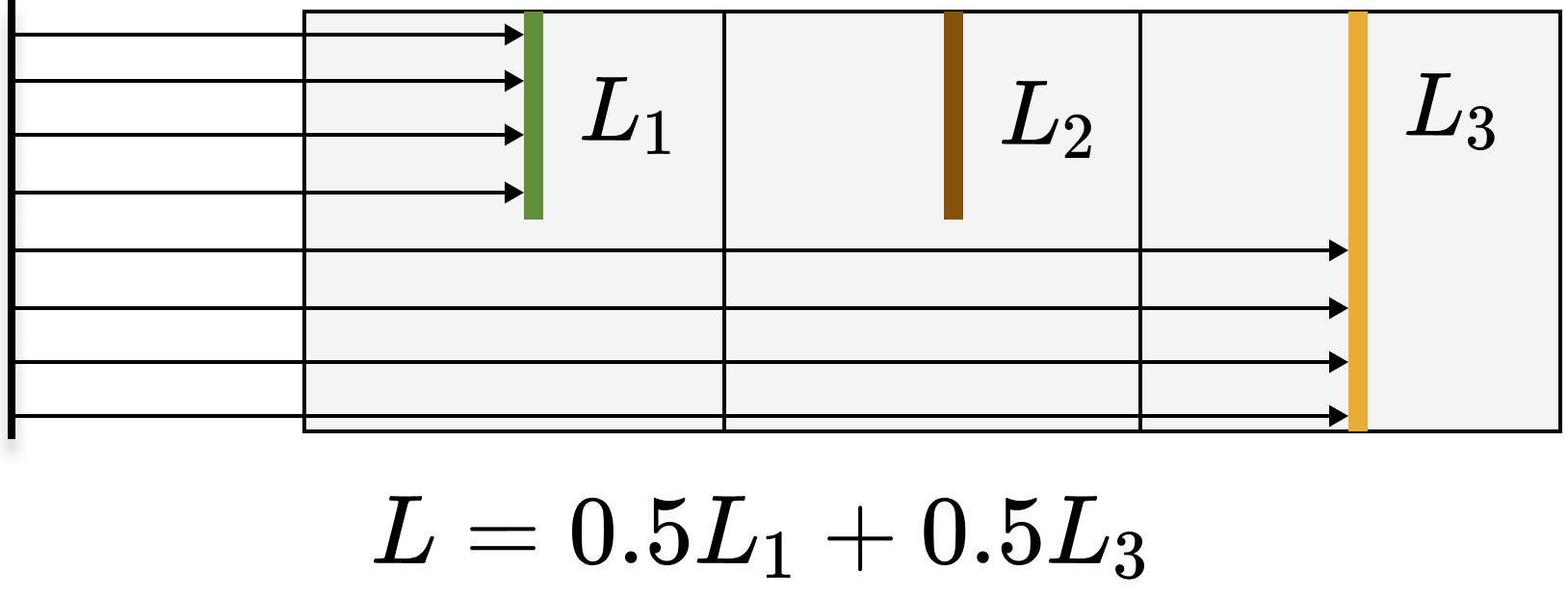} \\
        \small{\textsf{(a)}} & \small{\textsf{(b)}} \\
        \multicolumn{2}{c}{\includegraphics[width=3.1in]{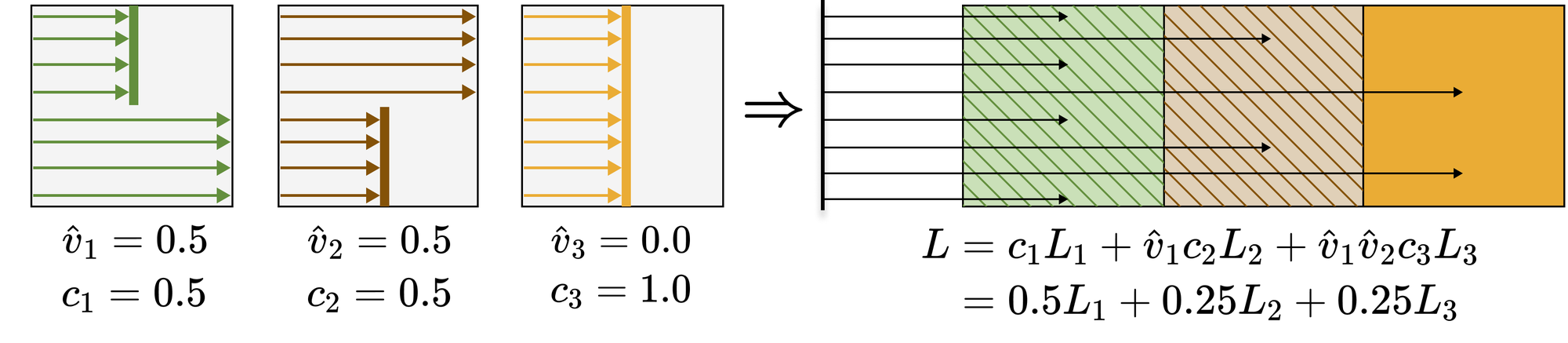}} \\
        \multicolumn{2}{c}{\small{\textsf{(c)}}} \\
        \includegraphics[width=1.5in]{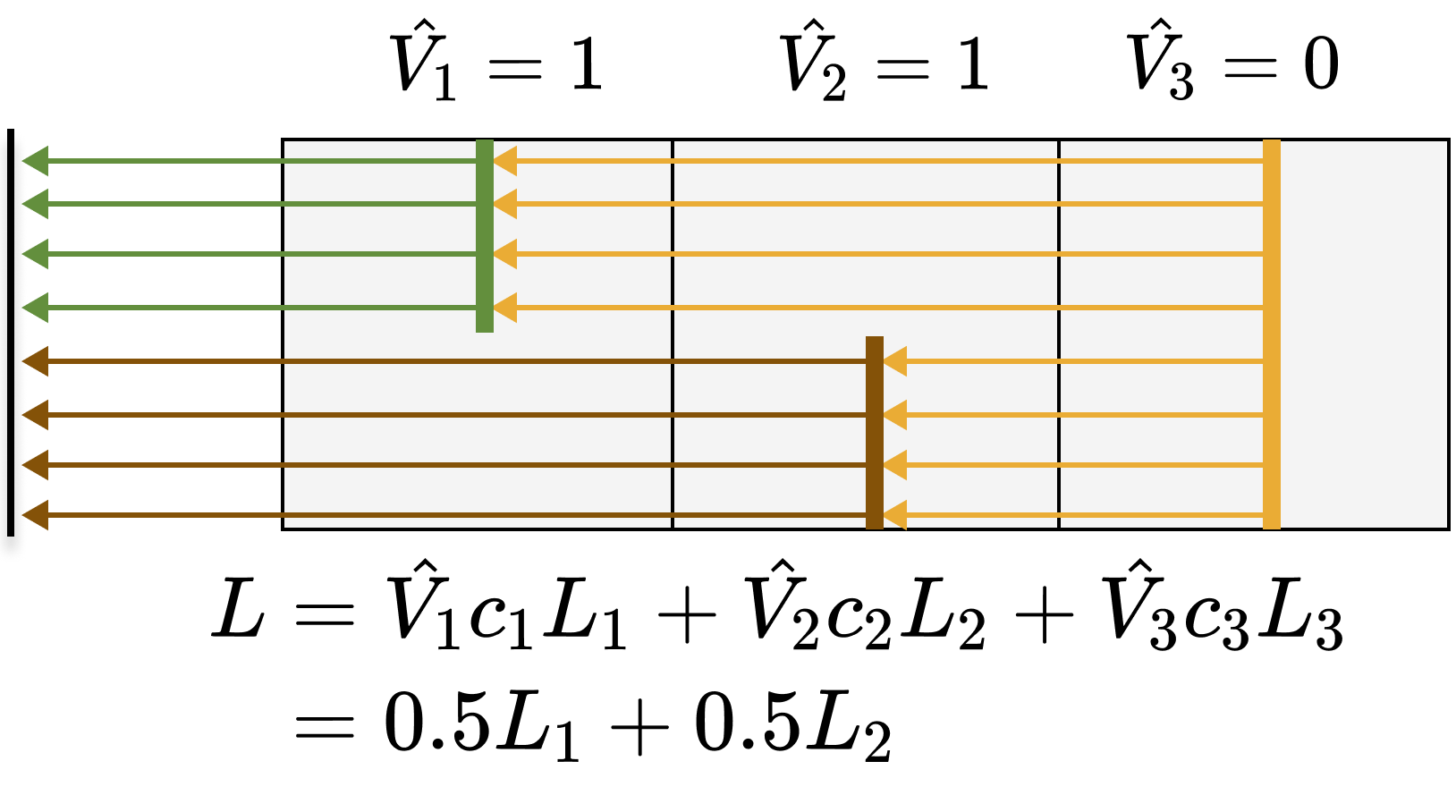}
        &
        \includegraphics[width=1.5in]{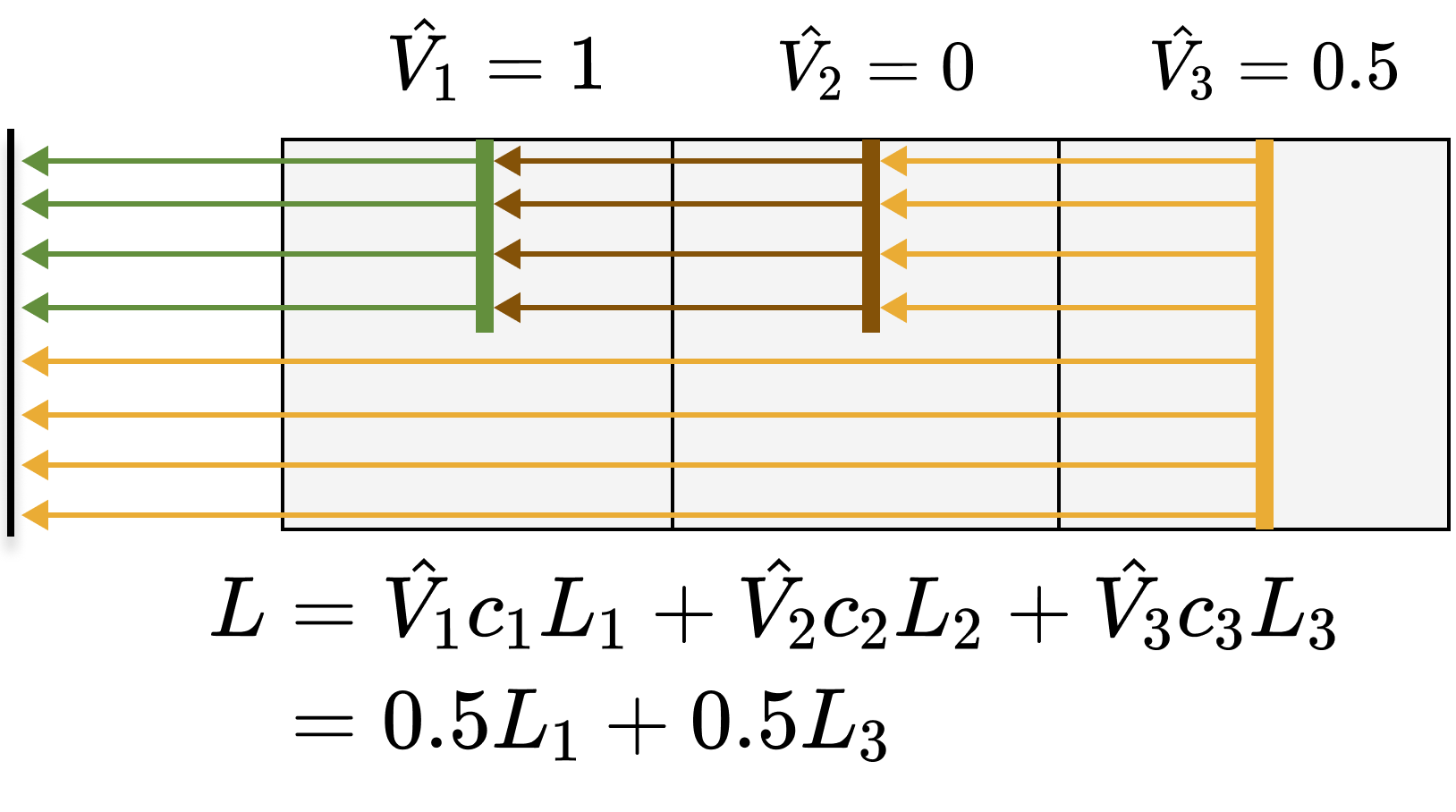} \\
        \small{\textsf{(d)}} & \small{\textsf{(e)}} \\
    \end{tabular}
    \caption{\label{fig:intervoxel_correlation}
        Even if voxels have identical local properties, different types of inter-voxel correlation leads to different accumulation results ((a) and (b)). While the
        traditional volumetric model (c) is unable to distinguish these two cases and produces wrong results, our method accounts for this correlation
        naturally by tracking global visibility ((d) and (e)).
    }
\end{figure}

Instead of modeling the visibility by local properties, we propose to model it as a global function. Recall that we produce the split visibility integral in
\autoref{eq:split_vis}. We further separate the visibility along incident and outgoing directions:
\begin{align} \label{eq:define_aiv}
\begin{split}
            &\frac{1}{|A|} \int_A V(x, \omega_i) V(x, \omega_o)  \,\D{x} \approx \hat{V}(\omega_i) \hat{V}(\omega_o), \\
            &\hat{V}(\omega) = \frac{1}{|A|} \int_A V(x, \omega) \,\D{x}, \\
\end{split}
\end{align}
where $\hat{V}(\omega)$ is the average visibility from points on the surfaces $A$ inside a voxel through the entire scene along $\omega$, and we name it
\emph{aggregated interior visibility} (AIV). We illustrate how modeling the global visibility naturally captures spatial correlation by a flatland example shown
in \autoref{fig:intervoxel_correlation}. We consider the accumulation of three voxels in two configurations where the results are different due to
different types of inter-voxel correlation. In (a), the first two voxels are negatively correlated, while in (b), they are positively correlated.
Both configurations have identical per-voxel visibilities (transmittance) $\hat{v}_i$ and coverages $c_i$ as denoted in (c).
The traditional volumetric model is thus not able to recognize the correlation and produces an incorrect result by applying the Beer-Lambert law.
By explicitly tracking the global AIV $\hat{V}_i$, our method naturally incorporates correlation and produces correct accumulation results.

A slightly different matter arises when a scene aggregate is placed in front of a background or other external objects. In order to correctly blend the
contribution from the scene aggregate and the external environment, we need to keep track of another type of aggregated visibility with origins not on surfaces
but in the free space. Let $P$ be a pixel footprint in world space observing the scene aggregate from direction $\omega$. Intuitively, we would like to know the average
visibility from points on $P$ through the entire scene along $\omega$. However, caution is needed as we should only count the subset of $P$, $P^+$, such that
rays originated from $P^+$ actually intersect the scene aggregate primitives, as illustrated in \autoref{fig:theory_illus1}. 
We define this \emph{aggregated boundary visibility} (ABV) as
\begin{equation} \label{eq:define_abv}
    \hat{V}_b(\omega; P^+) = \frac{1}{|P^+|}\int_{P^+} V(x, \omega) \,\D{x}.
\end{equation}
As the name implies, the ABV term only needs to be defined on the 2D ``boundary'' that encloses the 3D scene aggregate. The bound can be any suitable manifold 
that is
reasonably tight such that no external objects intersect it. In our implementation, the ABV term is precomputed and stored on the boundary faces of the voxels.
The aggregated visibility functions, AIV and ABV, represent high-dimensional (5D and 4D, respectively) and all-frequency signals. Therefore, we choose to
represent them in the Haar wavelet basis. We discuss the details of truncation and further compression strategies in \autoref{subsec:compression}.

\rev{
\paragraph{Discussion}
In their work, \citet{weier2023neural} propose to learn binary visibility by a visibility network and optimize the binary threshold by a weighted 
F-Measure. The network is trained for \emph{per-voxel} visibility query given a pair of vertices on the boundary of a voxel. During rendering, a ray is 
partitioned into multiple segments that intersect different voxels and the visibility of each segment is queried separately. 
We argue that this approach is theoretically limited to handle aggregated inter-voxel correlation due to two reasons. First, there is no aggregation of 
visibility at all as the network only supports \emph{point-to-point query} without any consideration of filter footprint. Second, this approach essentially 
assumes no correlation between voxels as both training and inference is performed in a \emph{per-voxel} manner. As discussed and illustrated in 
\autoref{fig:intervoxel_correlation}, even when individual voxels produce identical statistics, different combinations can still lead to different accumulation 
visibility. In addition, in their global illumination rendering, an indirect ray is simply spawned from the entry point of current voxel on its boundary with 
the new scattering direction, which is then used to query the network. This ignores the fact that scattering could happen anywhere inside the voxel and lead 
to a distribution over exiting positions.
}

\subsection{Evaluating Pixel Intensity} \label{subsec:evaluate_pixel}
We are now ready to present how to evaluate pixel intensity by accumulating the outgoing radiance of voxels under the far-field assumption. Assuming a pinhole
camera and a box pixel reconstruction filter, the intensity of a pixel is the integration of the receiving radiance over the footprint $P$ along its direction
$\omega_\mathrm{p}$:
\begin{equation} \label{eq:integrate_footprint}
    I = \frac{1}{|P|} \int_{P} L_i(x_\mathrm{p}, \omega_\mathrm{p}) \,\D{x_\mathrm{p}}.
\end{equation}
We first consider the case of a single voxel and background as illustrated in \autoref{fig:theory_illus1}.
The value of the integrand in \autoref{eq:integrate_footprint} depends on whether
$x_\mathrm{p}$ is ``covered'' by the truncated ellipsoid primitive $B$ of the voxel, defined as whether the ray spawned from $x_\mathrm{p}$ intersects $B$.
Let $P^+$ be the subset of $P$ that is covered, and $P^- \coloneq P \setminus P^+$:
\begin{enumerate}
    \item If $x_\mathrm{p} \in P^-$, $L_i(x_\mathrm{p}, \omega_\mathrm{p})$ simply evaluates to the background radiance $L_b(-\omega_\mathrm{p})$.
          Note that $L_b$ does not depend on position due to the far-field assumption.
    \item Otherwise, $x_\mathrm{p} \in P^+$ and $L_i(x_\mathrm{p}, \omega_\mathrm{p})$ is a blend between the outgoing radiance of the voxel,
    $L_o(-\omega_\mathrm{p})$ as defined in \autoref{eq:absdf_def}, and the background radiance by the primitive coverage $c(\omega_\mathrm{p})$ as defined
    in \autoref{eq:primitive_coverage}.
\end{enumerate}
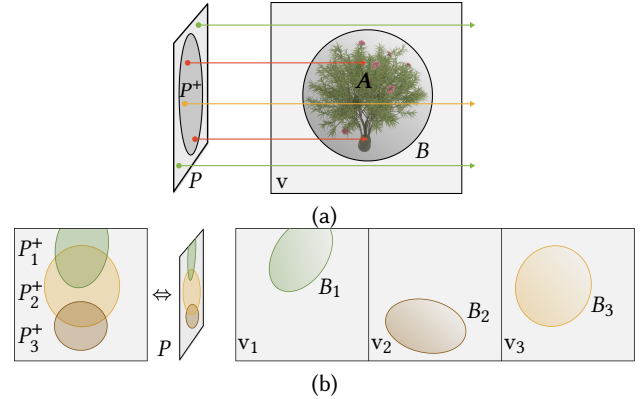
\begin{figure}[t]
	\newlength{\lenTheoryIllus}
	\setlength{\lenTheoryIllus}{1.0in}
	\centering
    \begin{tabular}{c}
        \begin{overpic}[height=\lenTheoryIllus,unit=1mm]{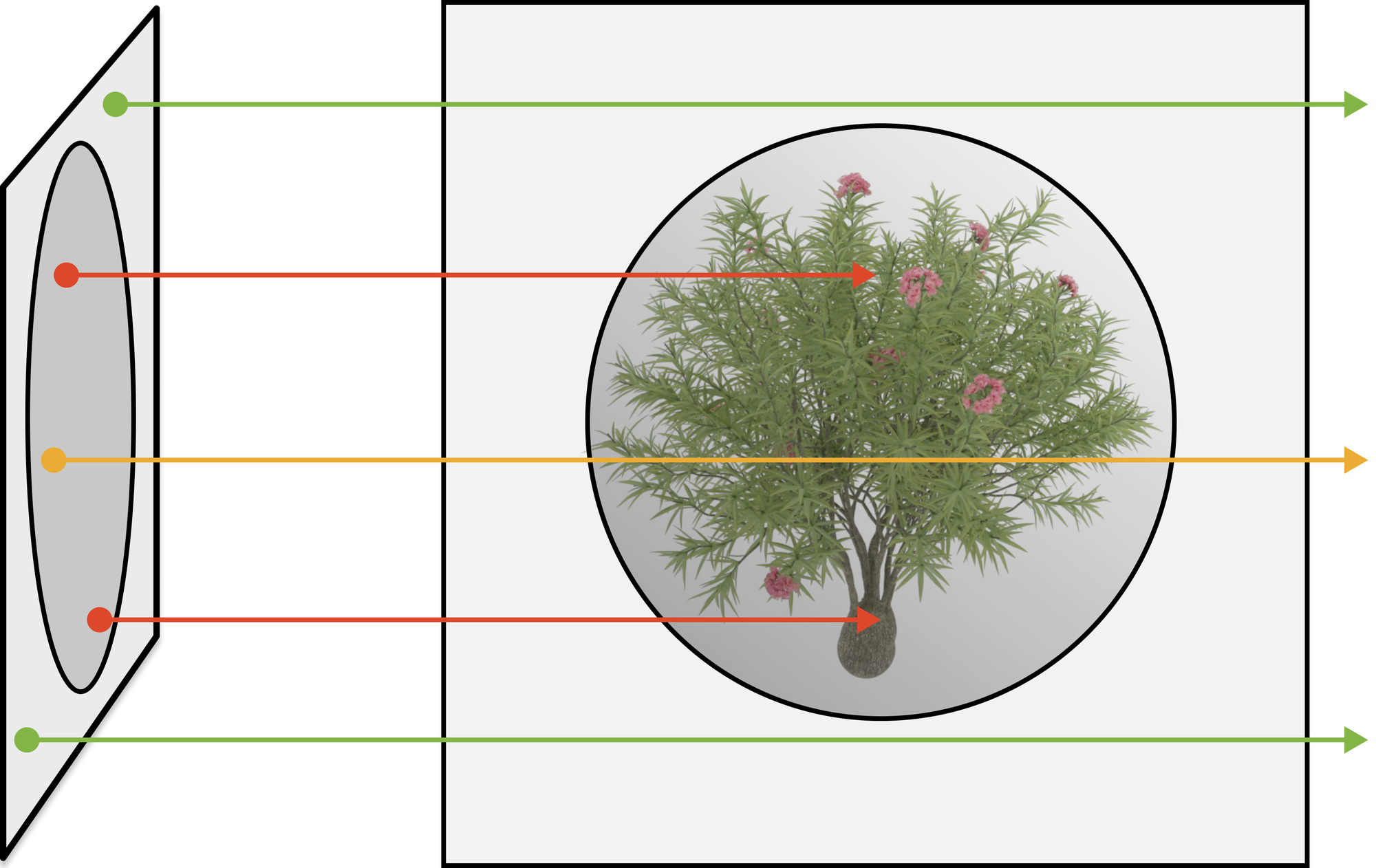}
            \put(5, 2){\normalsize $P$}
            \put(2, 32){\small $P^+$}
            \put(34, 2){\normalsize $\mathrm{v}$}
            \put(60, 35){\normalsize $\bm{A}$}
            \put(80, 12){\normalsize $B$}
        \end{overpic} \\
        \textsf{(a)} \\
        \begin{overpic}[height=0.7in,unit=1mm]{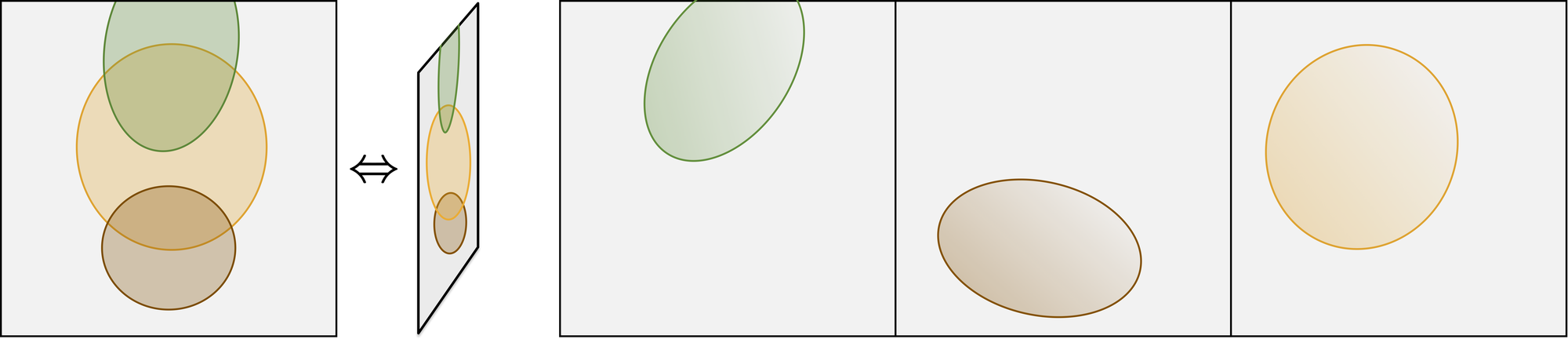}
            \put(23, 1){\normalsize $P$}
            \put(36, 2){\normalsize $\mathrm{v}_1$}
            \put(57.5, 2){\normalsize $\mathrm{v}_2$}
            \put(79, 2){\normalsize $\mathrm{v}_3$}
            \put(49, 11){\normalsize $B_1$}
            \put(73, 7){\normalsize $B_2$}
            \put(93, 8){\normalsize $B_3$}  
            \put(0.5, 17){\normalsize $P_1^+$}
            \put(0.5, 10){\normalsize $P_2^+$}
            \put(0.5, 3){\normalsize $P_3^+$}                        
        \end{overpic} \\
        \textsf{(b)} \\
    \end{tabular}
    \caption{\label{fig:theory_illus1} 
    (a) $P^+$ is the subset of a pixel footprint $P$ that is covered by a primitive $B$. A point outside $P^+$ is guaranteed not to 
    be covered by the underlying geometry (green). For a point inside $P^+$,
    it may (red) or may not (yellow) be covered by the underlying geometry. After the geometry is abstracted away, this notion of semi-transparency of $B$ is
    preserved by the primitive coverage $c(\omega_{\mathrm{p}})$.
    (b) When there are multiple voxels, each primitive covers a subset of the pixel footprint $P_k^+$ and their union 
    becomes $P^+$: $P^+ = \bigcup_k P^+_k$.    
    }
\end{figure}

Therefore, the pixel intensity becomes
\begin{align} \label{eq:accumulate_single_voxel}
\begin{split}
    I = \frac{1}{|P|} \Big( &\int_{P^+} L_o(-\omega_\mathrm{p}) c(\omega_\mathrm{p}) +
        L_b(-\omega_\mathrm{p}) (1-c(\omega_\mathrm{p})) \,\D{x_\mathrm{p}} \,+ \\
      &\int_{P^-} L_b(-\omega_\mathrm{p}) \,\D{x_\mathrm{p}} \Big).
\end{split}
\end{align}
Notice that the integrands do not depend on $x_\mathrm{p}$ at all because all the quantities are aggregated. We can thus collapse the integrals to
\begin{align} \label{eq:accumulate_single_voxel_collapsed}
\begin{split}
    I = &\frac{|P^+|}{|P|} \Big( L_o(-\omega_\mathrm{p}) c(\omega_\mathrm{p}) + L_b(-\omega_\mathrm{p}) (1-c(\omega_\mathrm{p})) \Big) + \\
        &\frac{|P^-|}{|P|} L_b(-\omega_\mathrm{p}),
\end{split}
\end{align}
where the fraction $|P^+|/|P|$ is the pixel coverage in the common sense (not to be confused with the primitive coverage).

\begin{figure}[t]
	\newlength{\lenTheoryIllusB}
	\setlength{\lenTheoryIllusB}{1.7in}
	\centering
    \begin{overpic}[height=\lenTheoryIllusB,unit=1mm]{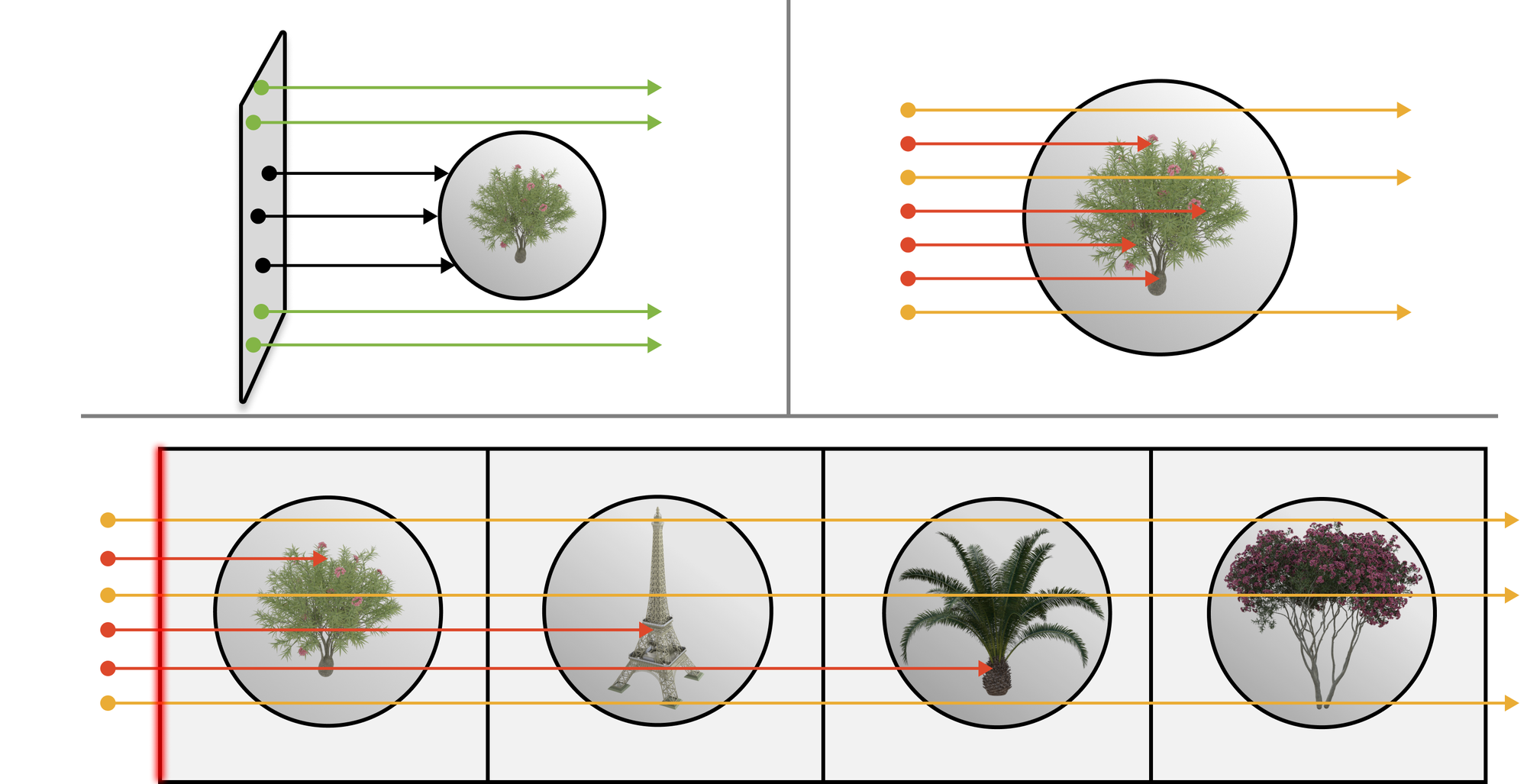}
        \put(53, 37){\normalsize $c_k$}
        \put(5, 36){\normalsize $\frac{|P_k^+|}{|P|}$}
        \put(1, 10){\normalsize $\hat{V}_b$}
        \put(11.5, 1){\normalsize $\mathrm{v}_1$}
        \put(33, 1){\normalsize $\mathrm{v}_2$}
        \put(55, 1){\normalsize $\mathrm{v}_3$}
        \put(77, 1){\normalsize $\mathrm{v}_4$}
    \end{overpic}
    \caption{\label{fig:theory_illus2} Illustrating several terms that appear in \autoref{eq:accumulate_multi_voxel}: The per-voxel pixel coverage (top left), 
    primitive coverage (top right), and the global ABV required to compose the background (bottom).}
\end{figure}
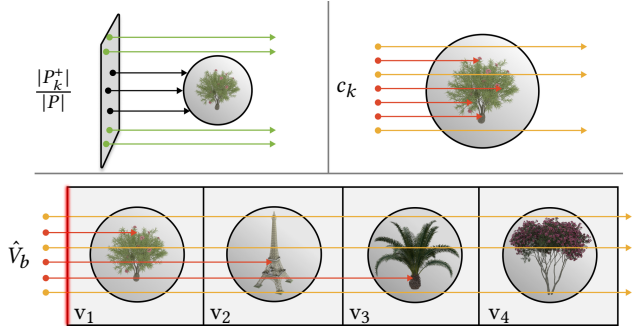

It is straightforward to extend the above formulation to support multiple voxels. Let $\{\mathrm{v}_k\}$ be a list of voxels with truncated ellipsoid primitives
$\{B_k\}$. The pixel intensity is the sum of the contributions from all voxels:
\begin{align} \label{eq:accumulate_multi_voxel}
\begin{split}
        I = &\sum_k \frac{|P^+_k|}{|P|} L_o^{k}(-\omega_\mathrm{p}) c_k(\omega_\mathrm{p}) \,+
        \frac{|P^+|}{|P|} L_b(-\omega_\mathrm{p})\hat{V_b}(\omega_\mathrm{p}) + \\
        &\frac{|P^-|}{|P|} L_b(-\omega_\mathrm{p}), \\
\end{split}
\end{align}
where $P^+_k$ is the subset of $P$ covered by $B_k$ and $P^+ = \bigcup_k P^+_k$.
\autoref{eq:accumulate_multi_voxel} is similar to \autoref{eq:accumulate_single_voxel_collapsed} except that the aggregated boundary visibility
$\hat{V}_b(\omega_\mathrm{p})$ replaces $(1-c(\omega_\mathrm{p}))$ in the second term. This is because now we require the average visibility from $P^+$ through
the entire scene to reconstruct the silhouette of the scene aggregate and compose the background. \autoref{fig:theory_illus2} illustrates several terms in
\autoref{eq:accumulate_multi_voxel}.
We note that the voxel accumulation described by \autoref{eq:accumulate_multi_voxel} is order-independent. Traditionally, 
ray marching or back-to-front alpha blending is required for resolving the occlusion (transmittance) between voxels. However, we have already done so in a 
correlation-aware manner as we have precomputed global visibility as aggregated visibility functions. Therefore, the accumulation during rendering reduces to 
a simple summation where each voxel modulates its contribution by its AIV.
This order-independent property allows efficient parallel implementations.

\section{Scene Aggregation Pipeline} \label{sec:pipeline}
In this section, we describe the pipeline of converting an input scene to our aggregated representation and rendering the scene aggregate with suitable
LoD selection. In particular, we discuss the practical strategy to compress visibility data, which occupies the majority of the memory footprint.

\subsection{Precomputation} \label{subsec:precompute}
The precomputation first performs discretization of the input scene. We utilize the sparsity of typical scenes and discretize the input scene into a hierarchy
consisting of multiple levels of sparse grids. Each level doubles the spatial resolution of the one in the previous level. Due to the sparsity, the actual growth
rate of non-empty voxel count is only quadratic instead of cubic as a function of resolution. We report the sparsity of all scenes used in \autoref{tab:scenes}.
In our current implementation, each level is precomputed separately. However, it is possible to cache and reuse collected data across levels, which we leave for
future optimization.

For each level, the precomputation involves two stages. The first stage precomputes the ``interior'' of the scene. For each non-empty voxel, we need to acquire
the following information:
\begin{enumerate}
    \setlength{\itemsep}{0pt}
    \setlength{\parskip}{1pt}
    \item The total surface area $|A|$ within the voxel (for normalization).
    \item The first two moments of roughness $\alpha$.
    \item Directional moments of material parameters $\beta^c$, $\beta^m$, and $\beta^s$ (\autoref{eq:directional_moments}) with angular resolution $d$.
    \item The surface normal distribution $p_N(n)$ function represented by one or a mixture of SGGX distributions with $k$ components (\autoref{eq:multi_lobe_ndf}).
    \item The ellipsoid of the truncated ellipsoid primitive as an affine transform.
    \item Wavelet basis coefficients for the aggregated visibility (\autoref{eq:define_aiv} and \autoref{eq:define_abv}).
\end{enumerate}
Apart from the surface area which can be computed analytically, the rest of the information is estimated via Monte Carlo sampling and ray tracing. We uniformly sample the
surfaces within the voxel. The sample budget is a tunable parameter. It should not be too low to avoid noisy estimation. As smaller voxels contain fewer surfaces,
we find that one suitable strategy is to allocate sample budget to be inversely proportional to the square of the resolution of current level. This strategy also
helps balance the computation cost across different levels. Each surface sample includes position, normal, and material parameters looked up from texture maps.
We then proceed to estimate each type of information respectively:

\paragraph{Roughness moments} These are straightforward to compute by moving averages.

\paragraph{Directional moments}
An easy way to estimate directional moments is to simply evaluate \autoref{eq:directional_moments} at the center direction of each angular grid cell.
However, this is prone to aliasing for highly glossy surface samples. Instead, for each surface sample, we warp a low discrepancy sequence to $\mathbb{S}^2$ by
the weight kernel $g$. Then we compute the numerator and the denominator of \autoref{eq:directional_moments} for each
sequence element and splat them to separate angular grids. We accumulate the contribution for all surface samples and perform the normalization (division) at
the end.

\paragraph{Surface normal distribution} For a single SGGX component, we follow the estimation method by Heitz et al.~\shortcite{heitz2015sggx}. For a mixture, we
perform a K-means clustering on the surface normal samples and fit each cluster as one component, similar to the process by Zhao et al.~\shortcite{zhao2016downsampling}.
The initial cluster centers are selected to be away from each other. Furthermore, to avoid undesirable homogeneous clusters, we repeat the fitting
for $1$ to a maximum of $k$ components and choose the result that yields the highest likelihood.
In practice, we find $k \leq 4$ are sufficient for most cases.

\paragraph{Truncated ellipsoid primitive} Finding the optimal minimum volume enclosing ellipsoid is a semidefinite programming problem that can be costly to solve
\cite{todd2016minimum}. Instead, we compute the approximate minimum bounding ellipsoid for a voxel with a simple heuristic. We first perform Principal Component
Analysis (PCA) on the sampled positions and transform them to the unit cube by the eigenvectors. We then compute a bounding sphere and transform it back to world space
to obtain a bounding ellipsoid. The resulting ellipsoid is tight enough for our purposes.

\begin{figure}[tb]
	\newlength{\lenAIVPlot}
	\setlength{\lenAIVPlot}{0.65in}
    \addtolength{\tabcolsep}{-4.5pt}
    \renewcommand{\arraystretch}{0.5}
    \centering
    \begin{tabular}{ccccc}
        \textsf{(a) Setup} & \textsf{(b) Ref.} & \textsf{(c) Ours} & \textsf{(d) Ref.} & \textsf{(e) Ours}
		\\
        \frame{\includegraphics[height=\lenAIVPlot]{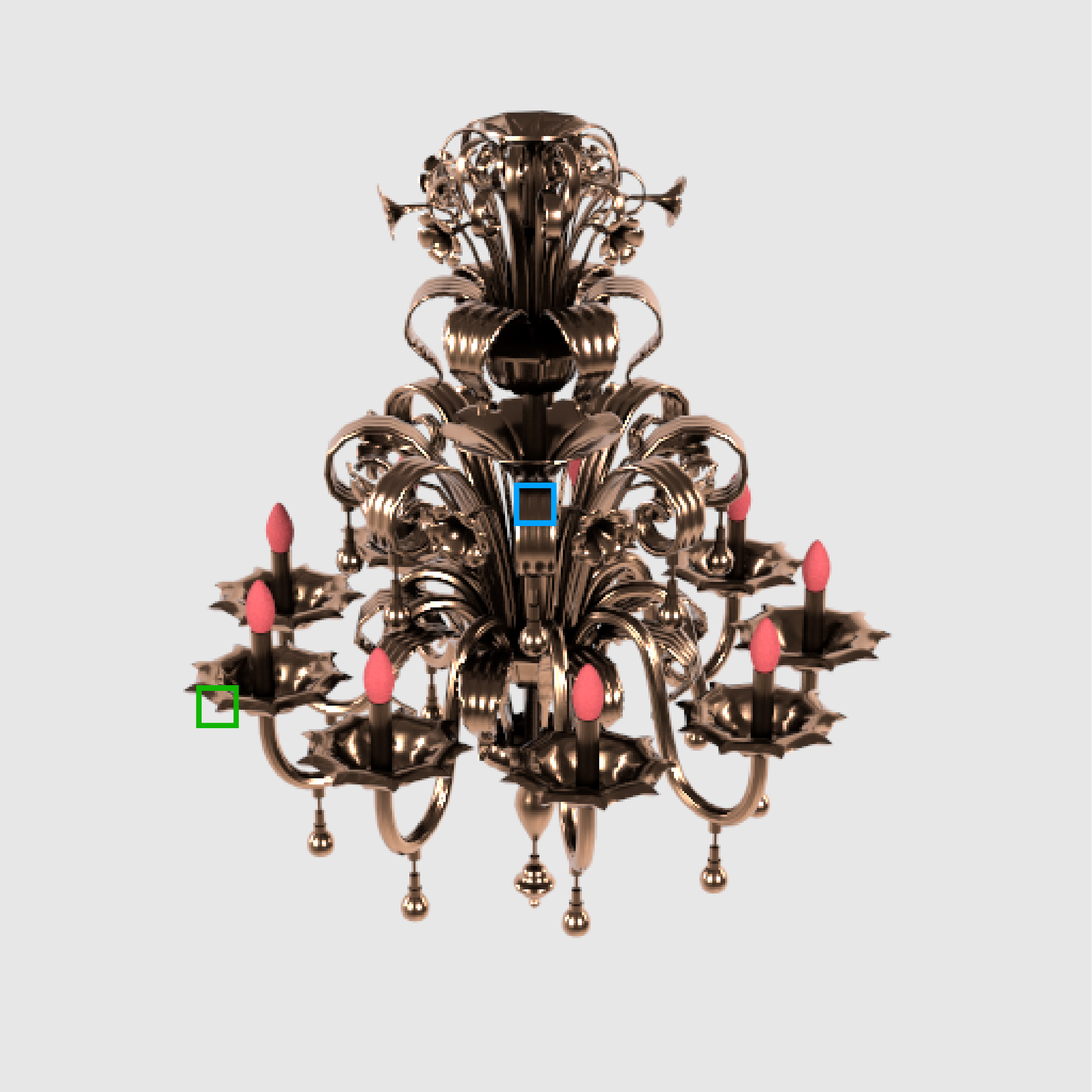}}
        &
        \begin{adjustbox}{margin=0.4pt, bgcolor=lightgreen}{\includegraphics[height=\lenAIVPlot]{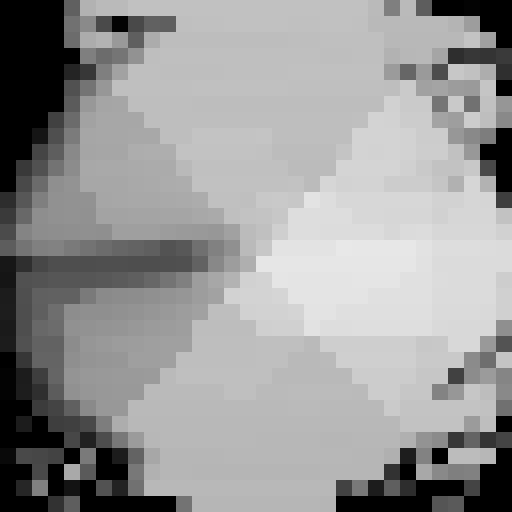}}\end{adjustbox}
        &
        \begin{adjustbox}{margin=0.4pt, bgcolor=lightgreen}{\includegraphics[height=\lenAIVPlot]{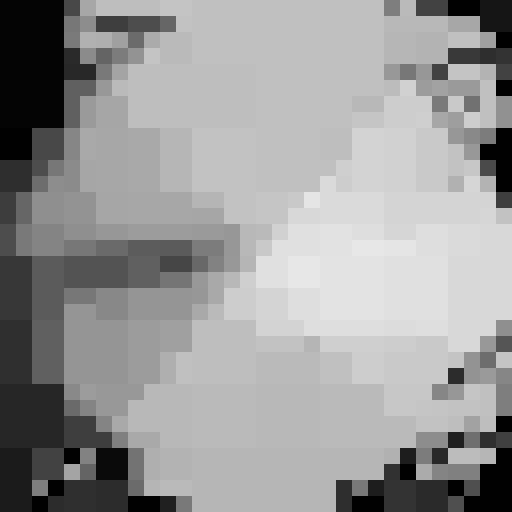}}\end{adjustbox}
        &
        \begin{adjustbox}{margin=0.4pt, bgcolor=lightblue}{\includegraphics[height=\lenAIVPlot]{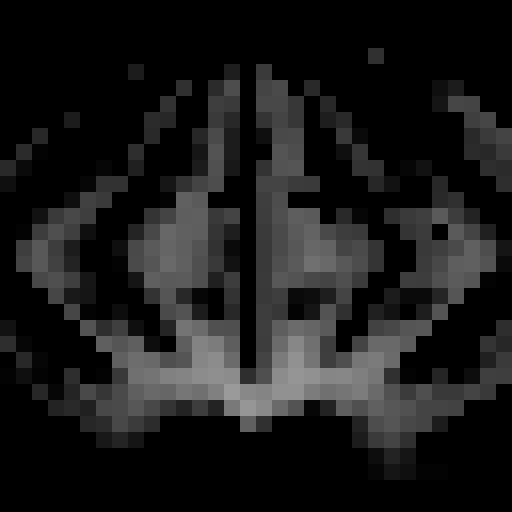}}\end{adjustbox}
        &
        \begin{adjustbox}{margin=0.4pt, bgcolor=lightblue}{\includegraphics[height=\lenAIVPlot]{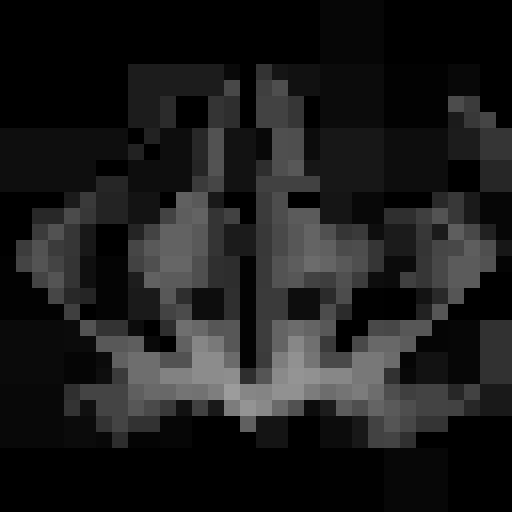}}\end{adjustbox}
        \\
        \textsf{PSNR (dB):} & & 32.84 & & 42.63
		\\
        \frame{\includegraphics[height=\lenAIVPlot]{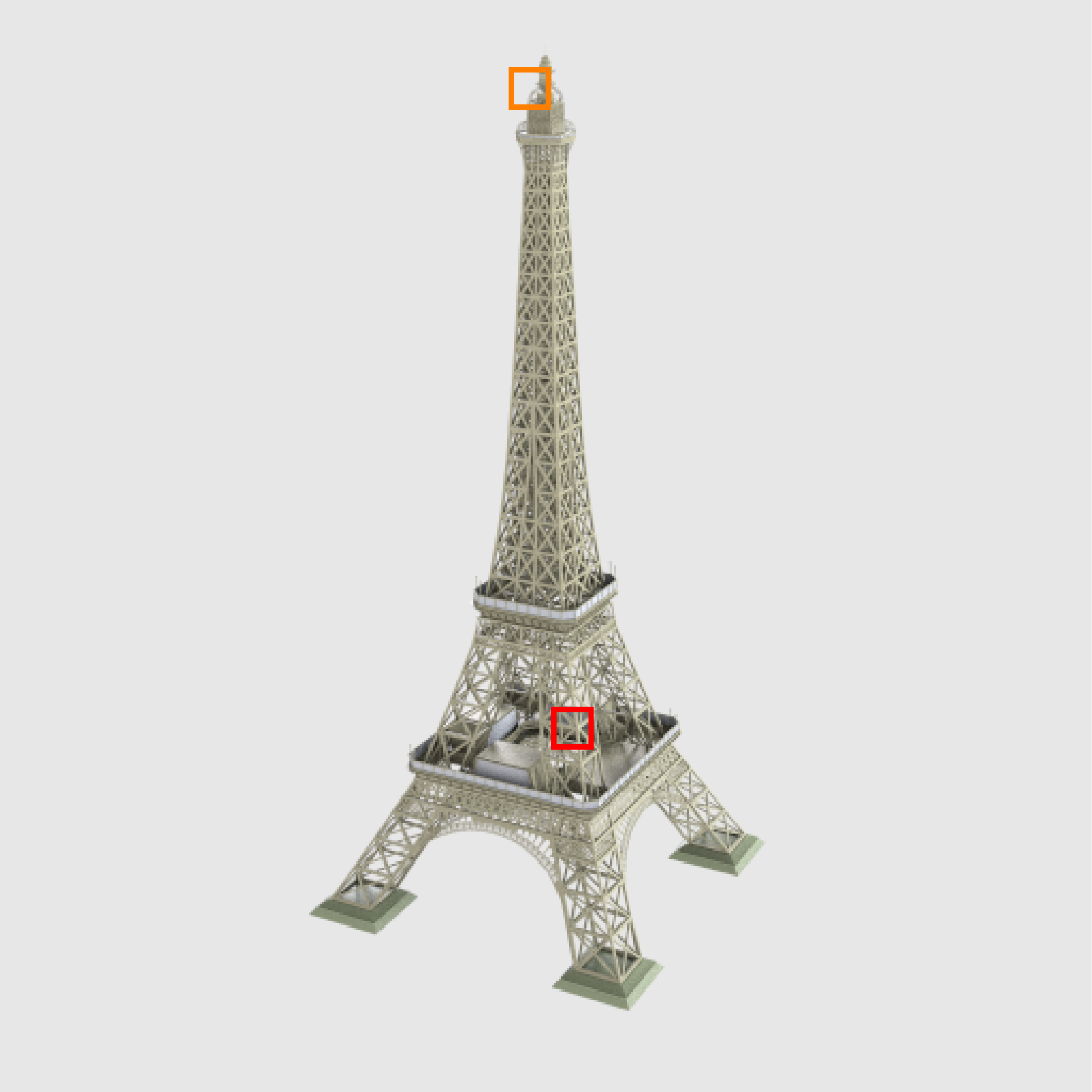}}
        &
        \begin{adjustbox}{margin=0.4pt, bgcolor=red}{\includegraphics[height=\lenAIVPlot]{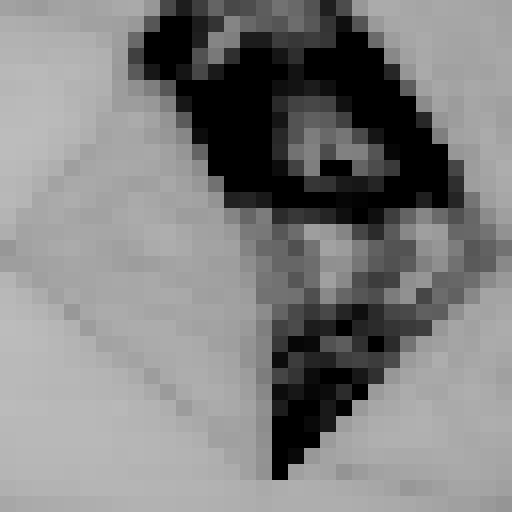}}\end{adjustbox}
        &
        \begin{adjustbox}{margin=0.4pt, bgcolor=red}{\includegraphics[height=\lenAIVPlot]{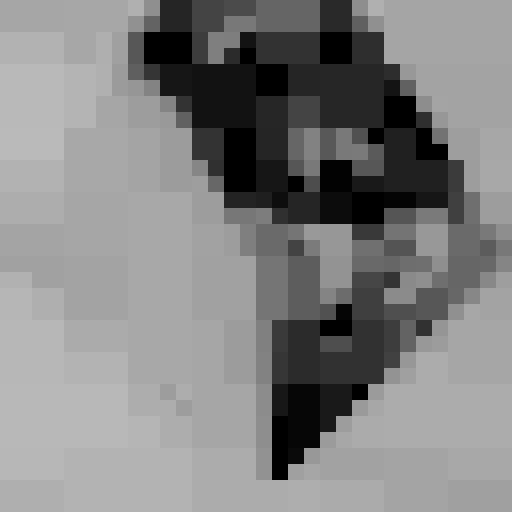}}\end{adjustbox}
        &
        \begin{adjustbox}{margin=0.4pt, bgcolor=orange}{\includegraphics[height=\lenAIVPlot]{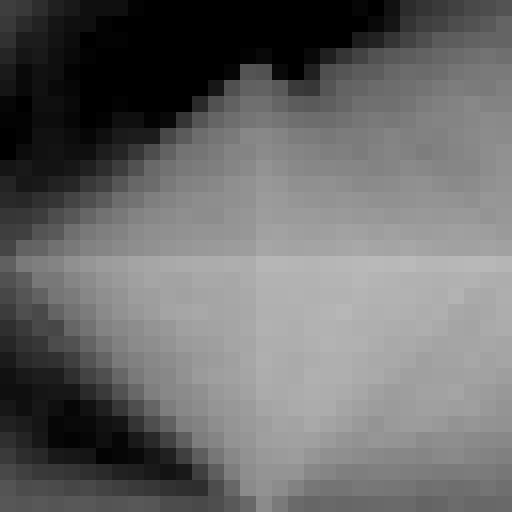}}\end{adjustbox}
        &
        \begin{adjustbox}{margin=0.4pt, bgcolor=orange}{\includegraphics[height=\lenAIVPlot]{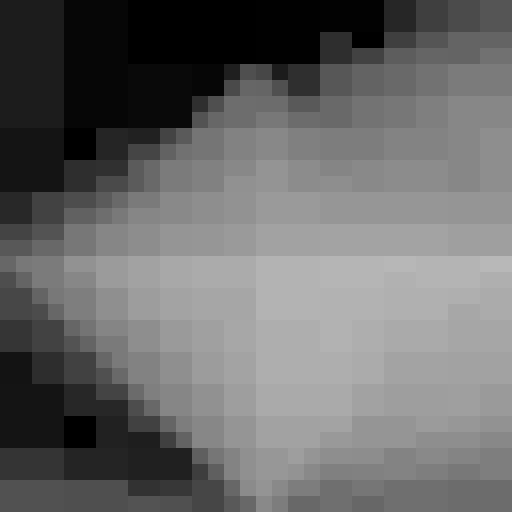}}\end{adjustbox}
        \\
        \textsf{PSNR (dB):} & & 32.17 & & 37.72
		\\
        \frame{\includegraphics[height=\lenAIVPlot]{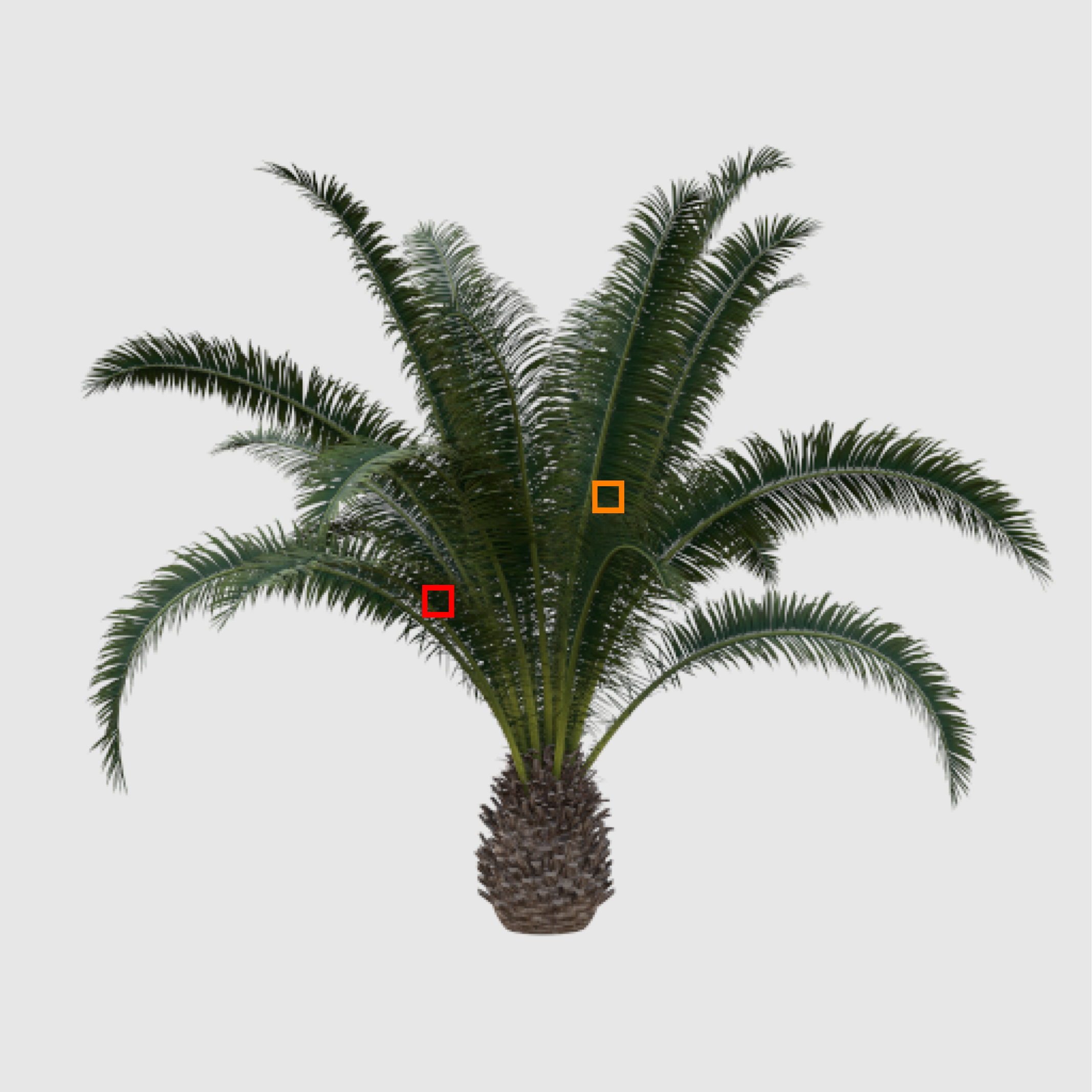}}
        &
        \begin{adjustbox}{margin=0.4pt, bgcolor=red}{\includegraphics[height=\lenAIVPlot]{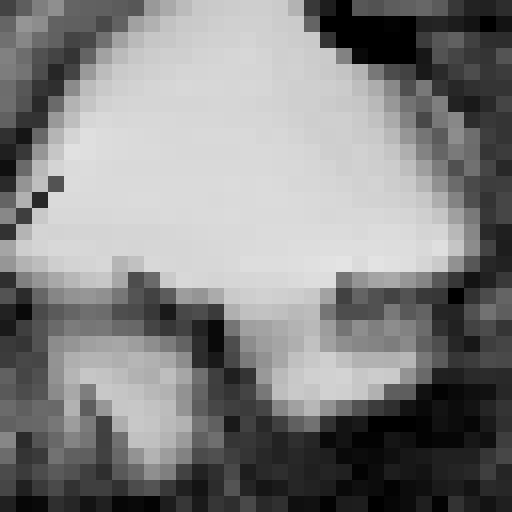}}\end{adjustbox}
        &
        \begin{adjustbox}{margin=0.4pt, bgcolor=red}{\includegraphics[height=\lenAIVPlot]{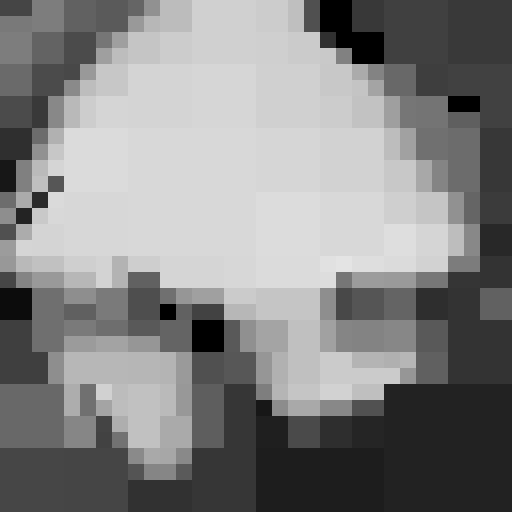}}\end{adjustbox}
        &
        \begin{adjustbox}{margin=0.4pt, bgcolor=orange}{\includegraphics[height=\lenAIVPlot]{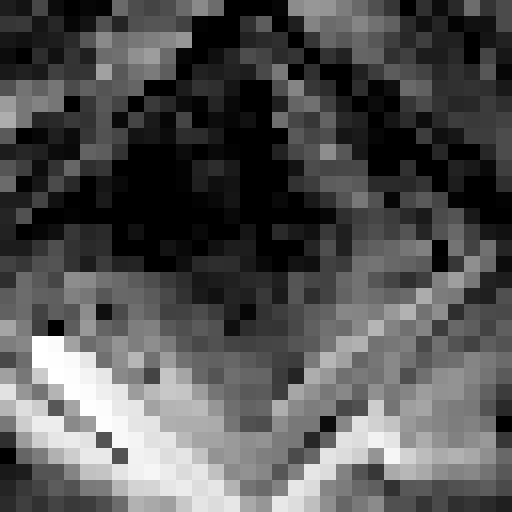}}\end{adjustbox}
        &
        \begin{adjustbox}{margin=0.4pt, bgcolor=orange}{\includegraphics[height=\lenAIVPlot]{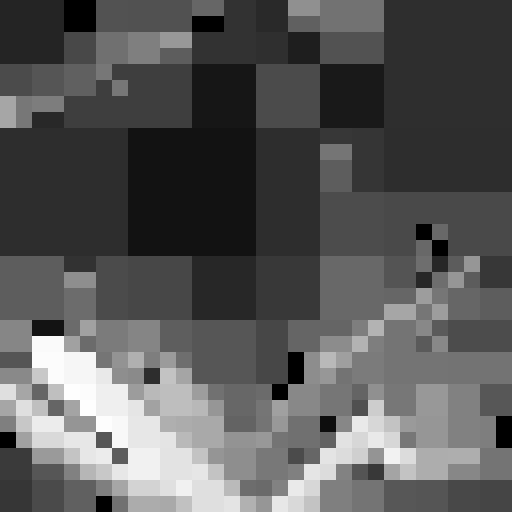}}\end{adjustbox}
        \\
        \textsf{PSNR (dB):} & & 28.76 & & 26.99
        \\
    \end{tabular}
    \caption{\label{fig:aiv_plot}
			Visualizing the AIV terms of selected voxels. Each spherical plot is parameterized by the equal-area mapping~\cite{clarberg2008fast}. The
			wavelet-based projection and compression is able to preserve the high-frequency visibility.
        }
\end{figure}

\paragraph{Aggregated visibility coefficients}
We trace visibility rays with uniformly sampled directions starting from each surface sample and project the results to the Haar wavelet basis. We use the equal-area
mapping \cite{clarberg2008fast} to parameterize the spherical domain.
Again, the visibility sample rate should not be too low to avoid noisy estimation. However, when there are a sufficient number of surface samples, the cost of
tracing visibility can be amortized. In practice, we find $16$-$64$ rays per surface sample is enough. In \autoref{fig:aiv_plot}, we visualize the AIV terms and
compare our compressed terms to the references.

\begin{figure}[t]
	\newlength{\lenCoverage}
	\setlength{\lenCoverage}{0.65in}
    \addtolength{\tabcolsep}{-4.5pt}
    \renewcommand{\arraystretch}{0.5}
    \centering
    \begin{tabular}{cccccc}
        \raisebox{5pt}{\rotatebox{90}{\textsf{Ours ($64^3$)}}}
        &
        \includegraphics[height=\lenCoverage]{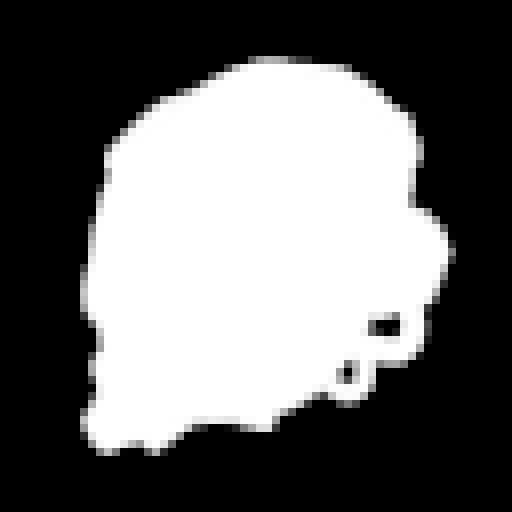}
		&
		\includegraphics[height=\lenCoverage]{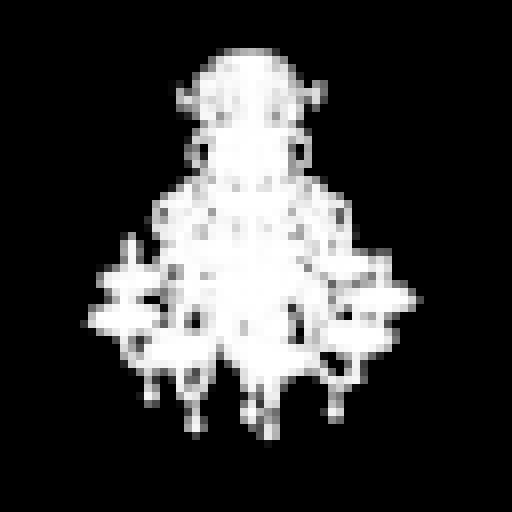}
		&
		\includegraphics[height=\lenCoverage]{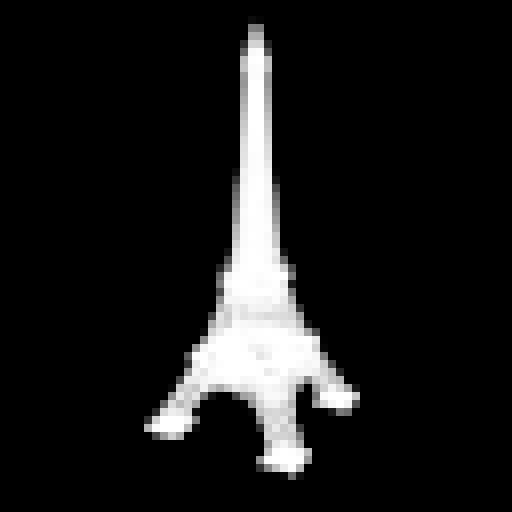}
		&
		\includegraphics[height=\lenCoverage]{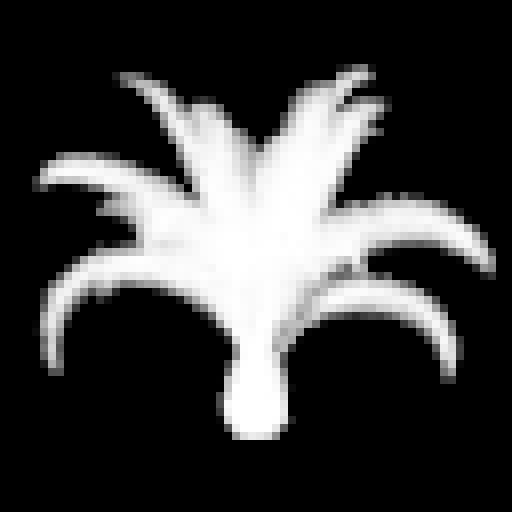}
		&
		\includegraphics[height=\lenCoverage]{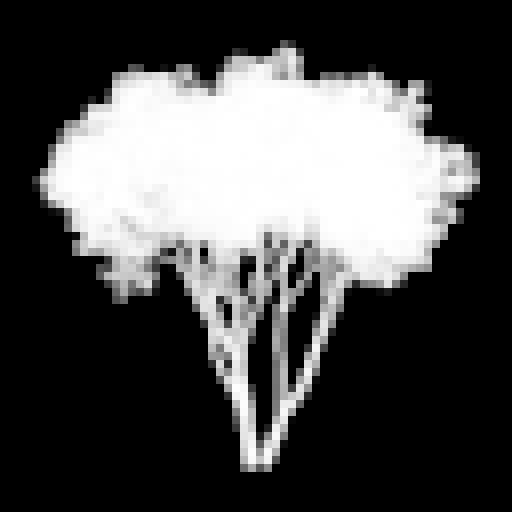}
		\\
        \raisebox{10pt}{\rotatebox{90}{\textsf{PT Ref.}}}
        &
		\includegraphics[height=\lenCoverage]{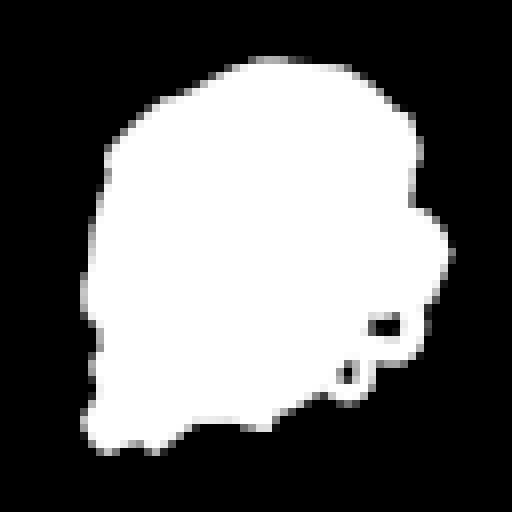}
		&
		\includegraphics[height=\lenCoverage]{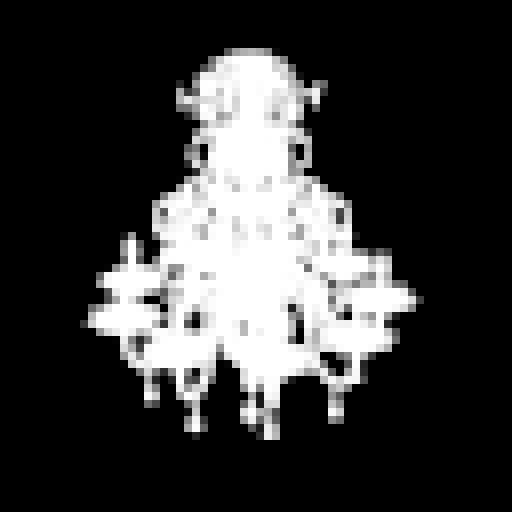}
		&
		\includegraphics[height=\lenCoverage]{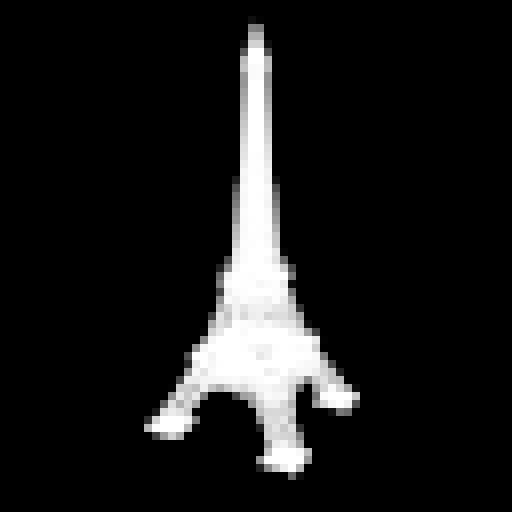}
		&
		\includegraphics[height=\lenCoverage]{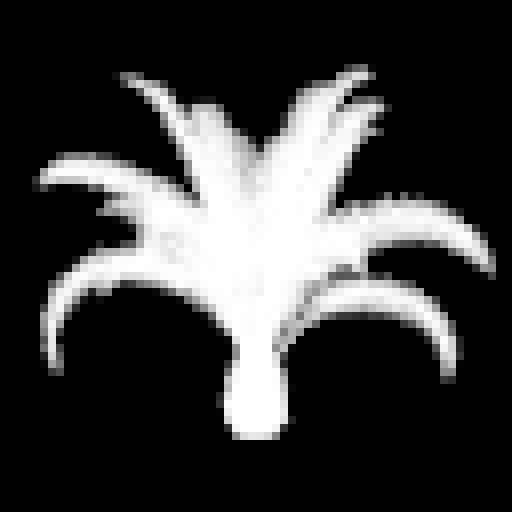}
		&
		\includegraphics[height=\lenCoverage]{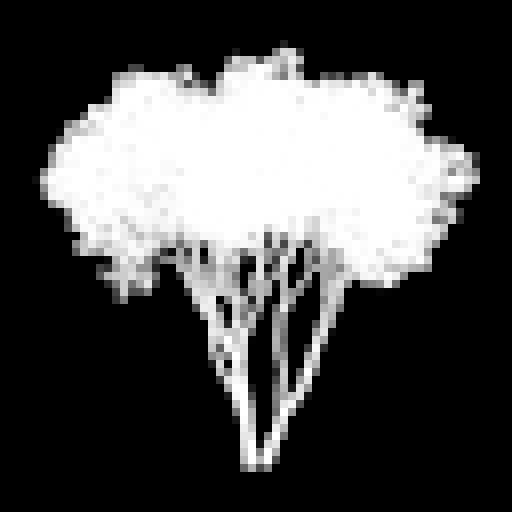}
    \end{tabular}
    \caption{\label{fig:coverage}
			Visualizing coverage for all five scenes in \autoref{fig:main_comparison}. Our ABV term accurately reconstructs partial
    		coverage (transmittance) for different types of scenes after compression.}
\end{figure}

The second stage of the precomputation handles the ABV term. Recall that this term is only defined
on the boundary of the entire scene. Therefore, after scene discretization, we precompute it for the boundary faces of the voxels. The list of boundary faces can
be determined by one simple flood fill iteration. For each boundary face, we consider all directions in its inward facing hemisphere. For each direction, we
cast visibility rays with origin uniformly sampled on the face to estimate the average visibility. We use concentric mapping \cite{shirley1997low} to parameterize
the hemispherical domain and obtain a 2D visibility map. The map is then projected to the Haar wavelet basis.
The ABV contains a high-frequency signal as it is responsible for reconstructing the silhouette of the scene. In practice, we choose a relatively high angular
resolution of $64^2$ for accurate reconstruction. \autoref{fig:coverage} shows the accurate coverage reconstruction with our ABV term.

\subsection{Compression Strategy for Visibility Data} \label{subsec:compression}
In order to preserve all-frequency information, we represent both types of visibility: the aggregated interior visibility (AIV), and the aggregated boundary
visibility (ABV) by wavelet coefficients. Typically, we are able to perform nonlinear approximation and truncate a large number of coefficients while preserving
good quality. However, this is not enough when the angular resolution is high. We typically use $32^2$ resolution for AIV and $64^2$ resolution for ABV as it is
responsible for reconstructing a sharp silhouette. Even with a typical $90\%$ to $95\%$ truncation rate, the memory cost can still be high as the spatial resolution
grows. Therefore, we further apply Clustered Principal Component Analysis (CPCA), which is proven to be effective at compressing basis coefficients
~\cite{sloan2003,LiuSSS04}. One performance issue for CPCA is that PCA has cubic time complexity and quadratic space complexity with respect to input
data matrix size. Thus, it becomes impractical to directly apply it to a fine LoD level. We apply a simple heuristic by dividing a level into individual
blocks of no more than $64^3$ and applying CPCA to each block separately. This works well in practice, since a large extent of spatial locality is still
preserved in each block that can be exploited by CPCA. It is possible to develop more sophisticated methods to scale CPCA or compress coefficients which is left
for future work.

\begin{figure}[t]
	\newlength{\lenCPCASmall}
	\setlength{\lenCPCASmall}{0.65in}
	\newlength{\lenCPCABig}
	\setlength{\lenCPCABig}{1.32in}
    \addtolength{\tabcolsep}{-4.5pt}
    \renewcommand{\arraystretch}{0.5}
    \centering
    \begin{tabular}{cccccc}
        \multicolumn{2}{c}{}
		&
        \includegraphics[height=\lenCPCASmall]{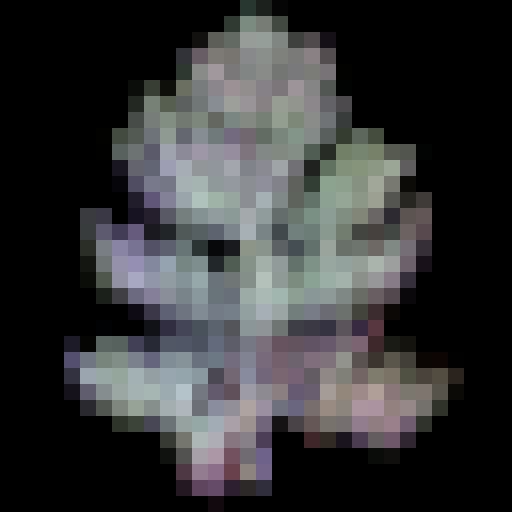}
		&
		\includegraphics[height=\lenCPCASmall]{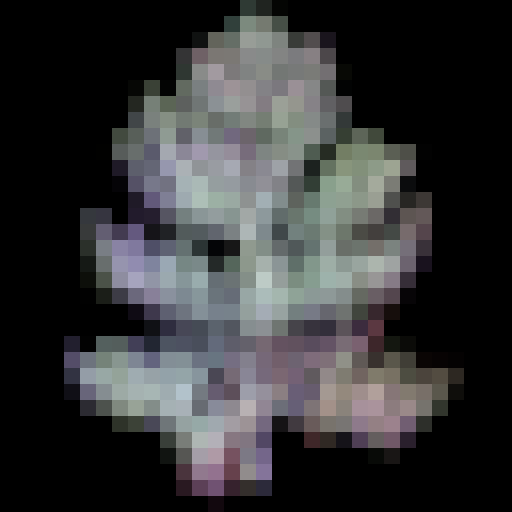}
		&
		\includegraphics[height=\lenCPCASmall]{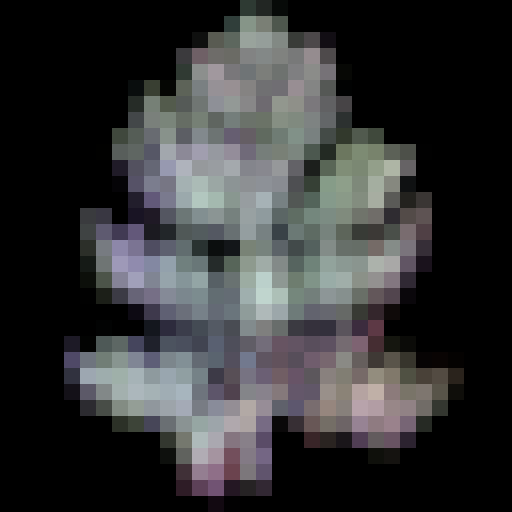}
		&
        \raisebox{28pt}{\rotatebox{-90}{$32^3$}}
		\\
		\multicolumn{2}{c}{\multirow[t]{2}{*}{\frame{\includegraphics[height=\lenCPCABig]{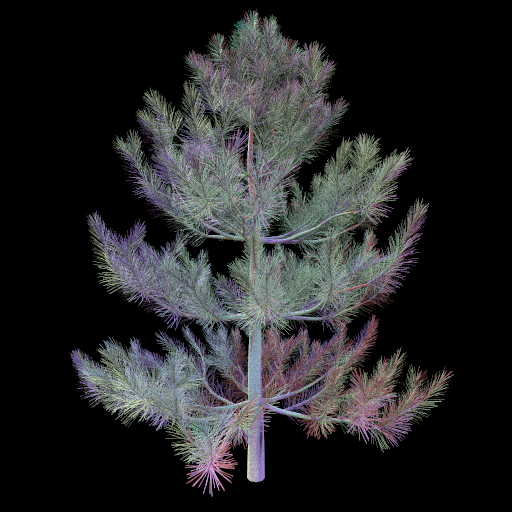}}}}
        &
		\includegraphics[height=\lenCPCASmall]{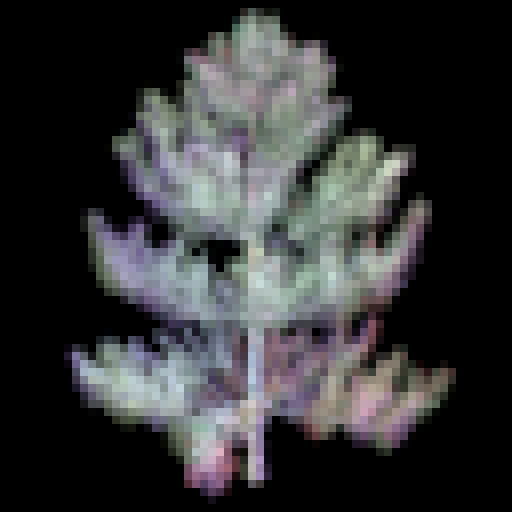}
		&
		\includegraphics[height=\lenCPCASmall]{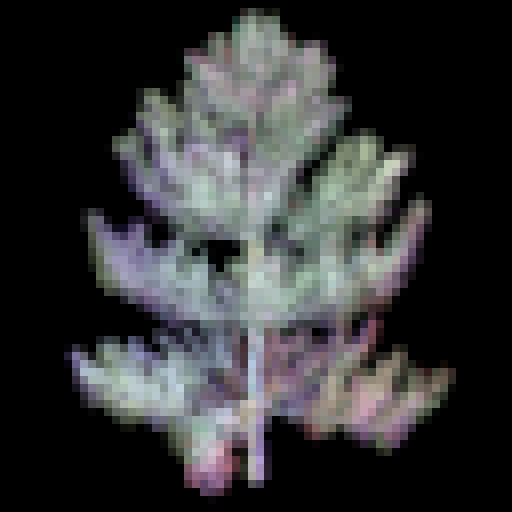}
		&
		\includegraphics[height=\lenCPCASmall]{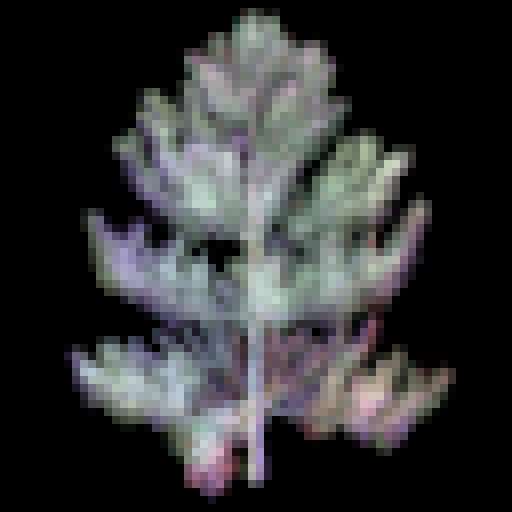}
		&
		\raisebox{28pt}{\rotatebox{-90}{$64^3$}}
		\\
		\multicolumn{2}{c}{\textsf{(a) PT Ref. (High Res.)}} & \textsf{(b) No CPCA} & \textsf{(c) ABV} & \textsf{(d) ABV+AIV} &
    \end{tabular}
    \caption{\label{fig:cpca} We apply CPCA to compress both the aggregated boundary (ABV) and interior visibility (AIV) data. In this example, we reach a
					$\sim4\times$ compression ratio while having little impact on the visual quality.}
\end{figure}

In \autoref{fig:cpca}, we validate the effectiveness of our current CPCA-based compression. For this \emph{Colortree} scene, we compress AIV to 30 clusters
each with 10 representatives and ABV to 30 clusters each with 60 representatives. Each representative still goes through coefficient truncation after CPCA.
Overall, we gain an extra $\sim 4 \times$ compression ratio without negatively impacting the visual quality.

\subsection{Rendering with LoD Selection}
The rendering of our scene aggregate follows \autoref{eq:accumulate_multi_voxel}. For each pixel, we need to determine the list of voxels $\{\mathrm{v}_k\}$ whose
primitives $\{B_k\}$ cover the pixel footprint $P$ and compute the pixel coverage $|P_k^+|/|P|$ for each $k$. In our current implementation, we choose to compute it by
multi-sampled ray casting. Each ray traverses the discretized scene by a digital differential
analyzer (DDA). For each encountered voxel, we compute and accumulate its outgoing radiance $L_o^{k}$ with the primitive coverage
$c_k$ (\autoref{eq:primitive_coverage}). Note that the traversal can be in arbitrary order, which enables possible rasterization-based approaches.

To utilize the different LoDs included in the hierarchy, we can enhance the above procedure by associating each ray with a cone aperture that covers the pixel
footprint, akin to ray differentials \cite{igehy1999tracing}. During the traversal of each ray, we determine the LoD level by the cross section size of
the cone. In our current implementation, we switch to a coarser LoD only at the boundary of the coarser voxels for efficiency. A continuous LoD blending scheme is
possible but more costly. \autoref{algo:rendering} provides pseudocode for the rendering procedure.

\begin{algorithm}[t]
	\caption{\label{algo:rendering}
		Rendering a scene aggregate $S$ given a cone with center ray $r$ and aperture $\theta$.
	}
	\begin{algorithmic}[1]
		\Function{Render}{$S$, $r$, $\theta$}
		\State $t, t_{\text{max}}$ = $\text{intersect}(S\text{.bound}, r)$
		\State $s$ = $\text{base voxel size of } S$
		\State $\omega$ = $\text{direction of } r$
		\State $I = 0$, $\hat{V}_b = 1$
		\While {$t < t_{\text{max}}$}
			\State level = $\text{floor}(\text{log2}(\tan(0.5\cdot\theta$)$\cdot t / s))$
			\State $\mathrm{v}_k, \triangle t = \text{DDAToNextVoxel}(r, t, \text{S[level].grid})$
			\If {$\mathrm{v}_k$ is on the boundary of $S$}
				\State $\hat{V}_b$ = EvaluateABV($t$, $\omega$)
			\EndIf
			\If {$\text{intersect}(B_k, r)$}
				\State $L_o = 0$
				\For {$i = 1 \text{ to } m$}
					\State $L_o \pluseq (1/m)\cdot \text{MISDirectLighting}(\mathrm{v}_k)$
					\State \Comment{incident AIV by $\hat{V}_i = \text{EvaluateAIV}(\mathrm{v}_k, \omega_i)$}
				\EndFor
				\State $\hat{V}_o = \text{EvaluateAIV}(\mathrm{v}_k, -\omega)$
				\State $I \pluseq L_o \cdot \hat{V}_o \cdot |A_k| \,/\, |B_k|_{\omega}$
				\State \Comment{See supplemental document}
			\EndIf
			\State $t \pluseq \triangle t$
		\EndWhile
		\State $\hat{L}_b = \text{EvaluateBackground}(-\omega)$
		\State $I \pluseq \hat{L}_b \cdot \hat{V}_b$
		\State \Return{$I$}
		\EndFunction
	\end{algorithmic}
\end{algorithm}

To enable next event estimation (NEE) with multiple importance sampling (\autoref{algo:rendering}, line 15), we develop a straightforward importance sampling
routine for our factored ABSDF $\hat{f}_\mathrm{novis}$ (\autoref{eq:absdf_distrib}) as follows:
\begin{enumerate}
    \item Pick one component between the specular and diffuse components. This can be done simply by uniform sampling.
    \item For the specular component, we first pick one convolved lobe from \autoref{eq:sggx_conv_2} based on the lobe weights $w_j$. Then we sample the corresponding
          SGGX distribution.
    \item For the diffuse component, note that we cannot directly sample a convolved lobe from \autoref{eq:diffuse_sg} because $\kappa_{ij}$ cannot be determined
          without $\omega_i$. Therefore, we resort to a simple strategy by assuming a fixed $\kappa$ during importance sampling. The rest is
          similar to the specular case: we pick one convolved lobe (but with the fixed $\kappa$) and sample the corresponding SGGX distribution.
\end{enumerate}
The corresponding PDF computation is also straightforward. The sample budget for NEE is decoupled from the ray casting sample budget.

\section{Results and Discussion} \label{sec:results}
In this section, we provide rendering results produced by our scene aggregation pipeline and detailed comparison to existing techniques. We implement our method
in a custom CPU renderer using Embree \citep{wald2014embree} as the ray tracing backend for precomputation and reference generation. The sparse hierarchical
data structure is implemented using OpenVDB/NanoVDB \citep{museth2013vdb, museth2021nanovdb}. All timings are measured on a desktop machine with an Intel i9-13900K
CPU and 64 GB of main memory. Unless otherwise stated, we use path-traced images with direct illumination as reference.

We compare our method to three current state-of-the-art methods: the hybrid mesh-volume LoD method (HybridLoD) \citep{loubet2017hybrid}, the non-exponential
transmittance volumetric model (NonExp) \citep{vicini2021non}, and the deep appearance prefiltering (DAP) \citep{bako2023deep}. For HybridLoD, we use the official
implementation provided by authors with modifications for asset loading purposes. For NonExp, we re-implemented the method based on the paper as the source code is
not available. For DAP, we used the authors' pre-trained results as training is prohibitively expensive and requires a GPU cluster. We provide different images
from references and provide root mean squared error (RMSE) to evaluate the quality of each method. In addition, we provide a supplementary video with varying
magnification levels, camera rotation, and lighting conditions to demonstrate the temporal stability of our method.

\begin{table*}[tb]
	\centering
	\caption{Scene Configuration. For all scenes: Surface NDF mixture component count is set to $k\leq4$; AIV is recorded at $32^2$ angular resolution;
	ABV is recorded at  $64^2$ angular resolution. At most top $10\%$ basis coefficients are kept. More are truncated as long as the reconstruction preserves
	more than $95\%$ accuracy. \footnotesize{*Instanced triangle count.} \footnotesize{${}^{\dagger}$This scene is composed of instances of multiple aggregated
	objects. We only report the largest object.}}
	\begin{tabular}{r|r|r|r|r|r|r}
		\Xhline{1pt}
		\textbf{Scene} & \multicolumn{2}{c|}{\textbf{Original}} & \multicolumn{4}{c}{\textbf{Ours}} \\
		\cline{2-7}
							& \#Tris 			& Memory 	& Max Res. 	& Occupancy & Total Mem. & Precomp. Time \\
		\hline
		\emph{Helmet} 		& 15K 				& 7.7 MB 	& $256^3$ 	& $2.36\%$ 	& 155.1 MB 	& 1583 sec \\
		\emph{Chandelier} 	& 106K 				& 11.7 MB 	& $256^3$ 	& $1.39\%$ 	& 180.6 MB 	& 748 sec \\
		\emph{Tower} 		& 453K 				& 45.7 MB 	& $256^3$ 	& $1.66\%$ 	& 255.8 MB 	& 1108 sec \\
		\emph{Palm} 		& 2.2M 				& 349.6 MB 	& $256^3$ 	& $1.25\%$ 	& 216.2 MB	& 1345 sec \\
		\emph{Oleander} 	& 2.7M 				& 398.8 MB 	& $256^3$ 	& $3.71\%$	& 327.6 MB 	& 3537 sec \\
		\emph{Coral Reef} 	& 4.1M 				& 513.0 MB 	& $256^3$ 	& $3.18\%$	& 209.5 MB 	& 1624 sec \\
		\emph{Forest} 		& 16.2M (*175.2M) 	& 2.5 GB 	& $512^3$	& $2.86\%$	& 1.06 GB	& 16064 sec \\
		\rev{\emph{Metropolis}}	& \rev{88.6M (*301.8M)}	& \rev{46.9 GB}	& \rev{${}^{\dagger}256^3$}	& \rev{${}^{\dagger}7.01\%$}	& \rev{5.33 GB}	& \rev{18254 sec}\\
		\Xhline{1pt}
	\end{tabular}
	\label{tab:scenes}
\end{table*}

\begin{figure*}[ht!]
	\newlength{\lenRenderingComparisonSmall}
	\setlength{\lenRenderingComparisonSmall}{0.71in}
	\newlength{\lenRenderingComparisonBig}
	\setlength{\lenRenderingComparisonBig}{1.45in}
    \addtolength{\tabcolsep}{-4pt}
    \renewcommand{\arraystretch}{0.5}
    \centering
    \begin{tabular}{ccccccccccc}
        &
        \multicolumn{2}{c}{\textsf{PT Ref. (High Res.)}}
        &
        \textsf{PT Ref.}
        &
        \textsf{Ours} & \textsf{Difference}
        &
        \textsf{HybridLoD} & \textsf{Difference}
        &
        \textsf{NonExp} & \textsf{Difference}
        &
        \\
        &&&&
        &
        \frame{\begin{overpic}[width=\lenRenderingComparisonSmall]{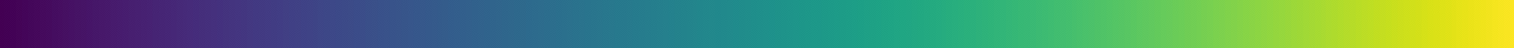}
            \put(-19, 0){\normalsize \small{0.0}}
            \put(101, 0){\normalsize \small{1.0}}
        \end{overpic}}
        &&
        \frame{\begin{overpic}[width=\lenRenderingComparisonSmall]{imgs/colorbar_hori.png}
            \put(-19, 0){\normalsize \small{0.0}}
            \put(101, 0){\normalsize \small{1.0}}
        \end{overpic}}
        &&
        \frame{\begin{overpic}[width=\lenRenderingComparisonSmall]{imgs/colorbar_hori.png}
            \put(-19, 0){\normalsize \small{0.0}}
            \put(101, 0){\normalsize \small{1.0}}
        \end{overpic}}
        &
        \\
        &
        \multicolumn{2}{c}{}
        &
        \frame{\includegraphics[height=\lenRenderingComparisonSmall]{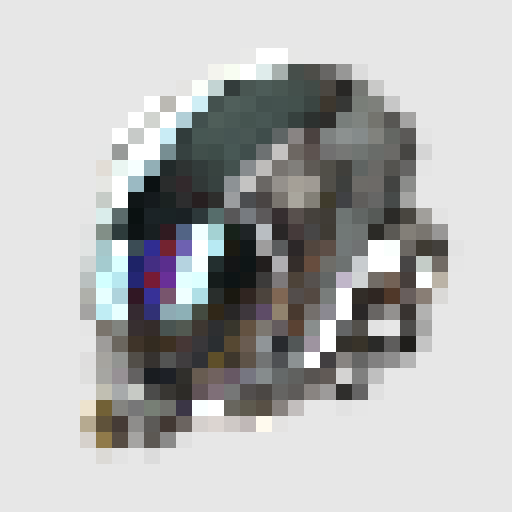}}
        &
        \frame{\includegraphics[height=\lenRenderingComparisonSmall]{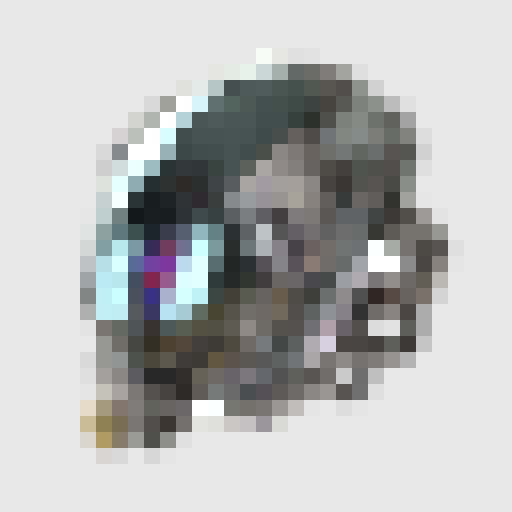}}
        &
        \frame{\includegraphics[height=\lenRenderingComparisonSmall]{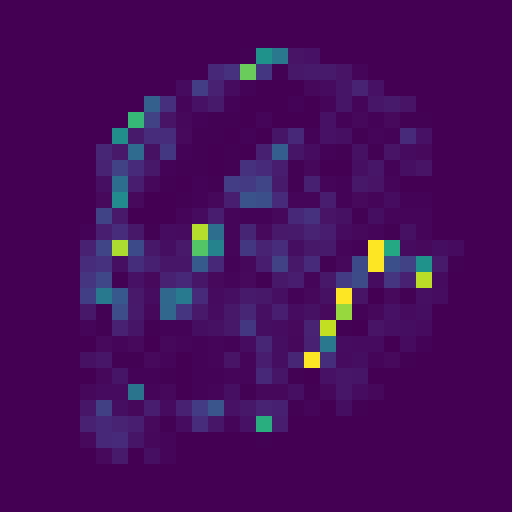}}
        &
        \frame{\includegraphics[height=\lenRenderingComparisonSmall]{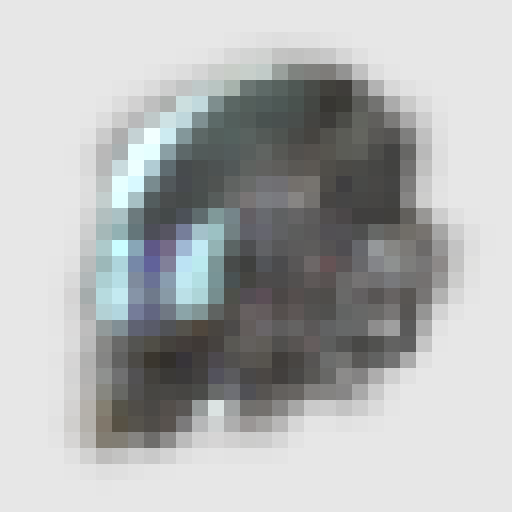}}
        &
        \frame{\includegraphics[height=\lenRenderingComparisonSmall]{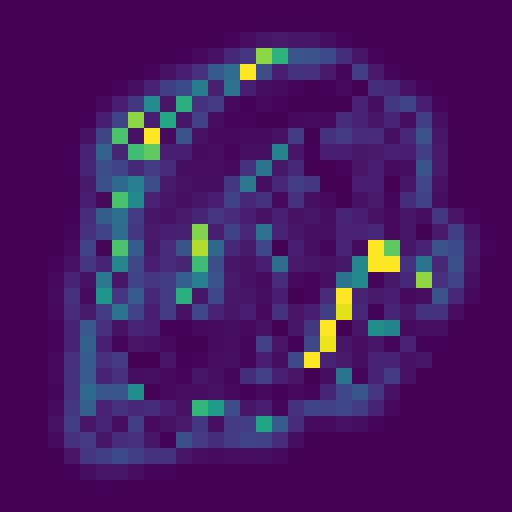}}
        &
        \frame{\includegraphics[height=\lenRenderingComparisonSmall]{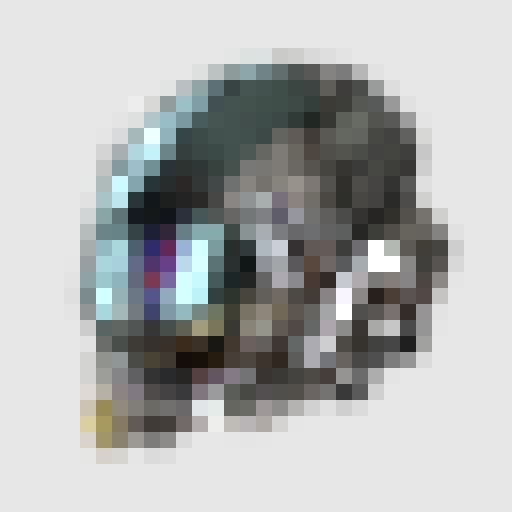}}
        &
        \frame{\includegraphics[height=\lenRenderingComparisonSmall]{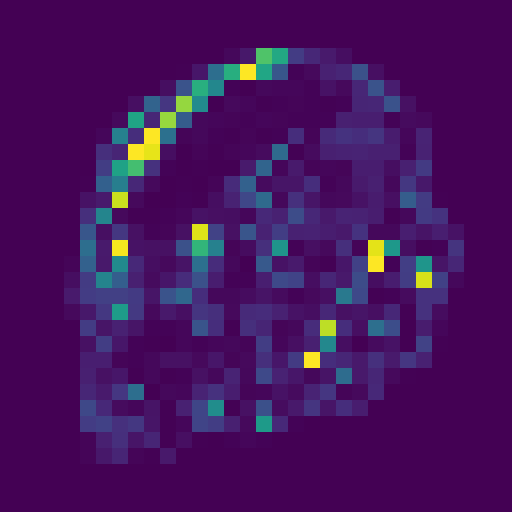}}
        &
        \raisebox{35pt}{\rotatebox{-90}{$32^3$}}
        \\
        \multirow[t]{2}{*}{\raisebox{42pt}{\rotatebox{90}{\emph{Helmet}}}}
        &
        \multicolumn{2}{c}{\multirow[t]{2}{*}{\frame{\includegraphics[height=\lenRenderingComparisonBig]{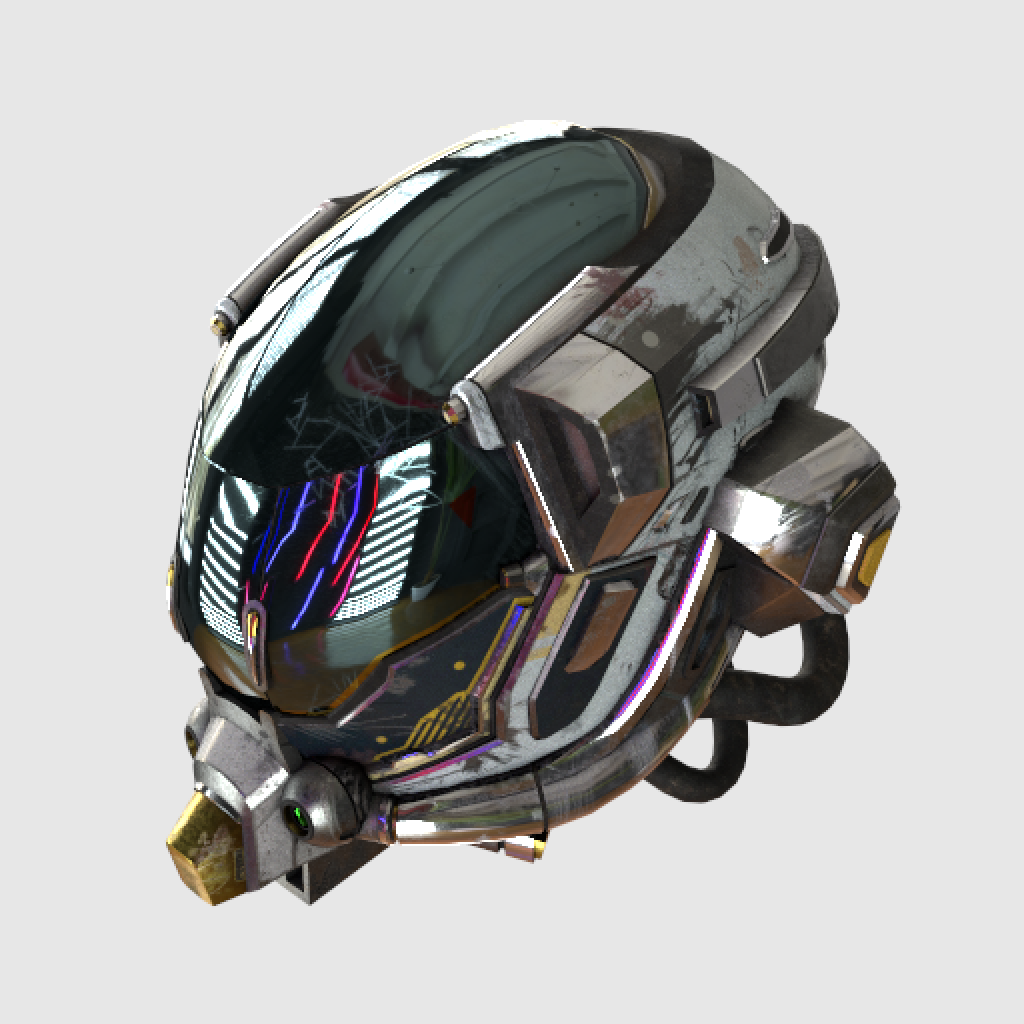}}}}
        &
        \frame{\includegraphics[height=\lenRenderingComparisonSmall]{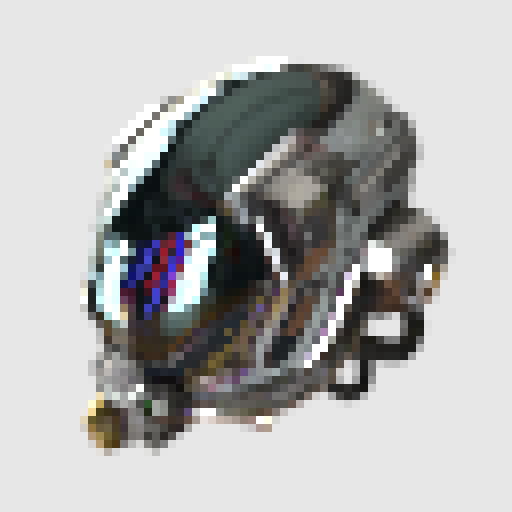}}
        &
        \frame{\includegraphics[height=\lenRenderingComparisonSmall]{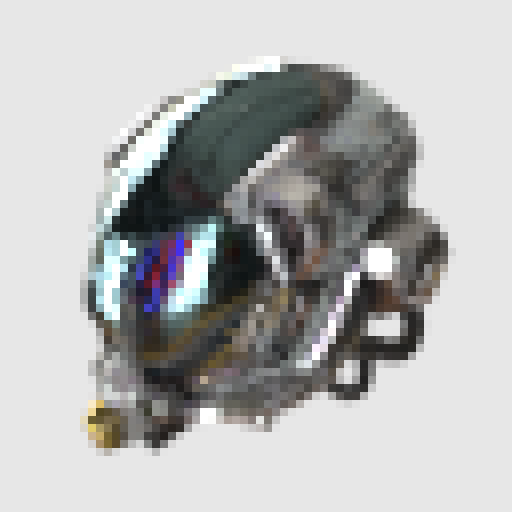}}
        &
        \frame{\includegraphics[height=\lenRenderingComparisonSmall]{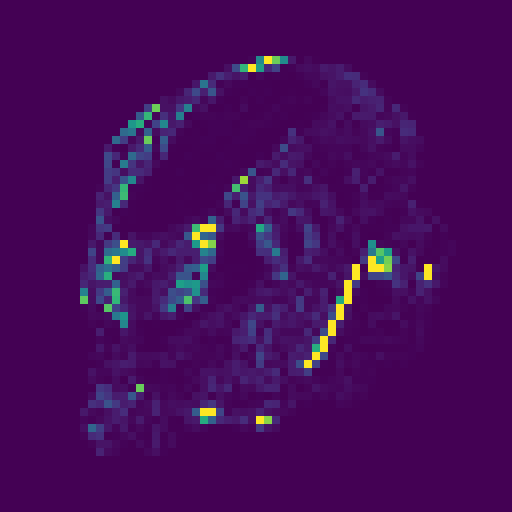}}
        &
        \frame{\includegraphics[height=\lenRenderingComparisonSmall]{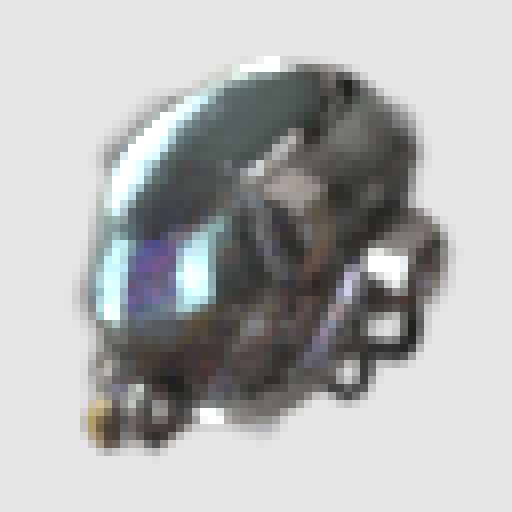}}
        &
        \frame{\includegraphics[height=\lenRenderingComparisonSmall]{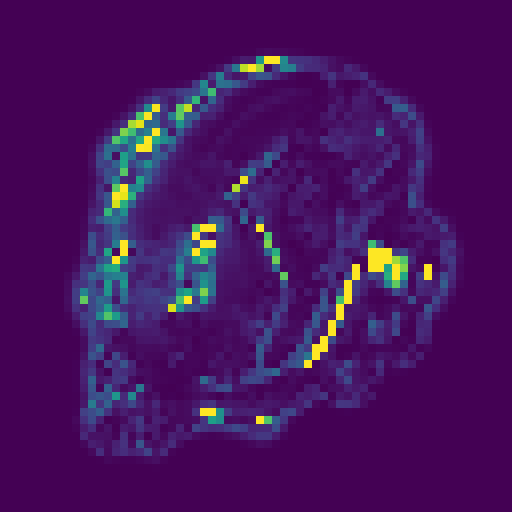}}
        &
        \frame{\includegraphics[height=\lenRenderingComparisonSmall]{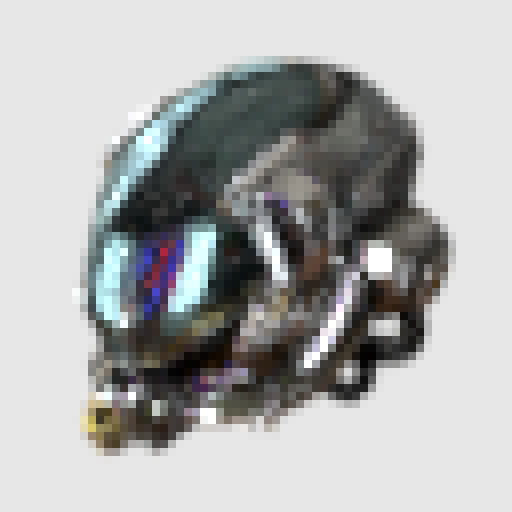}}
        &
        \frame{\includegraphics[height=\lenRenderingComparisonSmall]{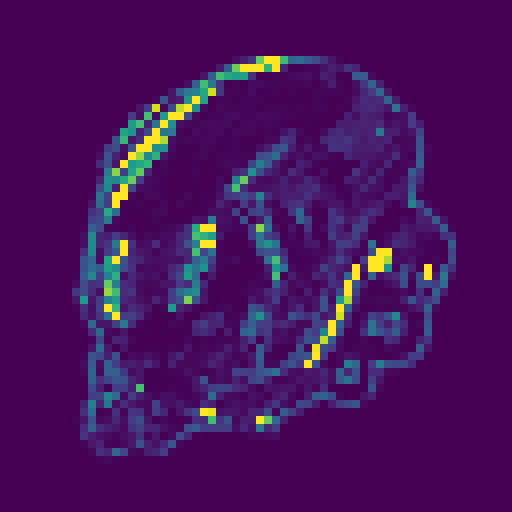}}
        &
        \raisebox{35pt}{\rotatebox{-90}{$64^3$}}
        \\
        &
        \multicolumn{2}{c}{\textsf{RMSE:}} & & & \textbf{0.165} / \textbf{0.158} & & {0.266} / {0.302} & & {0.203} / {0.238} &
        \\
        &
        \multicolumn{2}{c}{}
        &
        \frame{\includegraphics[height=\lenRenderingComparisonSmall]{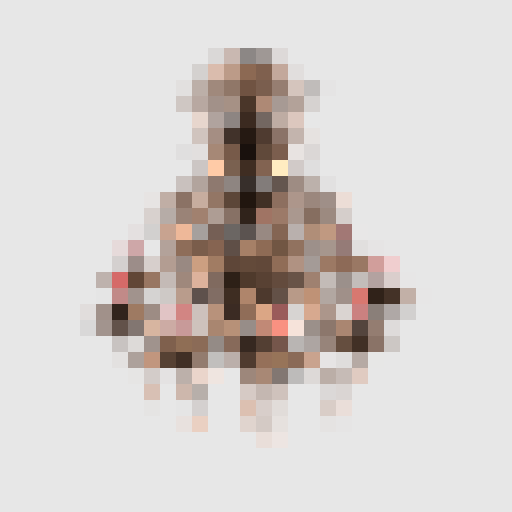}}
        &
        \frame{\includegraphics[height=\lenRenderingComparisonSmall]{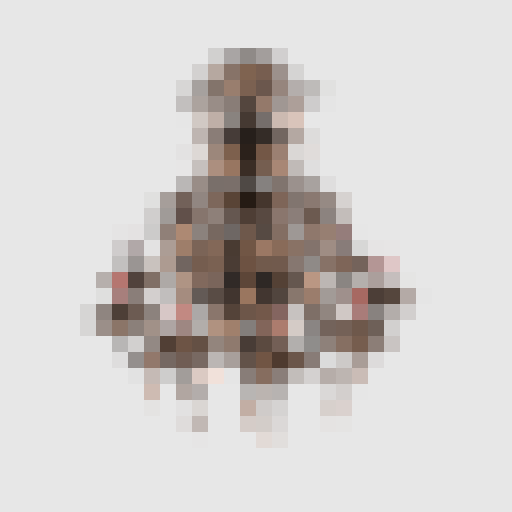}}
        &
        \frame{\includegraphics[height=\lenRenderingComparisonSmall]{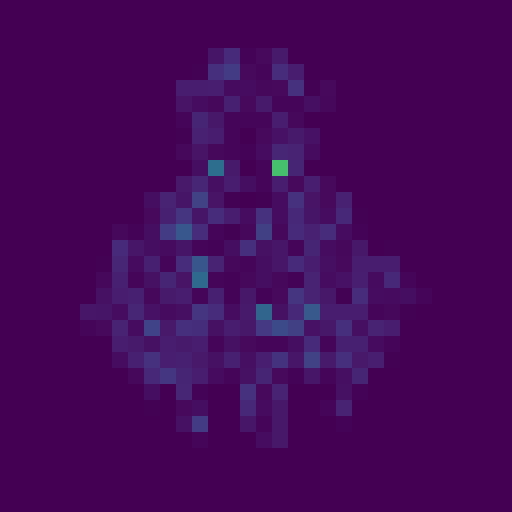}}
        &
        \frame{\includegraphics[height=\lenRenderingComparisonSmall]{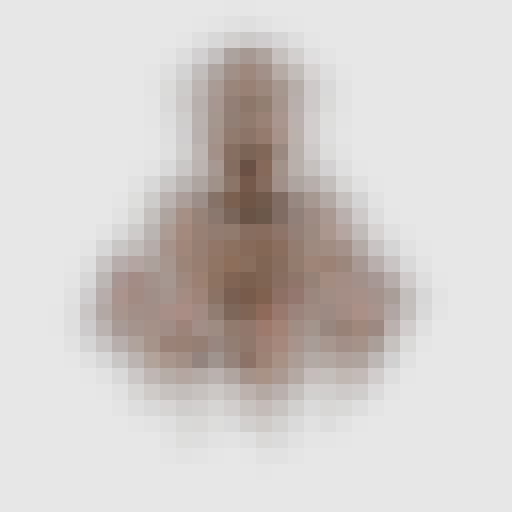}}
        &
        \frame{\includegraphics[height=\lenRenderingComparisonSmall]{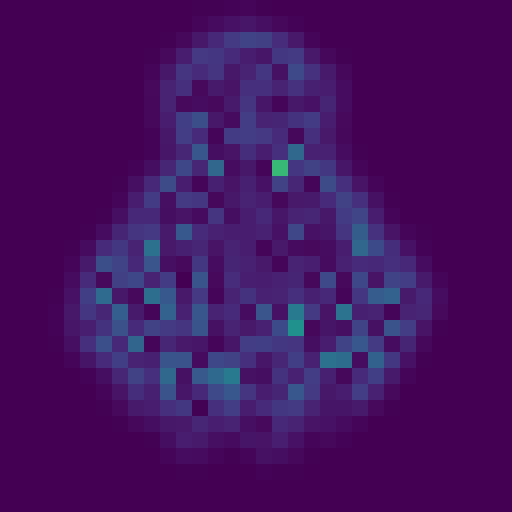}}
        &
        \frame{\includegraphics[height=\lenRenderingComparisonSmall]{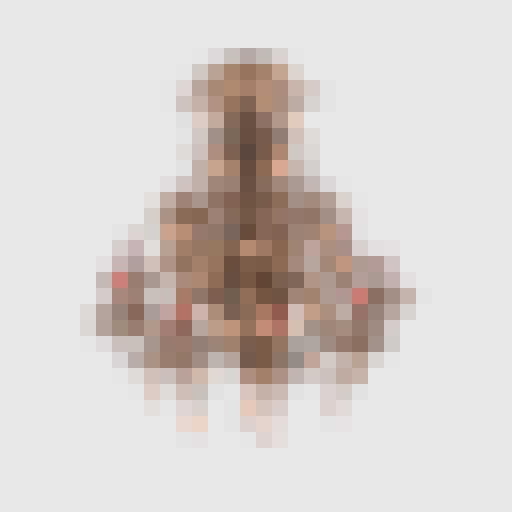}}
        &
        \frame{\includegraphics[height=\lenRenderingComparisonSmall]{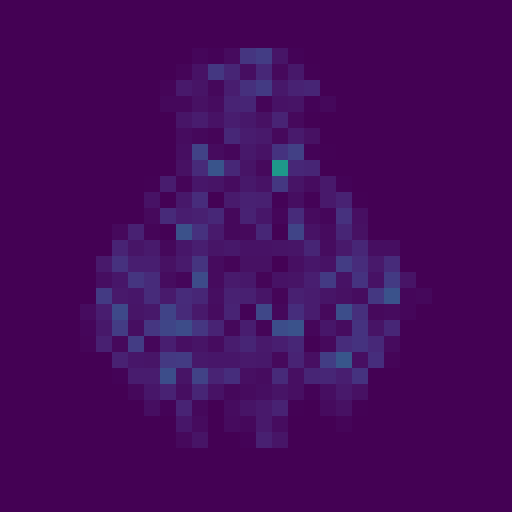}}
        &
        \raisebox{35pt}{\rotatebox{-90}{$32^3$}}
        \\
        \multirow[t]{2}{*}{\raisebox{42pt}{\rotatebox{90}{\emph{Chandelier}}}}
        &
        \multicolumn{2}{c}{\multirow[t]{2}{*}{\frame{\includegraphics[height=\lenRenderingComparisonBig]{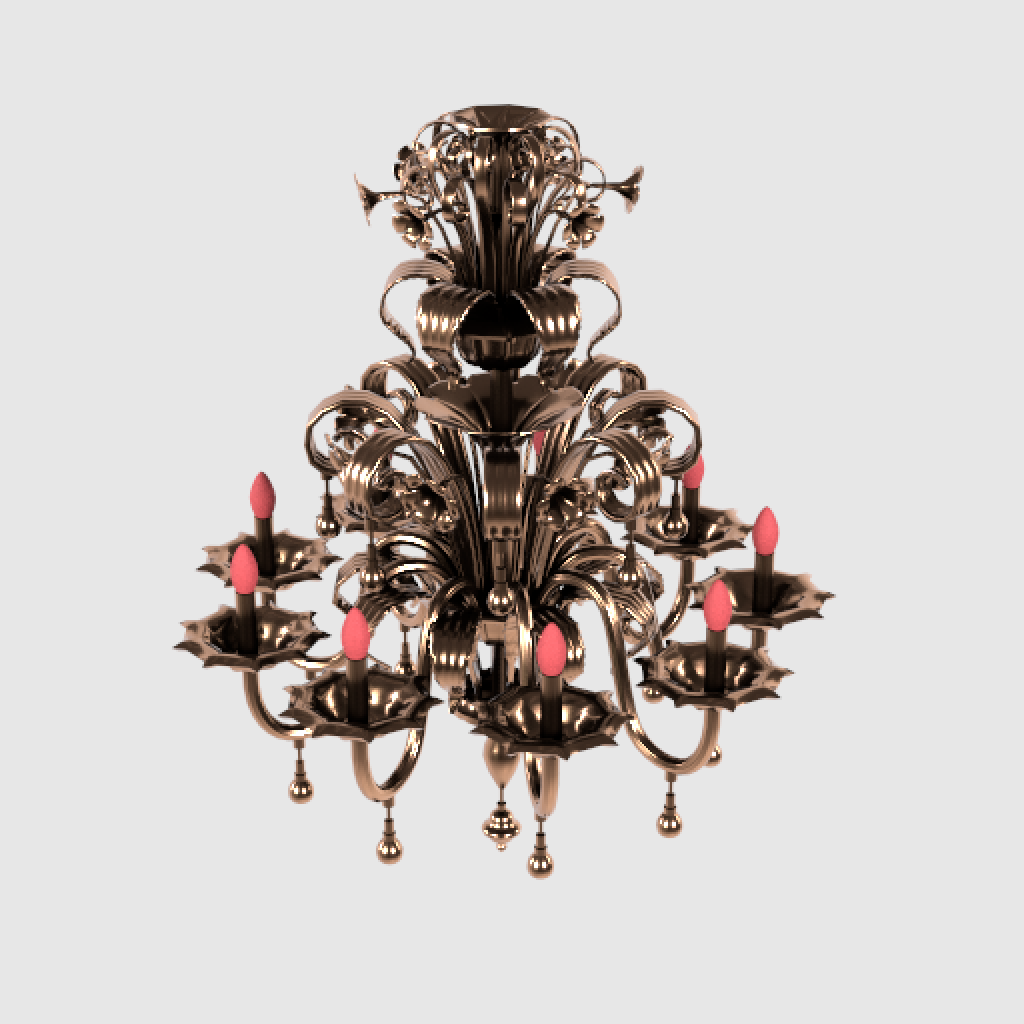}}}}
        &
        \frame{\includegraphics[height=\lenRenderingComparisonSmall]{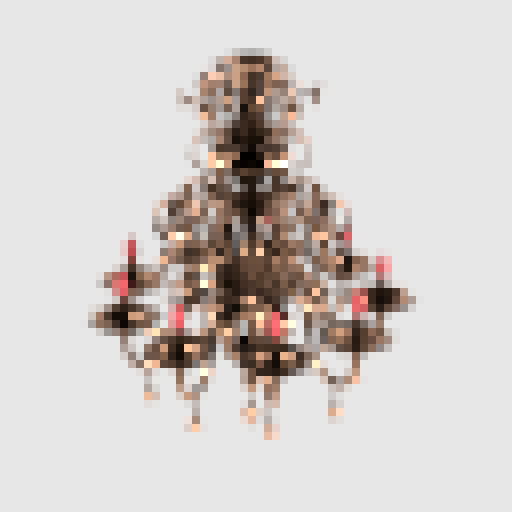}}
        &
        \frame{\includegraphics[height=\lenRenderingComparisonSmall]{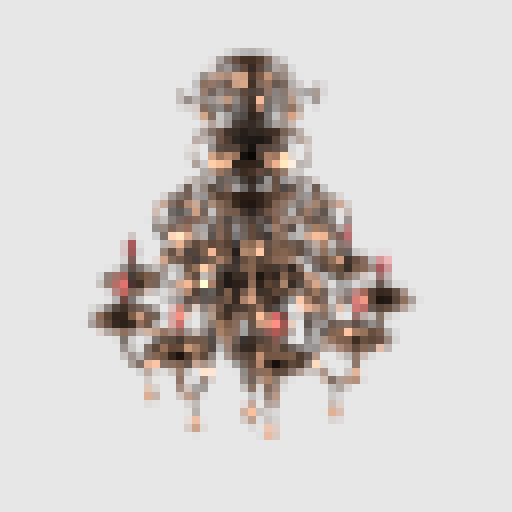}}
        &
        \frame{\includegraphics[height=\lenRenderingComparisonSmall]{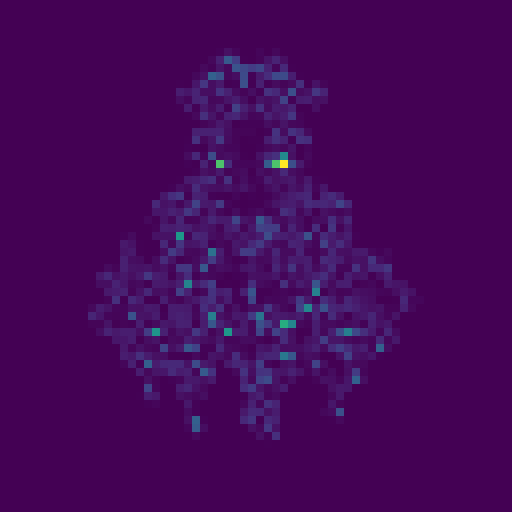}}
        &
        \frame{\includegraphics[height=\lenRenderingComparisonSmall]{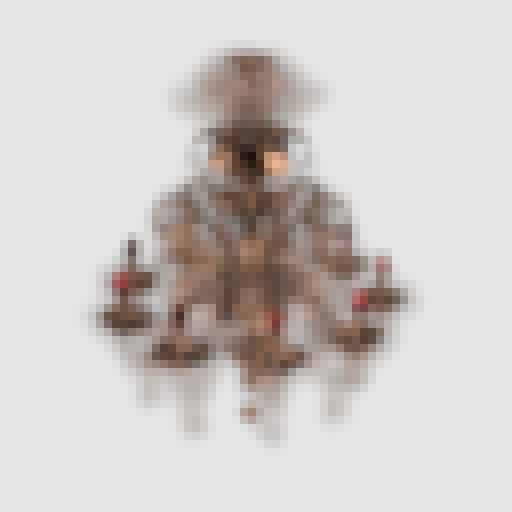}}
        &
        \frame{\includegraphics[height=\lenRenderingComparisonSmall]{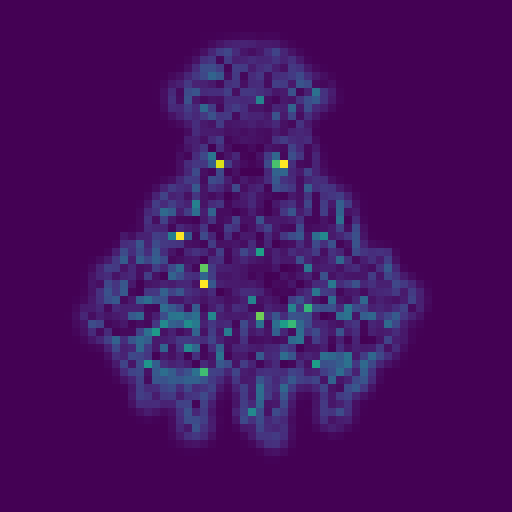}}
        &
        \frame{\includegraphics[height=\lenRenderingComparisonSmall]{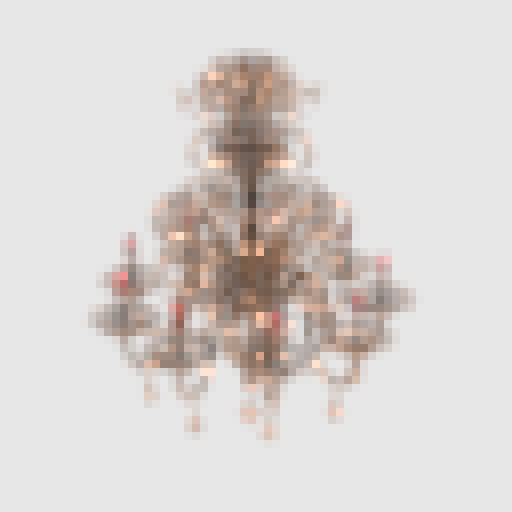}}
        &
        \frame{\includegraphics[height=\lenRenderingComparisonSmall]{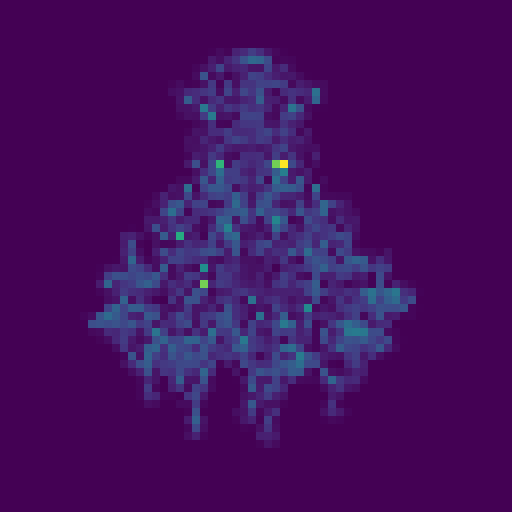}}
        &
        \raisebox{35pt}{\rotatebox{-90}{$64^3$}}
        \\
        &
        \multicolumn{2}{c}{\textsf{RMSE:}} & & & \textbf{0.059} / \textbf{0.075} & & {0.098} / {0.112} & & {0.064} / {0.099} &
        \\
        &
        \multicolumn{2}{c}{}
        &
        \frame{\includegraphics[height=\lenRenderingComparisonSmall]{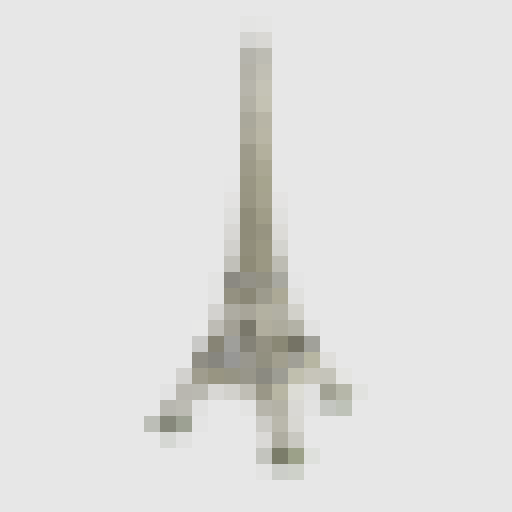}}
        &
        \frame{\includegraphics[height=\lenRenderingComparisonSmall]{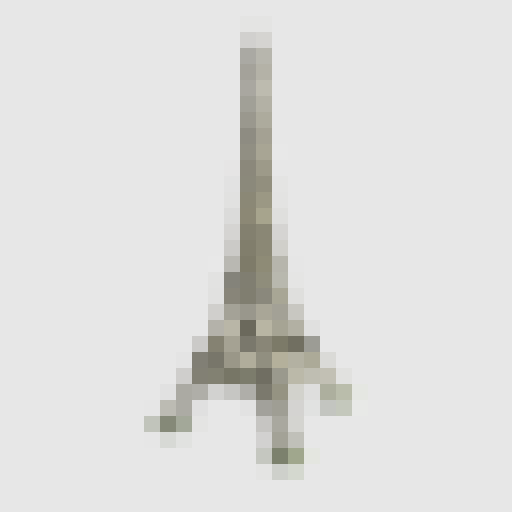}}
        &
        \frame{\includegraphics[height=\lenRenderingComparisonSmall]{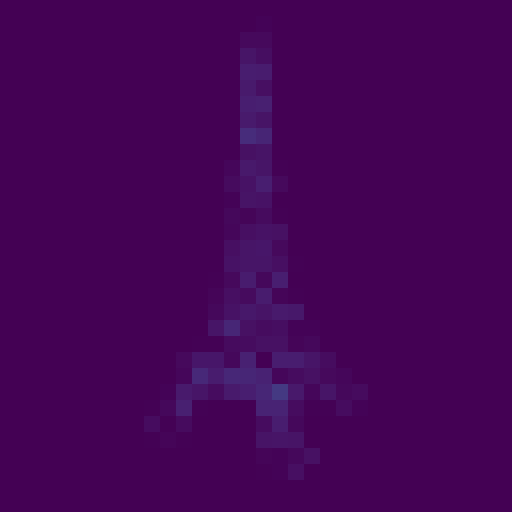}}
        &
        \frame{\includegraphics[height=\lenRenderingComparisonSmall]{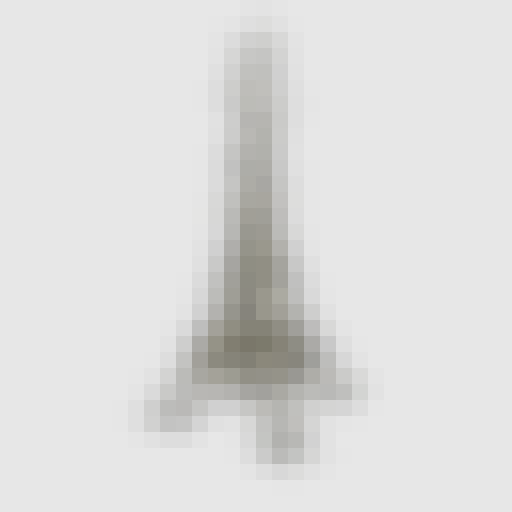}}
        &
        \frame{\includegraphics[height=\lenRenderingComparisonSmall]{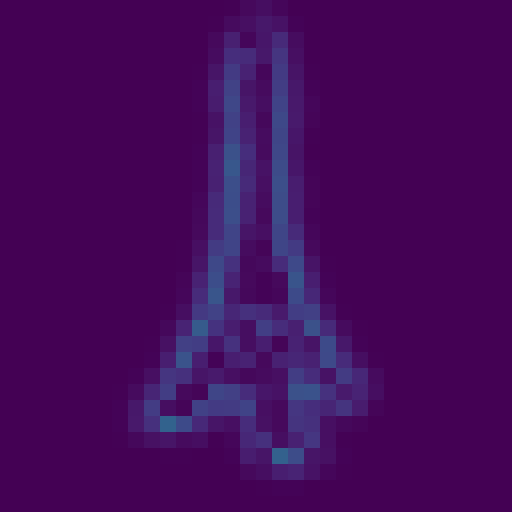}}
        &
        \frame{\includegraphics[height=\lenRenderingComparisonSmall]{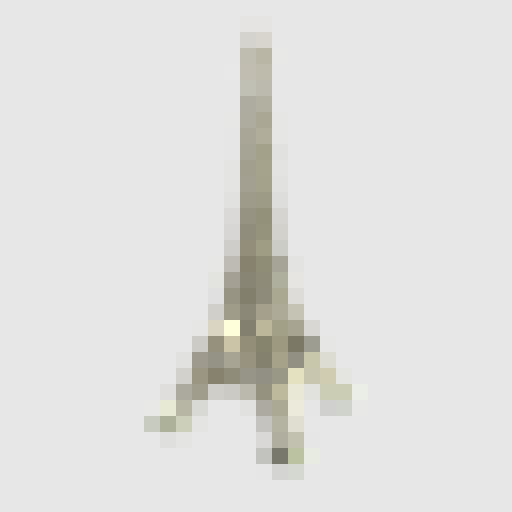}}
        &
        \frame{\includegraphics[height=\lenRenderingComparisonSmall]{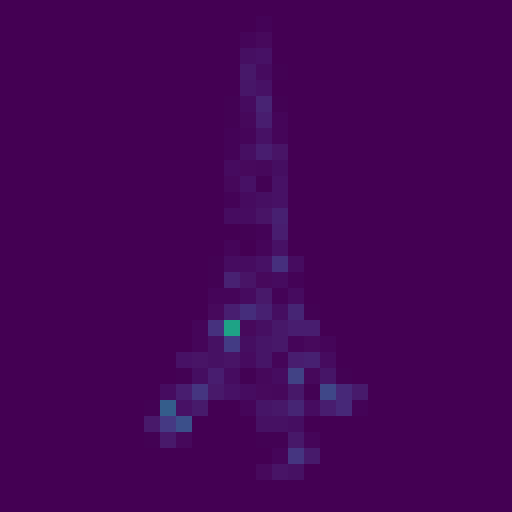}}
        &
        \raisebox{35pt}{\rotatebox{-90}{$32^3$}}
        \\
        \multirow[t]{2}{*}{\raisebox{42pt}{\rotatebox{90}{\emph{Tower}}}}
        &
        \multicolumn{2}{c}{\multirow[t]{2}{*}{\frame{\includegraphics[height=\lenRenderingComparisonBig]{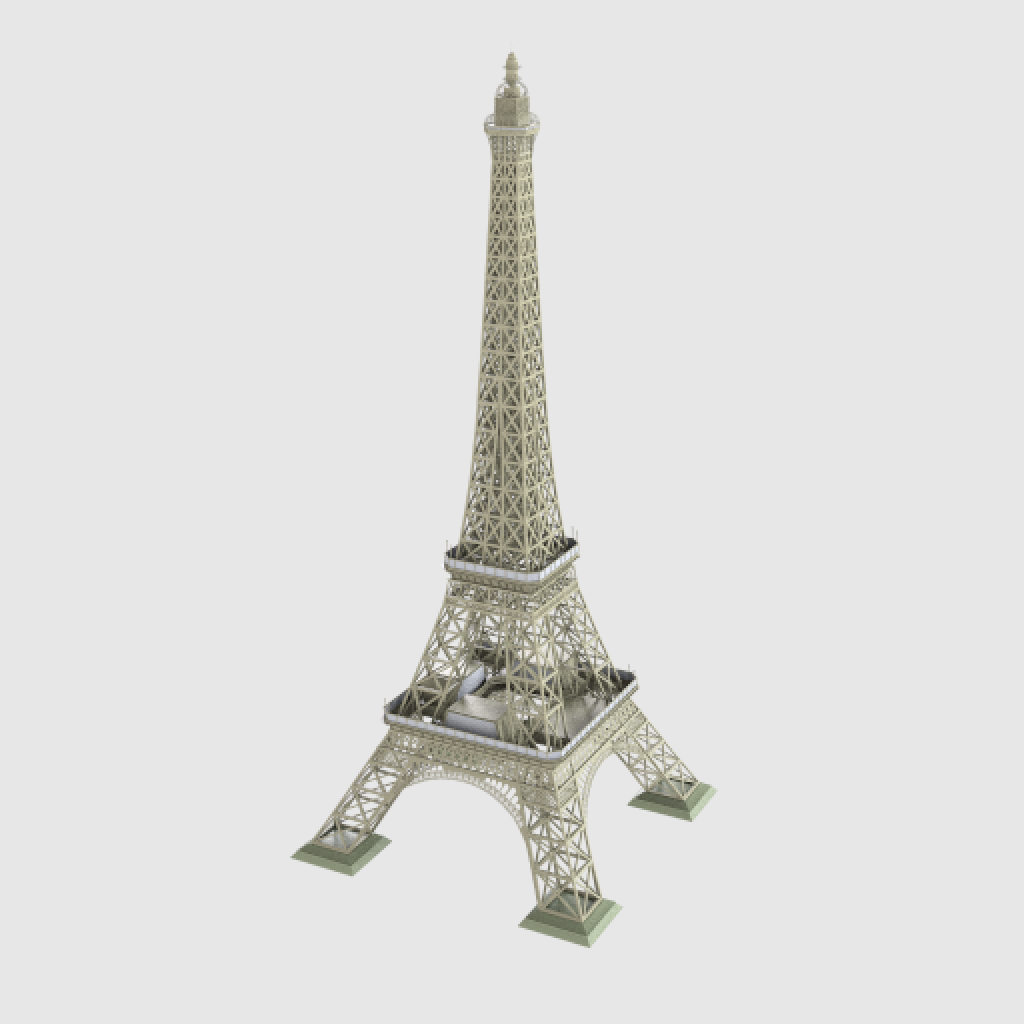}}}}
        &
        \frame{\includegraphics[height=\lenRenderingComparisonSmall]{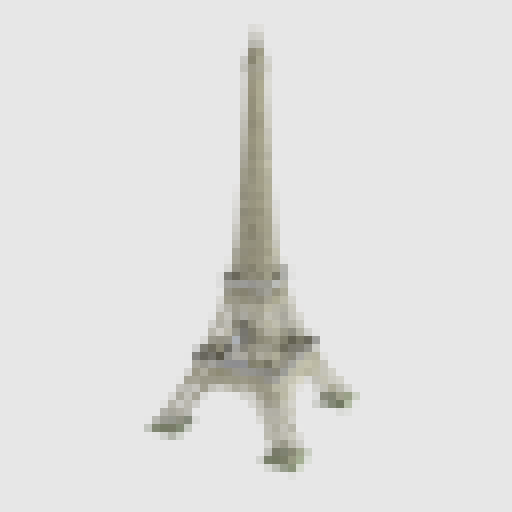}}
        &
        \frame{\includegraphics[height=\lenRenderingComparisonSmall]{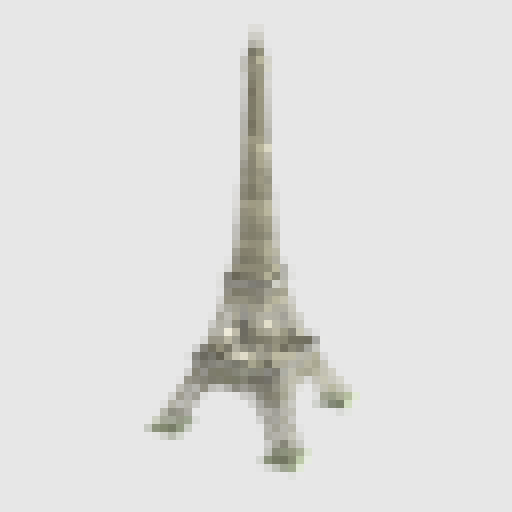}}
        &
        \frame{\includegraphics[height=\lenRenderingComparisonSmall]{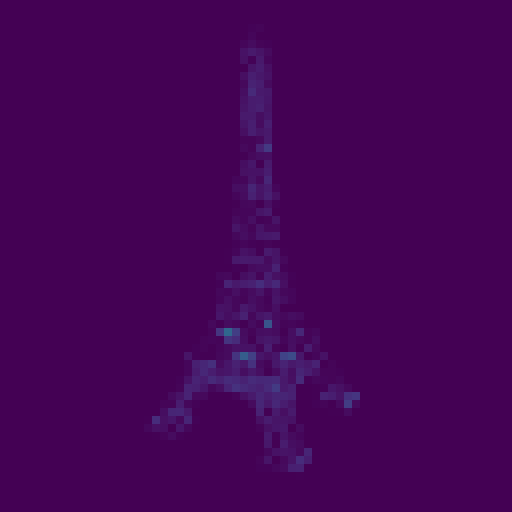}}
        &
        \frame{\includegraphics[height=\lenRenderingComparisonSmall]{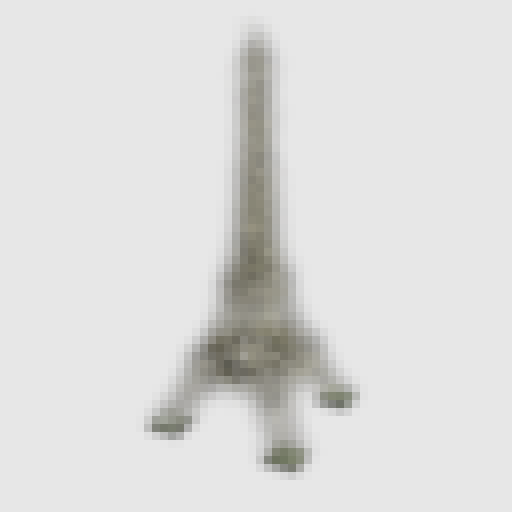}}
        &
        \frame{\includegraphics[height=\lenRenderingComparisonSmall]{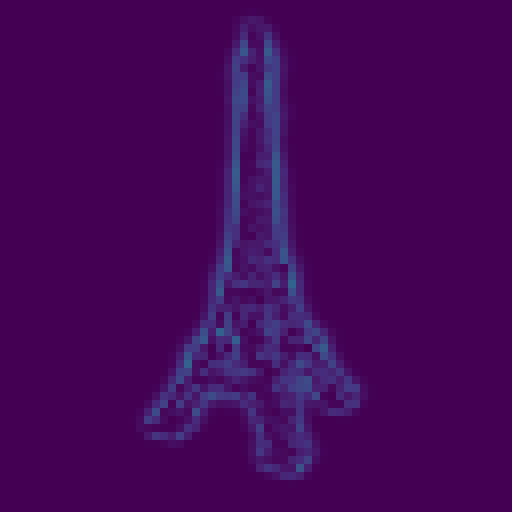}}
        &
        \frame{\includegraphics[height=\lenRenderingComparisonSmall]{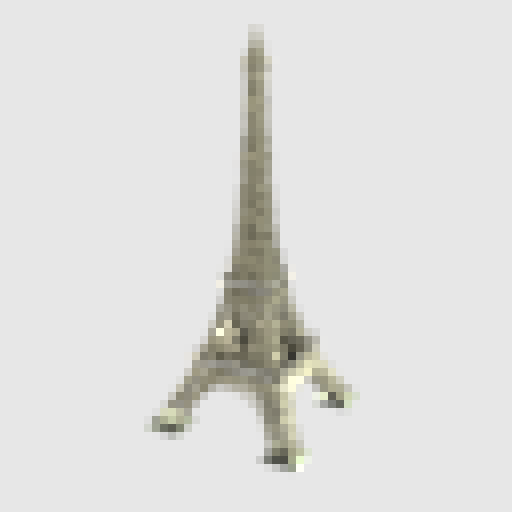}}
        &
        \frame{\includegraphics[height=\lenRenderingComparisonSmall]{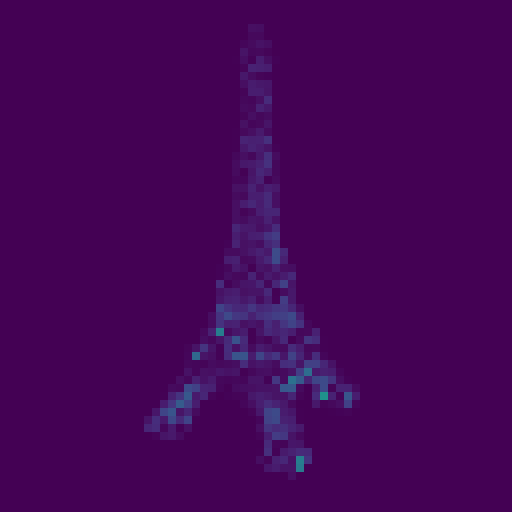}}
        &
        \raisebox{35pt}{\rotatebox{-90}{$64^3$}}
        \\
        &
        \multicolumn{2}{c}{\textsf{RMSE:}} & & & \textbf{0.020} / \textbf{0.022} & & {0.063} / {0.050} & & {0.033} / {0.038} &
        \\
        &
        \multicolumn{2}{c}{}
        &
        \frame{\includegraphics[height=\lenRenderingComparisonSmall]{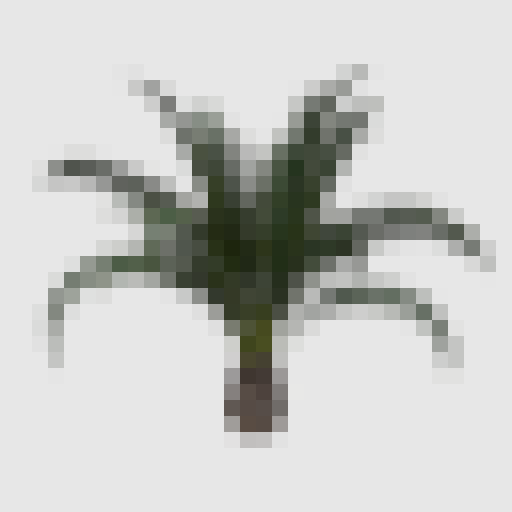}}
        &
        \frame{\includegraphics[height=\lenRenderingComparisonSmall]{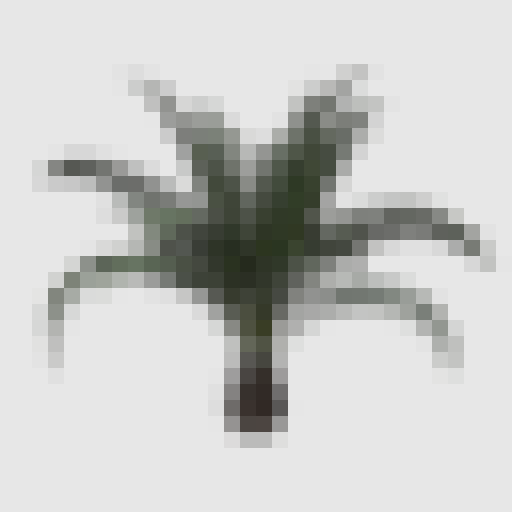}}
        &
        \frame{\includegraphics[height=\lenRenderingComparisonSmall]{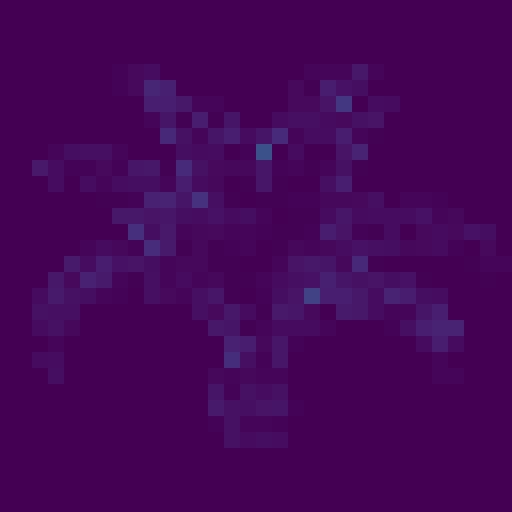}}
        &
        \frame{\includegraphics[height=\lenRenderingComparisonSmall]{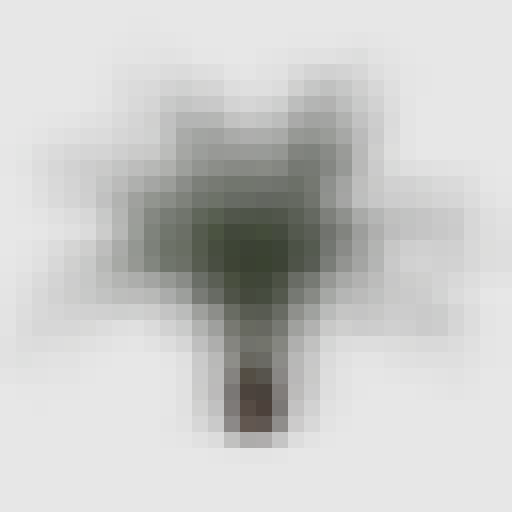}}
        &
        \frame{\includegraphics[height=\lenRenderingComparisonSmall]{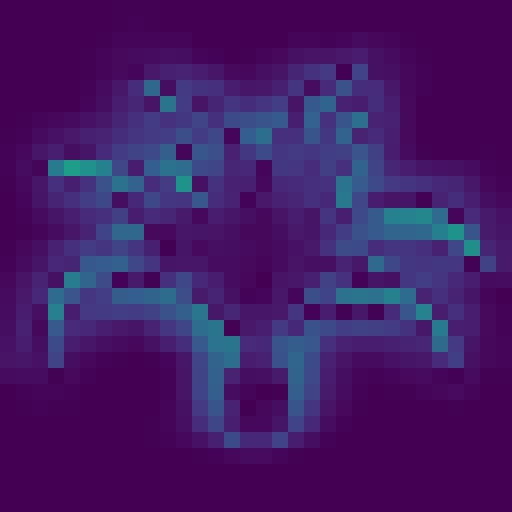}}
        &
        \frame{\includegraphics[height=\lenRenderingComparisonSmall]{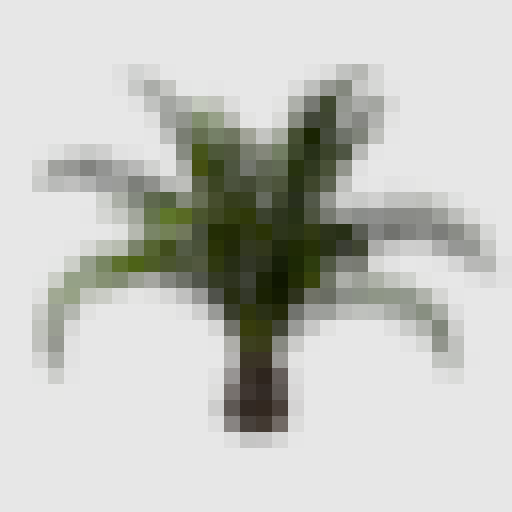}}
        &
        \frame{\includegraphics[height=\lenRenderingComparisonSmall]{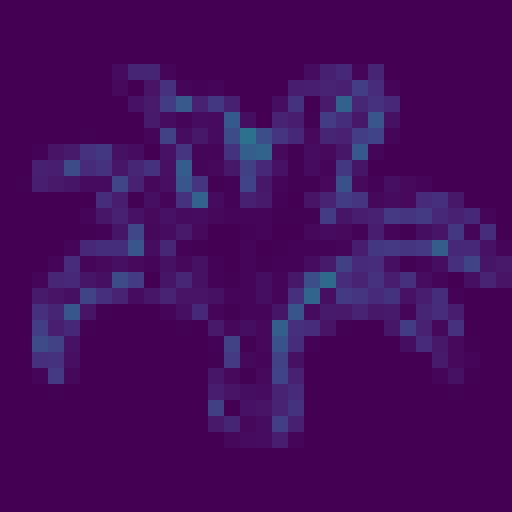}}
        &
        \raisebox{35pt}{\rotatebox{-90}{$32^3$}}
        \\
        \multirow[t]{2}{*}{\raisebox{42pt}{\rotatebox{90}{\emph{Palm}}}}
        &
        \multicolumn{2}{c}{\multirow[t]{2}{*}{\frame{\includegraphics[height=\lenRenderingComparisonBig]{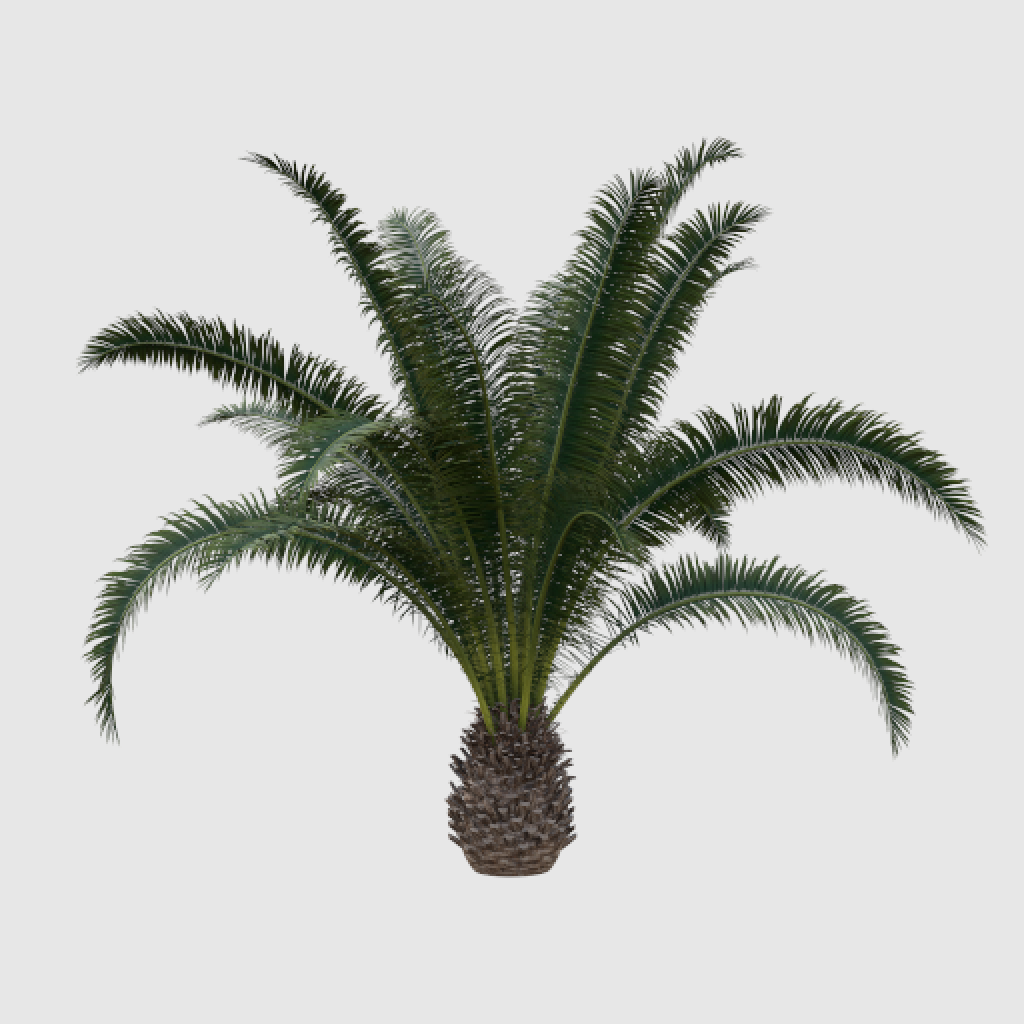}}}}
        &
        \frame{\includegraphics[height=\lenRenderingComparisonSmall]{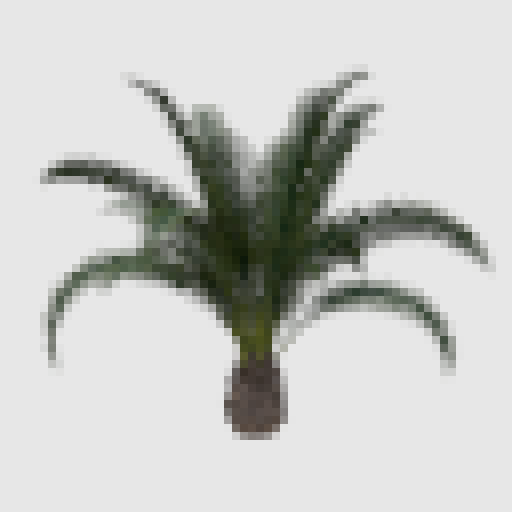}}
        &
        \frame{\includegraphics[height=\lenRenderingComparisonSmall]{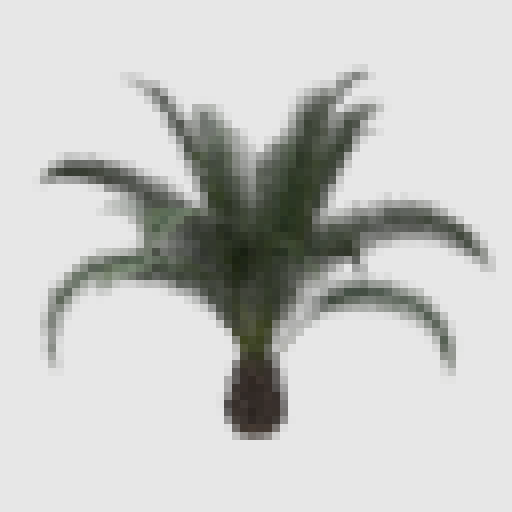}}
        &
        \frame{\includegraphics[height=\lenRenderingComparisonSmall]{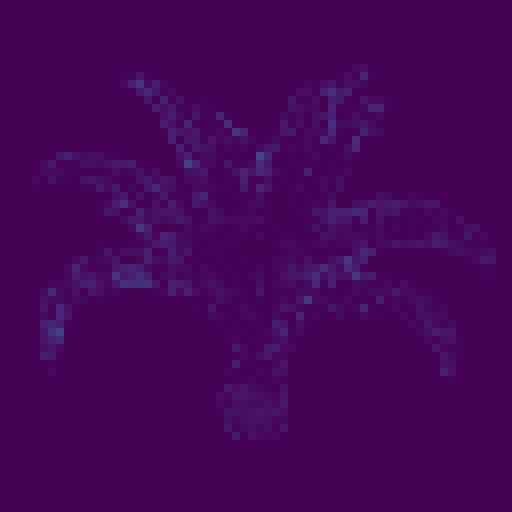}}
        &
        \frame{\includegraphics[height=\lenRenderingComparisonSmall]{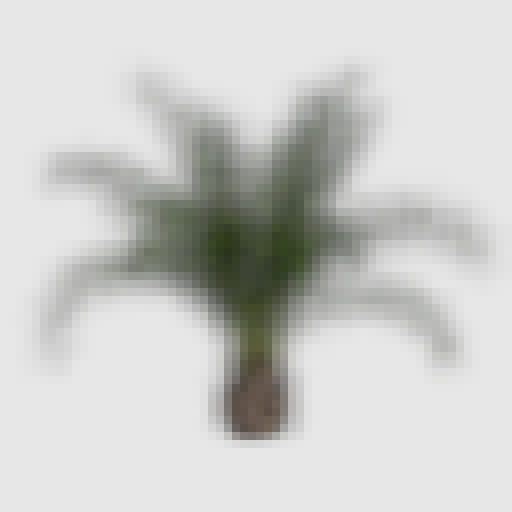}}
        &
        \frame{\includegraphics[height=\lenRenderingComparisonSmall]{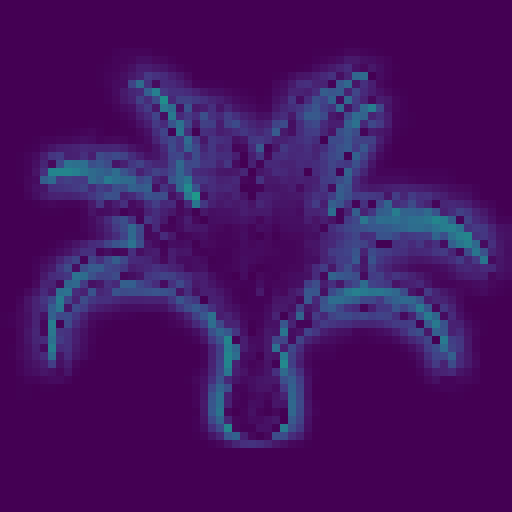}}
        &
        \frame{\includegraphics[height=\lenRenderingComparisonSmall]{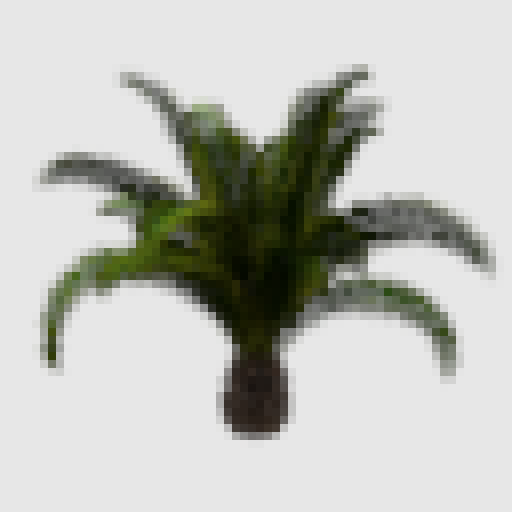}}
        &
        \frame{\includegraphics[height=\lenRenderingComparisonSmall]{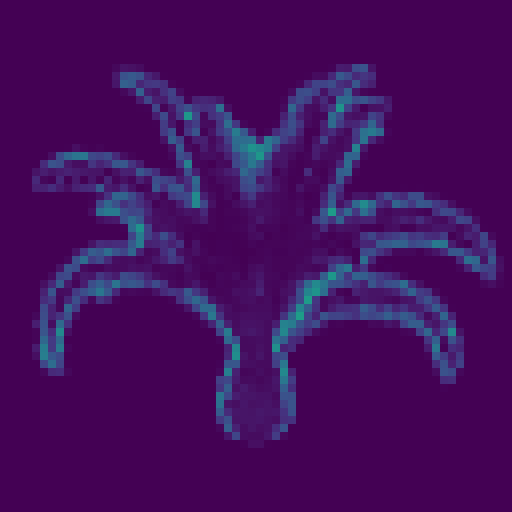}}
        &
        \raisebox{35pt}{\rotatebox{-90}{$64^3$}}
        \\
        &
        \multicolumn{2}{c}{\textsf{RMSE:}} & & & \textbf{0.029} / \textbf{0.024} & & {0.131} / {0.109} & & {0.070} / {0.098} &
        \\
        &
        \multicolumn{2}{c}{}
        &
        \frame{\includegraphics[height=\lenRenderingComparisonSmall]{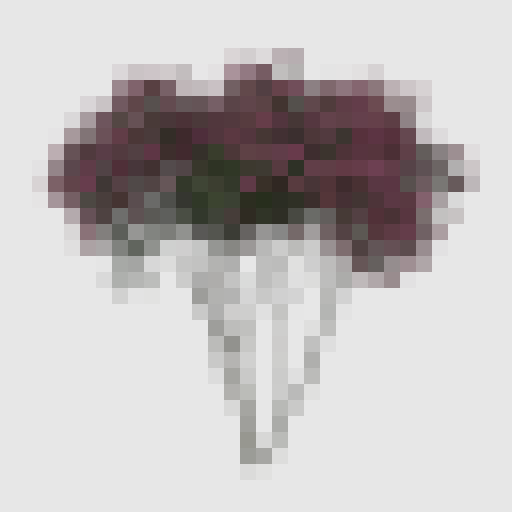}}
        &
        \frame{\includegraphics[height=\lenRenderingComparisonSmall]{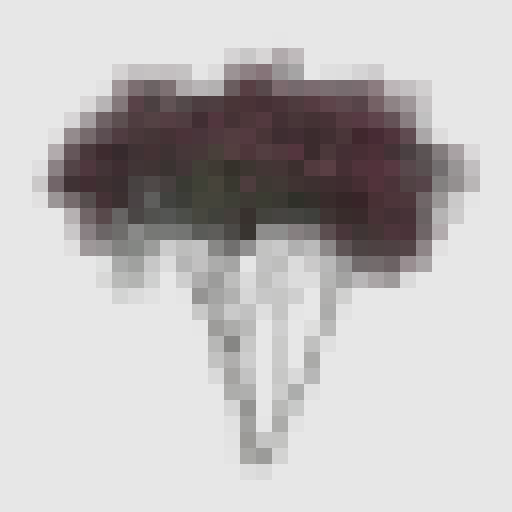}}
        &
        \frame{\includegraphics[height=\lenRenderingComparisonSmall]{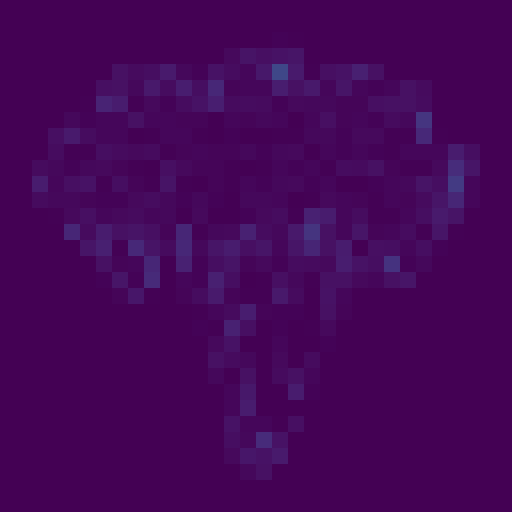}}
        &
        \frame{\includegraphics[height=\lenRenderingComparisonSmall]{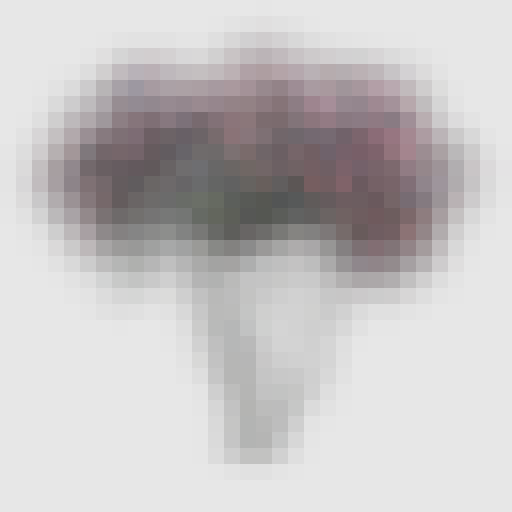}}
        &
        \frame{\includegraphics[height=\lenRenderingComparisonSmall]{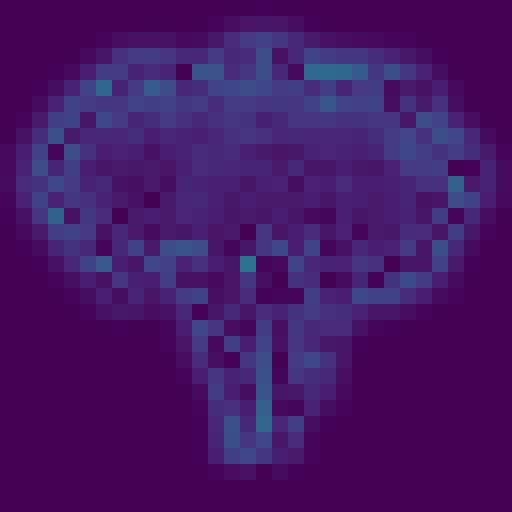}}
        &
        \frame{\includegraphics[height=\lenRenderingComparisonSmall]{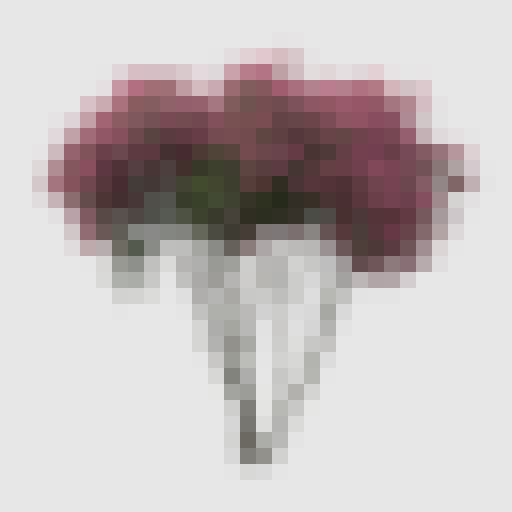}}
        &
        \frame{\includegraphics[height=\lenRenderingComparisonSmall]{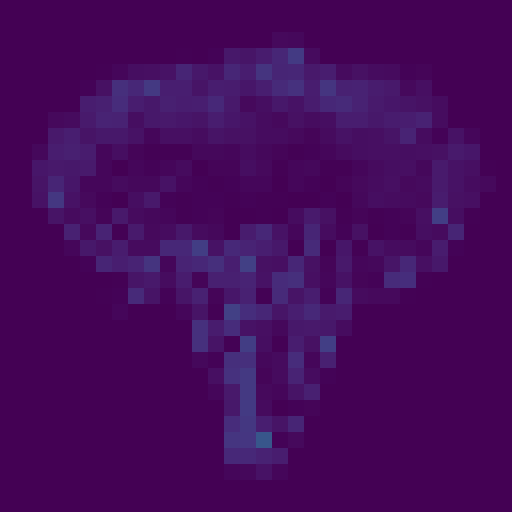}}
        &
        \raisebox{35pt}{\rotatebox{-90}{$32^3$}}
        \\
        \multirow[t]{2}{*}{\raisebox{42pt}{\rotatebox{90}{\emph{Oleander}}}}
        &
        \multicolumn{2}{c}{\multirow[t]{2}{*}{\frame{\includegraphics[height=\lenRenderingComparisonBig]{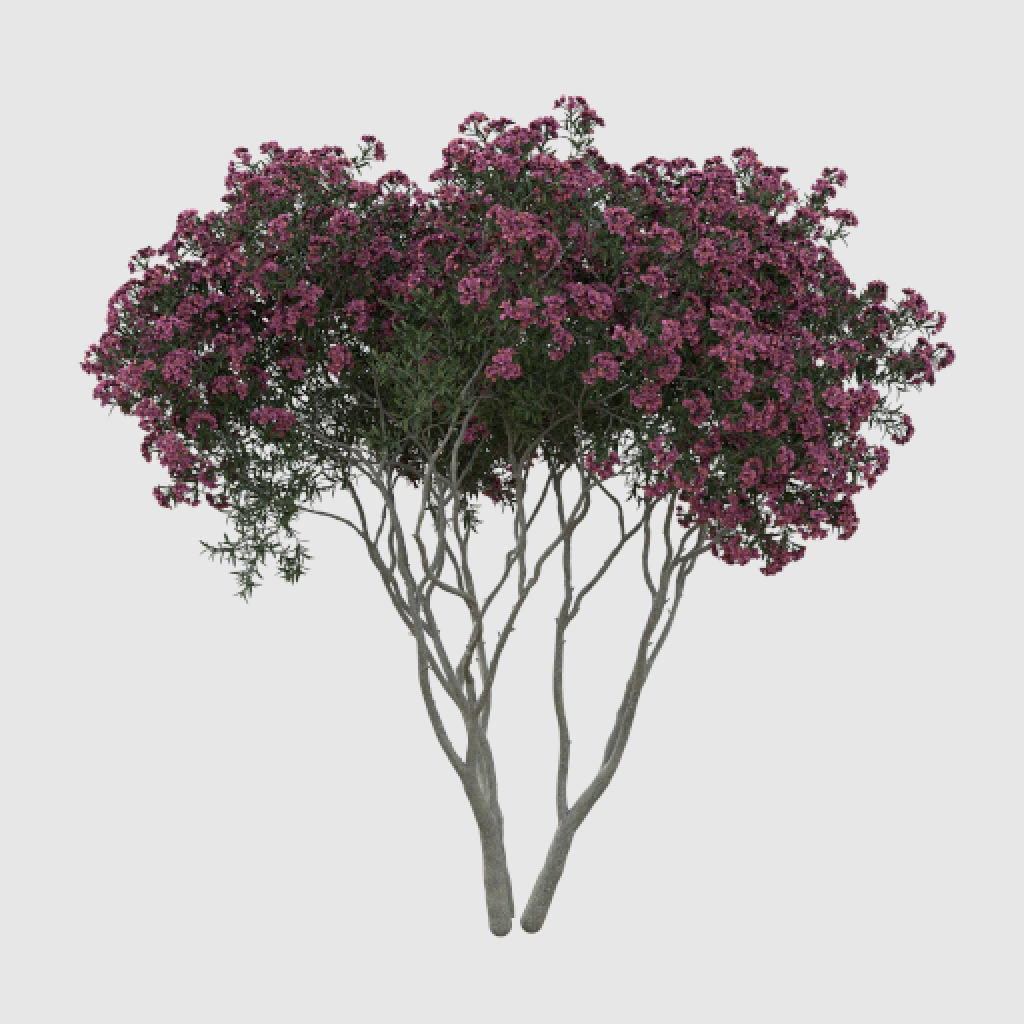}}}}
        &
        \frame{\includegraphics[height=\lenRenderingComparisonSmall]{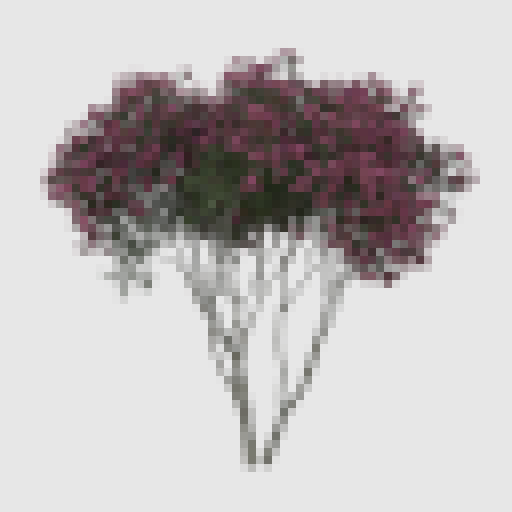}}
        &
        \frame{\includegraphics[height=\lenRenderingComparisonSmall]{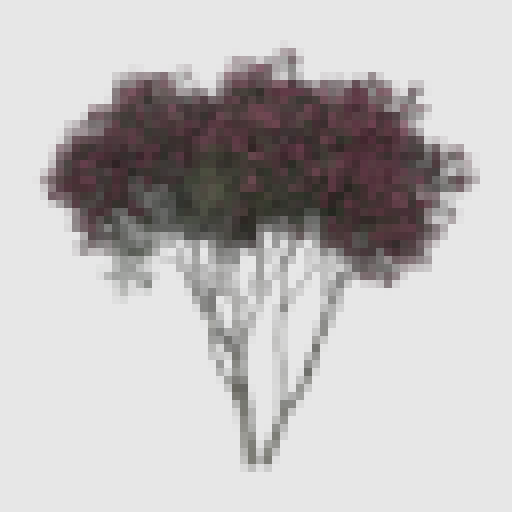}}
        &
        \frame{\includegraphics[height=\lenRenderingComparisonSmall]{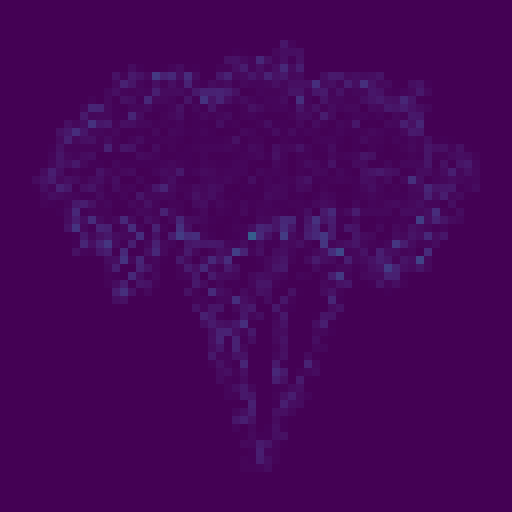}}
        &
        \frame{\includegraphics[height=\lenRenderingComparisonSmall]{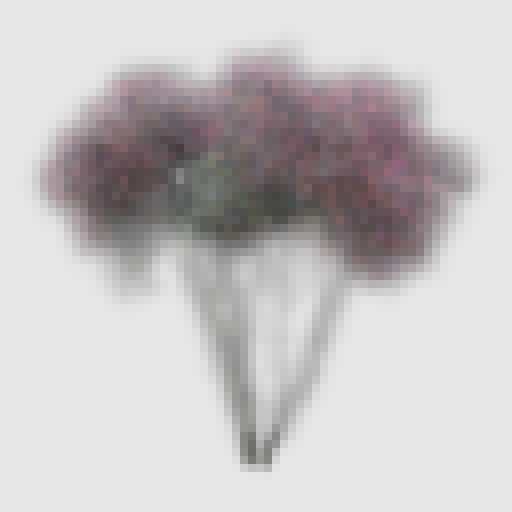}}
        &
        \frame{\includegraphics[height=\lenRenderingComparisonSmall]{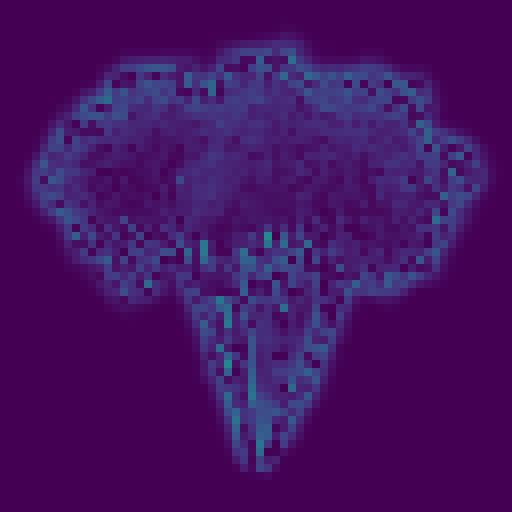}}
        &
        \frame{\includegraphics[height=\lenRenderingComparisonSmall]{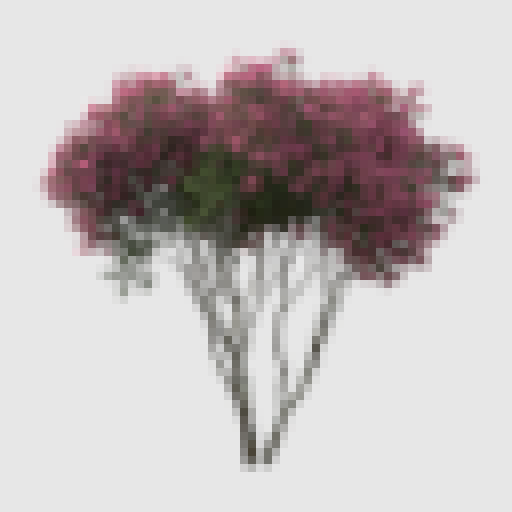}}
        &
        \frame{\includegraphics[height=\lenRenderingComparisonSmall]{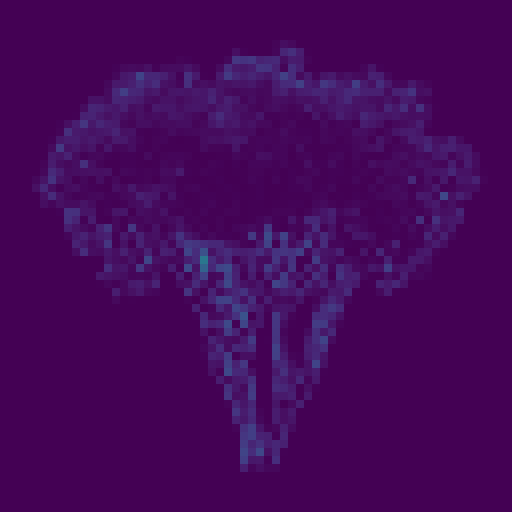}}
        &
        \raisebox{35pt}{\rotatebox{-90}{$64^3$}}
        \\
        &
        \multicolumn{2}{c}{\textsf{RMSE:}} & & & \textbf{0.028} / \textbf{0.027} & & {0.098} / {0.090} & & {0.049} / {0.045} &
    \end{tabular}
    \caption{\label{fig:main_comparison}
        Rendering results by our and existing methods on a variety of scenes. All images are rendered using 1024 samples per pixel. Each result is compared to
        the corresponding reference and the difference image is displayed on the side with RMSE provided. Our results achieve superior quality for all five scenes.
    }
\end{figure*}

\paragraph{Rendering Quality Comparison}
In \autoref{fig:main_comparison}, we compare our method to HybridLoD and NonExp on a set of scenes with varying geometric and material characteristics. For
each scene, we show the rendered results using 2 different LoD scales, $32^3$ and $64^4$. The image resolutions are $32^2$ and $64^2$, chosen such that a voxel
roughly projects to the footprint of a pixel. High resolution references are provided to better visualize the setup. The \emph{Helmet} scene has relatively low
geometric complexity but it consists of large specular surfaces which are traditionally challenging for LoD methods. The \emph{Chandelier} scene has intricate
geometric structure with varying degree of curvature that produces anisotropic highlights. The \emph{Tower} scene features organized thin structures that lead to
correlated partial occlusion. Finally, the \emph{Palm} and the \emph{Oleander} scene have larger complexity with both unstructured (leaves) and structured (trunk)
geometry.

For all scenes, our results achieve superior quality and produce closer matches to references, as can be verified by the difference images and the RMSE errors.
HybridLoD tends to produce bloated, over transparent results, which is especially
noticeable at coarser LoD resolution. This could be due to both misclassification (too much volume) and the neglect of correlation. Moreover, the mesh
simplification process could undesirably alter the curvature of the original geometry, causing loss of highlights (\emph{Helmet} and \emph{Chandelier}).
NonExp achieves better quality than HybridLoD in general, but still suffers from several issues. The transmittance optimization accounts for some correlation
but is usually not perfect, as shown in the \emph{Chandelier} renders (too leaky) and the \emph{Palm} renders (too opaque). The method ignores the complexity
in material and results in glossy appearance mismatch (\emph{Helmet} and \emph{Chandelier}). Ultimately, the empirical exponential-linear blending model is
unlikely to satisfy all constraints required to match transmittance for all directions.
In addition, we find that it is highly sensitive to the empirical ray offset parameter as a different value drastically alters brightness. We follow the authors'
suggestion and offset scattered rays by one voxel for all results.

\begin{figure}[h]
	\newlength{\lenCompareDAP}
	\setlength{\lenCompareDAP}{1.1in}
	\addtolength{\tabcolsep}{-4pt}
    \renewcommand{\arraystretch}{0.5}
	\centering
    \begin{tabular}{ccc}
		\includegraphics[height=\lenCompareDAP]{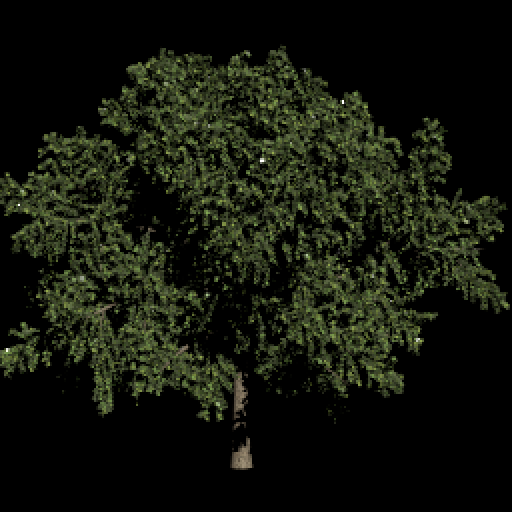}
		&
        \includegraphics[height=\lenCompareDAP]{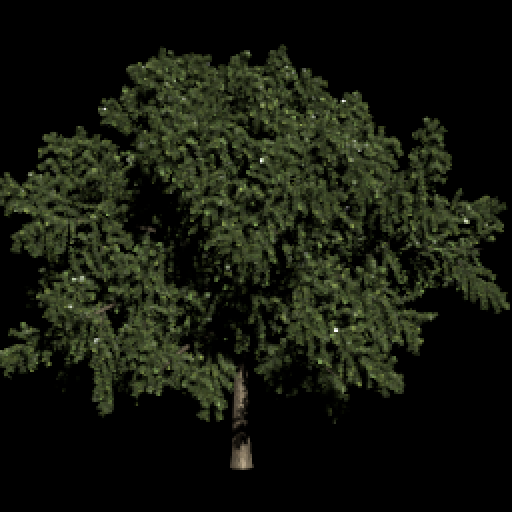}
		&
		\includegraphics[height=\lenCompareDAP]{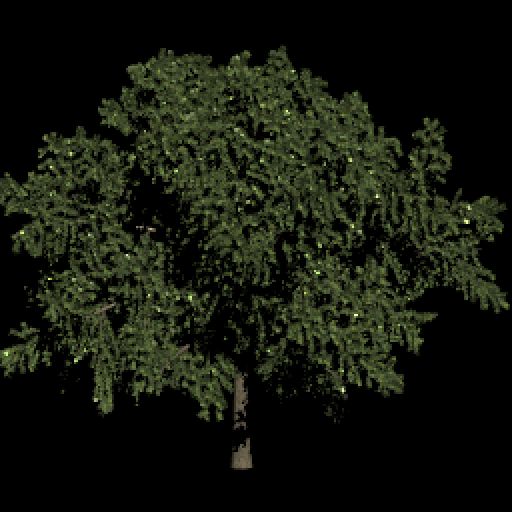}
		\\
		\multicolumn{1}{r}{\frame{\begin{overpic}[height=\lenCompareDAP]{imgs/colorbar.png}
			\put(-20, 92){\normalsize 0.25}
			\put(-15, 1){\normalsize 0.0}
		\end{overpic}}}
		&
		\begin{overpic}[height=\lenCompareDAP]{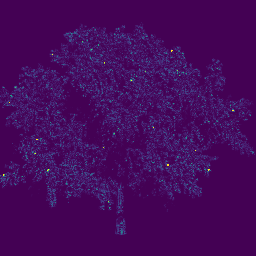}
			\put(2, 90){\normalsize \color{white} \textsf{RMSE:} 0.026}
		\end{overpic}
		&
		\begin{overpic}[height=\lenCompareDAP]{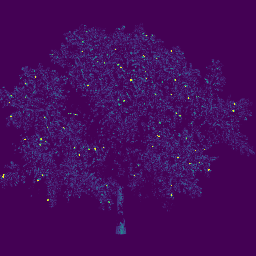}
			\put(2, 90){\normalsize \color{white} 0.031}
		\end{overpic}
		\\
		(a) \textsf{PT Reference} & \textsf{(b) Ours ($256^3$)} & \textsf{(c) DAP ($256^3$)}
    \end{tabular}
    \caption{\label{fig:dap_comparison} We compare our method to DAP with their pretrained \emph{Oak} scene. Despite the small imperfection of the hard shadow
			cast on the tree trunk due to visibility coefficient truncation and compression, our result overall reaches similar visual quality.}
\end{figure}

\paragraph{Comparison to DAP}
In \autoref{fig:dap_comparison}, we provide a separate comparison to DAP as we only use their pretrained asset. The \emph{Oak} scene presents two difficulties
including the glinty appearance from the highly glossy material and the hard shadow cast by a directional light. Our method is able to capture the highlight
accurately, but fails to reconstruct the hard shadow perfectly due to the coefficient truncation and compression error. This can be alleviated with more
conservative truncation/compression parameters at the cost of a larger memory footprint. Overall, our method is able to reach a comparable visual quality
(slightly better in terms of RMSE). We emphasize that our method only requires a fraction of precomputation time, memory cost, and rendering time to reach such
quality.

\begin{table}[h]
	\centering
	\caption{Statistics of the rendering results in \autoref{fig:main_comparison}. Memory consumption and render times are measured using $64^3$ LoD 
    resolution and 1024 samples per pixel for all methods.}
	\begin{tabular}{r|r|r|r|r}
		\Xhline{1pt}
        \multicolumn{1}{l|}{\emph{Helmet}} &
        \textbf{PT Ref.} & 
        \textbf{Ours} &
        \textbf{HybridLoD} &
        \textbf{NonExp}     
        \\
        \hline         
        \multicolumn{1}{l|}{Mem. (MB)}    & 7.7    & 9.7      & 5.1    & 1.0  \\
        \multicolumn{1}{l|}{Time (sec)}   & 1.37   & 1.68     & 23.78    & 57.36  \\           
        \multicolumn{1}{l|}{RMSE}         & ---   & 0.158     & 0.302   & 0.238 \\
        \hline
        \multicolumn{1}{l|}{\emph{Chandelier}} &
        \textbf{PT Ref.} & 
        \textbf{Ours} &
        \textbf{HybridLoD} &
        \textbf{NonExp}     
        \\         
        \hline
        \multicolumn{1}{l|}{Mem. (MB)}    & 11.7    & 6.4      & 6.6    & 0.5 \\
        \multicolumn{1}{l|}{Time (sec)}   & 1.46    & 1.74      & 32.33    & 51.08 \\       
        \multicolumn{1}{l|}{RMSE}         & ---   & 0.075     & 0.112   & 0.099 \\
        \hline
        \multicolumn{1}{l|}{\emph{Tower}} &
        \textbf{PT Ref.} & 
        \textbf{Ours} &
        \textbf{HybridLoD} &
        \textbf{NonExp}     
        \\         
        \hline
        \multicolumn{1}{l|}{Mem. (MB)}    & 45.7     & 9.6     & 4.3    & 0.7  \\
        \multicolumn{1}{l|}{Time (sec)}   & 1.66    & 1.20      & 19.72    & 37.13  \\                
        \multicolumn{1}{l|}{RMSE}         & ---     & 0.022     & 0.050   & 0.038 \\
        \hline
        \multicolumn{1}{l|}{\emph{Palm}} &
        \textbf{PT Ref.} & 
        \textbf{Ours} &
        \textbf{HybridLoD} &
        \textbf{NonExp}     
        \\         
        \hline
        \multicolumn{1}{l|}{Mem. (MB)}    & 349.6     & 7.3     & 4.0    & 0.5  \\
        \multicolumn{1}{l|}{Time (sec)}   & 2.93    & 1.81     & 30.50    & 56.79  \\                
        \multicolumn{1}{l|}{RMSE}         & ---     & 0.024     & 0.109   & 0.098 \\
        \hline
        \multicolumn{1}{l|}{\emph{Oleander}} &
        \textbf{PT Ref.} & 
        \textbf{Ours} &
        \textbf{HybridLoD} &
        \textbf{NonExp}     
        \\         
        \hline
        \multicolumn{1}{l|}{Mem. (MB)}    & 398.8    & 12.8     & 4.1     & 1.4  \\
        \multicolumn{1}{l|}{Time (sec)}   & 4.56    & 5.12      & 84.78    & 192.39  \\                
        \multicolumn{1}{l|}{RMSE}         & ---     & 0.027     & 0.090   & 0.045 \\                        		
		\Xhline{1pt}
	\end{tabular}
	\label{tab:main_comparison}
\end{table}

\paragraph{Complex Scenes}
We showcase the practicality of our method by demonstrating results on significantly more complex scenes. Each scene shown in this part features a collection
of assets with multiple geometric parts and materials. We compare our results to references but not to other methods, because they require either non-trivial
engineering effort or excessive precomputation budget to support assets at this scale.

The \emph{Coral Reef} scene in \autoref{fig:teaser} includes a variety of geometry (flat surfaces and unstructured details) and materials (glossy and diffuse).
The environment light features a dynamic range of 80,000:1 and produces strong highlights on glossy surfaces. Despite the challenging configuration, our results
accurately preserve the appearance across different scales. The insets show how our method deals with a particularly challenging part with numerous thin glossy
branches. At the coarsest scale, even our multi-lobe NDF does not have sufficient angular resolution to resolve all the highlights, resulting in a slightly
darker look. However, this is alleviated at finer scales.

\begin{figure}[h]
	\newlength{\lenBotanic}
	\setlength{\lenBotanic}{1.8in}
	\newlength{\lenBotanicDiff}
	\setlength{\lenBotanicDiff}{1.75in}
	\addtolength{\tabcolsep}{-5pt}
	\centering
    \begin{tabular}{cc}
		\multicolumn{2}{c}{
        \begin{overpic}[height=\lenBotanic]{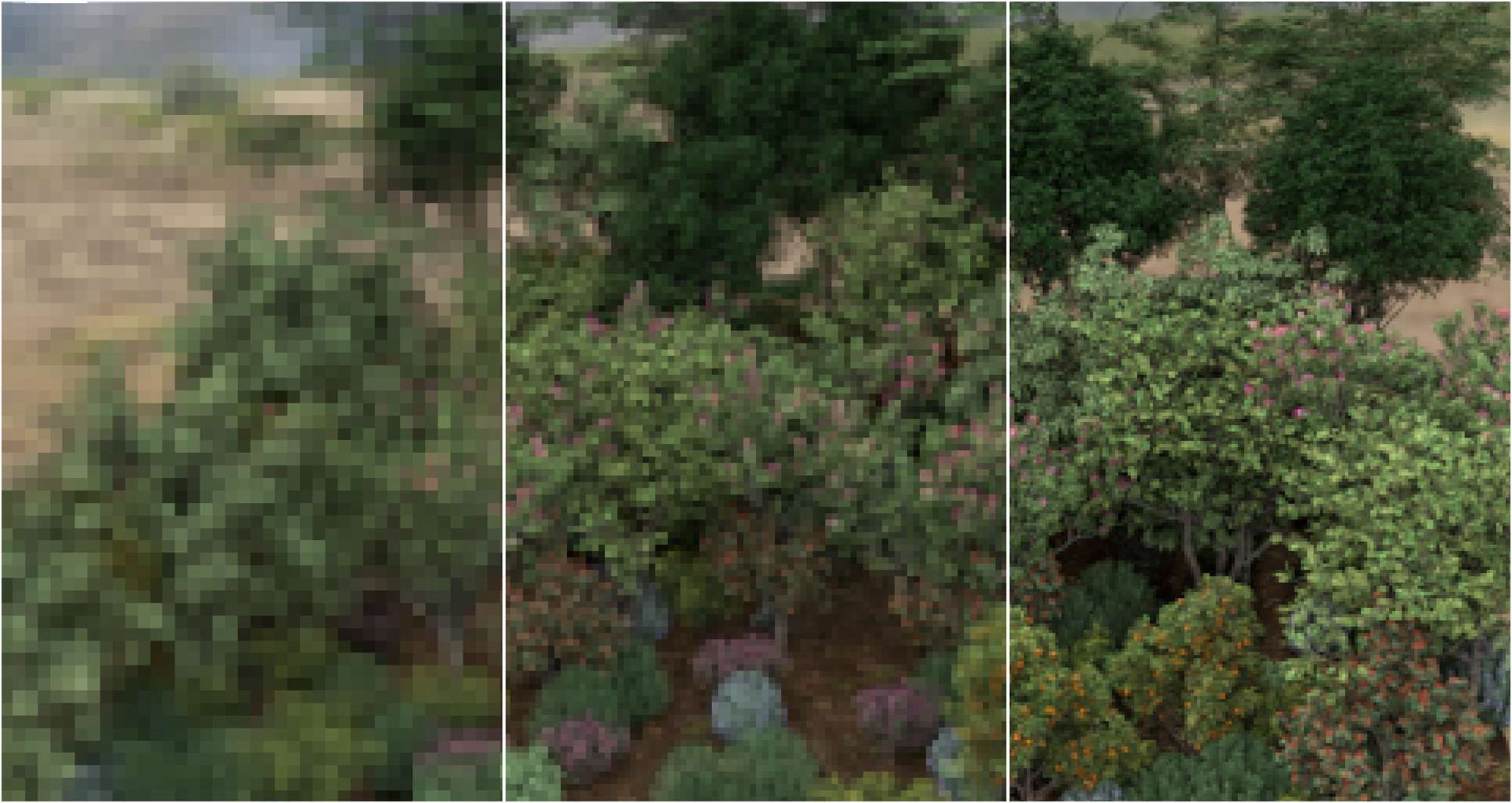}
			\put(1, 49){\normalsize \color{white}  $128^3$}
			\put(35, 49){\normalsize \color{white} $256^3$}
			\put(68, 49){\normalsize \color{white} $512^3$}
		\end{overpic}}
		\\
		\multicolumn{2}{c}{\textsf{(a) Ours}}\\
		\multicolumn{2}{c}{\includegraphics[height=\lenBotanic]{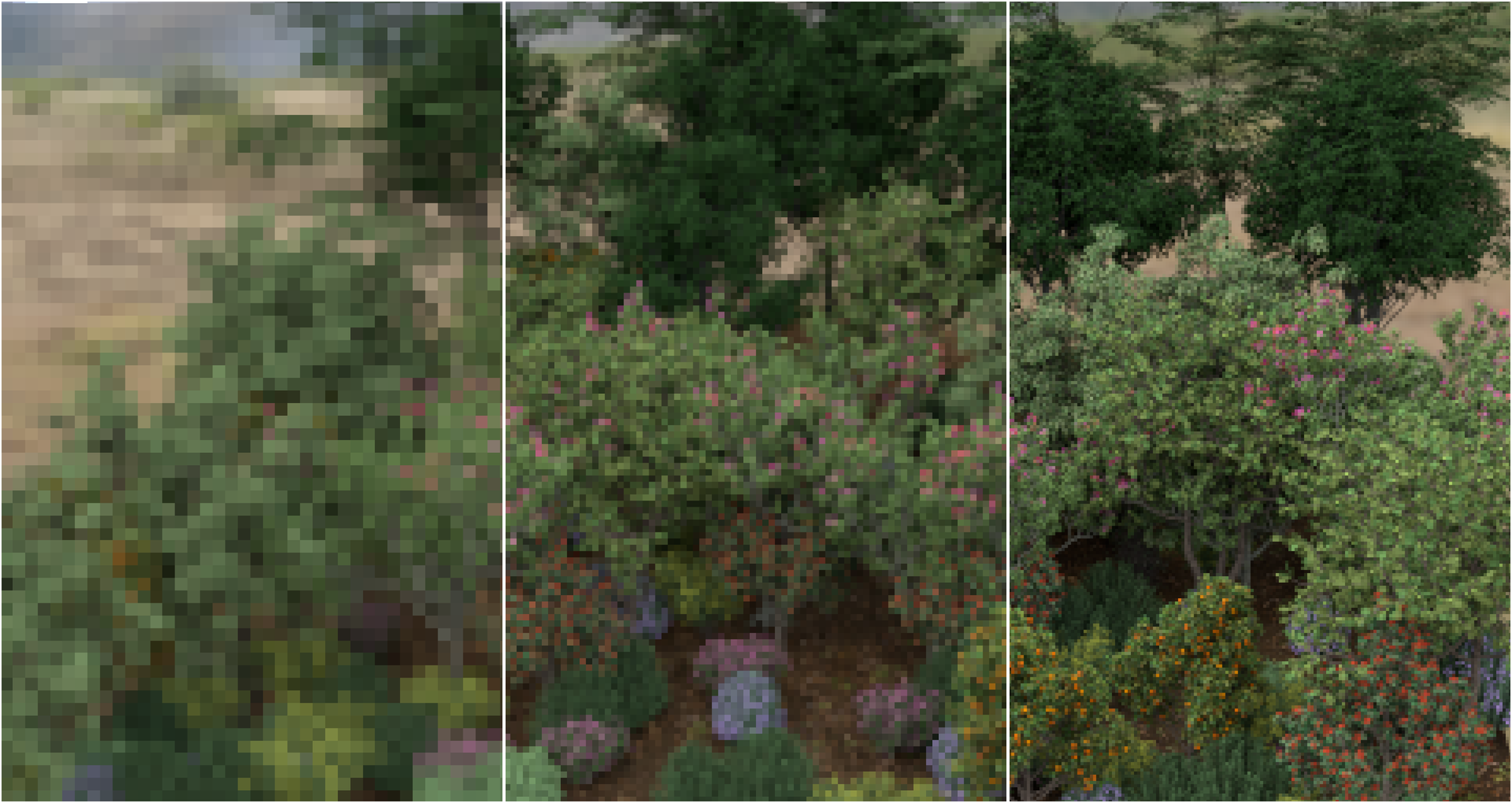}}\\
		\multicolumn{2}{c}{\textsf{(b) PT Reference}}\\
        \begin{overpic}[height=\lenBotanicDiff]{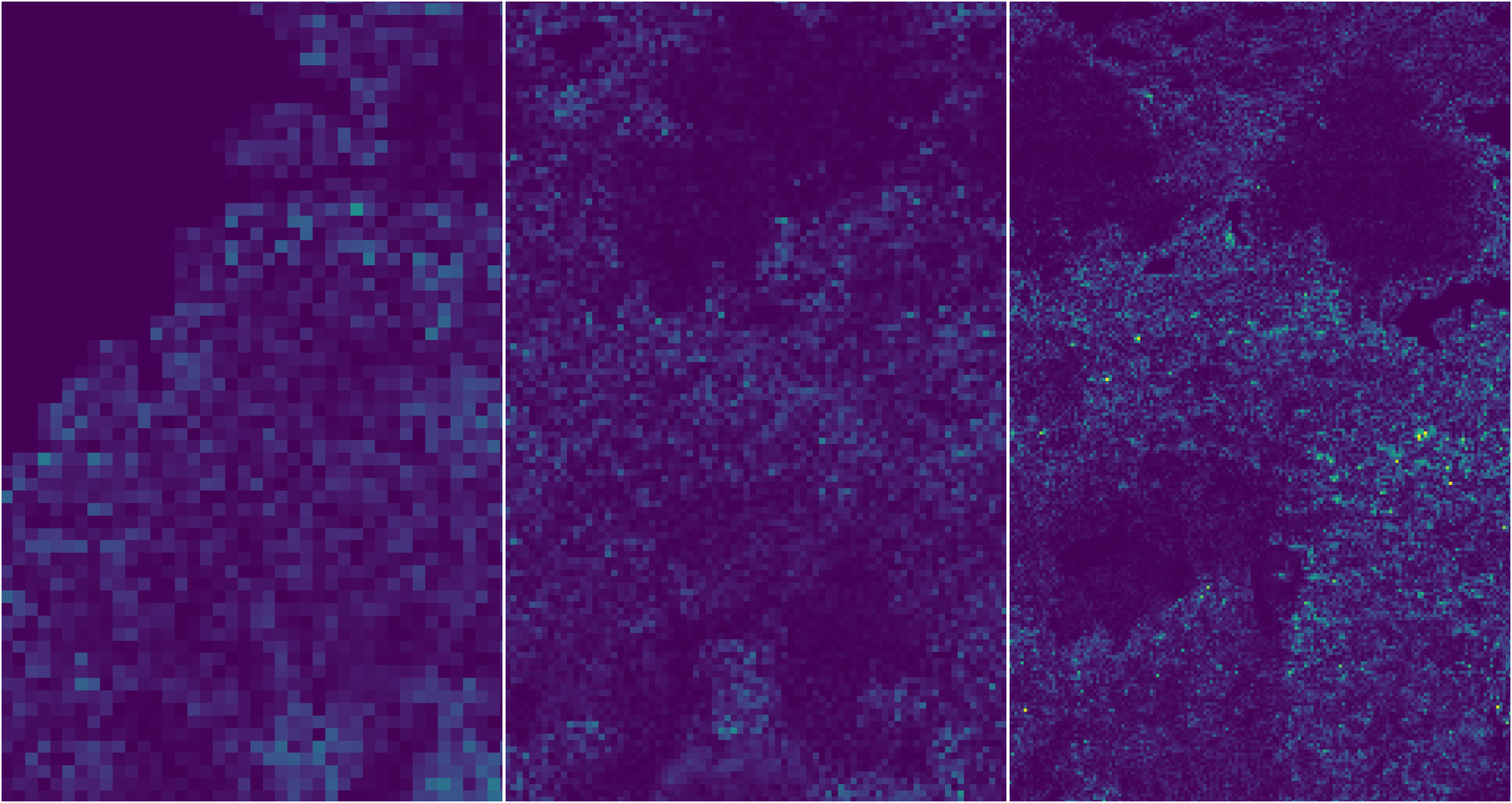}
			\put(1, 49){\normalsize \color{white} \textsf{RMSE:} 0.042}
			\put(35, 49){\normalsize \color{white} 0.036}
			\put(68, 49){\normalsize \color{white} 0.035}
		\end{overpic}
		&
        \frame{\begin{overpic}[height=\lenBotanicDiff]{imgs/colorbar.png}
			\put(4, 95){\normalsize 0.25}
			\put(4, 1){\normalsize 0.0}
		\end{overpic}}
		\\
		\multicolumn{2}{c}{\textsf{(c) Difference}}
    \end{tabular}
    \caption{\label{fig:botanic} Our results faithfully preserve the partial coverage, visibility, and appearance of the large \emph{Forest} scene at different
			scales. The difference images are scaled for clarity. }
\end{figure}

\rev{
The \emph{Forest} scene in \autoref{fig:botanic} features the largest single-object geometric complexity. It has $16.2$ million unique
triangles and $175.2$ million after instancing. Our results remain close to references at the three LoD scales shown ($128^3$, $256^3$, and $512^3$).
We note that the memory cost of our method is agnostic to whether or not the original scene contains instanced geometry.
}

\rev{
Finally, our method supports assembly of multiple aggregated objects and instancing for even larger scenes. The \emph{Metropolis} cityscape in
\autoref{fig:teaser} is composed of $82$ unique aggregated objects and $270$ instances. Each instance selects its level based on its screen-space
projection size so that content far away can be rendered with coarser LoDs accordingly (\autoref{fig:teaser}, right). Different instances are treated as
uncorrelated and ABVs are used to compute the partial occlusion between them. This is where the LoD approach truly shines as our representation provides nearly
an order-of-magnitude of memory saving. In fact, we are not even able to generate reference for this scene in our testing environment: The original scene costs
46.9 GB alone and when taking into account the auxiliary data structures such as mipmaps and the BVH, the total memory would have exceeded the 64 GB main
memory of our testing machine, requiring out-of-core rendering. Our representation only costs 5.33 GB and \autoref{fig:teaser} takes 1300 seconds to render at 
$1280 \times 540$ resolution with 2048 samples per pixel.
}

\paragraph{Performance}
In \autoref{tab:scenes}, we report the configuration for all scenes, including both the original version and our representation. It is clear that all scenes
have a large degree of sparsity. As expected, the memory footprint of our representation is largely independent of the original scene complexity. While our method
cannot beat the explicit representation for small scenes, the asymptotical benefit is evident for larger scenes. Recall that the memory footprint scales
quadratically with respect to LoD resolution. If a model only occupies a small portion of the rendered image, then only a coarser LoD level with much smaller
memory is needed. It is possible to further reduce the memory footprint with a more optimized implementation, e.g., by quantizing the stored data. GPU ray
tracing is likely to reduce the precomputation time by an order of magnitude.

\rev{
In \autoref{tab:main_comparison}, we compare the memory consumption and rendering times required by our method, HybridLoD, and NonExp. At a relatively
modest resolution of $64^3$, all methods reduce the memory consumption, especially for more complicated models. However, our method does require more memory
than HybridLoD and NonExp mainly due to the high-dimensional aggregated visibility data even after compression. Performance wise, none of the methods show
significant advantage over the PT baseline at equal samples. In fact, HybridLoD and NonExp are significantly slower, likely due to the complexity of (nested)
ray marching or sampling. We note that the baseline is backed by highly optimized ray tracing kernels from Embree, and our CPU re-implementation of NonExp is
relatively unoptimized. Our method is much more comparable to the baseline, as it is simpler with precomputed aggregated visibility
(\autoref{eq:accumulate_multi_voxel}). The equal-sample comparison could be unfavorable to our and other LoD methods because they are already prefiltered and 
should require fewer samples to reach the same or similar quality. In the following, we assess how our method improves efficiency under equal time.
}

\begin{figure}[ht!]
	\newlength{\lenBotanicEqTime}
	\setlength{\lenBotanicEqTime}{1.8in}
	\addtolength{\tabcolsep}{-5pt}
	\centering
    \begin{tabular}{c}
        \begin{overpic}[height=\lenBotanicEqTime]{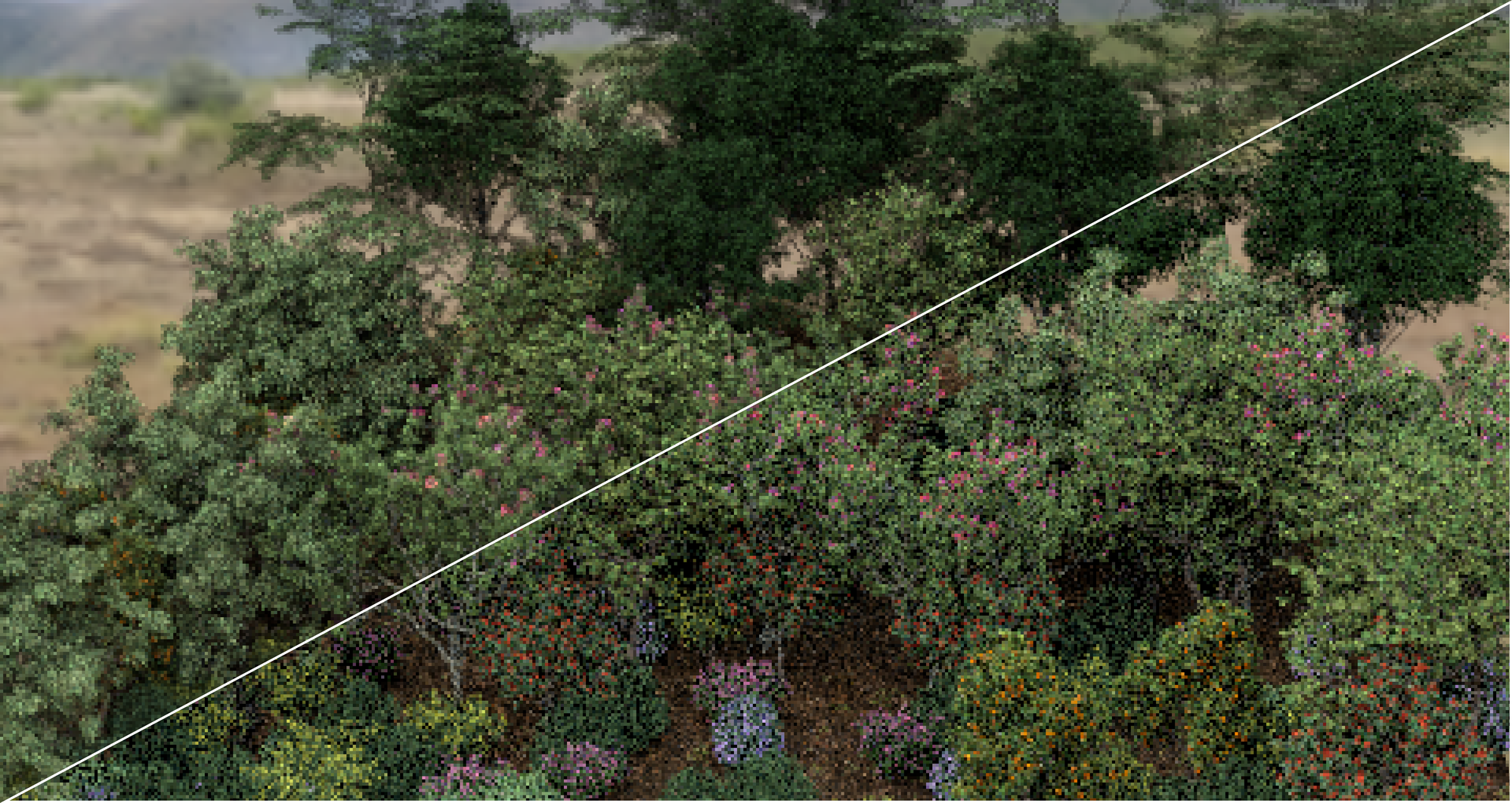}
			\put(1, 50){\normalsize \color{white}  \textsf{Ours}}
			\put(1, 47){\normalsize \color{white}  \textsf{2 ray spp}}
			\put(1, 44){\normalsize \color{white}  \textsf{5 nee spp}}
			\put(89, 5){\normalsize \color{white}  \textsf{PT}}
			\put(89, 2){\normalsize \color{white}  \textsf{18 spp}}
		\end{overpic}
    \end{tabular}
    \caption{\label{fig:botanic_eq_time} Equal-time rendering comparison. Both images are rendered for 1 second. Our method performs splitting to improve
	efficiency by allocating 5 NEE samples per ray sample.}
\end{figure}

Thanks to appearance aggregation, we are able to render complex scenes efficiently. \autoref{fig:botanic_eq_time} shows equal-time rendering comparison for
the \emph{Forest} scene. The performance improvement comes from two aspects: First, we avoid spending a large number of samples to trace explicit geometry as
our representation is already anti-aliased. This allows us to perform splitting and allocate more samples for lighting (\autoref{algo:rendering}, line 14).
Moreover, we do not need to trace shadow rays either, since the visibility information is readily available from the precomputed AIV. The variance
reduction is modest, as ray tracing is highly optimized by Embree. Again, our rendering speed can be further improved with a GPU implementation.

\begin{figure*}[ht!]
	\newlength{\lenAblationSmall}
	\setlength{\lenAblationSmall}{0.82in}
    \addtolength{\tabcolsep}{-4pt}
    \renewcommand{\arraystretch}{0.5}
    \centering    
    \begin{tabular}{ccccccccc}
        &
        \textsf{PT Ref.} & \textsf{Maximum} & 
        \textsf{1 NDF Lobe} &
        \textsf{10\% AIV Coeff.} & \textsf{1\% AIV Coeff.} & 
        \textsf{10\% ABV Coeff.} & \textsf{1\% ABV Coeff.} &
        \textsf{1 CPCA Rep.}
        \\
        \raisebox{2pt}{\rotatebox{90}{\emph{Chandelier} $64^3$}}
        &
        \frame{\includegraphics[height=\lenAblationSmall]{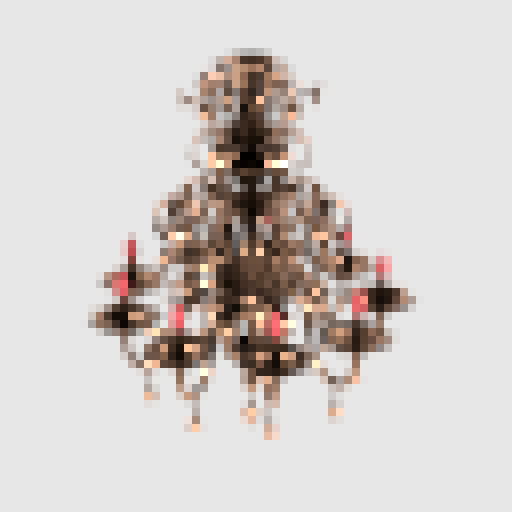}}
        &
        \frame{\includegraphics[height=\lenAblationSmall]{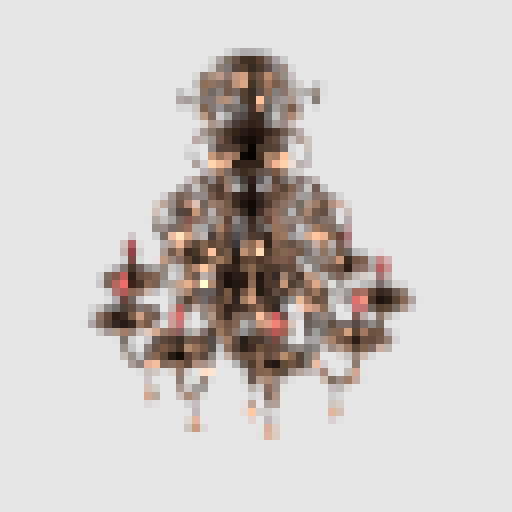}}
        &
        \frame{\includegraphics[height=\lenAblationSmall]{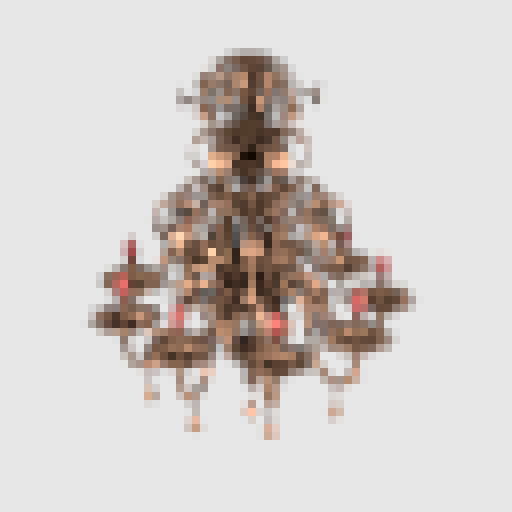}}
        &
        \frame{\includegraphics[height=\lenAblationSmall]{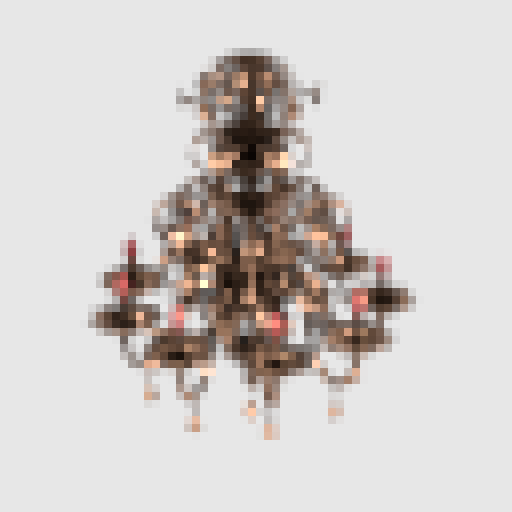}}
        &
        \frame{\includegraphics[height=\lenAblationSmall]{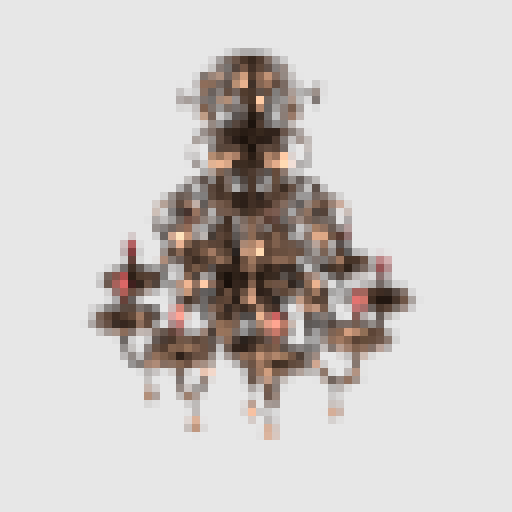}}
        &
        \frame{\includegraphics[height=\lenAblationSmall]{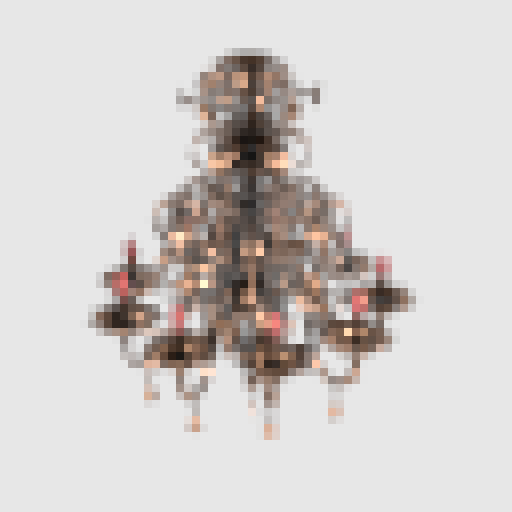}}
        &
        \frame{\includegraphics[height=\lenAblationSmall]{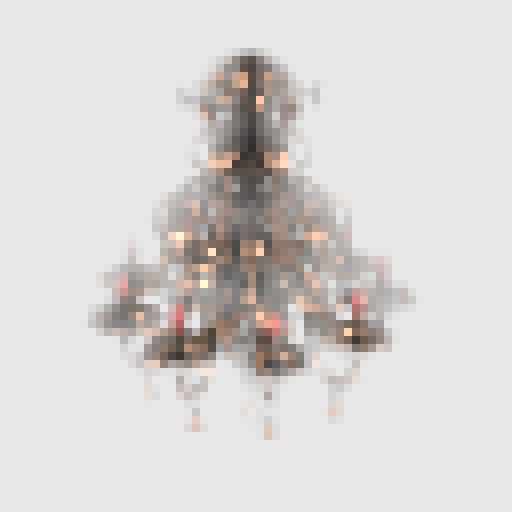}}
        &
        \frame{\includegraphics[height=\lenAblationSmall]{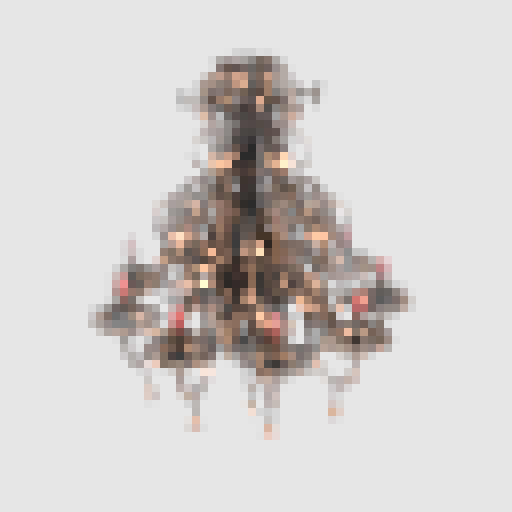}}
        \\
        \raisebox{10pt}{\rotatebox{90}{\textsf{Difference}}}
        &
		\multicolumn{1}{r}{\frame{\begin{overpic}[height=\lenAblationSmall]{imgs/colorbar.png}
			\put(-20, 92){\normalsize 1.0}
			\put(-20, 1){\normalsize 0.0}
		\end{overpic}}}
        &
        \frame{\includegraphics[height=\lenAblationSmall]{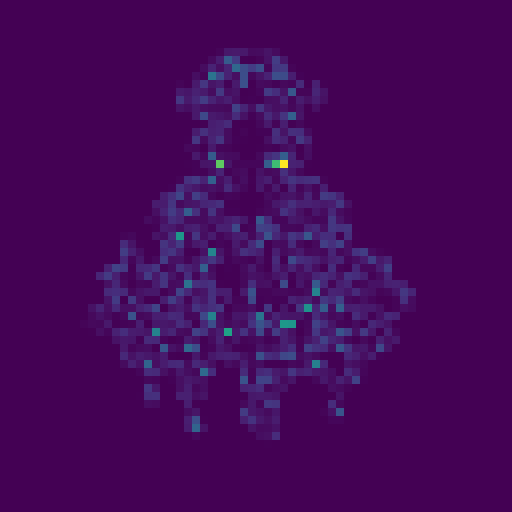}}
        &
        \frame{\includegraphics[height=\lenAblationSmall]{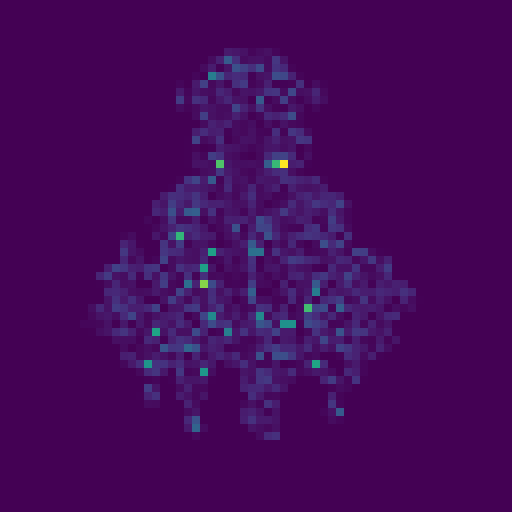}}
        &
        \frame{\includegraphics[height=\lenAblationSmall]{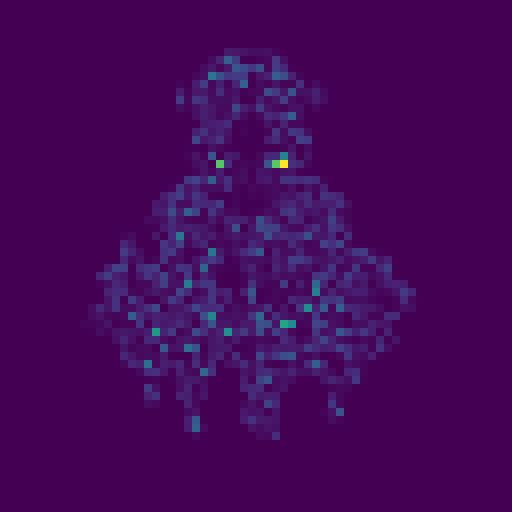}}
        &
        \frame{\includegraphics[height=\lenAblationSmall]{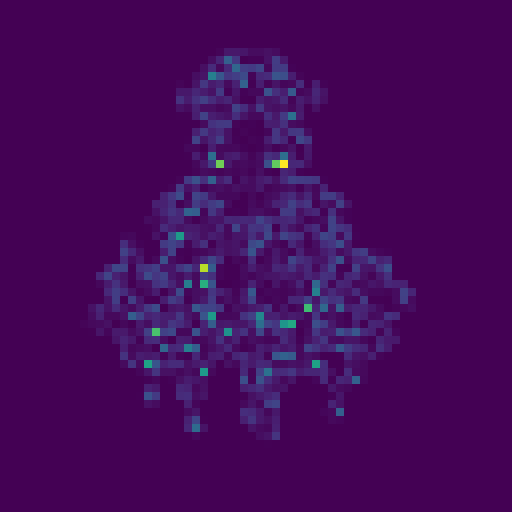}}
        &
        \frame{\includegraphics[height=\lenAblationSmall]{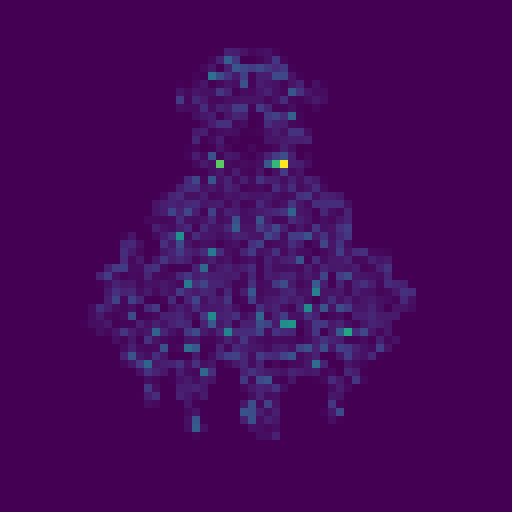}}
        &
        \frame{\includegraphics[height=\lenAblationSmall]{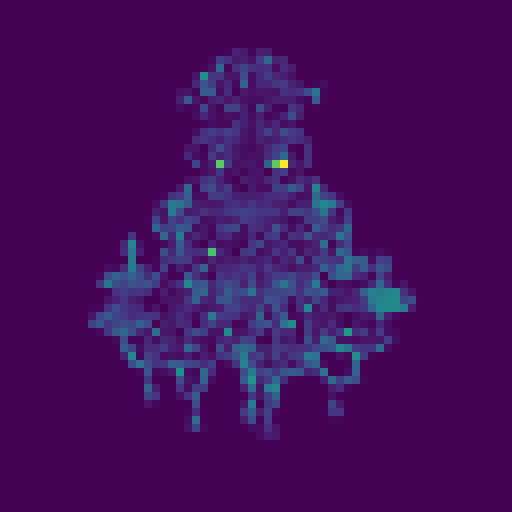}}
        &
        \frame{\includegraphics[height=\lenAblationSmall]{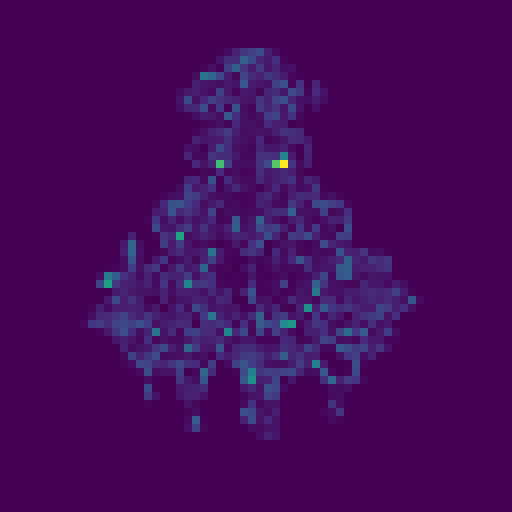}}
        \\  
        \raisebox{10pt}{\rotatebox{90}{\textsf{(Alt. View)}}}
        &
        \frame{\includegraphics[height=\lenAblationSmall]{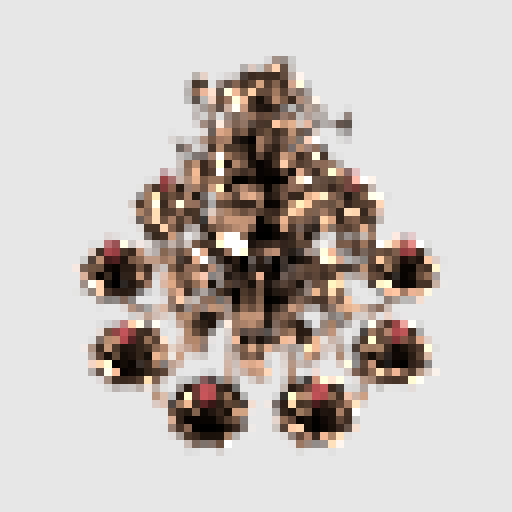}}
        &
        \frame{\includegraphics[height=\lenAblationSmall]{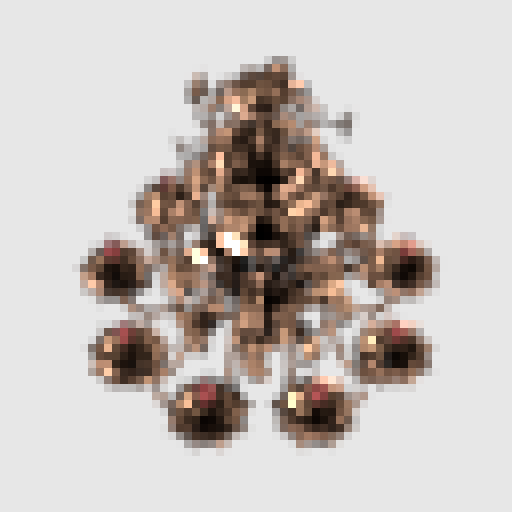}}
        &
        \frame{\includegraphics[height=\lenAblationSmall]{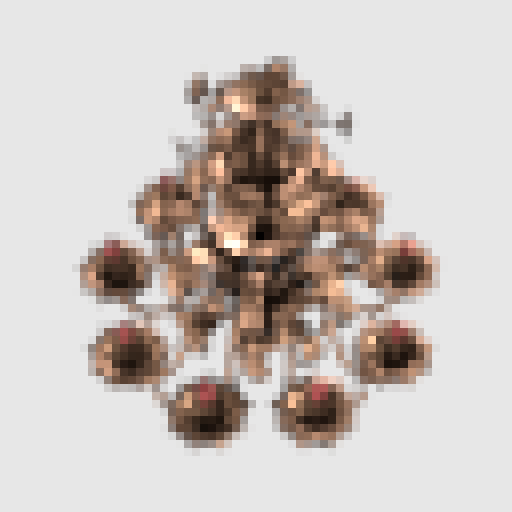}}
        &
        \frame{\includegraphics[height=\lenAblationSmall]{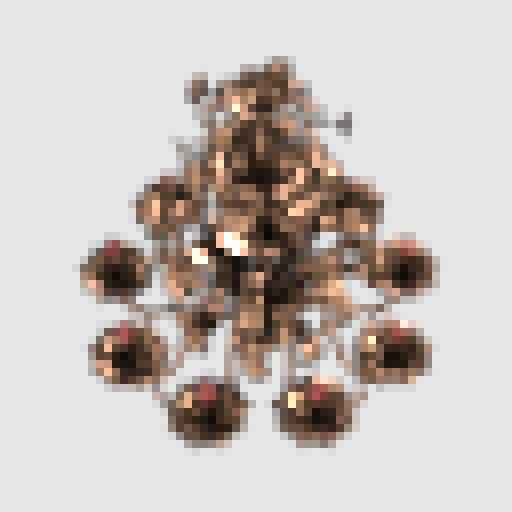}}
        &
        \frame{\includegraphics[height=\lenAblationSmall]{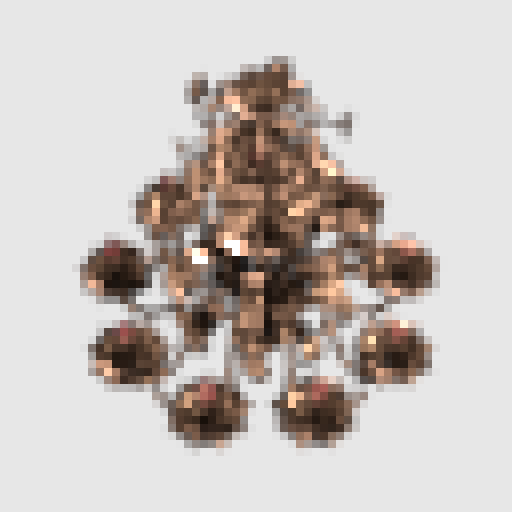}}
        &
        \frame{\includegraphics[height=\lenAblationSmall]{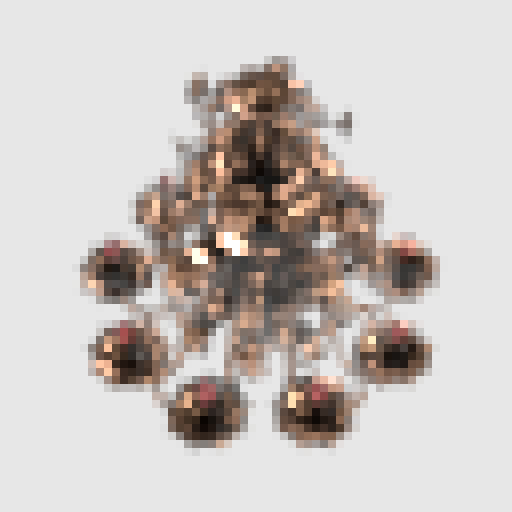}}
        &
        \frame{\includegraphics[height=\lenAblationSmall]{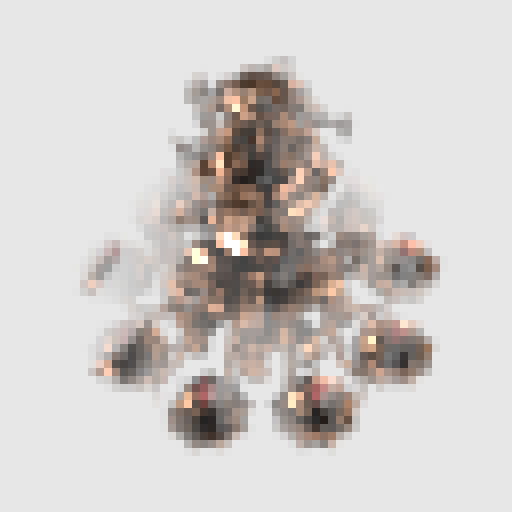}}
        &
        \frame{\includegraphics[height=\lenAblationSmall]{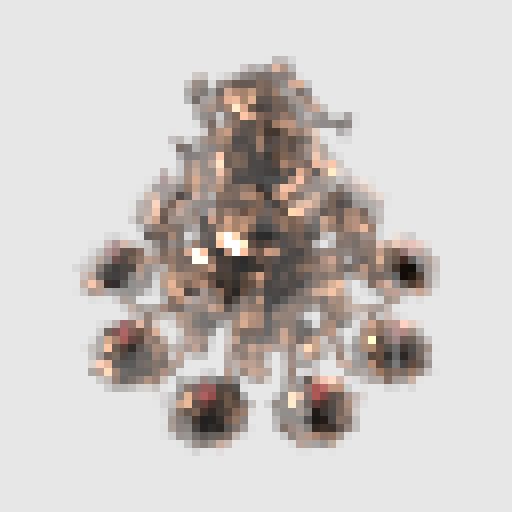}}
        \\
        \raisebox{10pt}{\rotatebox{90}{\textsf{Difference}}}
        &
		\multicolumn{1}{r}{\frame{\begin{overpic}[height=\lenAblationSmall]{imgs/colorbar.png}
			\put(-20, 92){\normalsize 1.0}
			\put(-20, 1){\normalsize 0.0}
		\end{overpic}}}
        &
        \frame{\includegraphics[height=\lenAblationSmall]{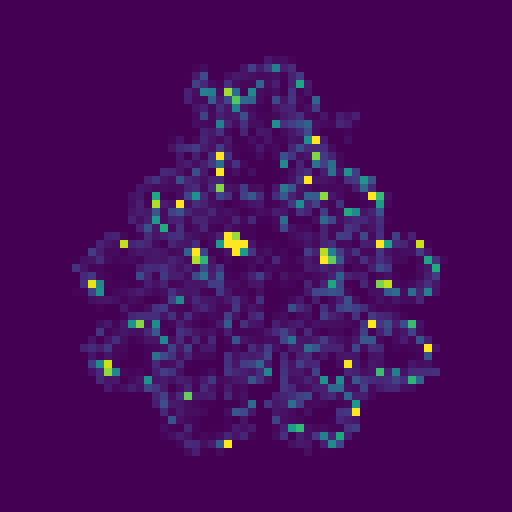}}
        &
        \frame{\includegraphics[height=\lenAblationSmall]{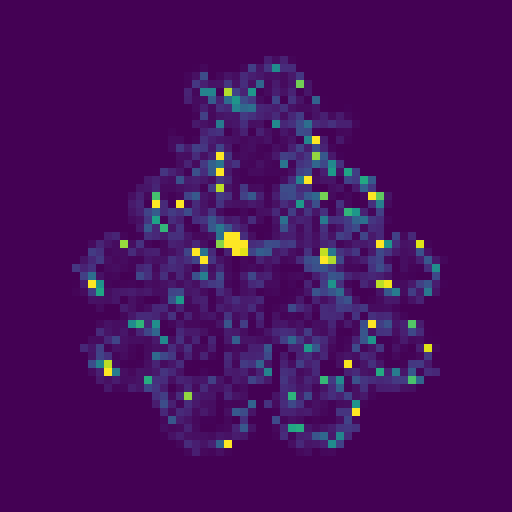}}
        &
        \frame{\includegraphics[height=\lenAblationSmall]{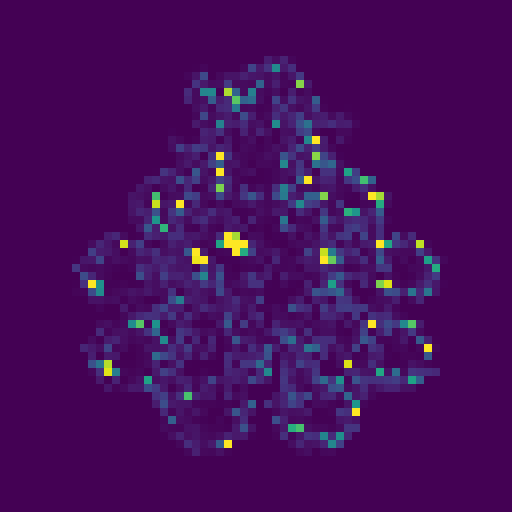}}
        &
        \frame{\includegraphics[height=\lenAblationSmall]{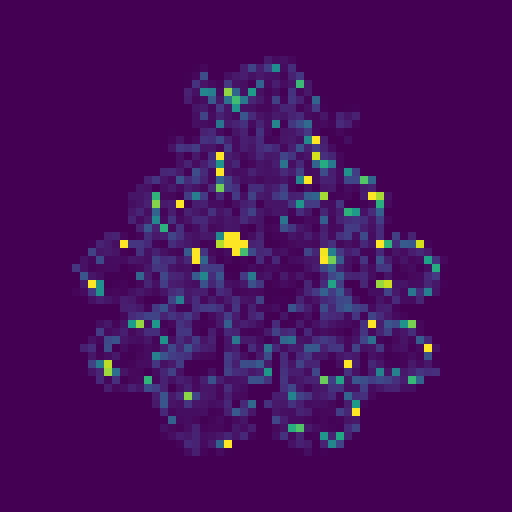}}
        &
        \frame{\includegraphics[height=\lenAblationSmall]{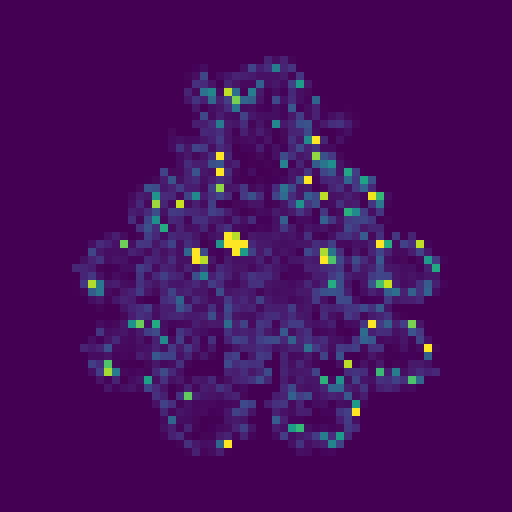}}
        &
        \frame{\includegraphics[height=\lenAblationSmall]{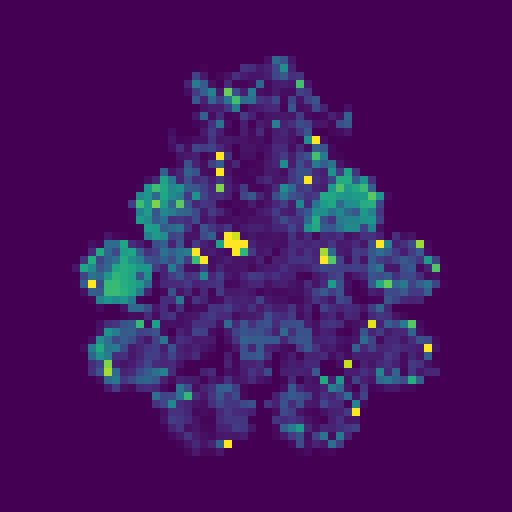}}
        &
        \frame{\includegraphics[height=\lenAblationSmall]{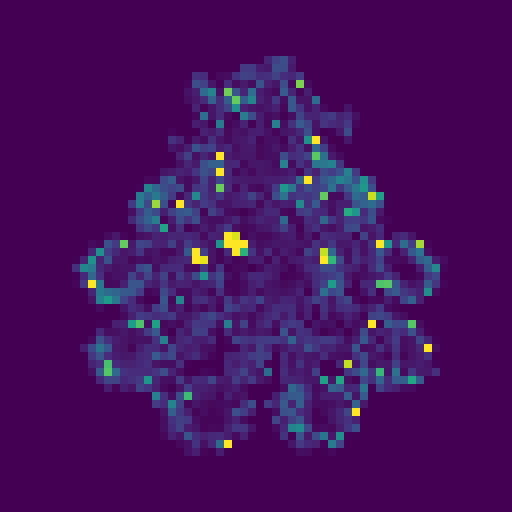}}
        \\  
        \raisebox{2pt}{\rotatebox{90}{\emph{Oleander} $64^3$}}
        &
        \frame{\includegraphics[height=\lenAblationSmall]{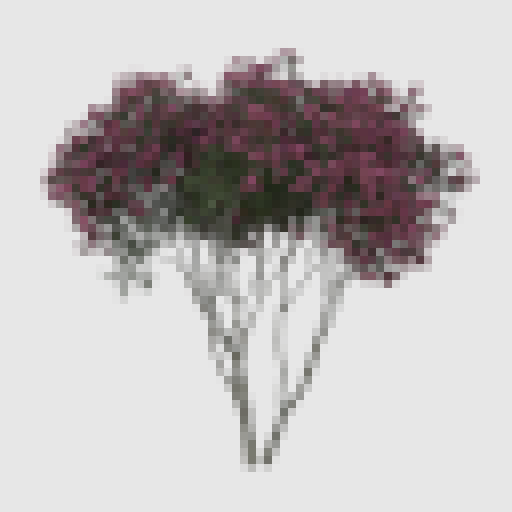}}
        &
        \frame{\includegraphics[height=\lenAblationSmall]{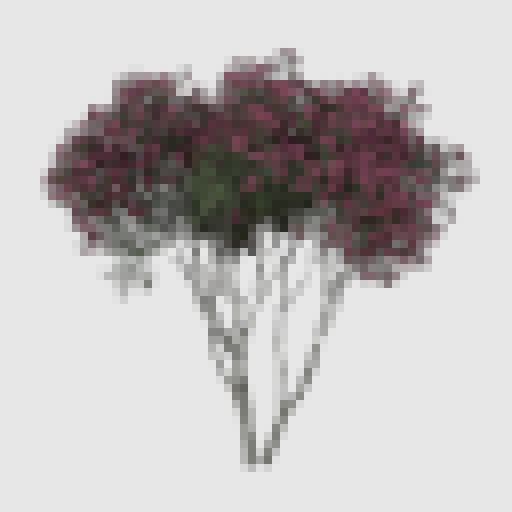}}
        &
        \frame{\includegraphics[height=\lenAblationSmall]{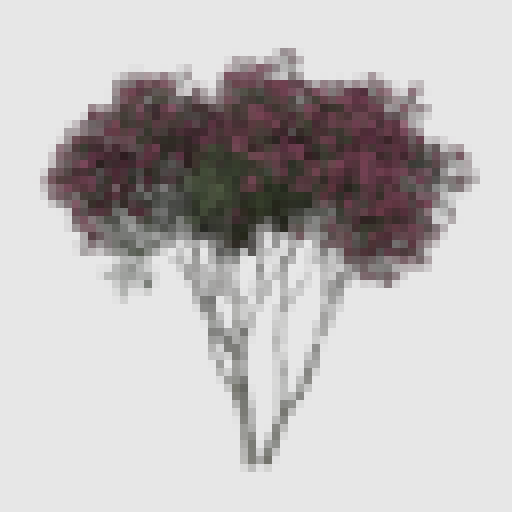}}
        &
        \frame{\includegraphics[height=\lenAblationSmall]{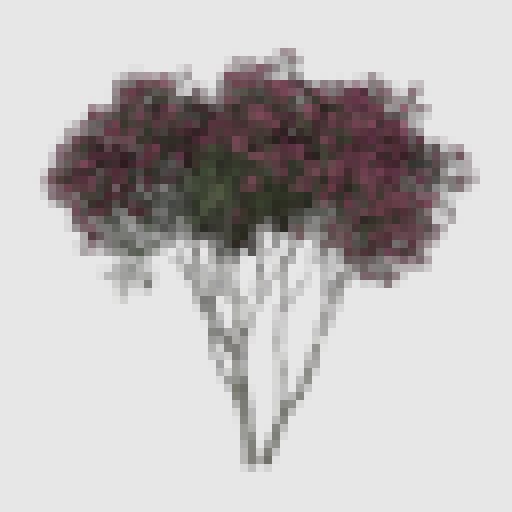}}
        &
        \frame{\includegraphics[height=\lenAblationSmall]{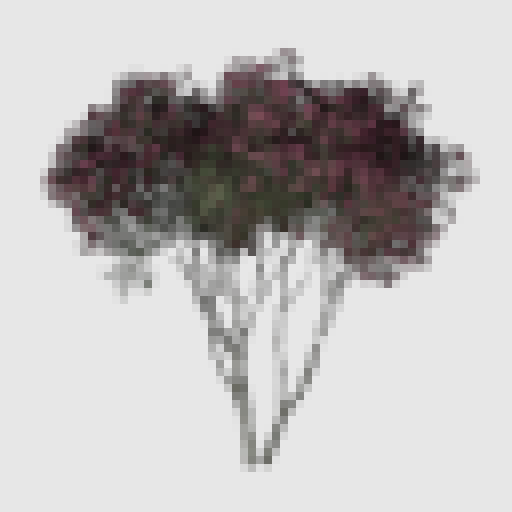}}
        &
        \frame{\includegraphics[height=\lenAblationSmall]{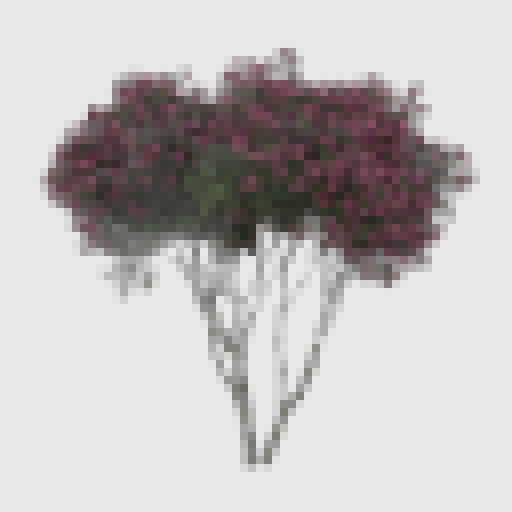}}
        &
        \frame{\includegraphics[height=\lenAblationSmall]{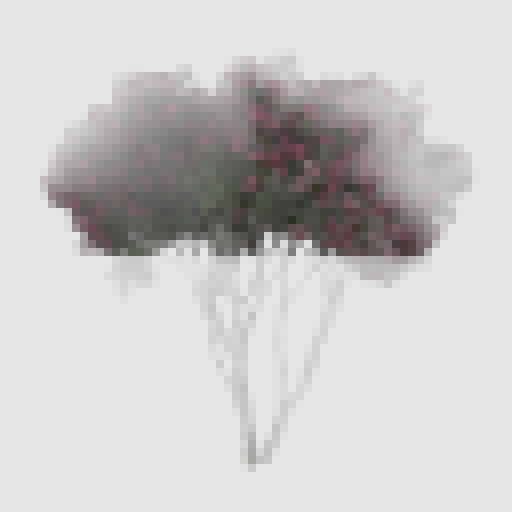}}
        &
        \frame{\includegraphics[height=\lenAblationSmall]{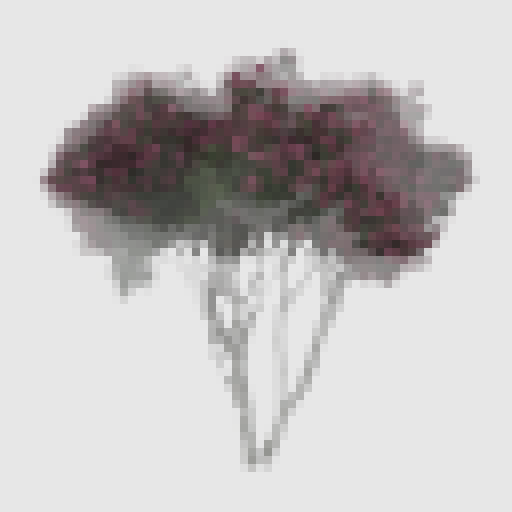}}
        \\
        \raisebox{10pt}{\rotatebox{90}{\textsf{Difference}}}
        &
		\multicolumn{1}{r}{\frame{\begin{overpic}[height=\lenAblationSmall]{imgs/colorbar.png}
			\put(-20, 92){\normalsize 0.4}
			\put(-20, 1){\normalsize 0.0}
		\end{overpic}}}
        &
        \frame{\includegraphics[height=\lenAblationSmall]{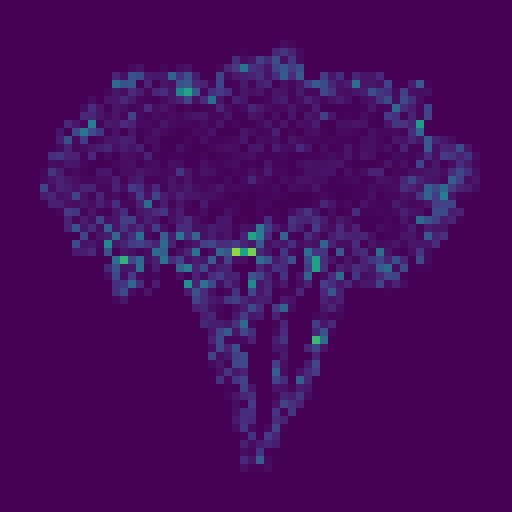}}
        &
        \frame{\includegraphics[height=\lenAblationSmall]{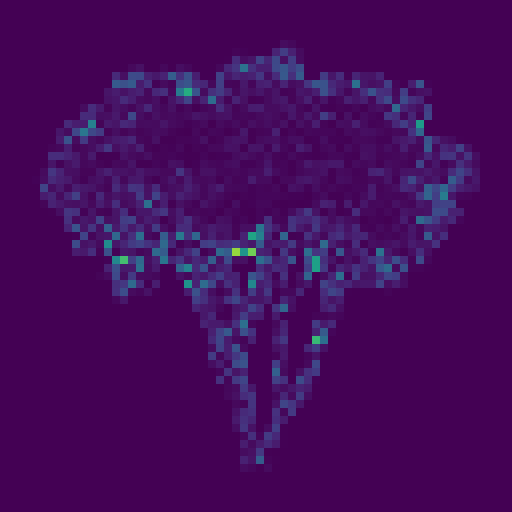}}
        &
        \frame{\includegraphics[height=\lenAblationSmall]{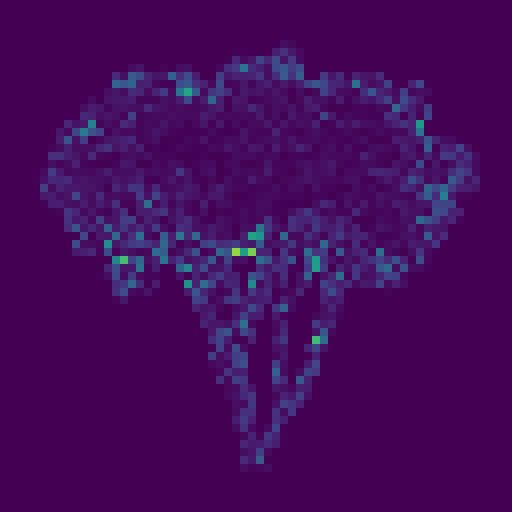}}
        &
        \frame{\includegraphics[height=\lenAblationSmall]{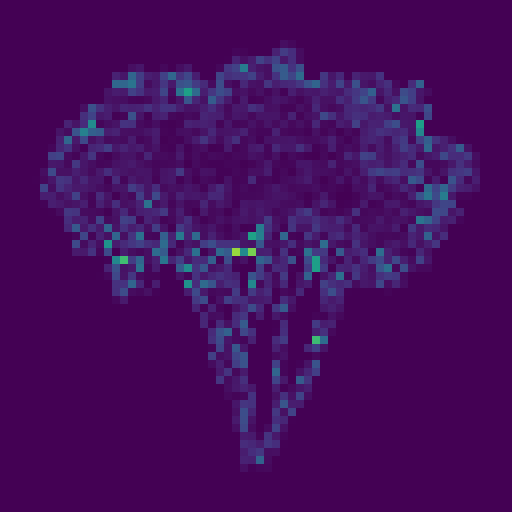}}
        &
        \frame{\includegraphics[height=\lenAblationSmall]{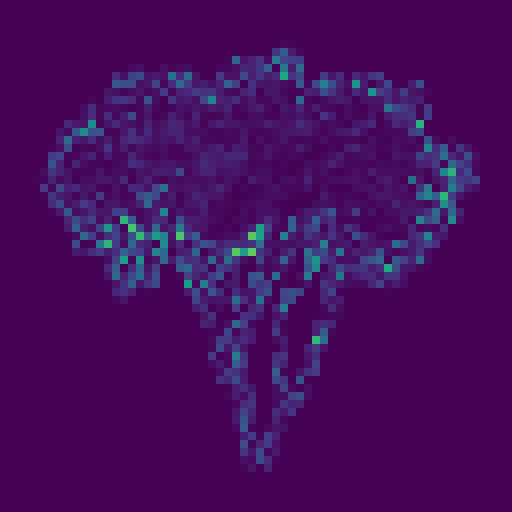}}
        &
        \frame{\includegraphics[height=\lenAblationSmall]{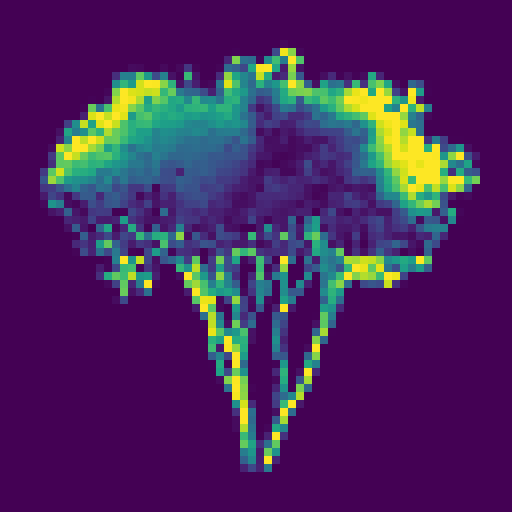}}
        &
        \frame{\includegraphics[height=\lenAblationSmall]{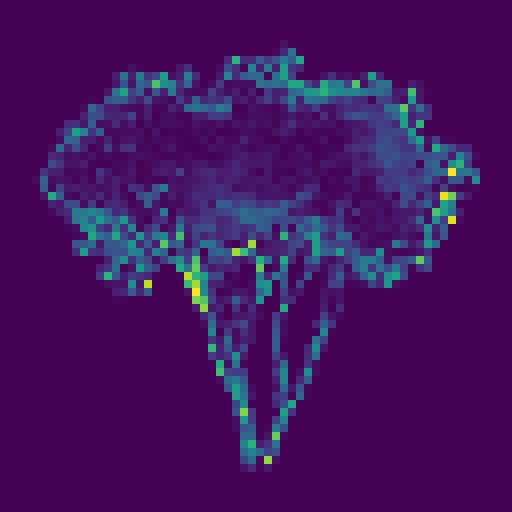}}
        \\  
        \raisebox{10pt}{\rotatebox{90}{\textsf{(Alt. View)}}}
        &
        \frame{\includegraphics[height=\lenAblationSmall]{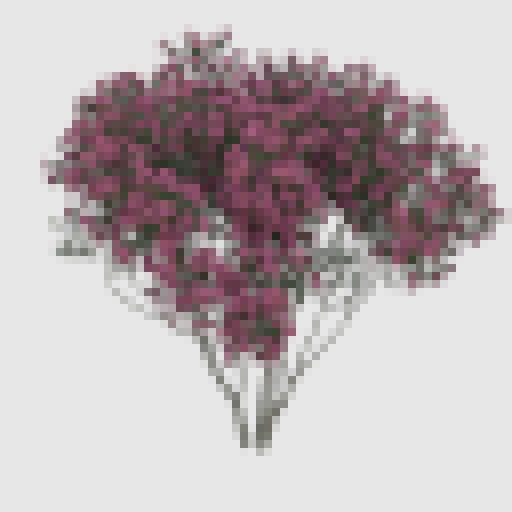}}
        &
        \frame{\includegraphics[height=\lenAblationSmall]{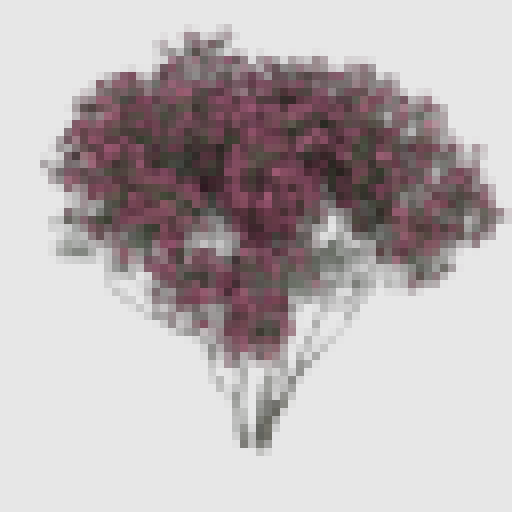}}
        &
        \frame{\includegraphics[height=\lenAblationSmall]{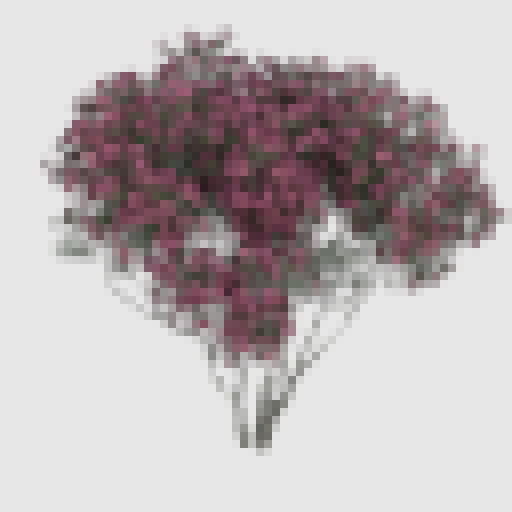}}
        &
        \frame{\includegraphics[height=\lenAblationSmall]{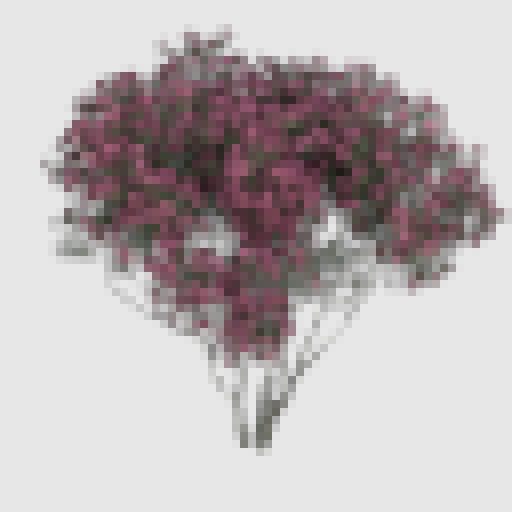}}
        &
        \frame{\includegraphics[height=\lenAblationSmall]{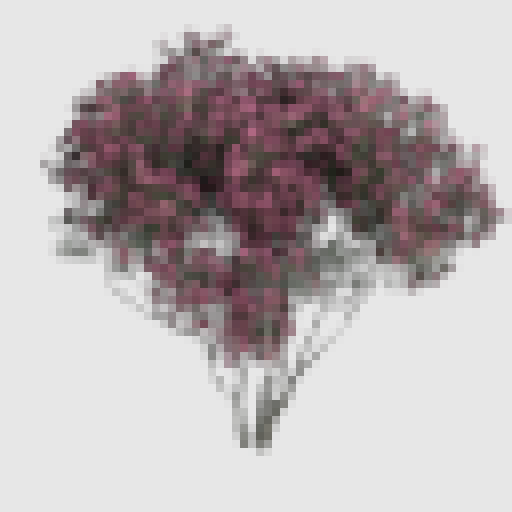}}
        &
        \frame{\includegraphics[height=\lenAblationSmall]{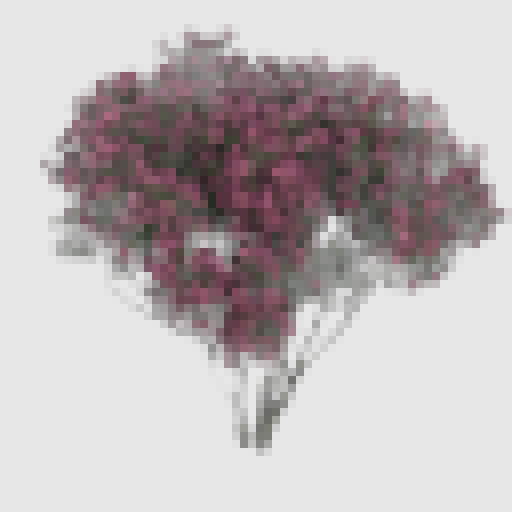}}
        &
        \frame{\includegraphics[height=\lenAblationSmall]{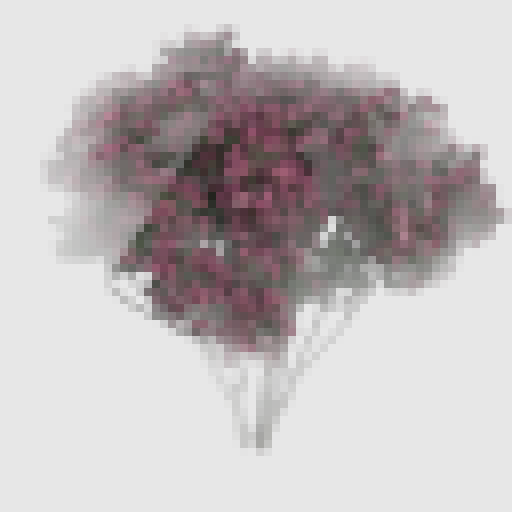}}
        &
        \frame{\includegraphics[height=\lenAblationSmall]{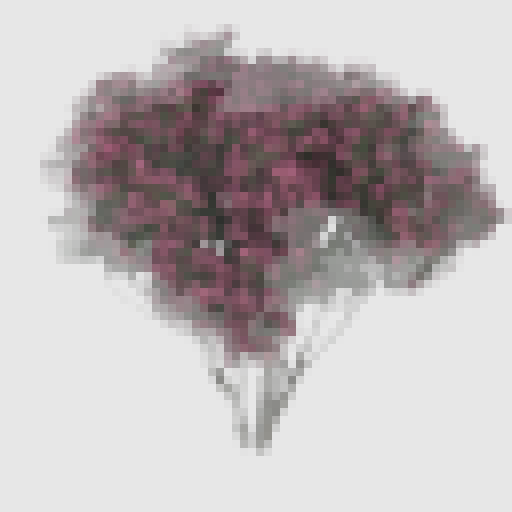}}
        \\
        \raisebox{10pt}{\rotatebox{90}{\textsf{Difference}}}
        &
		\multicolumn{1}{r}{\frame{\begin{overpic}[height=\lenAblationSmall]{imgs/colorbar.png}
			\put(-20, 92){\normalsize 0.4}
			\put(-20, 1){\normalsize 0.0}
		\end{overpic}}}
        &
        \frame{\includegraphics[height=\lenAblationSmall]{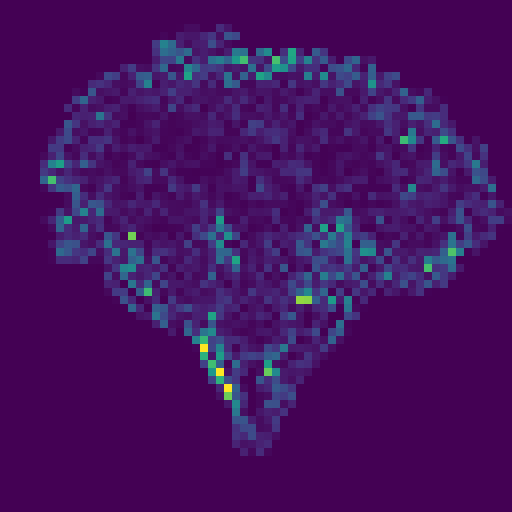}}
        &
        \frame{\includegraphics[height=\lenAblationSmall]{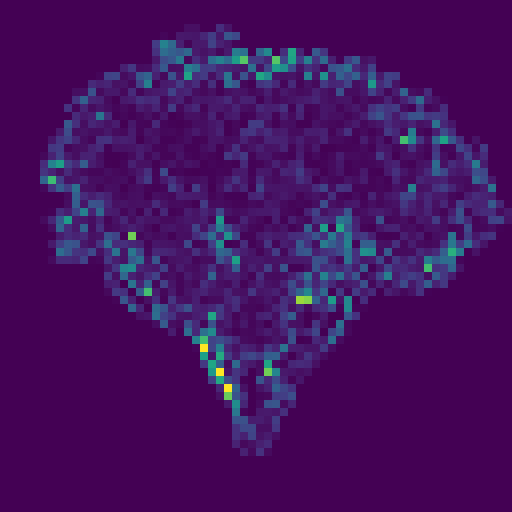}}
        &
        \frame{\includegraphics[height=\lenAblationSmall]{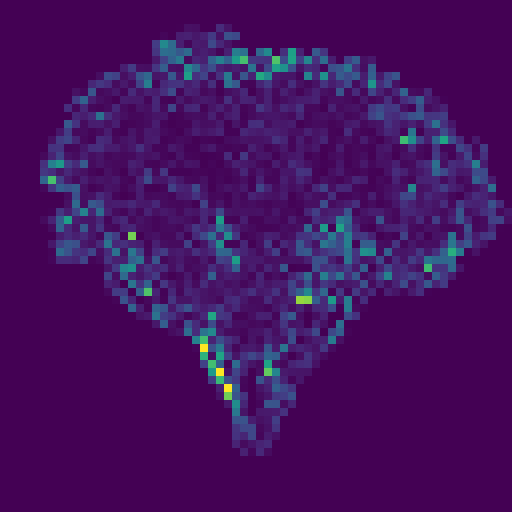}}
        &
        \frame{\includegraphics[height=\lenAblationSmall]{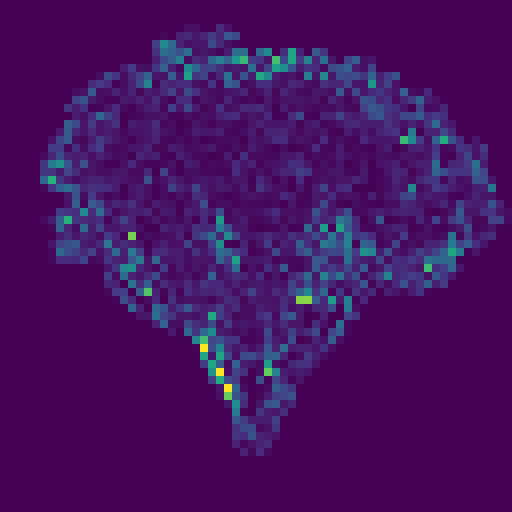}}
        &
        \frame{\includegraphics[height=\lenAblationSmall]{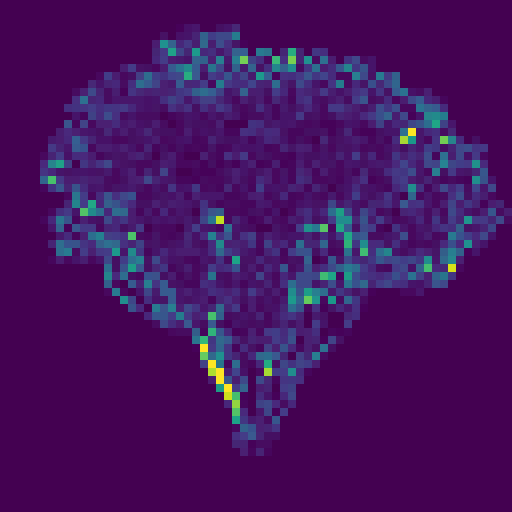}}
        &
        \frame{\includegraphics[height=\lenAblationSmall]{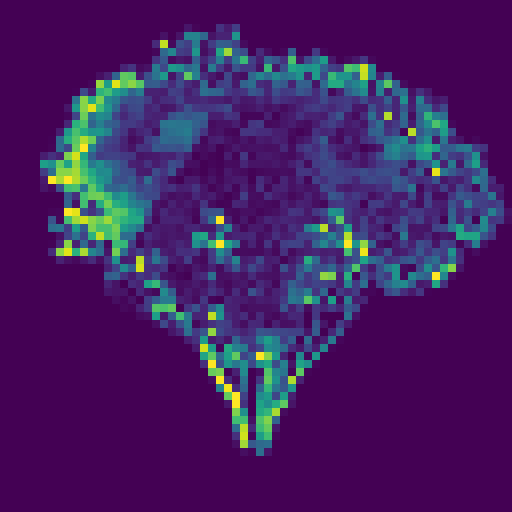}}
        &
        \frame{\includegraphics[height=\lenAblationSmall]{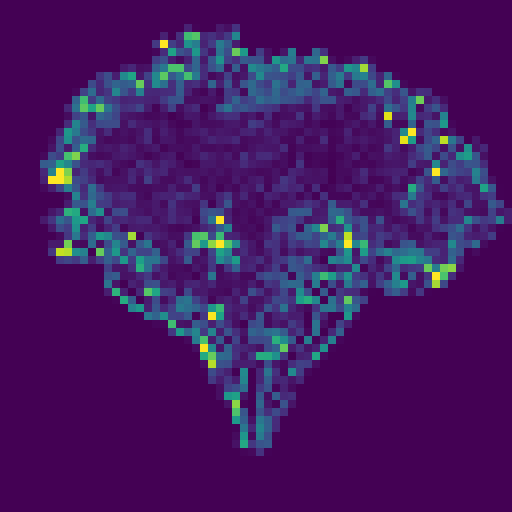}}   
    \end{tabular}
    \caption{\label{fig:ablation}
        Ablation study analyzing the impact of parameters on appearance and cost. All images are rendered using 1024 samples per pixel.
        Difference images between each configuration and the corresponding reference are provided. Please refer to \autoref{tab:ablation} for statistics.
    }    
\end{figure*}    

\begin{table*}[h]
	\centering
	\caption{Ablation study analyzing the impact of parameters on appearance and cost. Given the ``maximum'' configuration with up to 4 surface NDF lobes and \textbf{no} 
    compression, we change each paramter and show how it affects accuracy, memory requirement, and render time. Please refer to \autoref{fig:ablation} for rendered results.}
	\begin{tabular}{r|r|r|r|r|r|r|r|r}
		\Xhline{1pt}
		\multicolumn{2}{c|}{\textbf{Scene}} & 
        \textbf{Maximum} & 
        \textbf{1 NDF Lobe} &
        \textbf{10\% AIV} &
        \textbf{1\% AIV} &
        \textbf{10\% ABV} &
        \textbf{1\% ABV} &
        \textbf{1 CPCA Rep.}        
        \\
        \cline{1-9}
        \multirow[c]{3}{*}{\makecell{\emph{Chandelier} \\ $64^3$}}
        & \multicolumn{1}{l|}{Mem. (MB)}    & 70.1     & 69.5 (0.99$\times$)      & 48.9 (0.70$\times$)   & 46.8 (0.67$\times$)   & 31.2 (0.44$\times$)   & 27.3 (0.39$\times$)       & 27.2 (0.37$\times$) \\
        & \multicolumn{1}{l|}{Time (sec)}   & 1.91      & 1.87 (0.98$\times$)       & 1.91 (1.00$\times$)   & 1.62 (0.85$\times$)   & 1.81 (0.95$\times$)   & 1.78 (0.93$\times$)       & 1.85 (0.97$\times$) \\                
        & \multicolumn{1}{l|}{RMSE}         & 0.070     & 0.077                     & 0.070                 & 0.081                 & 0.071                 & 0.101                     & 0.078 \\
        \hline
        \multirow[c]{2}{*}{\makecell{\emph{Chandelier} \\ (alt. view)}} 
        & \multicolumn{1}{l|}{Time (sec)}   & 2.86      & 2.72 (0.95$\times$)       & 2.83 (0.99$\times$)   & 2.40 (0.84$\times$)   & 2.77 (0.97$\times$)   & 2.72 (0.95$\times$)       & 2.63 (0.92$\times$) \\        
        & \multicolumn{1}{l|}{RMSE}         & 0.242     & 0.322                     & 0.247                 & 0.290                 & 0.246                 & 0.266                     & 0.251 \\
        \hline
        \multirow[c]{3}{*}{\makecell{\emph{Oleander} \\ $64^3$}}
        & \multicolumn{1}{l|}{Mem. (MB)}    & 125.8     & 124.5 (0.99$\times$)      & 66.6 (0.53$\times$)  & 60.7 (0.48$\times$)  & 154.7 (0.62$\times$)  & 77.4 (0.58$\times$)      & 71.2 (0.57$\times$) \\
        & \multicolumn{1}{l|}{Time (sec)}   & 5.13      & 5.03 (0.98$\times$)       & 5.13 (1.00$\times$)   & 4.57 (0.89$\times$)   & 5.08 (0.99$\times$)   & 5.18 (1.01$\times$)       & 5.14 (1.00$\times$) \\        
        & \multicolumn{1}{l|}{RMSE}         & 0.025     & 0.025                     & 0.026                 & 0.028                 & 0.032                 & 0.116                     & 0.053 \\
        \hline
        \multirow[c]{2}{*}{\makecell{\emph{Oleander} \\ (alt. view)}}
        & \multicolumn{1}{l|}{Time (sec)}   & 8.06     & 7.90 (0.98$\times$)      & 7.93 (0.98$\times$)  & 6.77 (0.84$\times$)  & 7.66 (0.95$\times$)  & 7.90 (0.98$\times$)      & 7.81 (0.97$\times$) \\             
        & \multicolumn{1}{l|}{RMSE}         & 0.046     & 0.046                     & 0.047                 & 0.047                 & 0.055                 & 0.088                     & 0.070 \\   		
		\Xhline{1pt}
	\end{tabular}
	\label{tab:ablation}
\end{table*}

\rev{
\paragraph{Ablation Study}
In \autoref{tab:ablation} and \autoref{fig:ablation}, we conduct an ablation study on the impact of parameters to accuracy and cost. For each model, we
precompute it with the ``maximum'' configuration to serve as the control, where up to 4 surface NDF lobes are allowed and no compression is applied at all.
We then vary each parameter and assess its affect on rendering quality, memory requirement, and time compared to the control. The benefit of using multiple
NDF lobes is more prominent in the \emph{Chandelier} scene to capture the glossy base material together with curved surfaces; it is not obvious in the more
diffuse \emph{Oleander} scene. For both types of aggregated visibility, $10\%$ coefficient truncation only introduces barely recognizable error, while $1\%$
truncation results in visible inaccuracy. Finally, by keeping only 1 CPCA representative, CPCA is reduced to simple vector quantization and approximates
per-cluster subspaces poorly. The visibility data is high dimensional and thus compression parameters greatly affects the final memory requirement.
All parameters affect the shading cost but it is minor compared to voxel traversal and intersection test.
}

\paragraph{Limitations}
Our method has several limitations that could serve as fruitful topics for future research.
As prefaced, so far we have been focusing on direct illumination. In order for a scene aggregate to support global illumination, multiple scattering
between different parts of the scene should be modeled. This brings new challenges as discretizing and aggregating individual regions will inevitably lose the
information about how different regions interact with each other. One possible approach is to precompute and aggregate the entire transport from the external
environment to a given region. The definition of ABSDF should be extended accordingly in this case.

\begin{figure}[h]
	\newlength{\lenLimitationVisibility}
	\setlength{\lenLimitationVisibility}{0.75in}
	\centering
    \begin{tabular}{cccc}
		\frame{\includegraphics[height=\lenLimitationVisibility]{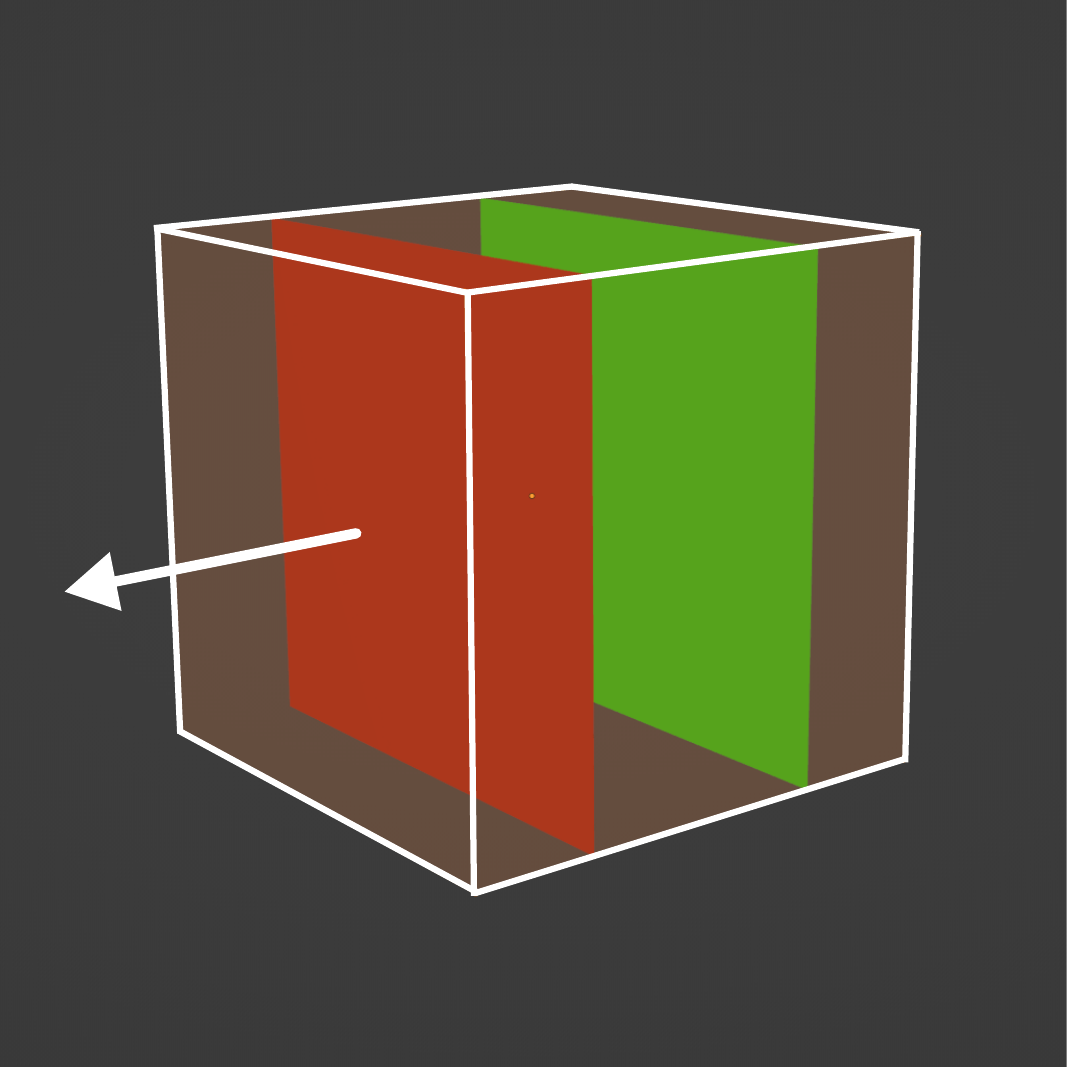}}
		&
		\frame{\includegraphics[height=\lenLimitationVisibility]{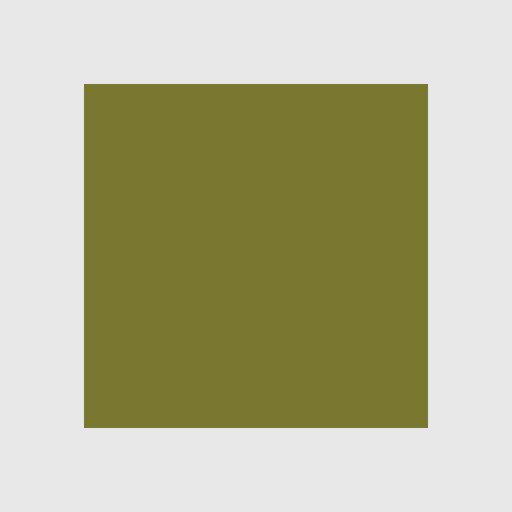}}
		&
		\frame{\includegraphics[height=\lenLimitationVisibility]{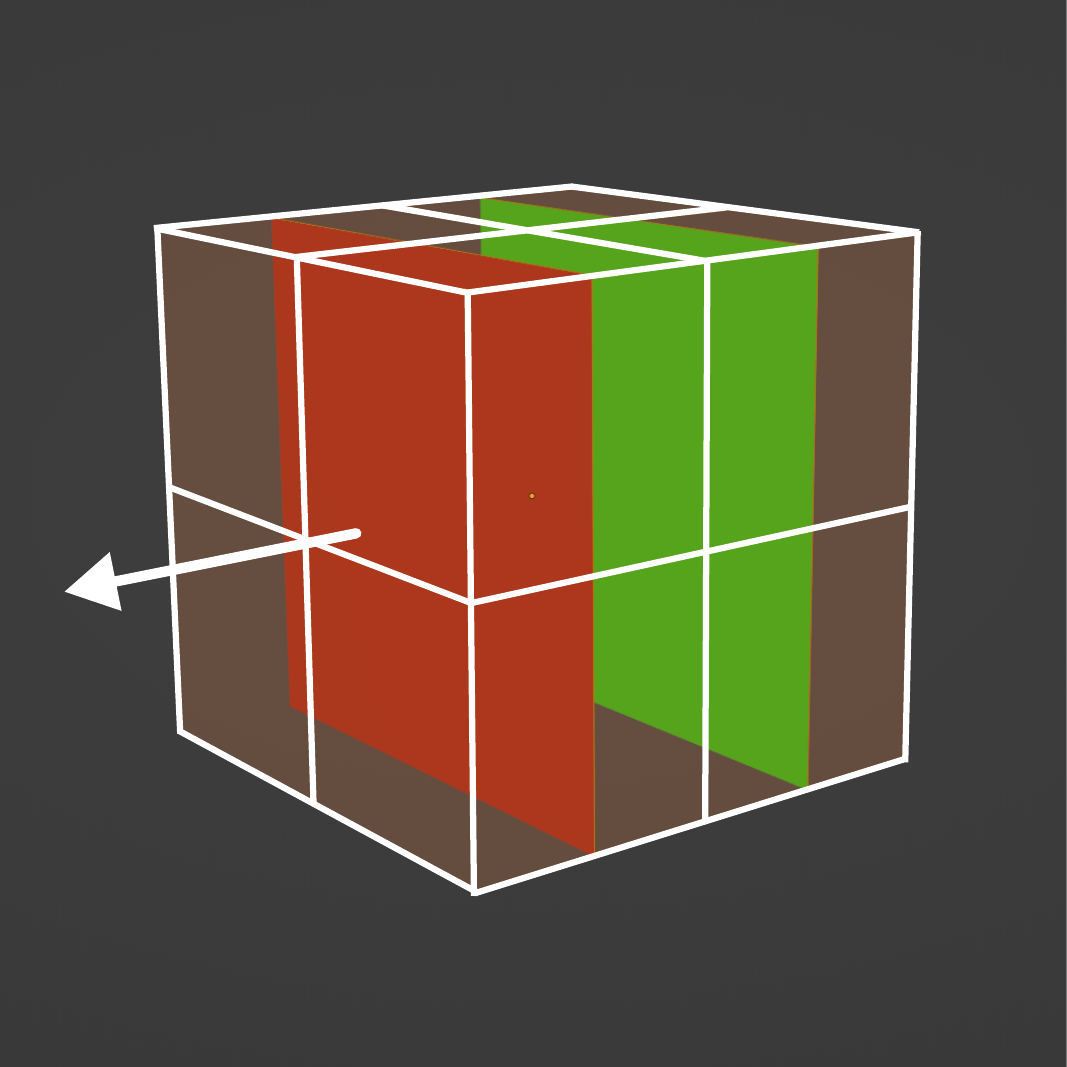}}
		&
		\frame{\includegraphics[height=\lenLimitationVisibility]{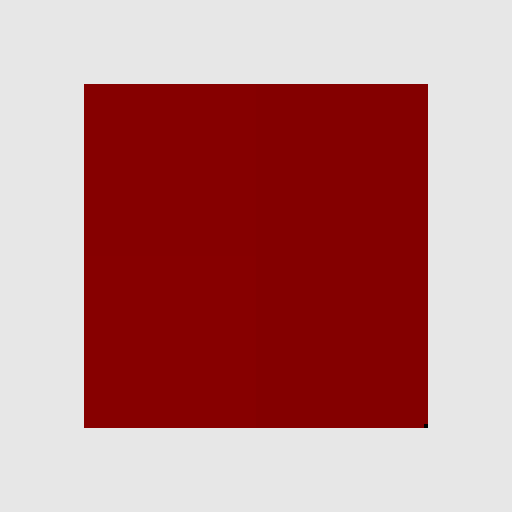}}
		\\
		\multicolumn{2}{c}{\textsf{(a) Render with $1$ voxel}}
		&
		\multicolumn{2}{c}{\textsf{(b) Render with $2^3$ voxels}}
    \end{tabular}
    \caption{\label{fig:limitation_visibility} (a) Due the separate visibility approximation (\autoref{eq:split_vis}), our method cannot handle correlation
	between visibility and material in a single voxel. The render averages the contribution from the red and green materials. (b) However, this limitation is
	mitigated as the voxelization resolution increases.}
\end{figure}

Two more limitations stem from the separate approximations made in \autoref{eq:split_vis} and \autoref{eq:define_aiv}. \autoref{eq:split_vis} assumes
independence between visibility and material in a single voxel which could lead to certain artifacts as illustrated in \autoref{fig:limitation_visibility}.
However, this is alleviated as the spatial resolution grows, since the correlated parts are more likely to be grouped into different voxels.
\autoref{eq:define_aiv} assumes independence between visibility along two directions. This could lead to incorrect occlusion when, for example, the camera
and the light source are collocated. To the best of our knowledge, compactly representing the general 4D correlated bidirectional visibility remains an open
problem for LoD techniques. In practice, we find that this assumption rarely causes noticeable artifacts.

Finally, it would be desirable to further support material models beyond the Disney BRDF. For example, foliage often exhibits non-negligible subsurface scattering
effects. One simple extension to our current ABSDF factorization could be to model an extra diffuse transmission component.

\section{Conclusion}
We present an efficient scene appearance aggregation method for LoD rendering.
Our method is based on a novel formulation for far-field scene aggregation with the definition of ABSDF, which captures the aggregated appearance of all 
surfaces within a volume. We develop a closed-form factorization of the ABSDF that supports all-frequency and view-dependent effects with handy evaluation and 
sampling procedures. Our representation naturally accounts for long-range correlation by recording two types of global visibility, the aggregated interior 
visibility and the aggregated boundary visibility. Our 
truncated ellipsoid primitive improves the preservation of local correlation compared to the na\"ive cubic primitive. 
We have demonstrated the accuracy of our method on a variety of scenes with different geometric and material characteristics and its scalability to large, 
complex scenes. 
Our results achieve higher quality than those from state-of-the-art LoD techniques.
While our implementation is far from optimized, we can already show the asymptotic advantages of our representation in terms of memory footprint and rendering 
speed compared to the original representation.
We believe our work is highly relevant to improving the scalability of physically based rendering, enabling the generation of richer, more realistic 3D content.

\bibliographystyle{ACM-Reference-Format}
\bibliography{main}

\end{document}


\maketitle

\section{Details and Validation in ABSDF Factorization}
We provide additional derivation details and numerical validation for different steps involved in our ABSDF factorization.

\subsection{SGGX-based Precomputed Convolution}
In the main article, we propose to represent the convolution of an SGGX distribution and an isotropic spherical distribution $g$ as another SGGX with the
same eigenvectors. We first describe our parameterization of SGGX for this purpose. Given an SGGX matrix $S$ with eigenbasis $(\omega_1, \omega_2, \omega_3)$
and projected area $\sigma(\omega_1) \leq \sigma(\omega_2) \leq \sigma(\omega_3)$, we can always scale the entire matrix by $1/\sigma^2(\omega_3)$.
The result is still a valid SGGX matrix:
\begin{equation}
    S' = \frac{S}{\sigma^2(\omega_3)} =
    (\omega_1, \omega_2, \omega_3)
    \begingroup %
    \setlength\arraycolsep{1pt}
    \begin{pmatrix}
    \frac{\sigma^2(\omega_1)}{\sigma^2(\omega_3)} & 0 & 0\\
    0 & \frac{\sigma^2(\omega_2)}{\sigma^2(\omega_3)} & 0\\
    0 & 0 & 1
    \end{pmatrix}
    \endgroup
    (\omega_1, \omega_2, \omega_3)^T,
\end{equation}
It is shown by Heitz et al.~\shortcite{heitz2015sggx} (supplemental document) that $S'$ is a double-sided GGX with roughness $\alpha_x = \sigma(\omega_1) / \sigma(\omega_3)$,
$\alpha_y = \sigma(\omega_1) / \sigma(\omega_3)$, and tangent frame $R = (\omega_1, \omega_2, \omega_3)$. Therefore, we can re-parameterize the SGGX as
$D_{\mathrm{sggx}}(n; R, \bm{\alpha})$, where $\bm{\alpha} = (\alpha_x, \alpha_y)$. 
Following Eq.~\ref{eq:per_sggx_lobe_conv} in the main article, the post-convolution distribution is a roughened SGGX
$D_{\mathrm{sggx}}(n; R, \bm{\alpha}_{+})$, where $\bm{\alpha}_{+}$ includes the additional roughness gained
from the convolution. We use nonlinear least-square fit to find the best mapping that gives the additional roughness from the original roughness the parameters
of $g$. The ground-truth convolution is computed by Monte Carlo integration with a large number of samples. In the following, we validate the accuracy of this
technique for the two encountered target distributions.

\begin{figure}[h]
	\centering
  \begin{tabular}{c}
    \includegraphics[height=2in]{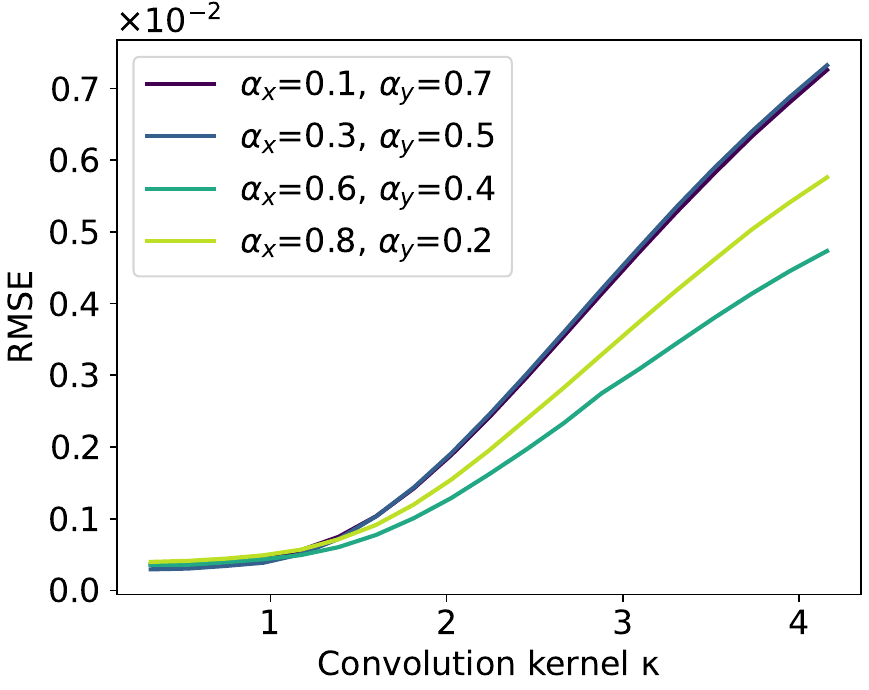}
    \\
    (a)
    \\
    \includegraphics[height=2in]{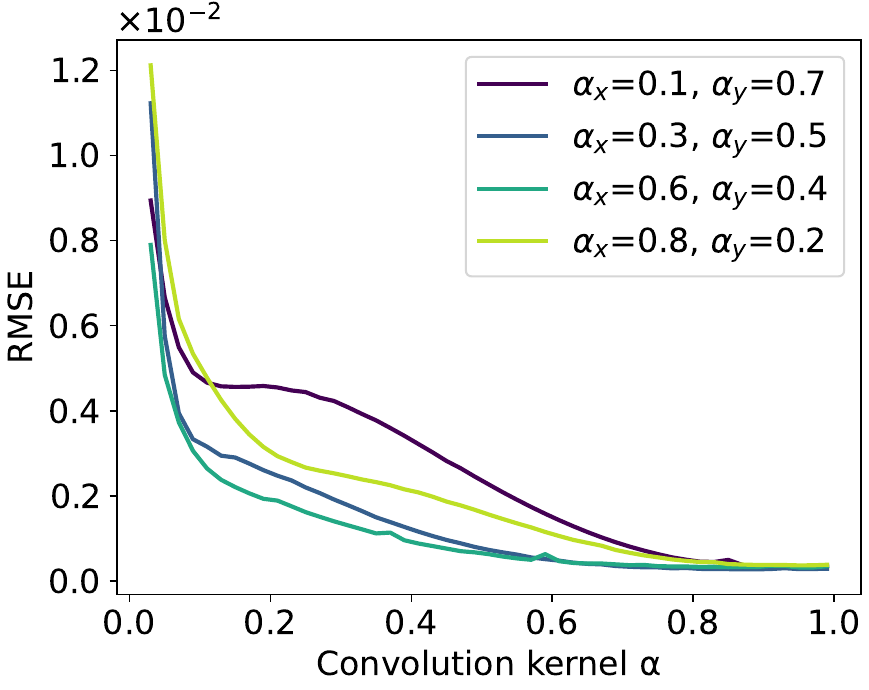}
    \\
    (b)
  \end{tabular}
    \caption{\label{fig:validate_conv_error}
        (a) RMSE of our SGGX-SG convolution fit with varying $\kappa$. Note that we only consider a small range of $\kappa$ because we only
        use SG to represent the clamped dot product term.
        (b) RMSE of our SGGX-GGX convolution fit with varying $\alpha$.
    }
\end{figure}

\begin{figure*}[t]
	\newlength{\lenValidateSGGXVMFConv}
	\setlength{\lenValidateSGGXVMFConv}{0.71in}
    \centering
    \begin{tabular}{cccc|cccc}
        Original & Kernel & Conv. Ours & Conv. Ref. & Original & Kernel & Conv. Ours & Conv. Ref. \\
        \frame{\begin{overpic}[width=0.6in]{imgs/colorbar_hori.png}
            \put(-23, 0){\normalsize \small{0.0}}
            \put(101, 0){\normalsize \small{1.0}}
        \end{overpic}} & & & & & & & \\
        \includegraphics[height=\lenValidateSGGXVMFConv]{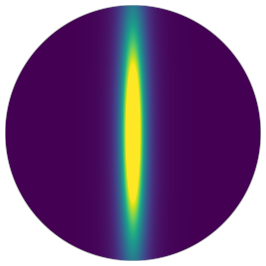}
        &
        \includegraphics[height=\lenValidateSGGXVMFConv]{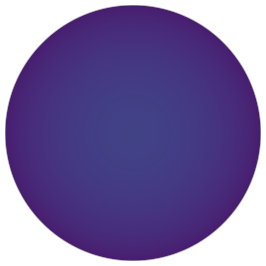}
        &
        \includegraphics[height=\lenValidateSGGXVMFConv]{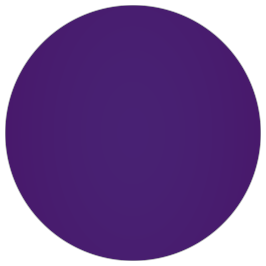}
        &
        \includegraphics[height=\lenValidateSGGXVMFConv]{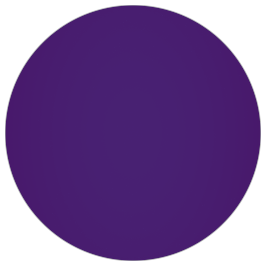}
        &
        \includegraphics[height=\lenValidateSGGXVMFConv]{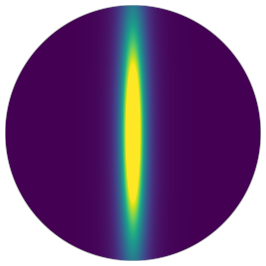}
        &
        \includegraphics[height=\lenValidateSGGXVMFConv]{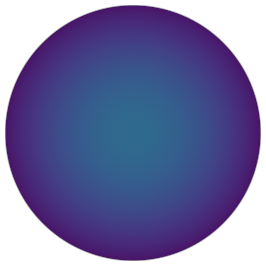}
        &
        \includegraphics[height=\lenValidateSGGXVMFConv]{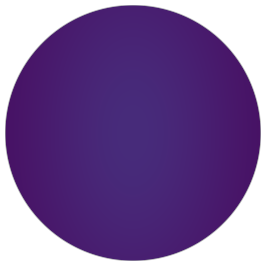}
        &
        \includegraphics[height=\lenValidateSGGXVMFConv]{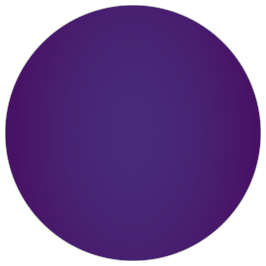}
        \\
        \multicolumn{4}{c}{$\alpha_x = 0.1$, $\alpha_y = 0.7$, $\kappa = 1.0$}
        &
        \multicolumn{4}{c}{$\alpha_x = 0.1$, $\alpha_y = 0.7$, $\kappa = 2.0$}
        \\
        \includegraphics[height=\lenValidateSGGXVMFConv]{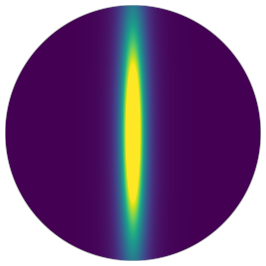}
        &
        \includegraphics[height=\lenValidateSGGXVMFConv]{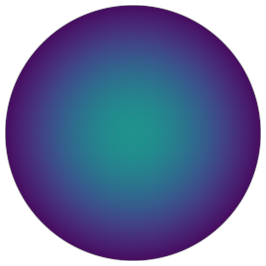}
        &
        \includegraphics[height=\lenValidateSGGXVMFConv]{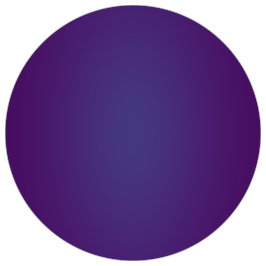}
        &
        \includegraphics[height=\lenValidateSGGXVMFConv]{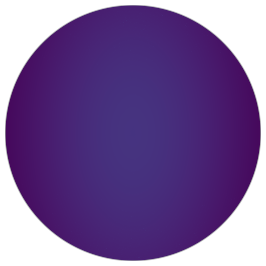}
        &
        \includegraphics[height=\lenValidateSGGXVMFConv]{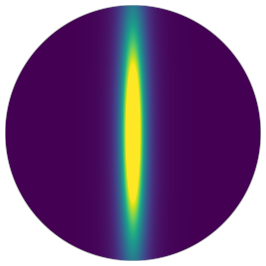}
        &
        \includegraphics[height=\lenValidateSGGXVMFConv]{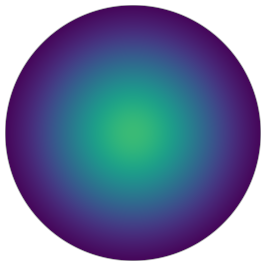}
        &
        \includegraphics[height=\lenValidateSGGXVMFConv]{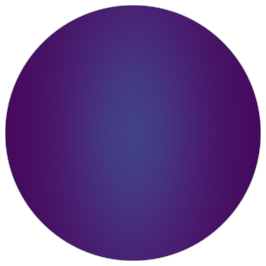}
        &
        \includegraphics[height=\lenValidateSGGXVMFConv]{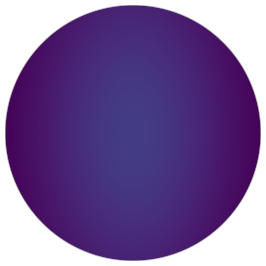}
        \\
        \multicolumn{4}{c}{$\alpha_x = 0.1$, $\alpha_y = 0.7$, $\kappa = 3.0$}
        &
        \multicolumn{4}{c}{$\alpha_x = 0.1$, $\alpha_y = 0.7$, $\kappa = 4.0$}
        \\
        \includegraphics[height=\lenValidateSGGXVMFConv]{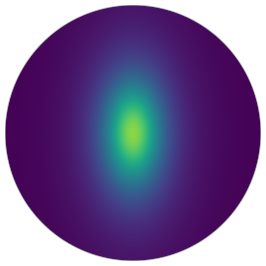}
        &
        \includegraphics[height=\lenValidateSGGXVMFConv]{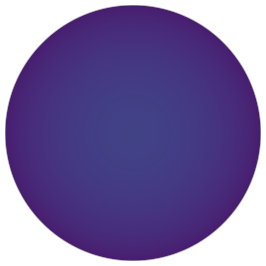}
        &
        \includegraphics[height=\lenValidateSGGXVMFConv]{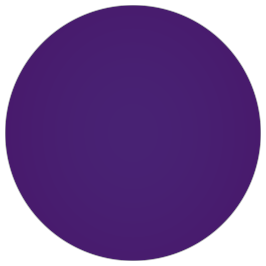}
        &
        \includegraphics[height=\lenValidateSGGXVMFConv]{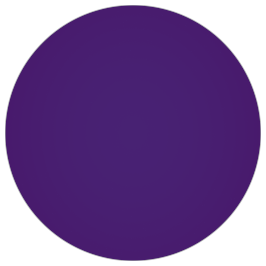}
        &
        \includegraphics[height=\lenValidateSGGXVMFConv]{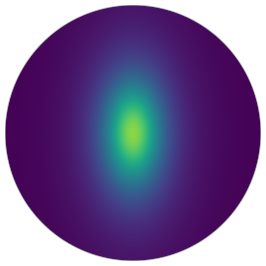}
        &
        \includegraphics[height=\lenValidateSGGXVMFConv]{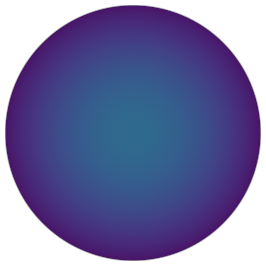}
        &
        \includegraphics[height=\lenValidateSGGXVMFConv]{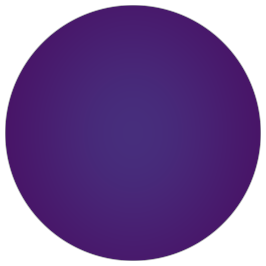}
        &
        \includegraphics[height=\lenValidateSGGXVMFConv]{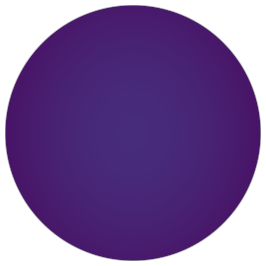}
        \\
        \multicolumn{4}{c}{$\alpha_x = 0.3$, $\alpha_y = 0.5$, $\kappa = 1.0$}
        &
        \multicolumn{4}{c}{$\alpha_x = 0.3$, $\alpha_y = 0.5$, $\kappa = 2.0$}
        \\
        \includegraphics[height=\lenValidateSGGXVMFConv]{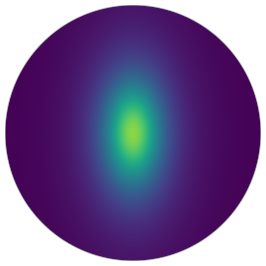}
        &
        \includegraphics[height=\lenValidateSGGXVMFConv]{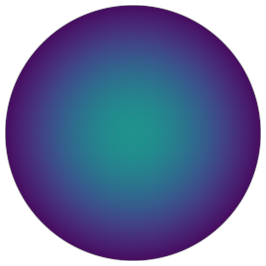}
        &
        \includegraphics[height=\lenValidateSGGXVMFConv]{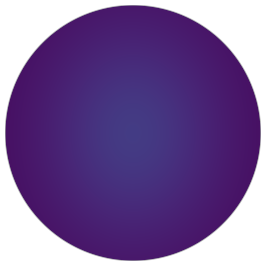}
        &
        \includegraphics[height=\lenValidateSGGXVMFConv]{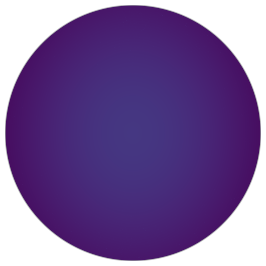}
        &
        \includegraphics[height=\lenValidateSGGXVMFConv]{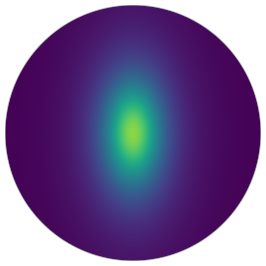}
        &
        \includegraphics[height=\lenValidateSGGXVMFConv]{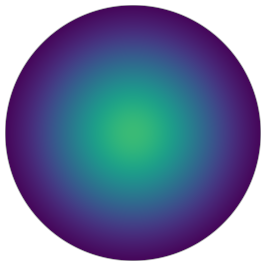}
        &
        \includegraphics[height=\lenValidateSGGXVMFConv]{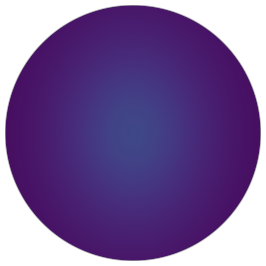}
        &
        \includegraphics[height=\lenValidateSGGXVMFConv]{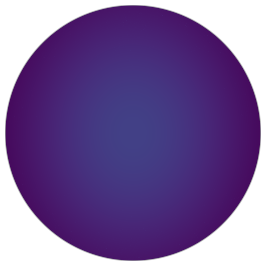}
        \\
        \multicolumn{4}{c}{$\alpha_x = 0.3$, $\alpha_y = 0.5$, $\kappa = 3.0$}
        &
        \multicolumn{4}{c}{$\alpha_x = 0.3$, $\alpha_y = 0.5$, $\kappa = 4.0$}
        \\
        \includegraphics[height=\lenValidateSGGXVMFConv]{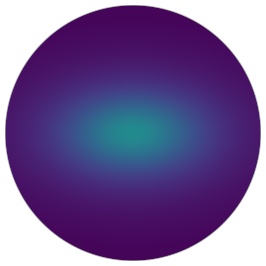}
        &
        \includegraphics[height=\lenValidateSGGXVMFConv]{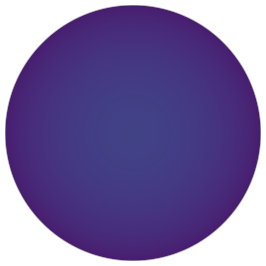}
        &
        \includegraphics[height=\lenValidateSGGXVMFConv]{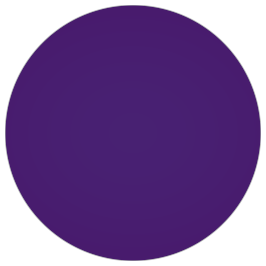}
        &
        \includegraphics[height=\lenValidateSGGXVMFConv]{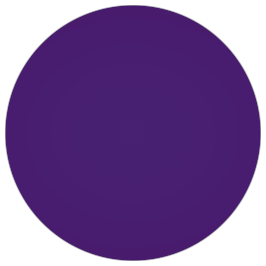}
        &
        \includegraphics[height=\lenValidateSGGXVMFConv]{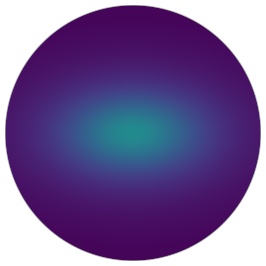}
        &
        \includegraphics[height=\lenValidateSGGXVMFConv]{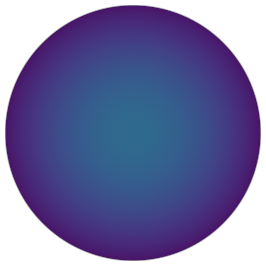}
        &
        \includegraphics[height=\lenValidateSGGXVMFConv]{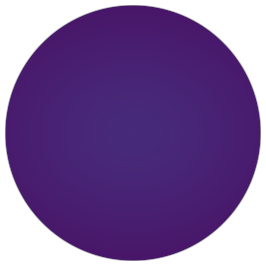}
        &
        \includegraphics[height=\lenValidateSGGXVMFConv]{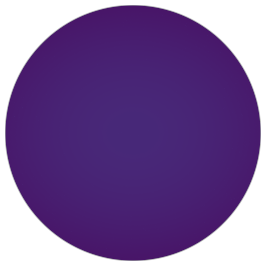}
        \\
        \multicolumn{4}{c}{$\alpha_x = 0.6$, $\alpha_y = 0.4$, $\kappa = 1.0$}
        &
        \multicolumn{4}{c}{$\alpha_x = 0.6$, $\alpha_y = 0.4$, $\kappa = 2.0$}
        \\
        \includegraphics[height=\lenValidateSGGXVMFConv]{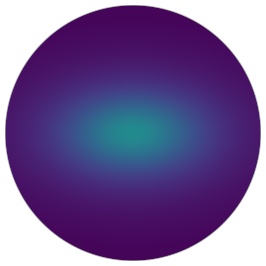}
        &
        \includegraphics[height=\lenValidateSGGXVMFConv]{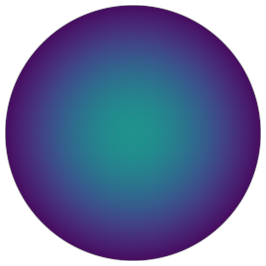}
        &
        \includegraphics[height=\lenValidateSGGXVMFConv]{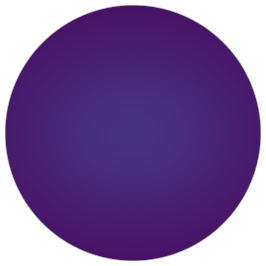}
        &
        \includegraphics[height=\lenValidateSGGXVMFConv]{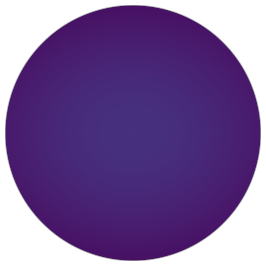}
        &
        \includegraphics[height=\lenValidateSGGXVMFConv]{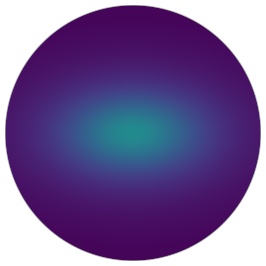}
        &
        \includegraphics[height=\lenValidateSGGXVMFConv]{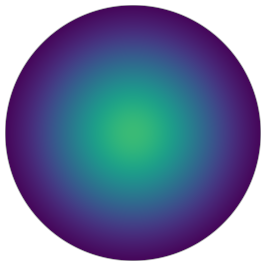}
        &
        \includegraphics[height=\lenValidateSGGXVMFConv]{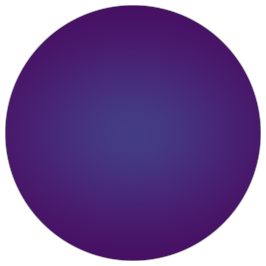}
        &
        \includegraphics[height=\lenValidateSGGXVMFConv]{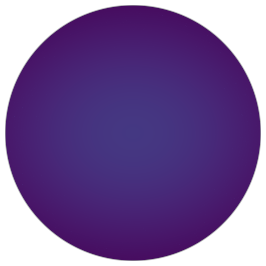}
        \\
        \multicolumn{4}{c}{$\alpha_x = 0.6$, $\alpha_y = 0.4$, $\kappa = 3.0$}
        &
        \multicolumn{4}{c}{$\alpha_x = 0.6$, $\alpha_y = 0.4$, $\kappa = 4.0$}
        \\
        \includegraphics[height=\lenValidateSGGXVMFConv]{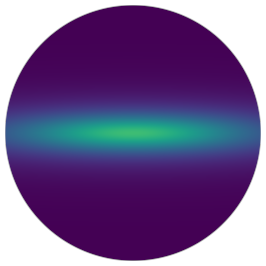}
        &
        \includegraphics[height=\lenValidateSGGXVMFConv]{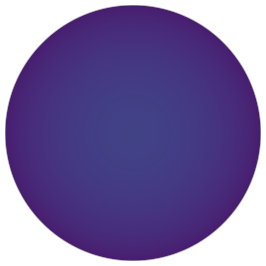}
        &
        \includegraphics[height=\lenValidateSGGXVMFConv]{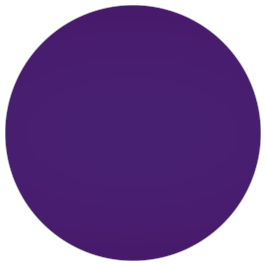}
        &
        \includegraphics[height=\lenValidateSGGXVMFConv]{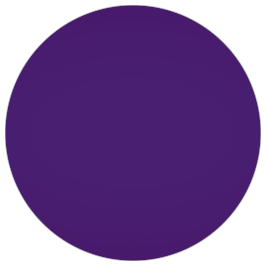}
        &
        \includegraphics[height=\lenValidateSGGXVMFConv]{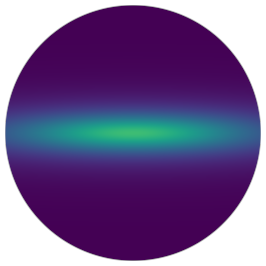}
        &
        \includegraphics[height=\lenValidateSGGXVMFConv]{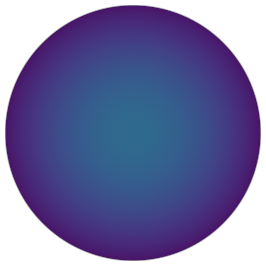}
        &
        \includegraphics[height=\lenValidateSGGXVMFConv]{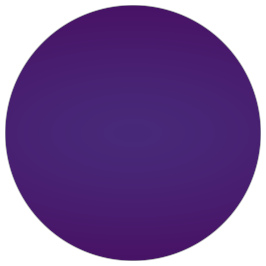}
        &
        \includegraphics[height=\lenValidateSGGXVMFConv]{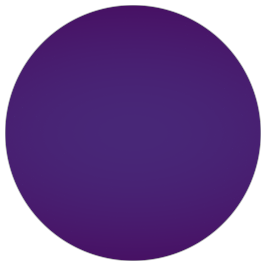}
        \\
        \multicolumn{4}{c}{$\alpha_x = 0.8$, $\alpha_y = 0.2$, $\kappa = 1.0$}
        &
        \multicolumn{4}{c}{$\alpha_x = 0.8$, $\alpha_y = 0.2$, $\kappa = 2.0$}
        \\
        \includegraphics[height=\lenValidateSGGXVMFConv]{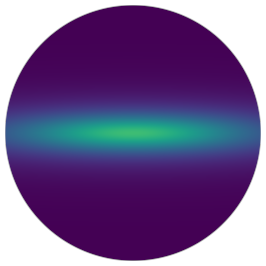}
        &
        \includegraphics[height=\lenValidateSGGXVMFConv]{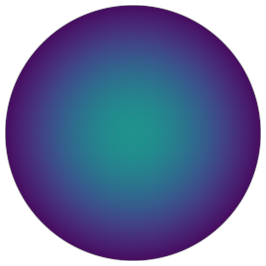}
        &
        \includegraphics[height=\lenValidateSGGXVMFConv]{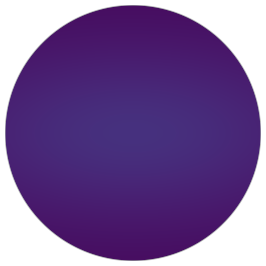}
        &
        \includegraphics[height=\lenValidateSGGXVMFConv]{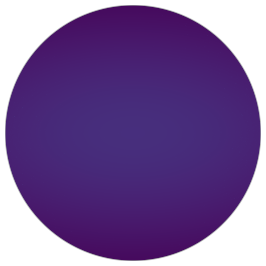}
        &
        \includegraphics[height=\lenValidateSGGXVMFConv]{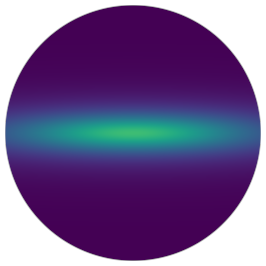}
        &
        \includegraphics[height=\lenValidateSGGXVMFConv]{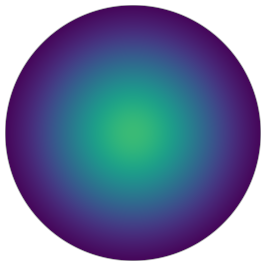}
        &
        \includegraphics[height=\lenValidateSGGXVMFConv]{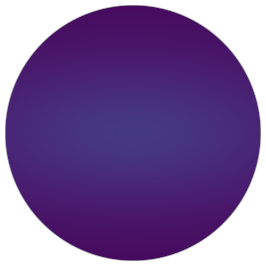}
        &
        \includegraphics[height=\lenValidateSGGXVMFConv]{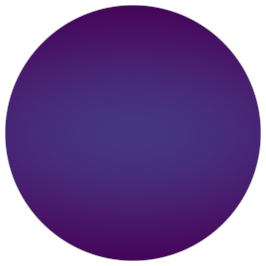}
        \\
        \multicolumn{4}{c}{$\alpha_x = 0.8$, $\alpha_y = 0.2$, $\kappa = 3.0$}
        &
        \multicolumn{4}{c}{$\alpha_x = 0.8$, $\alpha_y = 0.2$, $\kappa = 4.0$}
        \\

    \end{tabular}
    \caption{\label{fig:validate_sggx_vmf_conv_main}
        Convolving an SGGX with an SG kernel. Our fits closely matches the references in different configurations. Note that $\kappa$ is small because we only
        use SG to represent the clamped dot product term.
    }
\end{figure*}

In \S\ref{sec:factorize_diffuse}, we convolve an SGGX with a Spherical Gaussian (SG): $g(\omega) = \mathrm{SG}(\omega; n, \kappa)$ and compute a 3D mapping
$\bm{\alpha}_{+} = M_1(\bm{\alpha}, \kappa)$ (two channels). As we only consider a small range of $\kappa$, we use $20 \times 20 \times 20 \times 2$
for its resolution in practice.
In Fig~\ref{fig:validate_sggx_vmf_conv_main}, we visualize our fit using a single SGGX and compare it to the ground-truth convolution in different
configurations. In Fig~\ref{fig:validate_conv_error}, we plot the RMSE error between our fit and the ground truth with varying $\kappa$.

\begin{figure*}[t]
	\newlength{\lenValidateConv}
	\setlength{\lenValidateConv}{0.71in}
    \centering
    \begin{tabular}{cccc|cccc}
        Original & Kernel & Conv. Ours & Conv. Ref. & Original & Kernel & Conv. Ours & Conv. Ref. \\
        \frame{\begin{overpic}[width=0.6in]{imgs/colorbar_hori.png}
            \put(-23, 0){\normalsize \small{0.0}}
            \put(101, 0){\normalsize \small{1.0}}
        \end{overpic}} & & & & & & & \\
        \includegraphics[height=\lenValidateConv]{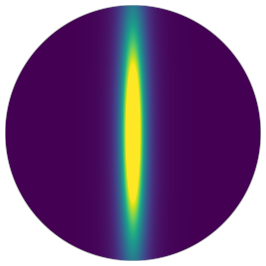}
        &
        \includegraphics[height=\lenValidateConv]{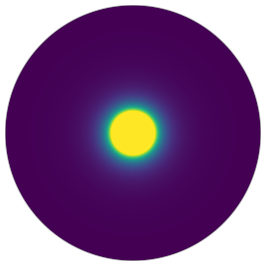}
        &
        \includegraphics[height=\lenValidateConv]{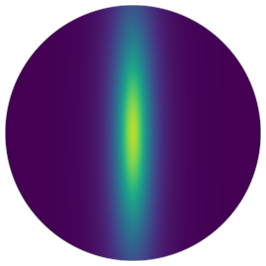}
        &
        \includegraphics[height=\lenValidateConv]{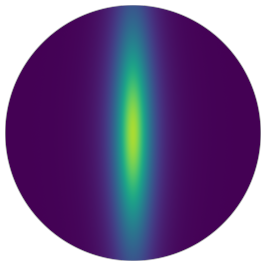}
        &
        \includegraphics[height=\lenValidateConv]{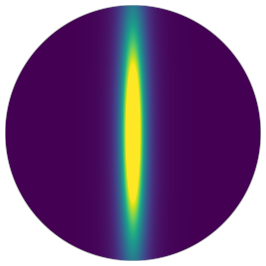}
        &
        \includegraphics[height=\lenValidateConv]{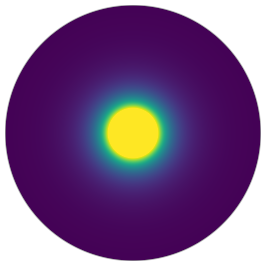}
        &
        \includegraphics[height=\lenValidateConv]{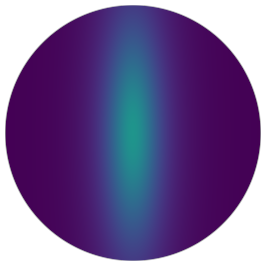}
        &
        \includegraphics[height=\lenValidateConv]{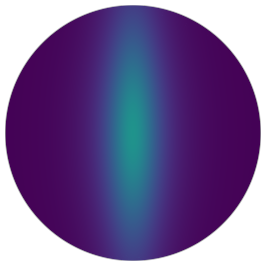}
        \\
        \multicolumn{4}{c}{$\alpha_x = 0.1$, $\alpha_y = 0.7$, $\alpha = 0.1$}
        &
        \multicolumn{4}{c}{$\alpha_x = 0.1$, $\alpha_y = 0.7$, $\alpha = 0.2$}
        \\
        \includegraphics[height=\lenValidateConv]{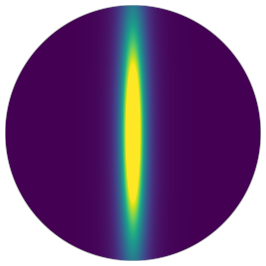}
        &
        \includegraphics[height=\lenValidateConv]{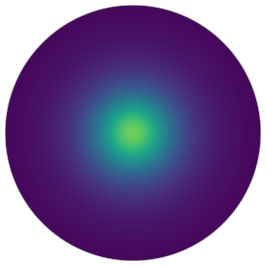}
        &
        \includegraphics[height=\lenValidateConv]{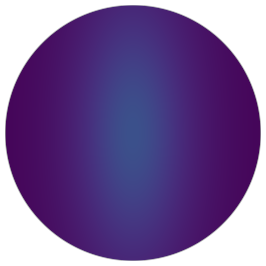}
        &
        \includegraphics[height=\lenValidateConv]{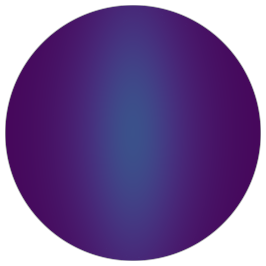}
        &
        \includegraphics[height=\lenValidateConv]{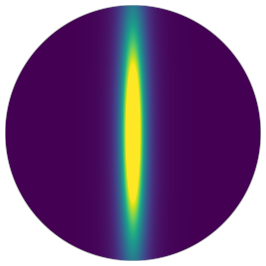}
        &
        \includegraphics[height=\lenValidateConv]{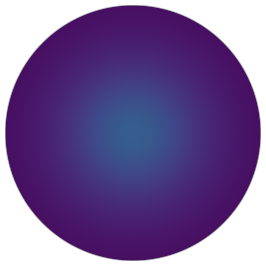}
        &
        \includegraphics[height=\lenValidateConv]{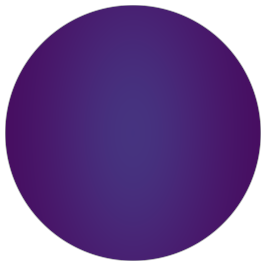}
        &
        \includegraphics[height=\lenValidateConv]{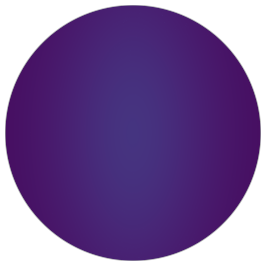}
        \\
        \multicolumn{4}{c}{$\alpha_x = 0.1$, $\alpha_y = 0.7$, $\alpha = 0.4$}
        &
        \multicolumn{4}{c}{$\alpha_x = 0.1$, $\alpha_y = 0.7$, $\alpha = 0.6$}
        \\
        \includegraphics[height=\lenValidateConv]{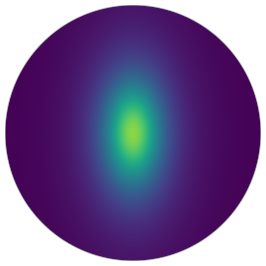}
        &
        \includegraphics[height=\lenValidateConv]{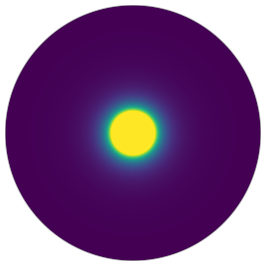}
        &
        \includegraphics[height=\lenValidateConv]{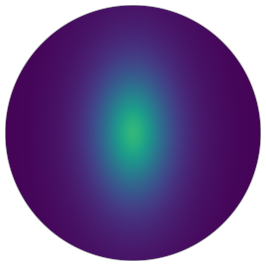}
        &
        \includegraphics[height=\lenValidateConv]{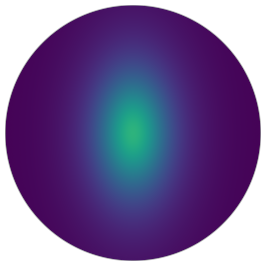}
        &
        \includegraphics[height=\lenValidateConv]{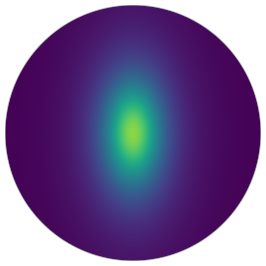}
        &
        \includegraphics[height=\lenValidateConv]{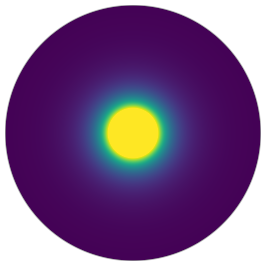}
        &
        \includegraphics[height=\lenValidateConv]{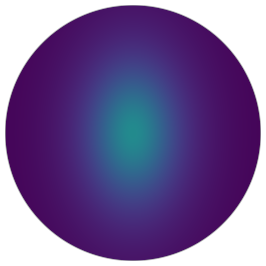}
        &
        \includegraphics[height=\lenValidateConv]{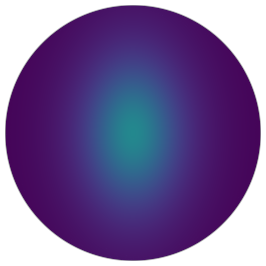}
        \\
        \multicolumn{4}{c}{$\alpha_x = 0.3$, $\alpha_y = 0.5$, $\alpha = 0.1$}
        &
        \multicolumn{4}{c}{$\alpha_x = 0.3$, $\alpha_y = 0.5$, $\alpha = 0.2$}
        \\
        \includegraphics[height=\lenValidateConv]{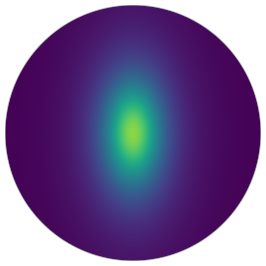}
        &
        \includegraphics[height=\lenValidateConv]{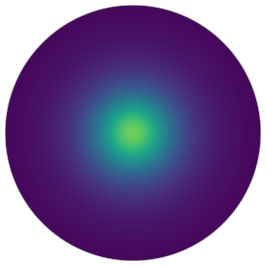}
        &
        \includegraphics[height=\lenValidateConv]{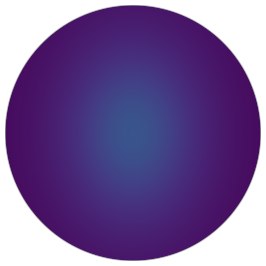}
        &
        \includegraphics[height=\lenValidateConv]{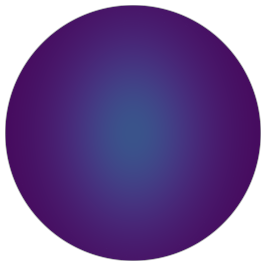}
        &
        \includegraphics[height=\lenValidateConv]{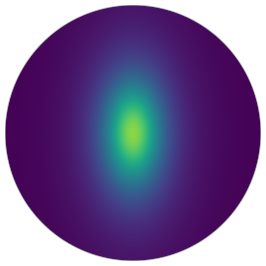}
        &
        \includegraphics[height=\lenValidateConv]{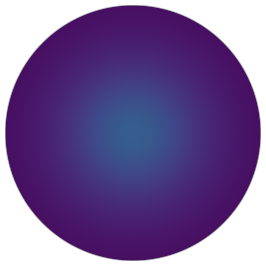}
        &
        \includegraphics[height=\lenValidateConv]{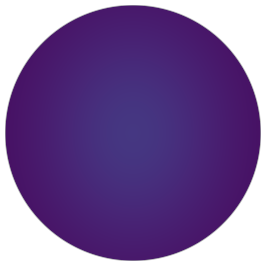}
        &
        \includegraphics[height=\lenValidateConv]{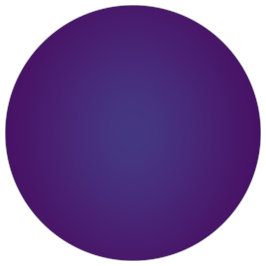}
        \\
        \multicolumn{4}{c}{$\alpha_x = 0.3$, $\alpha_y = 0.5$, $\alpha = 0.4$}
        &
        \multicolumn{4}{c}{$\alpha_x = 0.3$, $\alpha_y = 0.5$, $\alpha = 0.6$}
        \\
        \includegraphics[height=\lenValidateConv]{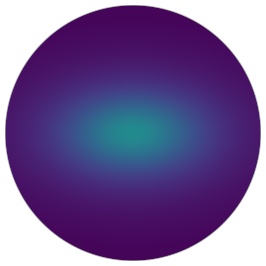}
        &
        \includegraphics[height=\lenValidateConv]{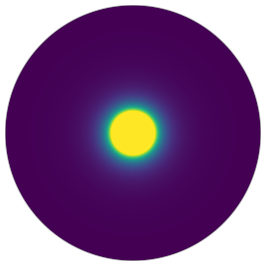}
        &
        \includegraphics[height=\lenValidateConv]{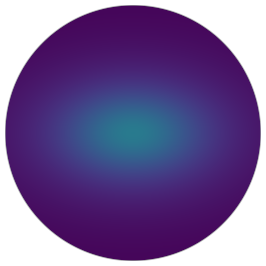}
        &
        \includegraphics[height=\lenValidateConv]{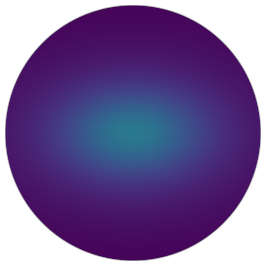}
        &
        \includegraphics[height=\lenValidateConv]{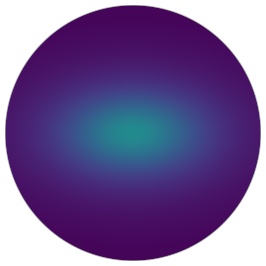}
        &
        \includegraphics[height=\lenValidateConv]{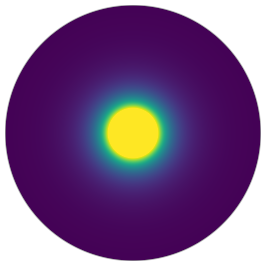}
        &
        \includegraphics[height=\lenValidateConv]{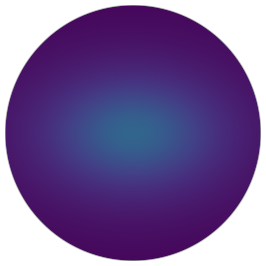}
        &
        \includegraphics[height=\lenValidateConv]{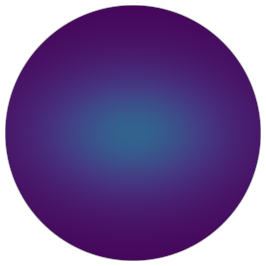}
        \\
        \multicolumn{4}{c}{$\alpha_x = 0.6$, $\alpha_y = 0.4$, $\alpha = 0.1$}
        &
        \multicolumn{4}{c}{$\alpha_x = 0.6$, $\alpha_y = 0.4$, $\alpha = 0.2$}
        \\
        \includegraphics[height=\lenValidateConv]{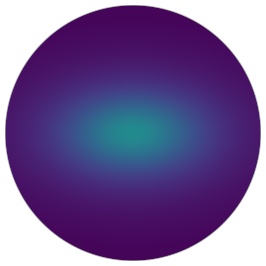}
        &
        \includegraphics[height=\lenValidateConv]{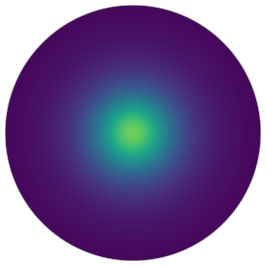}
        &
        \includegraphics[height=\lenValidateConv]{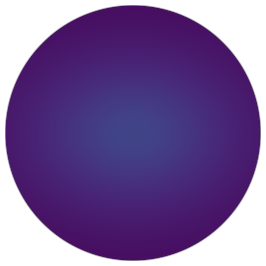}
        &
        \includegraphics[height=\lenValidateConv]{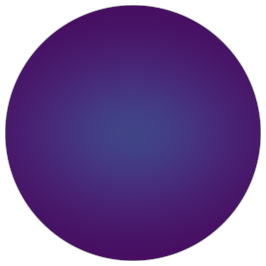}
        &
        \includegraphics[height=\lenValidateConv]{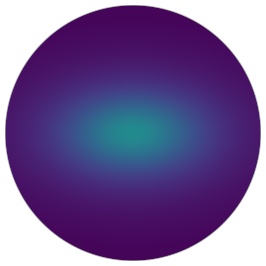}
        &
        \includegraphics[height=\lenValidateConv]{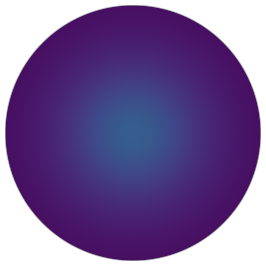}
        &
        \includegraphics[height=\lenValidateConv]{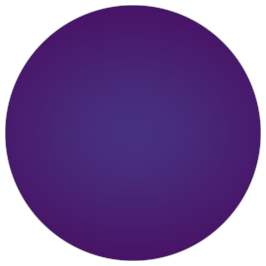}
        &
        \includegraphics[height=\lenValidateConv]{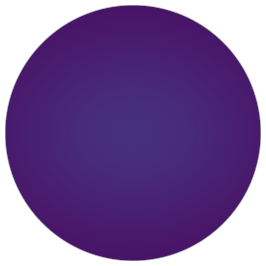}
        \\
        \multicolumn{4}{c}{$\alpha_x = 0.6$, $\alpha_y = 0.4$, $\alpha = 0.4$}
        &
        \multicolumn{4}{c}{$\alpha_x = 0.6$, $\alpha_y = 0.4$, $\alpha = 0.6$}
        \\
        \includegraphics[height=\lenValidateConv]{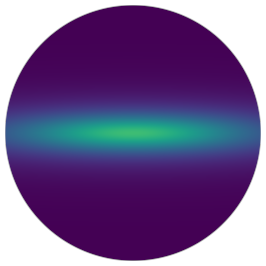}
        &
        \includegraphics[height=\lenValidateConv]{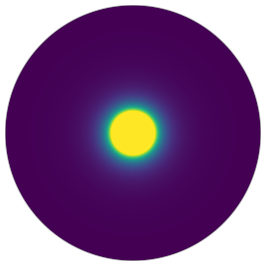}
        &
        \includegraphics[height=\lenValidateConv]{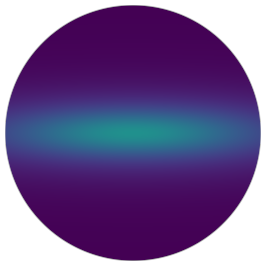}
        &
        \includegraphics[height=\lenValidateConv]{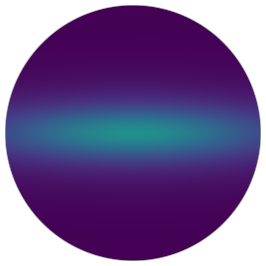}
        &
        \includegraphics[height=\lenValidateConv]{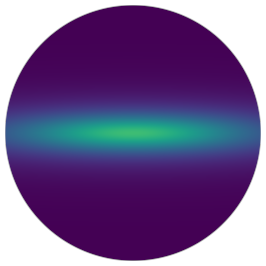}
        &
        \includegraphics[height=\lenValidateConv]{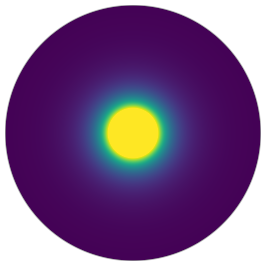}
        &
        \includegraphics[height=\lenValidateConv]{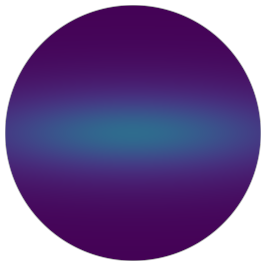}
        &
        \includegraphics[height=\lenValidateConv]{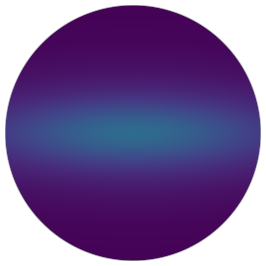}
        \\
        \multicolumn{4}{c}{$\alpha_x = 0.8$, $\alpha_y = 0.2$, $\alpha = 0.1$}
        &
        \multicolumn{4}{c}{$\alpha_x = 0.8$, $\alpha_y = 0.2$, $\alpha = 0.2$}
        \\
        \includegraphics[height=\lenValidateConv]{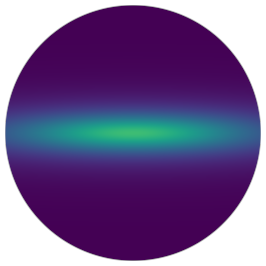}
        &
        \includegraphics[height=\lenValidateConv]{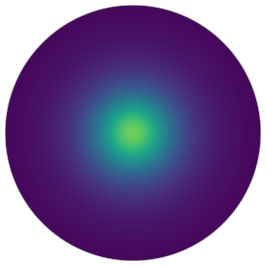}
        &
        \includegraphics[height=\lenValidateConv]{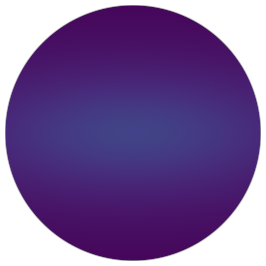}
        &
        \includegraphics[height=\lenValidateConv]{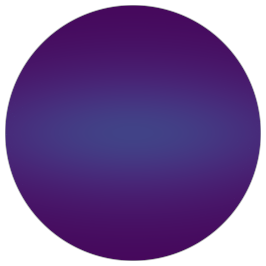}
        &
        \includegraphics[height=\lenValidateConv]{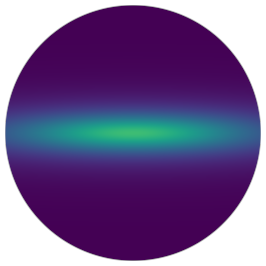}
        &
        \includegraphics[height=\lenValidateConv]{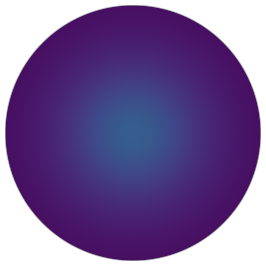}
        &
        \includegraphics[height=\lenValidateConv]{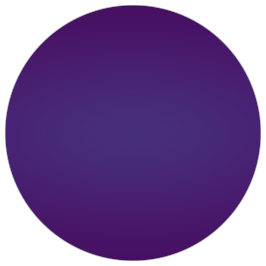}
        &
        \includegraphics[height=\lenValidateConv]{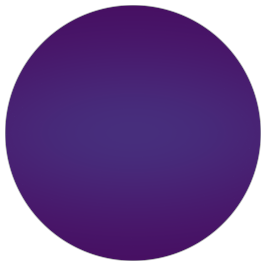}
        \\
        \multicolumn{4}{c}{$\alpha_x = 0.8$, $\alpha_y = 0.2$, $\alpha = 0.4$}
        &
        \multicolumn{4}{c}{$\alpha_x = 0.8$, $\alpha_y = 0.2$, $\alpha = 0.6$}
    \end{tabular}
    \caption{\label{fig:validate_sggx_ggx_conv_main}
        Convolving an SGGX with a GGX kernel. Our fits closely matches the references in different configurations.
    }
\end{figure*}

In \S\ref{subsec:conv_with_ndf}, we convolve an SGGX with a GGX microfacet distribution: $g(\omega_h) = D(\omega_h; n, \alpha)$ and compute a 3D mapping
$\bm{\alpha}_{+} = M_2(\bm{\alpha}, \alpha)$ (two channels).  We use $50 \times 50 \times 50 \times 2$ for its resolution in practice.
Similarly, we provide visualization and error plots to compare our fit to the ground truth in Fig~\ref{fig:validate_conv_error} and
Fig~\ref{fig:validate_sggx_ggx_conv_main}.

\subsection{Aggregated Microfacet Distribution}
\begin{figure*}[t]
	\newlength{\lenValidateBetaRoughness}
	\setlength{\lenValidateBetaRoughness}{1.3in}
    \addtolength{\tabcolsep}{-3pt}
    \centering
    \begin{tabular}{cccccc}
        & Map 1 & Map 2 & Map 3 & Map 4 & Map 5
        \\
        &
        \frame{\includegraphics[height=\lenValidateBetaRoughness]{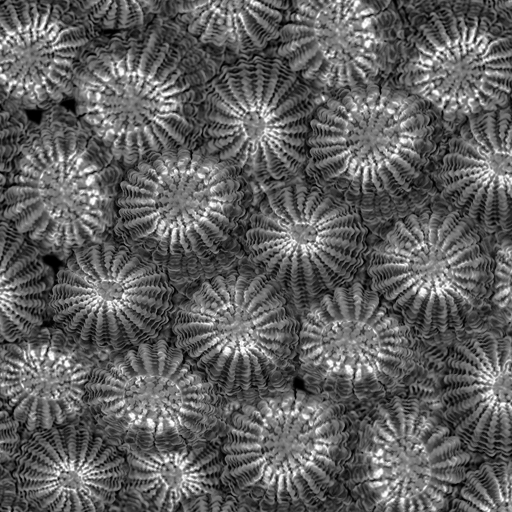}}
        &
        \frame{\includegraphics[height=\lenValidateBetaRoughness]{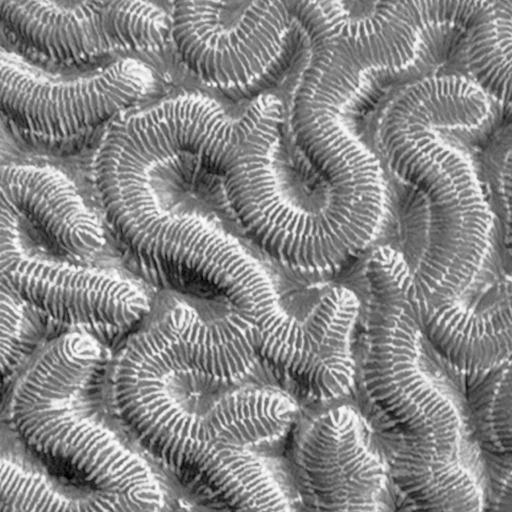}}
        &
        \frame{\includegraphics[height=\lenValidateBetaRoughness]{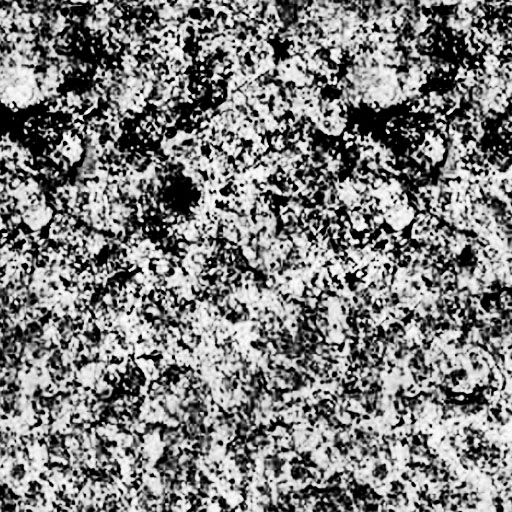}}
        &
        \frame{\includegraphics[height=\lenValidateBetaRoughness]{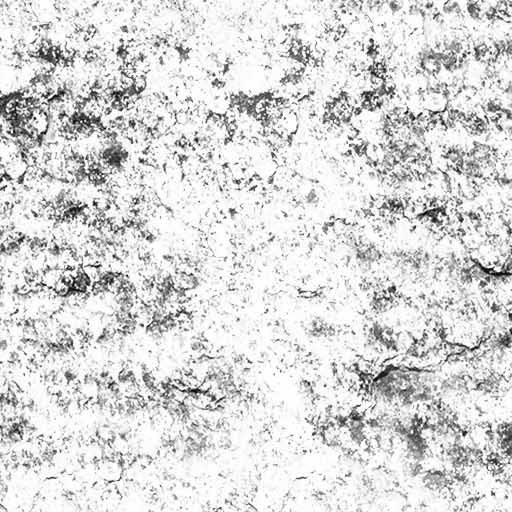}}
        &
        \frame{\includegraphics[height=\lenValidateBetaRoughness]{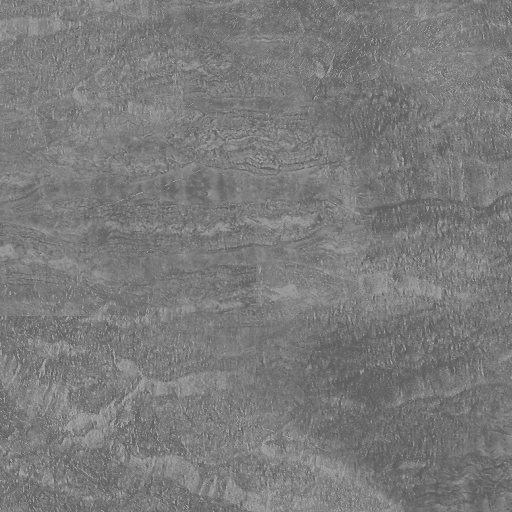}}
        \\
        \raisebox{45pt}{\rotatebox{90}{$p(\alpha)$}}
        &
        \includegraphics[height=\lenValidateBetaRoughness]{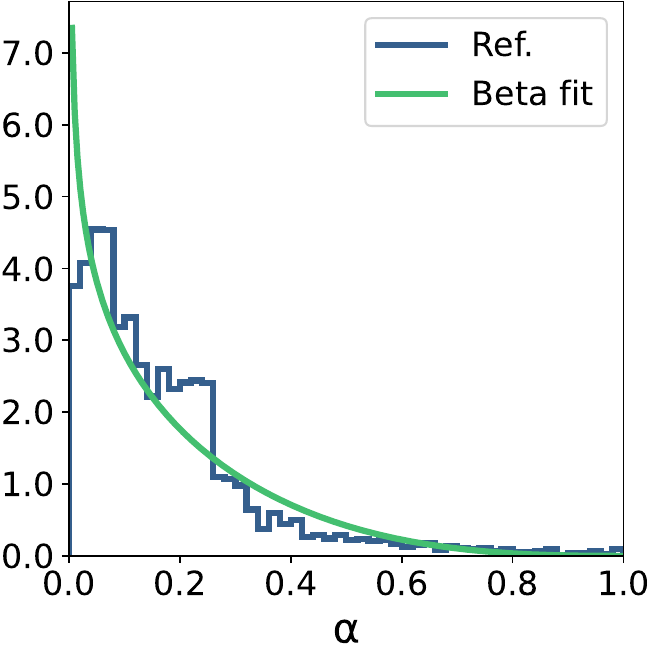}
        &
        \includegraphics[height=\lenValidateBetaRoughness]{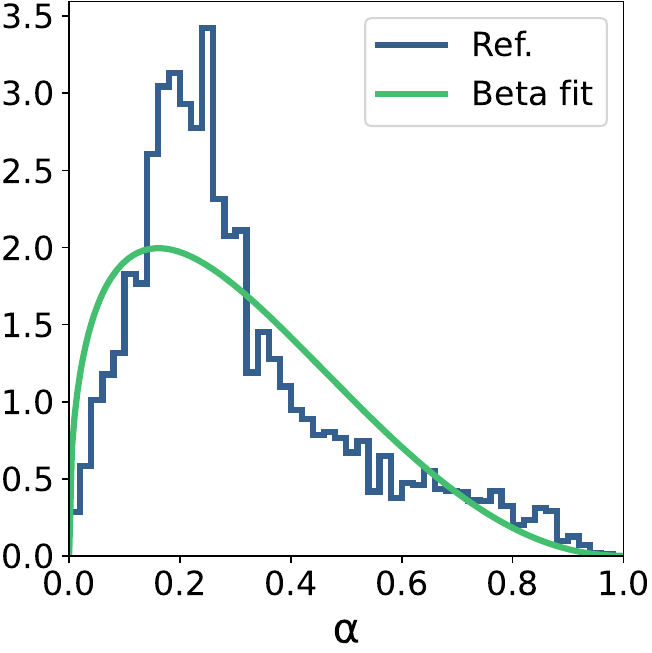}
        &
        \includegraphics[height=\lenValidateBetaRoughness]{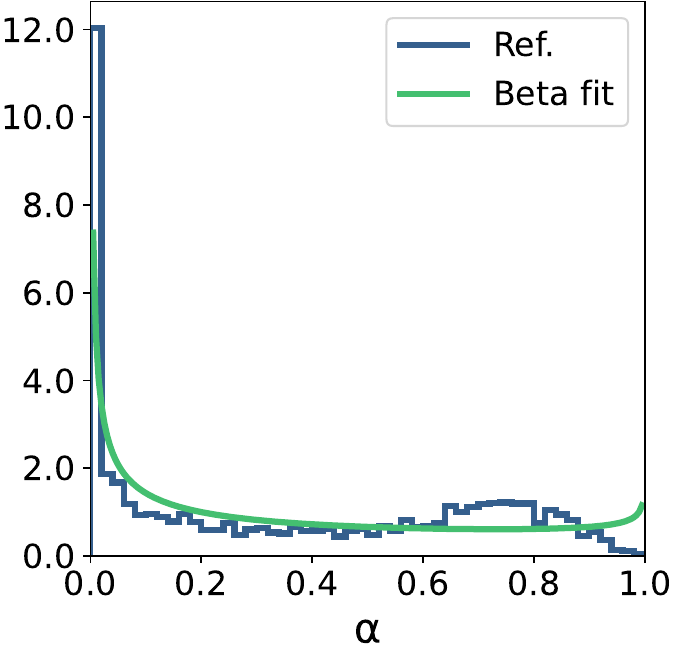}
        &
        \includegraphics[height=\lenValidateBetaRoughness]{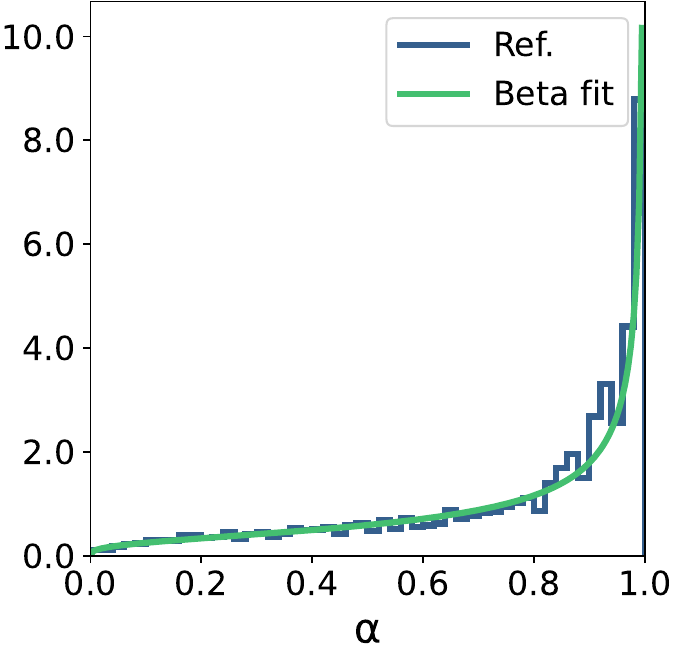}
        &
        \includegraphics[height=\lenValidateBetaRoughness]{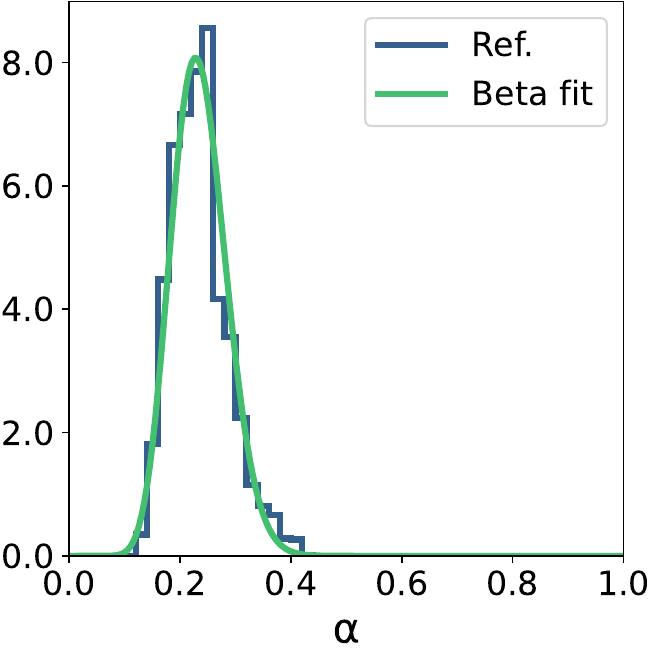}
        \\         
        \raisebox{45pt}{\rotatebox{90}{$D(\omega_m)$}}
        &
        \includegraphics[height=\lenValidateBetaRoughness]{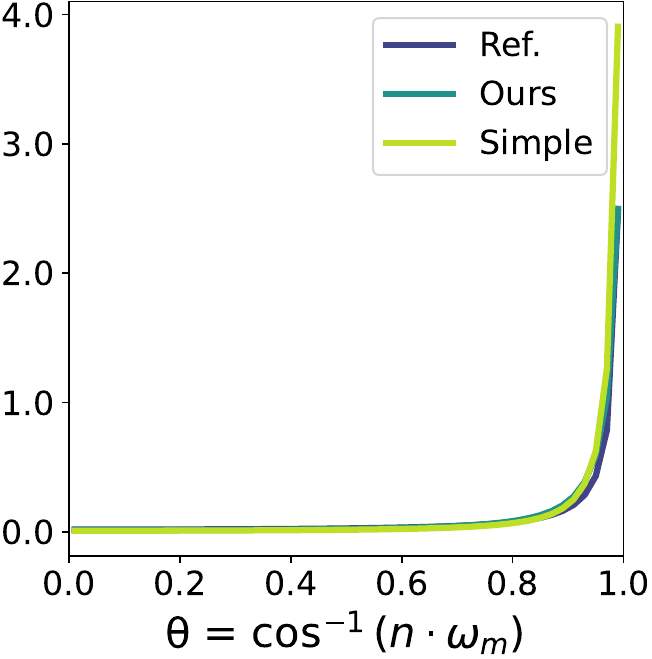}
        &
        \includegraphics[height=\lenValidateBetaRoughness]{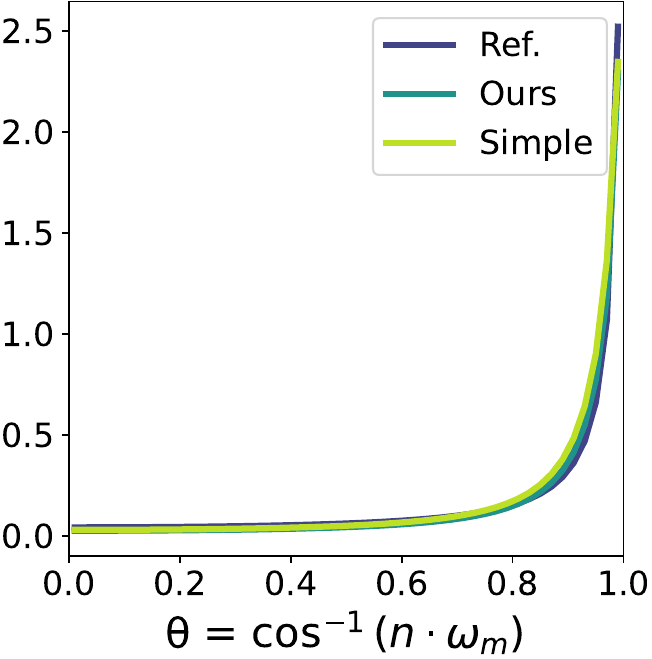}
        &
        \includegraphics[height=\lenValidateBetaRoughness]{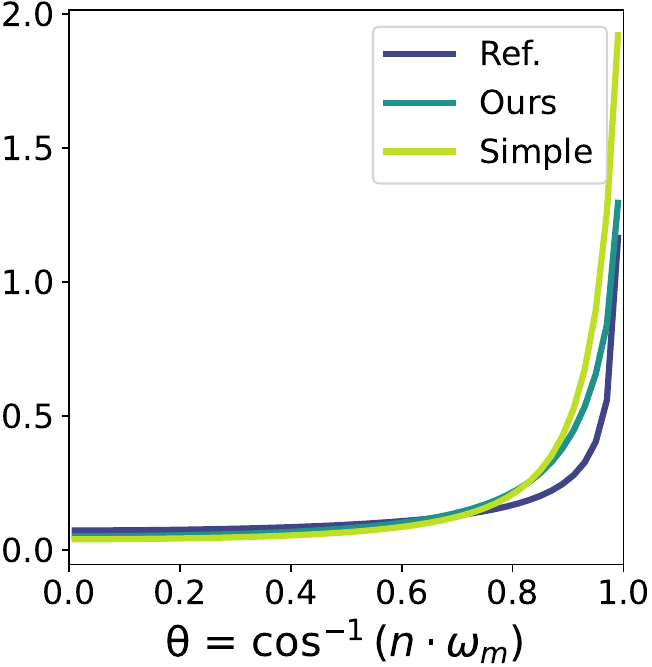}
        &
        \includegraphics[height=\lenValidateBetaRoughness]{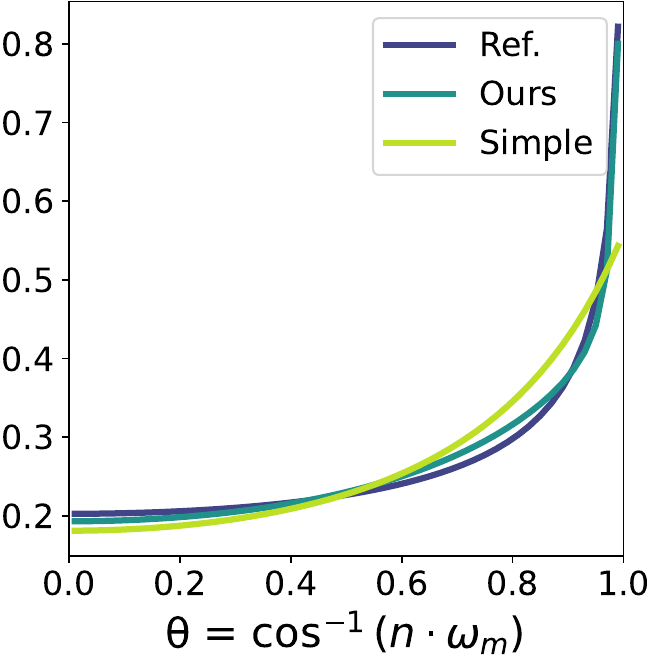}
        &
        \includegraphics[height=\lenValidateBetaRoughness]{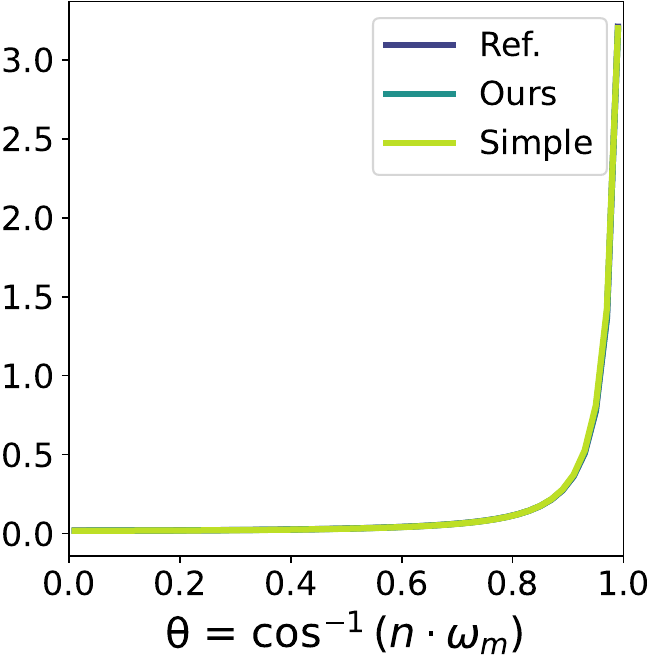}               
    \end{tabular}
    \caption{\label{fig:validate_beta_roughness}
        Fitting the distributions of a set of roughness maps as beta distributions and the aggregated microfacet distributions by our 2-lobe weighted sum.
    }
\end{figure*}

In \S\ref{subsec:conv_with_ndf}, we use a beta distribution to represent the roughness distribution $p_{\mathcal{A}}(\alpha) = \mathcal{B}(\alpha; a, b)$. Let $\mu$ and
$\sigma^2$ be the mean and the variance, the shape parameters of a beta distribution can be estimated as
\begin{align}
\begin{split}
  a &= -\mu (\sigma^2 + \mu^2 - \mu) / \sigma^2, \\
  b &= (\sigma^2 + \mu^2 - \mu) (\mu - 1) / \sigma^2. \\
\end{split}
\end{align}
We validate the effectiveness of this approach by estimating $p_{\mathcal{A}}(\alpha)$ from a set of roughness maps used for actual assets for the main article.
Fig~\ref{fig:validate_beta_roughness} shows the roughness maps, the ground-truth histograms and our estimated beta distributions. Our fits are reasonably
accurate and adapt to the different modes of real data. It is clear that a Gaussian distribution would have exceeded the valid $[0,1]$ range of $\alpha$
(e.g. map 1, 3, and 4). A truncated Gaussian can be restricted within the valid range and may produce similar quality, but the fitting process is more tedious.
We then validate the effectiveness of Eq.~\ref{eq:rough_var_two_lobe_fit} by comparing our 2-lobe fit to the ground-truth aggregated microfacet distributions as well as
a single-lobe fit with simple averaged roughness. Our fits are overall more accurate and guaranteed to be never worse than the simple fits.

\subsection{Correction for Conditioned Angular Domain}
\begin{figure*}[t]
	\newlength{\lenValidateRestricted}
	\setlength{\lenValidateRestricted}{0.8in}
    \centering
    \begin{tabular}{ccccccc}
        Kernel & \multicolumn{2}{c}{Clamped Orig.} & \multicolumn{2}{c}{Ref.} & \multicolumn{2}{c}{Ours} \\
        \frame{\begin{overpic}[width=0.6in]{imgs/colorbar_hori.png}
            \put(-23, 0){\normalsize \small{0.0}}
            \put(101, 0){\normalsize \small{0.5}}
        \end{overpic}} & & & & & & \\
        \includegraphics[height=\lenValidateRestricted]{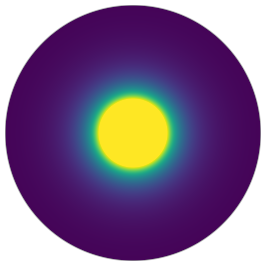}
        &
        \multicolumn{2}{c|}{\includegraphics[height=\lenValidateRestricted]{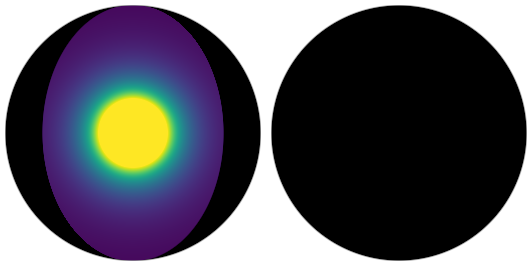}}
        &
        \multicolumn{2}{c|}{\includegraphics[height=\lenValidateRestricted]{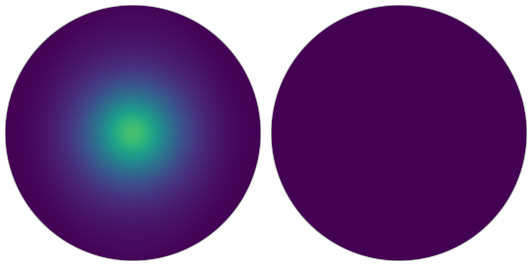}}
        &
        \multicolumn{2}{c}{\includegraphics[height=\lenValidateRestricted]{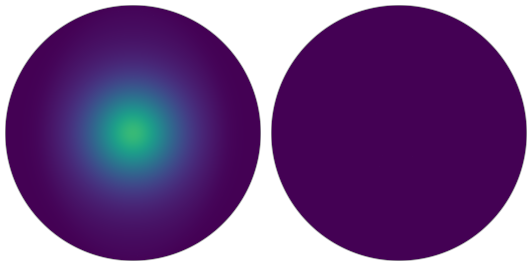}}
        \\
        &
        \multicolumn{2}{c|}{\includegraphics[height=\lenValidateRestricted]{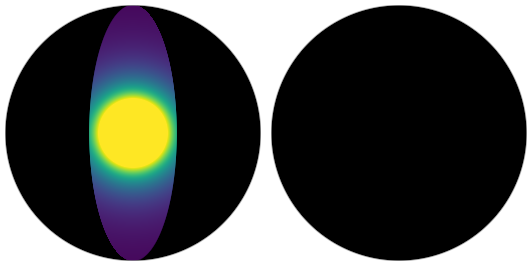}}
        &
        \multicolumn{2}{c|}{\includegraphics[height=\lenValidateRestricted]{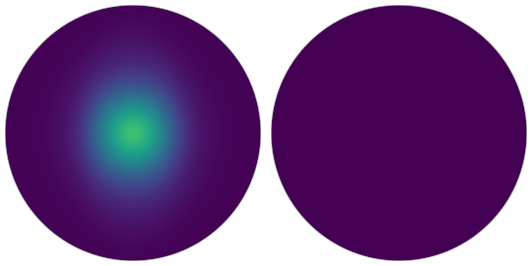}}
        &
        \multicolumn{2}{c}{\includegraphics[height=\lenValidateRestricted]{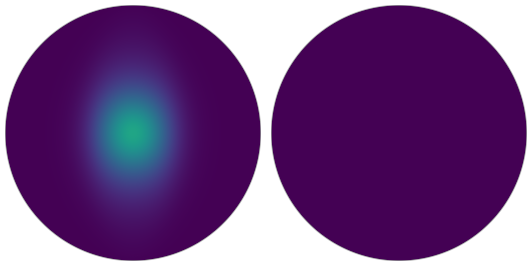}}
        \\
        &
        \multicolumn{2}{c|}{\includegraphics[height=\lenValidateRestricted]{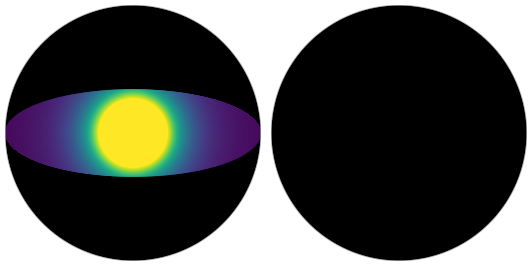}}
        &
        \multicolumn{2}{c|}{\includegraphics[height=\lenValidateRestricted]{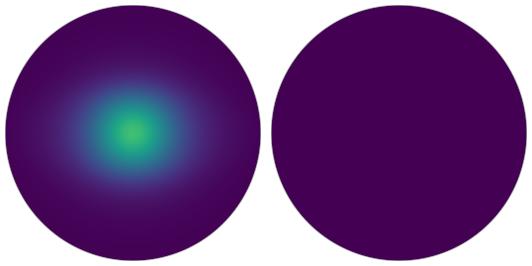}}
        &
        \multicolumn{2}{c}{\includegraphics[height=\lenValidateRestricted]{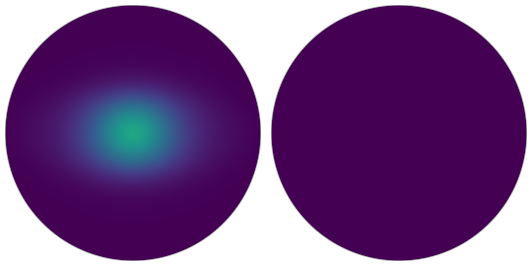}}
        \\
        &
        \multicolumn{2}{c|}{\includegraphics[height=\lenValidateRestricted]{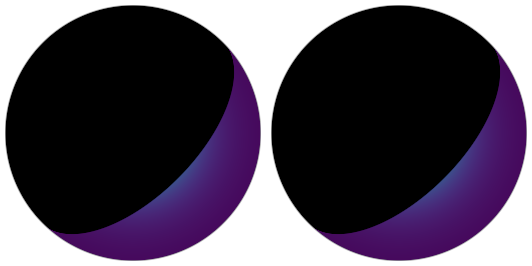}}
        &
        \multicolumn{2}{c|}{\includegraphics[height=\lenValidateRestricted]{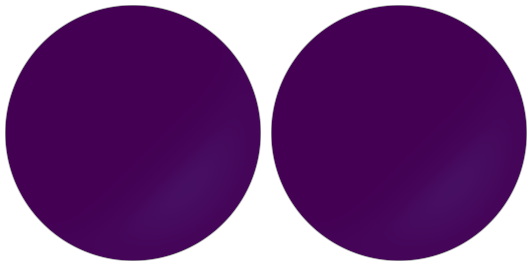}}
        &
        \multicolumn{2}{c}{\includegraphics[height=\lenValidateRestricted]{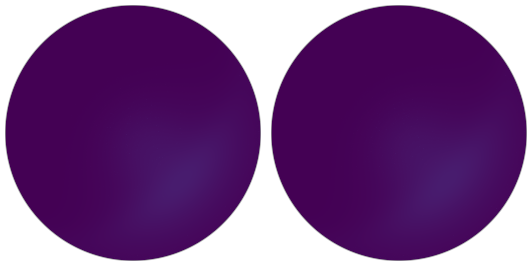}}
        \\
        &
        \multicolumn{2}{c|}{\includegraphics[height=\lenValidateRestricted]{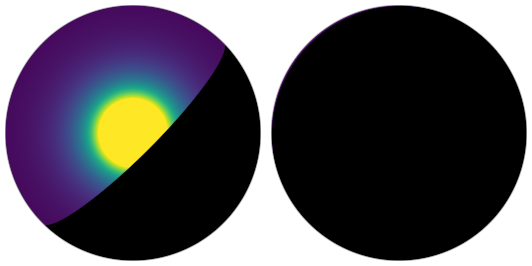}}
        &
        \multicolumn{2}{c|}{\includegraphics[height=\lenValidateRestricted]{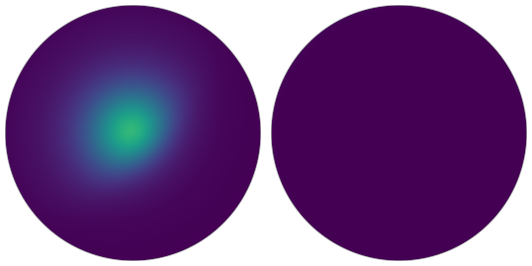}}
        &
        \multicolumn{2}{c}{\includegraphics[height=\lenValidateRestricted]{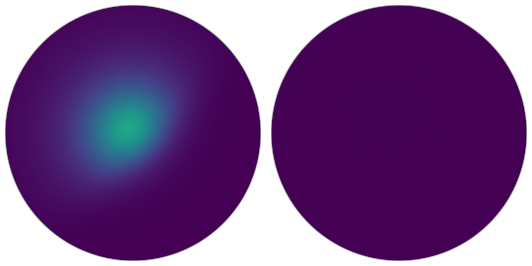}}
        \\
        &
        \multicolumn{2}{c|}{\includegraphics[height=\lenValidateRestricted]{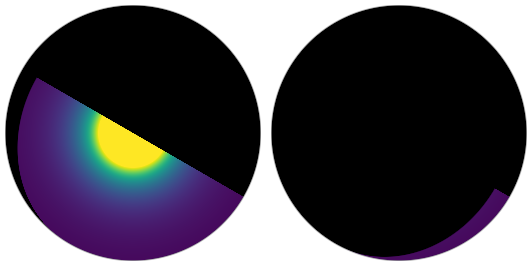}}
        &
        \multicolumn{2}{c|}{\includegraphics[height=\lenValidateRestricted]{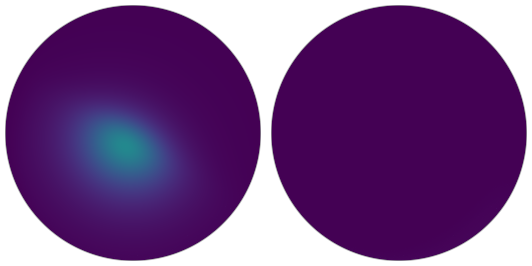}}
        &
        \multicolumn{2}{c}{\includegraphics[height=\lenValidateRestricted]{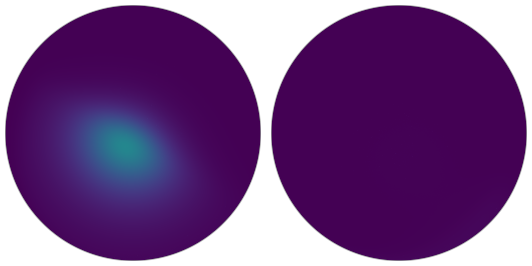}}
        \\
    \end{tabular}
    \caption{\label{fig:validate_restricted_angular_domain_main_1}
        Convolving an SGGX with a GGX over an restricted domain $\mathcal{X}$. The kernel has roughness $\alpha=0.2$ and the original SGGX has roughness
        $\alpha_x = 0.3, \alpha_y = 0.3$. The contribution of the original distribution is clamped to black by $\mathcal{X}$.
    }
\end{figure*}

\begin{figure*}[t]
    \setlength{\lenValidateRestricted}{0.8in}
    \centering
    \begin{tabular}{ccccccc}
        Kernel & \multicolumn{2}{c}{Clamped Orig.} & \multicolumn{2}{c}{Ref.} & \multicolumn{2}{c}{Ours} \\
        \frame{\begin{overpic}[width=0.6in]{imgs/colorbar_hori.png}
            \put(-23, 0){\normalsize \small{0.0}}
            \put(101, 0){\normalsize \small{0.5}}
        \end{overpic}} & & & & & & \\
        \includegraphics[height=\lenValidateRestricted]{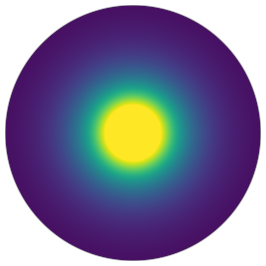}
        &
        \multicolumn{2}{c|}{\includegraphics[height=\lenValidateRestricted]{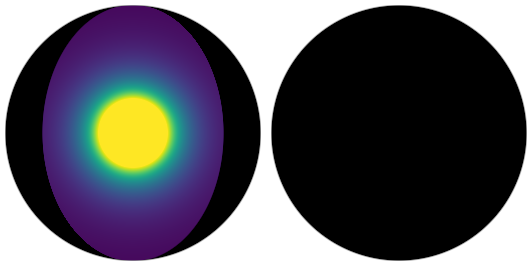}}
        &
        \multicolumn{2}{c|}{\includegraphics[height=\lenValidateRestricted]{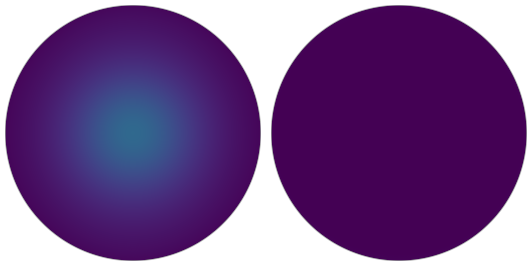}}
        &
        \multicolumn{2}{c}{\includegraphics[height=\lenValidateRestricted]{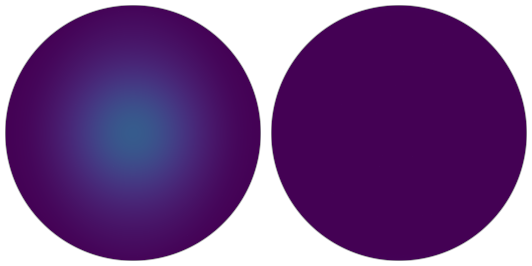}}
        \\
        &
        \multicolumn{2}{c|}{\includegraphics[height=\lenValidateRestricted]{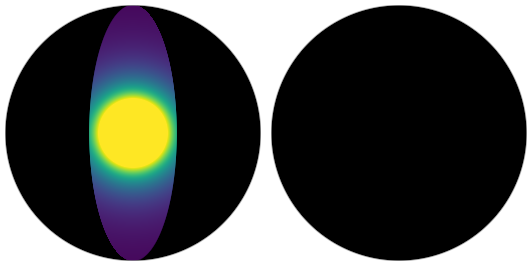}}
        &
        \multicolumn{2}{c|}{\includegraphics[height=\lenValidateRestricted]{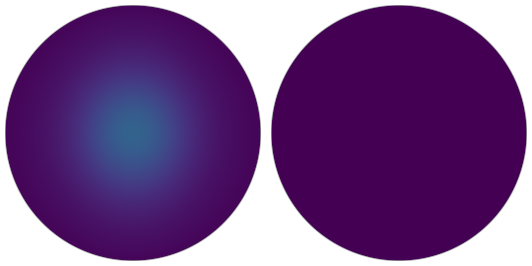}}
        &
        \multicolumn{2}{c}{\includegraphics[height=\lenValidateRestricted]{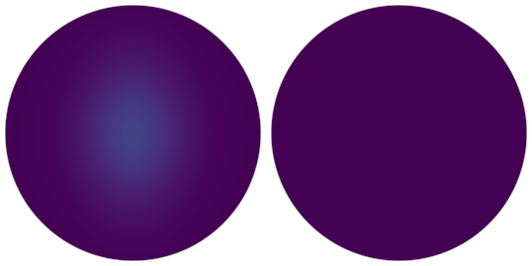}}
        \\
        &
        \multicolumn{2}{c|}{\includegraphics[height=\lenValidateRestricted]{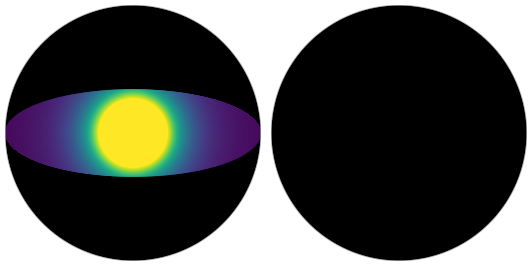}}
        &
        \multicolumn{2}{c|}{\includegraphics[height=\lenValidateRestricted]{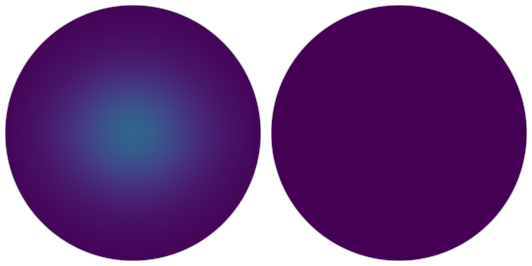}}
        &
        \multicolumn{2}{c}{\includegraphics[height=\lenValidateRestricted]{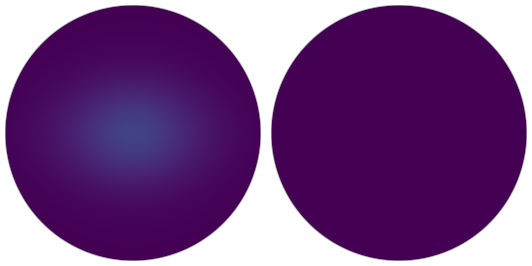}}
        \\
        &
        \multicolumn{2}{c|}{\includegraphics[height=\lenValidateRestricted]{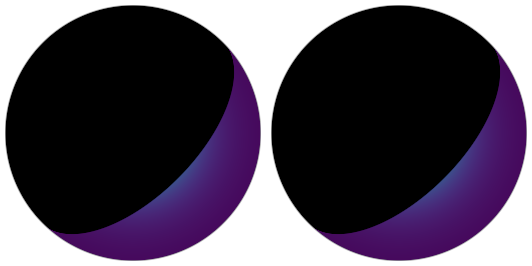}}
        &
        \multicolumn{2}{c|}{\includegraphics[height=\lenValidateRestricted]{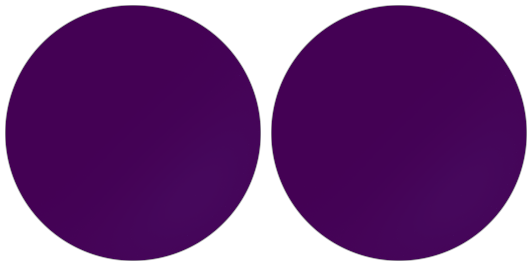}}
        &
        \multicolumn{2}{c}{\includegraphics[height=\lenValidateRestricted]{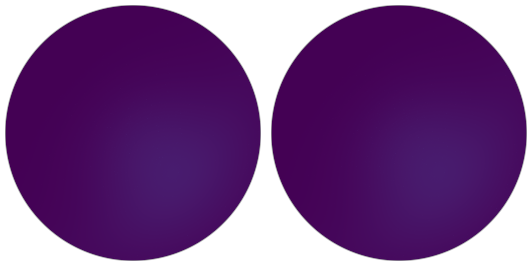}}
        \\
        &
        \multicolumn{2}{c|}{\includegraphics[height=\lenValidateRestricted]{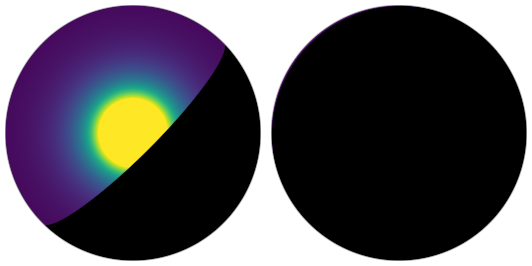}}
        &
        \multicolumn{2}{c|}{\includegraphics[height=\lenValidateRestricted]{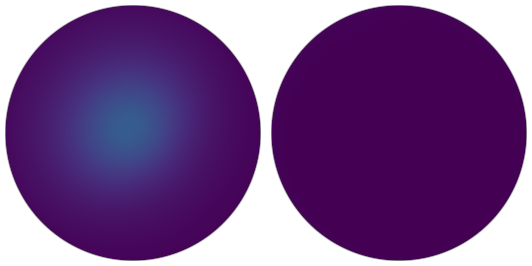}}
        &
        \multicolumn{2}{c}{\includegraphics[height=\lenValidateRestricted]{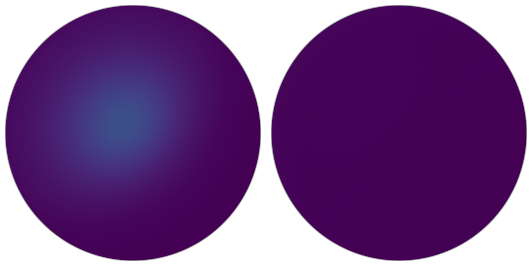}}
        \\
        &
        \multicolumn{2}{c|}{\includegraphics[height=\lenValidateRestricted]{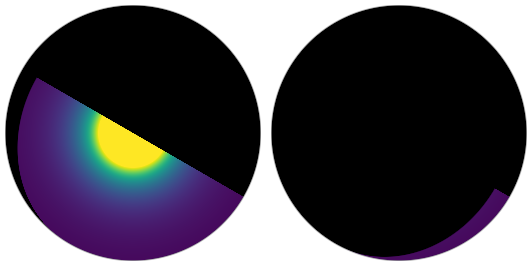}}
        &
        \multicolumn{2}{c|}{\includegraphics[height=\lenValidateRestricted]{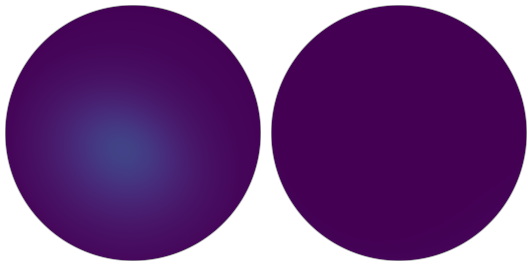}}
        &
        \multicolumn{2}{c}{\includegraphics[height=\lenValidateRestricted]{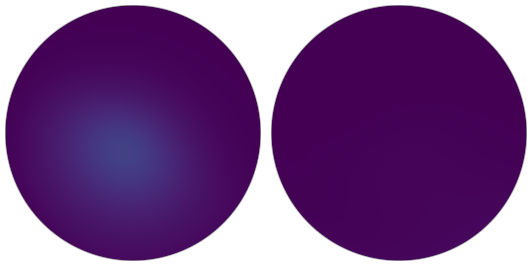}}
        \\
    \end{tabular}
    \caption{\label{fig:validate_restricted_angular_domain_main_2}
        Convolving an SGGX with a GGX over an restricted domain $\mathcal{X}$. The kernel has roughness $\alpha=0.4$ and the original SGGX has roughness
        $\alpha_x = 0.3, \alpha_y = 0.3$. The contribution of the original distribution is clamped to black by $\mathcal{X}$.
    }
\end{figure*}

\begin{figure*}[t]
    \setlength{\lenValidateRestricted}{0.8in}
    \centering
    \begin{tabular}{ccccccc}
        Kernel & \multicolumn{2}{c}{Clamped Orig.} & \multicolumn{2}{c}{Ref.} & \multicolumn{2}{c}{Ours} \\
        \frame{\begin{overpic}[width=0.6in]{imgs/colorbar_hori.png}
            \put(-23, 0){\normalsize \small{0.0}}
            \put(101, 0){\normalsize \small{0.5}}
        \end{overpic}} & & & & & & \\
        \includegraphics[height=\lenValidateRestricted]{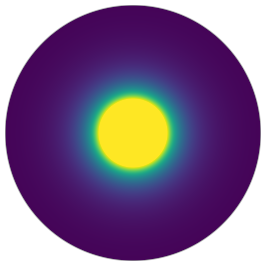}
        &
        \multicolumn{2}{c|}{\includegraphics[height=\lenValidateRestricted]{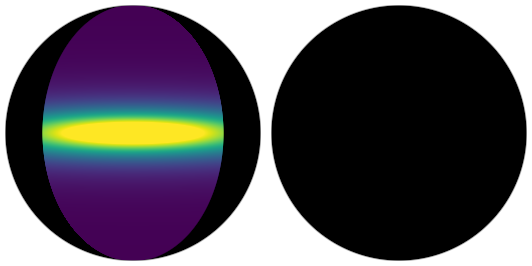}}
        &
        \multicolumn{2}{c|}{\includegraphics[height=\lenValidateRestricted]{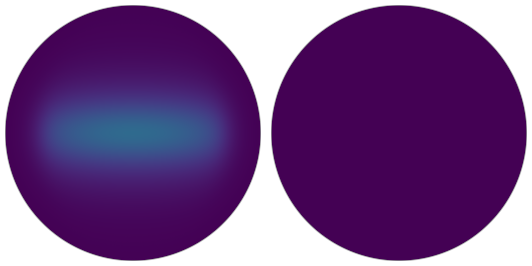}}
        &
        \multicolumn{2}{c}{\includegraphics[height=\lenValidateRestricted]{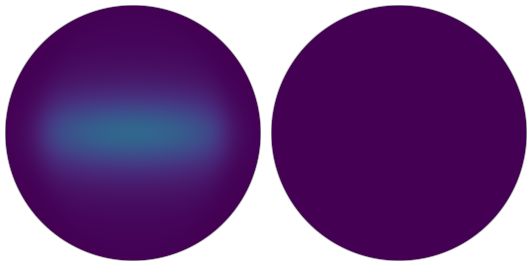}}
        \\
        &
        \multicolumn{2}{c|}{\includegraphics[height=\lenValidateRestricted]{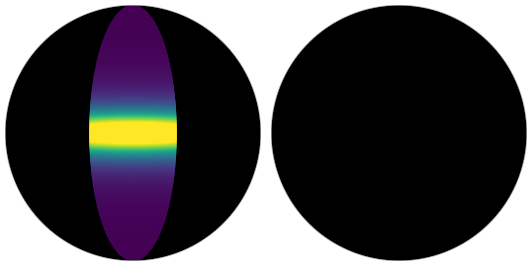}}
        &
        \multicolumn{2}{c|}{\includegraphics[height=\lenValidateRestricted]{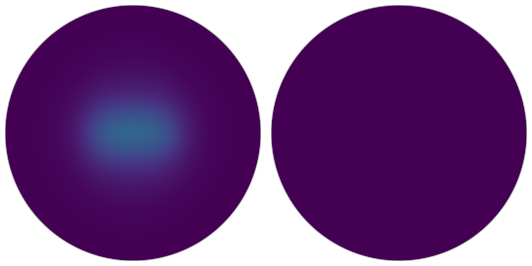}}
        &
        \multicolumn{2}{c}{\includegraphics[height=\lenValidateRestricted]{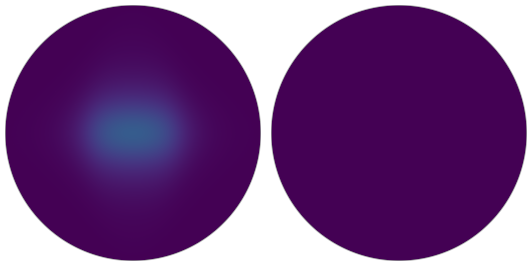}}
        \\
        &
        \multicolumn{2}{c|}{\includegraphics[height=\lenValidateRestricted]{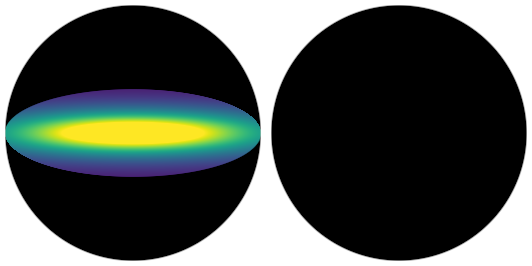}}
        &
        \multicolumn{2}{c|}{\includegraphics[height=\lenValidateRestricted]{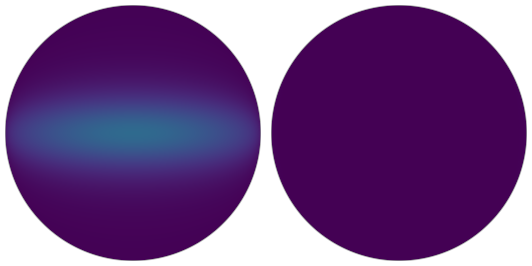}}
        &
        \multicolumn{2}{c}{\includegraphics[height=\lenValidateRestricted]{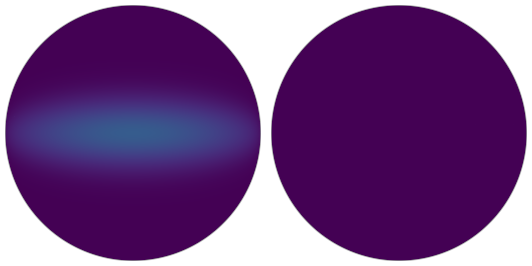}}
        \\
        &
        \multicolumn{2}{c|}{\includegraphics[height=\lenValidateRestricted]{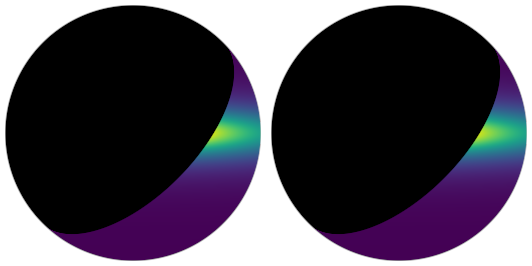}}
        &
        \multicolumn{2}{c|}{\includegraphics[height=\lenValidateRestricted]{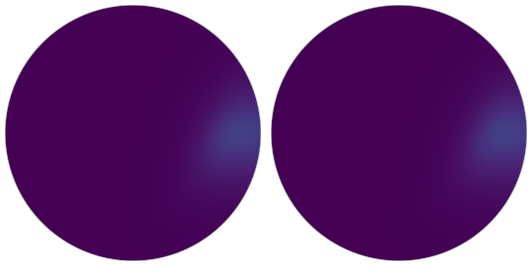}}
        &
        \multicolumn{2}{c}{\includegraphics[height=\lenValidateRestricted]{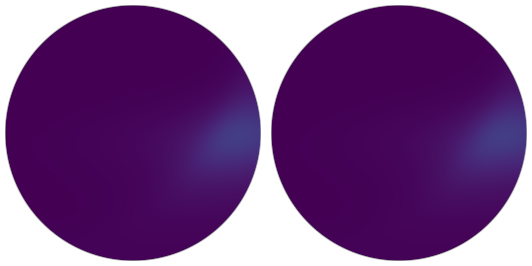}}
        \\
        &
        \multicolumn{2}{c|}{\includegraphics[height=\lenValidateRestricted]{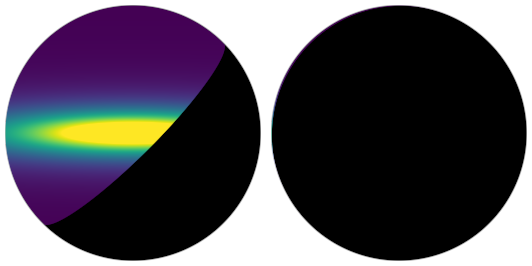}}
        &
        \multicolumn{2}{c|}{\includegraphics[height=\lenValidateRestricted]{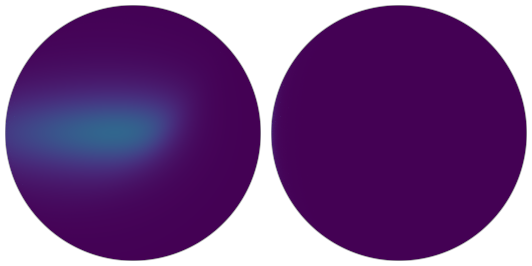}}
        &
        \multicolumn{2}{c}{\includegraphics[height=\lenValidateRestricted]{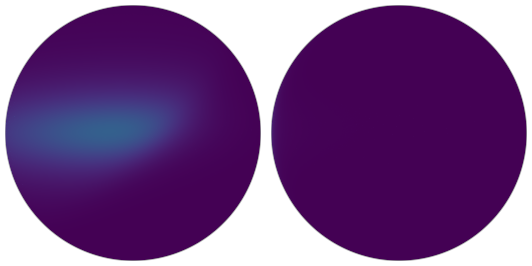}}
        \\
        &
        \multicolumn{2}{c|}{\includegraphics[height=\lenValidateRestricted]{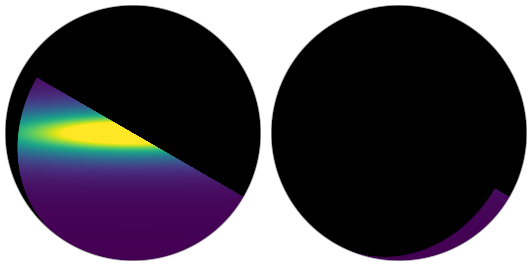}}
        &
        \multicolumn{2}{c|}{\includegraphics[height=\lenValidateRestricted]{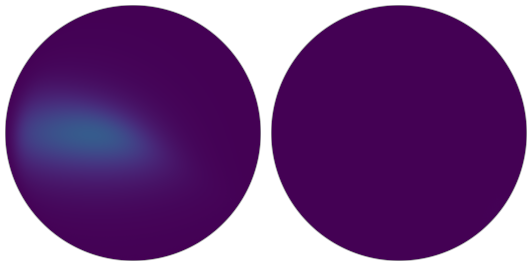}}
        &
        \multicolumn{2}{c}{\includegraphics[height=\lenValidateRestricted]{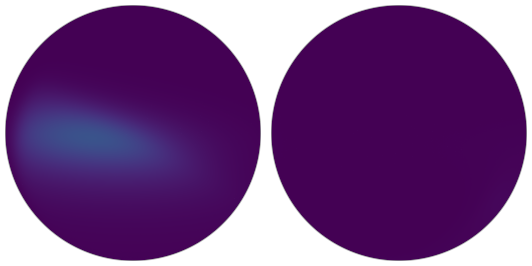}}
        \\
    \end{tabular}
    \caption{\label{fig:validate_restricted_angular_domain_main_3}
        Convolving an SGGX with a GGX over an restricted domain $\mathcal{X}$. The kernel has roughness $\alpha=0.2$ and the original SGGX has roughness
        $\alpha_x = 0.8, \alpha_y = 0.2$. The contribution of the original distribution is clamped to black by $\mathcal{X}$.
    }
\end{figure*}

\begin{figure*}[t]
    \setlength{\lenValidateRestricted}{0.8in}
    \centering
    \begin{tabular}{ccccccc}
        Kernel & \multicolumn{2}{c}{Clamped Orig.} & \multicolumn{2}{c}{Ref.} & \multicolumn{2}{c}{Ours} \\
        \frame{\begin{overpic}[width=0.6in]{imgs/colorbar_hori.png}
            \put(-23, 0){\normalsize \small{0.0}}
            \put(101, 0){\normalsize \small{0.5}}
        \end{overpic}} & & & & & & \\
        \includegraphics[height=\lenValidateRestricted]{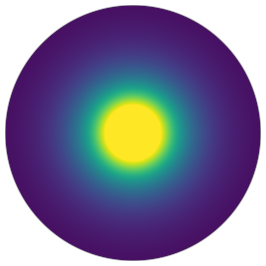}
        &
        \multicolumn{2}{c|}{\includegraphics[height=\lenValidateRestricted]{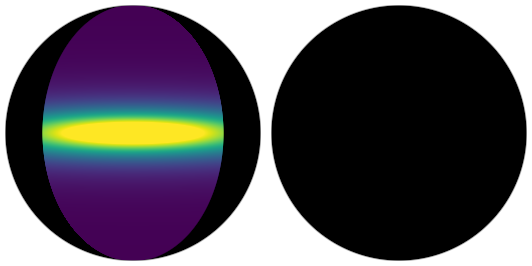}}
        &
        \multicolumn{2}{c|}{\includegraphics[height=\lenValidateRestricted]{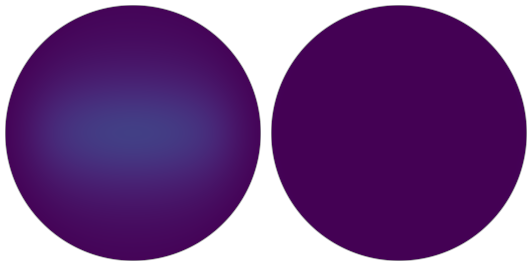}}
        &
        \multicolumn{2}{c}{\includegraphics[height=\lenValidateRestricted]{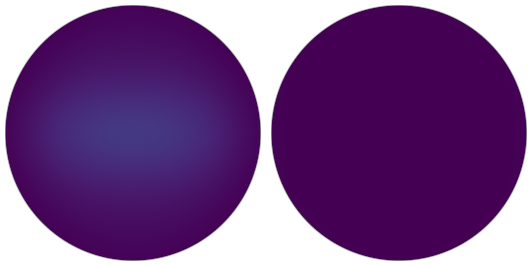}}
        \\
        &
        \multicolumn{2}{c|}{\includegraphics[height=\lenValidateRestricted]{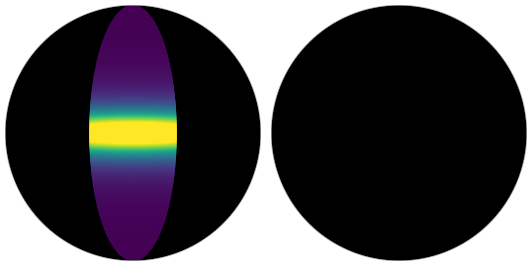}}
        &
        \multicolumn{2}{c|}{\includegraphics[height=\lenValidateRestricted]{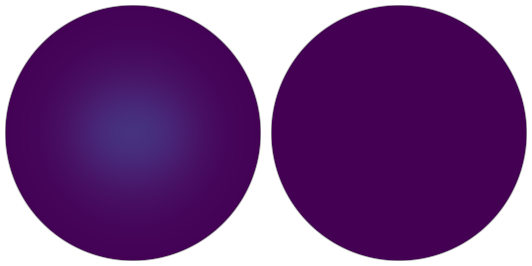}}
        &
        \multicolumn{2}{c}{\includegraphics[height=\lenValidateRestricted]{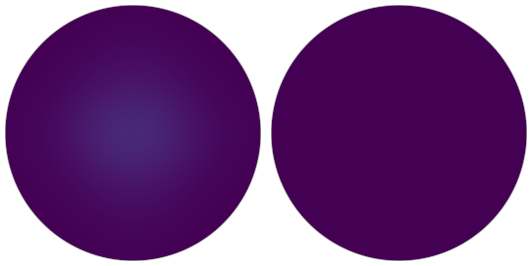}}
        \\
        &
        \multicolumn{2}{c|}{\includegraphics[height=\lenValidateRestricted]{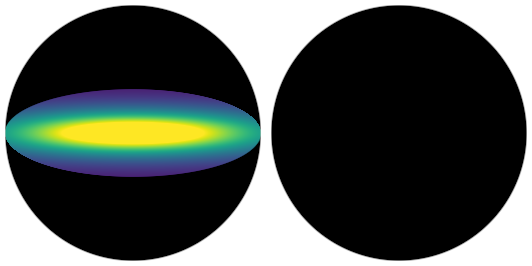}}
        &
        \multicolumn{2}{c|}{\includegraphics[height=\lenValidateRestricted]{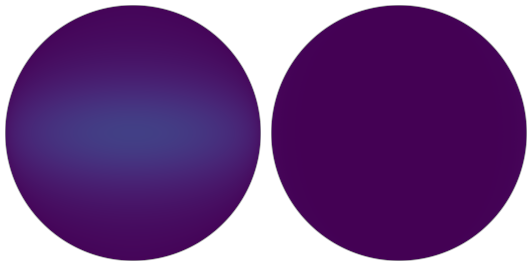}}
        &
        \multicolumn{2}{c}{\includegraphics[height=\lenValidateRestricted]{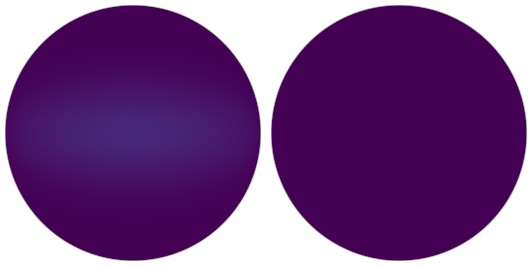}}
        \\
        &
        \multicolumn{2}{c|}{\includegraphics[height=\lenValidateRestricted]{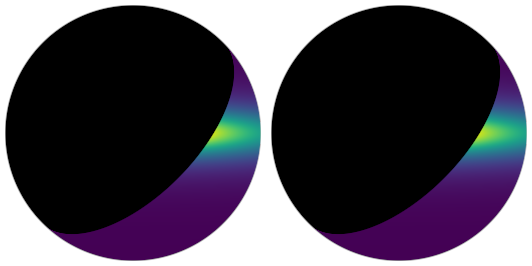}}
        &
        \multicolumn{2}{c|}{\includegraphics[height=\lenValidateRestricted]{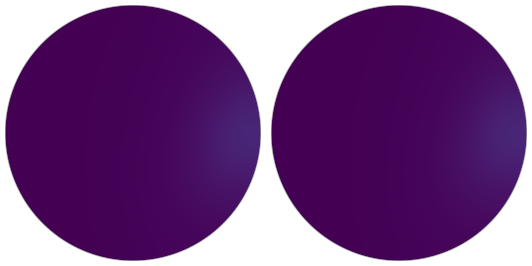}}
        &
        \multicolumn{2}{c}{\includegraphics[height=\lenValidateRestricted]{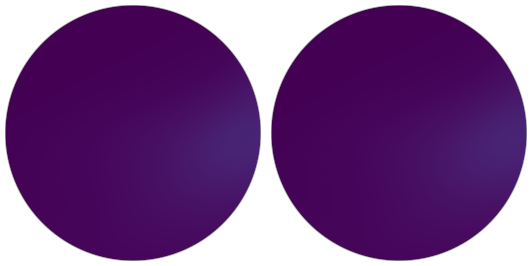}}
        \\
        &
        \multicolumn{2}{c|}{\includegraphics[height=\lenValidateRestricted]{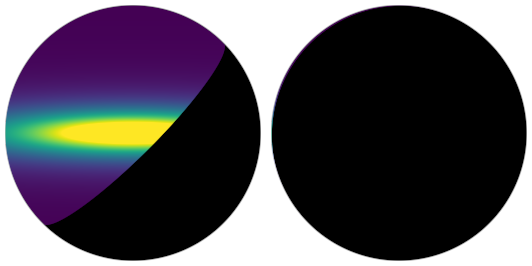}}
        &
        \multicolumn{2}{c|}{\includegraphics[height=\lenValidateRestricted]{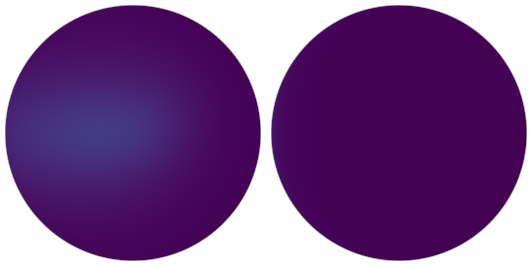}}
        &
        \multicolumn{2}{c}{\includegraphics[height=\lenValidateRestricted]{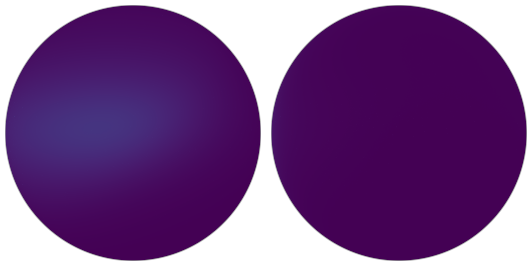}}
        \\
        &
        \multicolumn{2}{c|}{\includegraphics[height=\lenValidateRestricted]{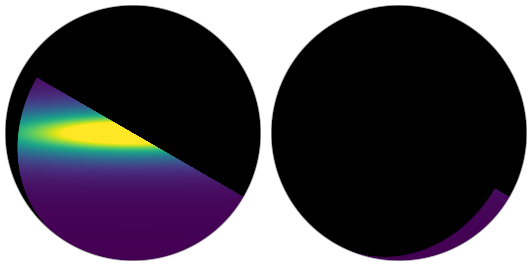}}
        &
        \multicolumn{2}{c|}{\includegraphics[height=\lenValidateRestricted]{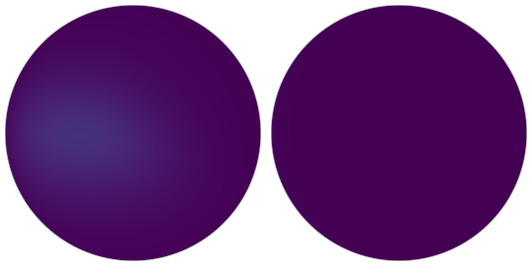}}
        &
        \multicolumn{2}{c}{\includegraphics[height=\lenValidateRestricted]{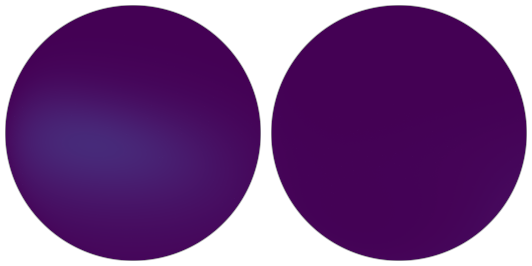}}
        \\
    \end{tabular}
    \caption{\label{fig:validate_restricted_angular_domain_main_4}
        Convolving an SGGX with a GGX over an restricted domain $\mathcal{X}$. The kernel has roughness $\alpha=0.4$ and the original SGGX has roughness
        $\alpha_x = 0.8, \alpha_y = 0.2$. The contribution of the original distribution is clamped to black by $\mathcal{X}$.
    }
\end{figure*}

We validate the approximation made in Eq.~\ref{eq:angular_domain} in a set of configurations including different shapes of the SGGX and the angular domain $\mathcal{X}$.
Fig~\ref{fig:validate_restricted_angular_domain_main_1} to \ref{fig:validate_restricted_angular_domain_main_4} visualizes our approximation and compare it to
the ground-truth. Our approximation qualitatively matches the geometric characteristics of the original distribution and the conditioned domain, and in general
achieves reasonable accuracy. The source of error comes from the constant contribution assumption in the \emph{shape term} in Eq.~\ref{eq:angular_domain}. 
This results in
under- or overestimates in some cases. For example, our approximation produces slightly darker results in Fig~\ref{fig:validate_restricted_angular_domain_main_2},
row 2 and 3, but slightly brighter results in row 4. In Fig~\ref{fig:validate_restricted_angular_domain_main_4}, row 5 and 6, our lobe shapes are slightly
stretched compared to references.

\section{Truncated Ellipsoid Primitive Projected Area}
We provide an efficient Monte Carlo estimator to calculate the projected area of the intersection of a cube and an ellipsoid.
We first uniformly sample points on the visible ellipsoid~\cite{heitz2015sggx}. Then, we cast a line from each point along $\omega$ to test if it
intersects the cube. The ratio of intersection times the full ellipsoid projected area provides an estimator to $|B|_{\omega}$.
Algorithm~\ref{algo:projected_area} provides the pseudocode. In practice, we optimize the implementation by using a fixed sample budget of $N=16$ and by
early-exiting if the ellipsoid is fully contained by the cube or vice versa.

\begin{algorithm}[h]
	\caption{\label{algo:projected_area}
		Computing the projected area along $\omega$ of a truncated ellipsoid primitive $B$.
	}
	\begin{algorithmic}[1]
		\Function{ProjectedArea}{$B$, $\omega$}
            \State $M \leftarrow$ the ellipsoid matrix, $c \leftarrow$ the ellipsoid center
            \State $Q \leftarrow$ the cube
            \State Compute an orthonormal basis $\Omega = (\omega_x, \omega_y, \omega)$ from $\omega$
            \State $M_{\omega}^{-1} \leftarrow \Omega^T M^{-1} \Omega$
            \State $M_{\omega xy}^{-1}$ $\leftarrow$ the upper left 2x2 block of $M_{\omega}^{-1}$
            \State Compute the Cholesky decomposition of $M_{\omega}^{-1}$: $M_{\omega}^{-1} = LL^T$
            \For {$i = 1$ to $N$}
                \State Sample a point on unit disk $(x, y)$ and lift it to the unit upper hemisphere: $p \leftarrow (x, y, \sqrt{1-x^2-y^2})$
                \State Warp $p$ to the visible ellipsoid surface: $p \leftarrow Lp$
                \State Transform $p$ back to world space: $p \leftarrow \Omega p + c$
                \State Cast a line $l$ that passes $p$ in direction $\omega$
                \If {$l$ intersects $Q$}
                    \State $m \leftarrow m+1$
                \EndIf
            \EndFor
            \State \Return{$\frac{m}{N} \cdot \pi \sqrt{\det(M_{\omega xy}^{-1})}$}
		\EndFunction
	\end{algorithmic}
\end{algorithm}

\section{Visibility Compression Details}
As described in \S\ref{subsec:compression}, we employ Clustered Principal Component Analysis (CPCA) to compress the aggregated visibility (both AIV and ABV). 
The method is similar to that of~\citet{sloan2003} and~\citet{LiuSSS04}, but we describe some of its details for completeness. 
Given a set of input visibility data, we first partition the angular domain: This is both to  
reduce the problem size and improve angular locality. Since we parameterize by the equal-area mapping~\citep{clarberg2008fast}, we simply partition the entire 
domain into $4 \times 4$ angular tiles. CPCA is applied individually to each group of tiles that cover the same subset of the domain. Note that many tiles are 
completely visible or occluded and can be culled away early. 

Let $C$ be the visibility matrix of a cluster of tiles where each column is one tile flattened to 1D and subtracted by the cluster mean. The standard principal 
component analysis (PCA) involves computing the (reduced) singular value decomposition (SVD) of $C$:
\begin{equation}
  C = U \Sigma V\trans.
\end{equation}
Next, we take the top $k$ \emph{left-singular vectors} from the $k$ leftmost columns of $U$ as the representative tiles (instead of the more common 
right-singular vectors from columns of $V$). For efficiency, we follow~\citet{sloan2003} and compute the eigendecomposition of either $CC \trans$ or 
$C \trans C$, whichever is smaller. Both produce eigenvalues equal to $\Sigma^2$. The eigenvectors of $CC \trans$ is $U$; The eigenvectors of $C \trans C$ is the 
right-singular vectors $V$ from which $U$ can be computed via $U = C V \Sigma^{-1}$. When possible, we only solve for the top eigenpairs by the Lanczos 
algorithm~\citep{golub2013matrix, qiuspectralib}. 

Finally, each representative tile goes through wavelet compression with non-standard 2D Haar wavelet. During rendering, it is unnecessary to decompress the 
entire tile. The wavelet coefficients can be arranged as an implicit sparse quadtree that supports random access~\citep{lalonde1997representations}. To 
evaluate visibility of a given direction, we can traverse a path down the tree, following a child if its basis supports the direction being evaluated.

\bibliographystyle{ACM-Reference-Format}
\bibliography{main}